\pdfoutput=1
\InputIfFileExists{./includeonly.tex}{}{}
\documentclass[11pt,letterpaper]{memoir}
\usepackage[british]{babel}
\usepackage[letterspace=75,final]{microtype}
\usepackage[T1]{fontenc}
\usepackage[utf8]{inputenc}
\usepackage{amsmath}
\usepackage{amsfonts}
\usepackage{amssymb}
\usepackage{amsthm}
\usepackage[donotfixamsmathbugs,allowspaces]{mathtools}
\mathtoolsset{mathic}
\usepackage{braket}
\usepackage[pdfusetitle]{hyperref}
\usepackage[all]{hypcap}
\usepackage{tensor}
\usepackage{picture}
\usepackage[final]{graphicx}
\usepackage{calc}
\usepackage[dvipsnames]{xcolor}
\usepackage{pifont}
\usepackage{pgfplots}
\pgfplotsset{compat=1.12}
\usepgfplotslibrary{external}
\usepackage{extdash}
\hypersetup{
	final,
	colorlinks,
	linkcolor={NavyBlue},
	citecolor={NavyBlue},
	urlcolor={RedViolet},
	bookmarksdepth=3,
	bookmarksnumbered=true
}
\usepackage[capitalise]{cleveref}
\crefname{equation}{equation}{equations}

\usepackage{csquotes}

\def\gettexliveversion#1(#2 #3 #4#5#6#7#8)#9\relax{#4#5#6#7}
\edef\texliveversion{\expandafter\gettexliveversion\pdftexbanner\relax}

\ifnum\texliveversion<2016
  \usepackage[backend=biber,style=authortitle,firstinits=true,uniquename=init,eprint=true,hyperref=true,sorting=nty]{biblatex}
  \usepackage{newunicodechar}
  \newunicodechar{ĭ}{\u{\i}}
  \newunicodechar{Ǧ}{\v{G}}
\else
  \usepackage[backend=biber,style=authortitle,giveninits=true,uniquename=init,eprint=true,hyperref=true,sorting=nty]{biblatex}
\fi

\newsubfloat{table}
\newsubfloat{figure}

\newcommand{\chapsymbol}{
\IfFileExists{MinionPro.styy}{
	\Pisymbol{MinionPro-Extra}{120}
	}{}
}

\setlrmarginsandblock{1in}{2.3in}{*}
\setulmarginsandblock{1.2in}{1.5in}{*}
\setmarginnotes{5mm}{2.3in-20mm}{\onelineskip}
\marginparmargin{outer}
\footnotesinmargin
\strictpagecheck 
\setheadfoot{\baselineskip}{0.45\lowermargin}
\makeevenfoot{ruled}{}{\thepage}{}
\makeoddfoot{ruled}{}{\thepage}{}
\copypagestyle{part}{empty} 
\setsecnumdepth{subsection}

\makeatletter
\makechapterstyle{mystyle}{

\renewcommand*{\chapnamefont}{\color{CadetBlue}\normalfont\Large\flushleft}

\renewcommand*{\printchaptername}{{\chapnamefont \chapsymbol \@chapapp}}

}
\makeatother
\chapterstyle{mystyle}

\checkandfixthelayout

\makeatletter
\g@addto@macro\bfseries{\boldmath}
\g@addto@macro\rmfamily{\mathversion{normal}}
\g@addto@macro\upshape{\mathversion{normal}}
\g@addto@macro\normalfont{\mathversion{normal}}
\makeatother

\makeatletter
\AtBeginDocument{
\check@mathfonts
\fontdimen13\textfont2=5.25pt
\fontdimen14\textfont2=4.75pt
\fontdimen16\textfont2=2.75pt
}
\makeatother

\ifdraftdoc
\makeevenfoot{plain}{}{\thepage}{\color{gray}\textit{draft: \today}}
\makeoddfoot{plain}{\color{gray}\textit{draft: \today}}{\thepage}{}
\makeevenfoot{ruled}{}{\thepage}{\color{gray}\textit{draft: \today}}
\makeoddfoot{ruled}{\color{gray}\textit{draft: \today}}{\thepage}{}
\fi

\newcommand{\draftnote}[1]{
\ifdraftdoc
\marginpar{\footnotesize\color{Cyan}#1}
\fi
}

\newenvironment{myquote}%
{\list{}{\leftmargin=0in\rightmargin=0in}\item[]}%
{\endlist}

\newcommand{\conj}{\widebar}
\renewcommand{\vec}{\mathbf}
\newcommand{\vecsymbol}{\boldsymbol}
\newcommand{\eder}{\mathrm{d}}
\newcommand{\der}[3][]{\frac{\mathrm{d}^{#1}{#2}}{{\mathrm{d}{#3}}^{#1}}}
\newcommand{\pder}[2]{\frac{\partial{#1}}{\partial{#2}}}
\newcommand{\transpose}{\mathrm{t}}
\DeclareMathOperator{\tr}{tr}
\DeclareMathOperator{\Span}{span}
\DeclareMathOperator{\sgn}{sgn}

\DeclareMathOperator{\rank}{rank}
\DeclareMathOperator{\Var}{Var}
\DeclareMathOperator{\Cov}{Cov}

\makeatletter
\DeclareRobustCommand\Dagger{\@ifstar\@@Dagger\@@Dagger}
\def\@Dagger{\mathrlap{^\dagger}}
\def\@@Dagger{^\dagger}
\DeclareRobustCommand\Star{\@ifstar\@@Star\@@Star}
\def\@Star{\mathrlap{^*}}
\def\@@Star{^*}
\DeclareRobustCommand\Transpose{\@ifstar\@@Transpose\@@Transpose}
\def\@Transpose{\protect\mathrlap{^\transpose}}
\def\@@Transpose{\protect^\transpose}
\DeclareRobustCommand\Prime{\@ifstar\@@Prime\@@Prime}
\def\@Prime{\mathrlap{^\prime}}
\def\@@Prime{^\prime}
\makeatother

\newcommand{\half}[1][1]{\tfrac{#1}{2}}
\newcommand{\shalf}[1][1]{\frac{#1}{2}} 
\newcommand{\Rho}{\mathrm{P}}
\newcommand{\casesif}{\quad\textnormal{if }\,} 
\newcommand{\casestext}[1]{\quad\textnormal{#1 }\,} 
\newcommand{\casestextn}[1]{\quad\textnormal{#1}} 
\newcommand{\setst}{\mid} 
\newcommand{\setstx}{\:\middle\vert\:} 

\newcommand{\CG}{Clebsch--Gordan}
\newcommand{\JS}{Jordan--Schwinger}
\newcommand{\WE}{Wigner--Eckart}

\IfFileExists{ubbextended.fd}{
	\DeclareSymbolFont{bbextended}{U}{bbextended}{m}{n}
	\DeclareMathSymbol{\ii}{0}{bbextended}{"69}
	\DeclareMathSymbol{\1}{0}{bbextended}{"31}
}{
	\usepackage{bbm}
	\newcommand{\ii}{\mathbbm{i}}
	\newcommand{\1}{\mathbbm{1}}
}
\newcommand{\R}{\mathbb{R}}
\newcommand{\C}{\mathbb{C}}
\newcommand{\K}{\mathbb{K}}
\newcommand{\Z}{\mathbb{Z}}
\newcommand{\N}{\mathbb{N}}
\newcommand{\0}{0}
\newcommand{\cC}{\mathcal{C}}
\newcommand{\cJ}{\mathcal{J}}
\newcommand{\cM}{\mathcal{M}}
\newcommand{\GL}{\mathrm{GL}}
\newcommand{\SL}{\mathrm{SL}}
\newcommand{\SO}{\mathrm{SO}}
\newcommand{\SU}{\mathrm{SU}}
\newcommand{\Spin}{\mathrm{Spin}}
\newcommand{\Sp}{\mathrm{Sp}}

\newcommand{\so}{\mathfrak{so}}
\newcommand{\su}{\mathfrak{su}}
\newcommand{\spin}{\mathfrak{spin}}
\newcommand{\g}{\mathfrak{g}}

\newcommand{\bigboxplus}{
  \mathop{
    \vphantom{\bigoplus} 
    \mathchoice
      {\vcenter{\hbox{\resizebox{\widthof{$\displaystyle\bigoplus$}}{!}{$\boxplus$}}}}
      {\vcenter{\hbox{\resizebox{\widthof{$\bigoplus$}}{!}{$\boxplus$}}}}
      {\vcenter{\hbox{\resizebox{\widthof{$\scriptstyle\oplus$}}{!}{$\boxplus$}}}}
      {\vcenter{\hbox{\resizebox{\widthof{$\scriptscriptstyle\oplus$}}{!}{$\boxplus$}}}}
  }\displaylimits 
}

\makeatletter
\let\save@mathaccent\mathaccent
\newcommand*\if@single[3]{%
  \setbox0\hbox{\({\mathaccent"0362{#1}}^H\)}%
  \setbox2\hbox{\({\mathaccent"0362{\kern0pt#1}}^H\)}%
  \ifdim\ht0=\ht2 #3\else #2\fi
  }
\newcommand*\rel@kern[1]{\kern#1\dimexpr\macc@kerna}
\newcommand*\widebar[1]{\@ifnextchar^{\wide@bar{#1}{0}}{\wide@bar{#1}{1}}}
\newcommand*\wide@bar[2]{\if@single{#1}{\wide@bar@{#1}{#2}{1}}{\wide@bar@{#1}{#2}{2}}}
\newcommand*\wide@bar@[3]{%
  \begingroup
  \def\mathaccent##1##2{%
    \let\mathaccent\save@mathaccent
    \if#32 \let\macc@nucleus\first@char \fi
    \setbox\z@\hbox{\(\macc@style{\macc@nucleus}_{}\)}%
    \setbox\tw@\hbox{\(\macc@style{\macc@nucleus}{}_{}\)}%
    \dimen@\wd\tw@
    \advance\dimen@-\wd\z@
    \divide\dimen@ 3
    \@tempdima\wd\tw@
    \advance\@tempdima-\scriptspace
    \divide\@tempdima 10
    \advance\dimen@-\@tempdima
    \ifdim\dimen@>\z@ \dimen@0pt\fi
    \rel@kern{0.6}\kern-\dimen@
    \if#31
      \overline{\rel@kern{-0.6}\kern\dimen@\macc@nucleus\rel@kern{0.4}\kern\dimen@}%
      \advance\dimen@0.4\dimexpr\macc@kerna
      \let\final@kern#2%
      \ifdim\dimen@<\z@ \let\final@kern1\fi
      \if\final@kern1 \kern-\dimen@\fi
    \else
      \overline{\rel@kern{-0.6}\kern\dimen@#1}%
    \fi
  }%
  \macc@depth\@ne
  \let\math@bgroup\@empty \let\math@egroup\macc@set@skewchar
  \mathsurround\z@ \frozen@everymath{\mathgroup\macc@group\relax}%
  \macc@set@skewchar\relax
  \let\mathaccentV\macc@nested@a
  \if#31
    \macc@nested@a\relax111{#1}%
  \else
    \def\gobble@till@marker##1\endmarker{}%
    \futurelet\first@char\gobble@till@marker#1\endmarker
    \ifcat\noexpand\first@char A\else
      \def\first@char{}%
    \fi
    \macc@nested@a\relax111{\first@char}%
  \fi
  \endgroup
}
\makeatother
\newenvironment{smallpmatrix}{\left(\begin{smallmatrix}}{\end{smallmatrix}\right)}
\newenvironment{smallmatrixDiv}[1]
	{
		\everymath={\scriptstyle}
		\arraycolsep=0.3\arraycolsep\ensuremath
		\begin{array}{@{}#1@{}}}
	{\end{array}}
\newenvironment{smallpmatrixDiv}[1]
	{\left(\begin{smallmatrixDiv}{#1}}{\end{smallmatrixDiv}\right)}
\newenvironment{pmatrixDiv}[1]
	{\left(\begin{array}{@{}#1@{}}}{\end{array}\right)}

\theoremstyle{plain}
\newtheorem{proposition}{Proposition}[chapter]
\newtheorem{definition}{Definition}[chapter]
\newtheorem{theorem}{Theorem}[chapter]
\newtheorem*{theorem*}{Theorem}
\newtheorem*{remark}{Remark}
\newtheorem{corollary}{Corollary}[chapter]
\newtheorem{lemma}{Lemma}[chapter]

\DeclarePairedDelimiter\abs{\lvert}{\rvert}
\DeclarePairedDelimiter\card{\lvert}{\rvert}

\DeclarePairedDelimiter\floor{\lfloor}{\rfloor}
\DeclarePairedDelimiter\paren{\lparen}{\rparen}
\DeclarePairedDelimiter\braces{\lbrace}{\rbrace}
\DeclarePairedDelimiter\bracks{\lbrack}{\rbrack}
\DeclarePairedDelimiter\commutator{\lbrack}{\rbrack}
\DeclarePairedDelimiterXPP{\eval}[2]{}{.}{\rvert}{_{#2}}{#1}
\makeatletter
\let\oldeval\eval
\def\eval{\@ifstar{\oldeval}{\oldeval*}}
\makeatother
\let\set\relax
\DeclarePairedDelimiter\set{\lbrace}{\rbrace}
\let\ket\relax
\DeclarePairedDelimiter{\ket}{\lvert}{\rangle}
\let\bra\relax
\DeclarePairedDelimiter{\bra}{\langle}{\rvert}
\let\braket\relax
\DeclarePairedDelimiter{\braket}{\langle}{\rangle}
\DeclarePairedDelimiter{\ketb}{\lvert}{\rbrack}
\DeclarePairedDelimiter{\brab}{\lbrack}{\rvert}
\DeclarePairedDelimiter{\braketlb}{\lbrack}{\rangle}
\DeclarePairedDelimiter{\braketrb}{\langle}{\rbrack}
\DeclarePairedDelimiter{\braketbb}{\lbrack}{\rbrack}
\DeclarePairedDelimiter{\praket}{\lparen}{\rparen}
\let\originalleft\left
\let\originalright\right
\renewcommand{\left}{\mathopen{}\mathclose\bgroup\originalleft}
\renewcommand{\right}{\aftergroup\egroup\originalright}

\usepackage{enumitem}
\newlist{proofenum}{enumerate}{1}
\setlist[proofenum,1]{wide, labelwidth=!, labelindent=0pt,label={(\roman*)}}
\newenvironment{proofenumerate}{\begin{proofenum}}{\qedhere\end{proofenum}}

\usepackage{tikz}

\tikzset{
vcenter/.style={
execute at end picture={
\path[draw] ([rotate around={180:#1}]perpendicular cs: vertical line through={#1}, horizontal line through={(current bounding box.north)})     ([rotate around={180:#1}]perpendicular cs: vertical line through={#1}, horizontal line through={(current bounding box.south)});}}}

\usetikzlibrary{decorations.markings,arrows,calc}

\newcommand{\trivalentlr}[3]{
\mathord{\vcenter{\hbox{
\tikzsetnextfilename{trivalentlr#1#2#3}
\footnotesize
\begin{tikzpicture}[vcenter=(c), x=1.85em,y=1.85em, line width=1.1pt, every node/.style={inner sep = 0pt, outer sep = 2pt}]
\coordinate (j1) at (0,{sqrt(2)});
\coordinate (j2) at (0,0);
\coordinate (j3) at ({1+0.5*sqrt(2)},{0.5*sqrt(2)});
\coordinate (c)  at ({0.5*sqrt(2)},{0.5*sqrt(2)});
\draw (j1) -- (c);
\draw (j2) -- (c);
\draw (j3) -- (c);
\node[anchor=east] at (j1) {\( #1 \)};
\node[anchor=east] at (j2) {\( #2 \)};
\node[anchor=west] at (j3) {\( #3 \)};
\end{tikzpicture}
}}}
}

\newcommand{\trivalentrl}[3]{
\tikzsetnextfilename{trivalentrl#1#2#3}
\mathord{\vcenter{\hbox{
\footnotesize
\begin{tikzpicture}[vcenter=(c), x=-1.85em,y=1.85em, line width=1.1pt, every node/.style={inner sep = 0pt, outer sep = 2pt}]
\coordinate (j1) at (0,{sqrt(2)});
\coordinate (j2) at (0,0);
\coordinate (j3) at ({1+0.5*sqrt(2)},{0.5*sqrt(2)});
\coordinate (c)  at ({0.5*sqrt(2)},{0.5*sqrt(2)});
\draw (j1) -- (c);
\draw (j2) -- (c);
\draw (j3) -- (c);
\node[anchor=west] at (j1) {\( #1 \)};
\node[anchor=west] at (j2) {\( #2 \)};
\node[anchor=east] at (j3) {\( #3 \)};
\end{tikzpicture}
}}}
}

\newcommand{\intidA}[2]{
\tikzsetnextfilename{intidA#1#2}
\mathord{\vcenter{\hbox{
\footnotesize
\begin{tikzpicture}[vcenter=(j1), x=1.8em,y=2em, line width=1.1pt, every node/.style={inner sep = 0pt, outer sep = 2pt}]
\coordinate (j1) at (0,0);
\coordinate (j2) at (2,0);
\draw (j1) -- (j2);
\node[anchor=east] at (j1) {\( #1 \)};
\node[anchor=west] at (j2) {\( #2 \)};
\end{tikzpicture}
}}}
}

\newcommand{\intidB}[3]{
\tikzsetnextfilename{intidB#1#2}
\mathord{\vcenter{\hbox{
\footnotesize
\begin{tikzpicture}[vcenter=(j2a), x=1.85em,y=1.85em, line width=1.1pt, every node/.style={inner sep = 0pt, outer sep = 2pt}]
\coordinate (j1a) at (0,1);
\coordinate (j1b) at (2,1);
\coordinate (j2a) at (0,0);
\coordinate (j2b) at (2,0);
\coordinate (j3a) at (0,-1);
\coordinate (j3b) at (2,-1);
\draw (j1a) -- (j1b);
\draw (j2a) -- (j2b);
\draw (j3a) -- (j3b);
\node[anchor=east] at (j1a) {\( #1 \)};
\node[anchor=west] at (j1b) {\( #1 \)};
\node[anchor=east] at (j2a) {\( #2 \)};
\node[anchor=west] at (j2b) {\( #2 \)};
\node[anchor=east] at (j3a) {\( #3 \)};
\node[anchor=west] at (j3b) {\( #3 \)};
\end{tikzpicture}
}}}
}

\newcommand{\racahintA}[5]{
\tikzsetnextfilename{racahintA#1#2#3#4#5}
\mathord{\vcenter{\hbox{
\footnotesize
\begin{tikzpicture}[vcenter=(c), x=1.85em,y=1.85em, line width=1.1pt, every node/.style={inner sep = 0pt, outer sep = 2pt}]
\coordinate (j1) at (0,2);
\coordinate (j2) at (0,1);
\coordinate (j3) at (0,0);
\coordinate (j) at (2.2,1);
\coordinate (c)  at (1,1);
\coordinate (j12) at ($(j1)!0.5!(c)$);
\draw (j1) -- (c);
\draw (j2) -- (j12);
\draw (j) -- (c);
\draw (j3) -- (c);
\node[anchor=east] at (j1) {\( #1 \)};
\node[anchor=east] at (j2) {\( #2 \)};
\node[anchor=east] at (j3) {\( #3 \)};
\node[anchor=west] at (j) {\( #4 \)};
\node[anchor=west, inner sep=1.5ex] at (j12) {\( #5 \)};
\end{tikzpicture}
}}}
}

\newcommand{\racahintB}[5]{
\tikzsetnextfilename{racahintB#1#2#3#4#5}
\mathord{\vcenter{\hbox{
\footnotesize
\begin{tikzpicture}[vcenter=(c), x=1.85em,y=1.85em, line width=1.1pt, every node/.style={inner sep = 0pt, outer sep = 2pt}]
\coordinate (j1) at (0,2);
\coordinate (j2) at (0,1);
\coordinate (j3) at (0,0);
\coordinate (j) at (2.2,1);
\coordinate (c)  at (1,1);
\coordinate (j12) at ($(j3)!0.5!(c)$);
\draw (j1) -- (c);
\draw (j2) -- (j12);
\draw (j) -- (c);
\draw (j3) -- (c);
\node[anchor=east] at (j1) {\( #1 \)};
\node[anchor=east] at (j2) {\( #2 \)};
\node[anchor=east] at (j3) {\( #3 \)};
\node[anchor=west] at (j) {\( #4 \)};
\node[anchor=west, outer sep=1.5ex] at (j12) {\( #5 \)};
\end{tikzpicture}
}}}
}

\newcommand{\intspiderA}[5]{
\tikzsetnextfilename{intspiderA#1#2#3#4#5}
\mathord{\vcenter{\hbox{
\footnotesize
\begin{tikzpicture}[vcenter=(j), x=1.85em,y=1.85em, line width=1.1pt, every node/.style={inner sep = 0pt, outer sep = 2pt}]
\coordinate (j1a) at (0,2);
\coordinate (j1b) at (3.2,2);
\coordinate (j2a) at (0,1);
\coordinate (j2b) at (3.2,1);
\coordinate (j3a) at (0,0);
\coordinate (j3b) at (3.2,0);
\coordinate (ca)  at (1,1);
\coordinate (cb) at (2.2,1);
\coordinate (j) at ($(ca)!0.5!(cb)$);
\coordinate (j12a) at ($(j1a)!0.5!(ca)$);
\coordinate (j12b) at ($(j1b)!0.5!(cb)$);
\draw (j1a) -- (ca);
\draw (j2a) -- (j12a);
\draw (ca) -- (cb);
\draw (j3a) -- (ca);
\draw (j1b) -- (cb);
\draw (j2b) -- (j12b);
\draw (j3b) -- (cb);
\node[anchor=east] at (j1a) {\( #1 \)};
\node[anchor=east] at (j2a) {\( #2 \)};
\node[anchor=east] at (j3a) {\( #3 \)};
\node[anchor=west, inner sep=1ex] at (j12a) {\( #5 \)};
\node[anchor=west] at (j1b) {\( #1 \)};
\node[anchor=west] at (j2b) {\( #2 \)};
\node[anchor=west] at (j3b) {\( #3 \)};
\node[anchor=east, inner sep=1ex] at (j12b) {\( #5 \)};
\node[anchor=north] at (j) {\( #4 \)};
\end{tikzpicture}
}}}
}

\newcommand{\intspiderB}[5]{
\tikzsetnextfilename{intspiderB#1#2#3#4#5}
\mathord{\vcenter{\hbox{
\footnotesize
\begin{tikzpicture}[vcenter=(j), x=1.85em,y=1.85em, line width=1.1pt, every node/.style={inner sep = 0pt, outer sep = 2pt}]
\coordinate (j1a) at (0,2);
\coordinate (j1b) at (3.2,2);
\coordinate (j2a) at (0,1);
\coordinate (j2b) at (3.2,1);
\coordinate (j3a) at (0,0);
\coordinate (j3b) at (3.2,0);
\coordinate (ca)  at (1,1);
\coordinate (cb) at (2.2,1);
\coordinate (j) at ($(ca)!0.5!(cb)$);
\coordinate (j12a) at ($(j3a)!0.5!(ca)$);
\coordinate (j12b) at ($(j3b)!0.5!(cb)$);
\draw (j1a) -- (ca);
\draw (j2a) -- (j12a);
\draw (ca) -- (cb);
\draw (j3a) -- (ca);
\draw (j1b) -- (cb);
\draw (j2b) -- (j12b);
\draw (j3b) -- (cb);
\node[anchor=east] at (j1a) {\( #1 \)};
\node[anchor=east] at (j2a) {\( #2 \)};
\node[anchor=east] at (j3a) {\( #3 \)};
\node[anchor=west, outer sep=1ex] at (j12a) {\( #5 \)};
\node[anchor=west] at (j1b) {\( #1 \)};
\node[anchor=west] at (j2b) {\( #2 \)};
\node[anchor=west] at (j3b) {\( #3 \)};
\node[anchor=east, outer sep=1ex] at (j12b) {\( #5 \)};
\node[anchor=south] at (j) {\( #4 \)};
\end{tikzpicture}
}}}
}

\newcommand{\intsquareA}[7]{
\tikzsetnextfilename{intsquareA#1#2#3#4#5#6#7}
\mathord{\vcenter{\hbox{
\footnotesize
\begin{tikzpicture}[vcenter=(ca), x=1.85em,y=1.85em, line width=1.1pt, every node/.style={inner sep = 0pt, outer sep = 2pt}]
\coordinate (c1) at (2.2,2);
\coordinate (c2) at (2.2,1);
\coordinate (c3) at (2.2,0);
\coordinate (j') at (0,1);
\coordinate (j) at (4.4,1);
\coordinate (ca)  at (1.2,1);
\coordinate (cb)  at (3.2,1);
\coordinate (j12) at ($(c1)!0.5!(cb)$);
\coordinate (j23) at ($(c3)!0.5!(ca)$);
\draw (ca) -- (c1) -- (cb);
\draw (ca) -- (c3) -- (cb);
\draw (j12) -- (j23);
\draw (j') -- (ca);
\draw (j) -- (cb);
\node[anchor=south east, outer sep=0pt] at ($(ca)!0.5!(c1)$) {\( #1 \)};
\node[anchor=south east, outer sep=0pt] at ($(j12)!0.5!(j23)$) {\( #2 \)};
\node[anchor=north west, outer sep=0pt] at ($(cb)!0.5!(c3)$) {\( #3 \)};
\node[anchor=west] at (j) {\( #6 \)};
\node[anchor=east] at (j') {\( #7 \)};
\node[anchor=west, inner sep=1ex] at (j12) {\( #4 \)};
\node[anchor=east] at (j23) {\( #5 \)};
\end{tikzpicture}
}}}
}

\newcommand{\intsquareB}[7]{
\tikzsetnextfilename{intsquareB#1#2#3#4#5#6#7}
\mathord{\vcenter{\hbox{
\footnotesize
\begin{tikzpicture}[vcenter=(ca), x=1.85em,y=1.85em, line width=1.1pt, every node/.style={inner sep = 0pt, outer sep = 2pt}]
\coordinate (c1) at (2.2,2);
\coordinate (c2) at (2.2,1);
\coordinate (c3) at (2.2,0);
\coordinate (j') at (0,1);
\coordinate (j) at (4.4,1);
\coordinate (ca)  at (1.2,1);
\coordinate (cb)  at (3.2,1);
\coordinate (j12) at ($(c1)!0.5!(ca)$);
\coordinate (j23) at ($(c3)!0.5!(cb)$);
\draw (ca) -- (c1) -- (cb);
\draw (ca) -- (c3) -- (cb);
\draw (j12) -- (j23);
\draw (j') -- (ca);
\draw (j) -- (cb);
\node[anchor=south west, outer sep=1pt] at ($(cb)!0.5!(c1)$) {\( #1 \)};
\node[anchor=south west, outer sep=1pt] at ($(j12)!0.5!(j23)$) {\( #2 \)};
\node[anchor=north east, outer sep=0pt] at ($(ca)!0.5!(c3)$) {\( #3 \)};
\node[anchor=west] at (j) {\( #6 \)};
\node[anchor=east] at (j') {\( #7 \)};
\node[anchor=east, inner sep=2pt] at (j12) {\( #4 \)};
\node[anchor=west, outer sep=1ex] at (j23) {\( #5 \)};
\end{tikzpicture}
}}}
}

\newcommand{\intracahsquare}[7]{
\tikzsetnextfilename{intracahsquare#1#2#3#4#5#6#7}
\mathord{\vcenter{\hbox{
\footnotesize
\begin{tikzpicture}[vcenter=(ca), x=1.85em,y=1.85em, line width=1.1pt, every node/.style={inner sep = 0pt, outer sep = 2pt}]
\coordinate (j1) at (-1,2);
\coordinate (j2) at (-1,1);
\coordinate (j3) at (-1,0);
\coordinate (c1) at (2.2,2);
\coordinate (c2) at (2.2,1);
\coordinate (c3) at (2.2,0);
\coordinate (c0) at (0,1);
\coordinate (j) at (4.4,1);
\coordinate (ca)  at (1.2,1);
\coordinate (cb)  at (3.2,1);
\coordinate (j12) at ($(c1)!0.5!(cb)$);
\coordinate (j23) at ($(c3)!0.5!(ca)$);
\coordinate (j23b) at ($(j3)!0.5!(c0)$);
\draw (ca) -- (c1) -- (cb);
\draw (ca) -- (c3) -- (cb);
\draw (j1) -- (c0) -- (j3);
\draw (j2) -- (j23b);
\draw (j12) -- (j23);
\draw (c0) -- (ca);
\draw (j) -- (cb);
\node[anchor=south east, outer sep=0pt] at ($(ca)!0.5!(c1)$) {\( #1 \)};
\node[anchor=south east, outer sep=0pt] at ($(j12)!0.5!(j23)$) {\( #2 \)};
\node[anchor=north west, outer sep=0pt] at ($(cb)!0.5!(c3)$) {\( #3 \)};
\node[anchor=west] at (j) {\( #6 \)};
\node[anchor=south] at ($(c0)!0.5!(ca)$) {\( #7 \)};
\node[anchor=west, inner sep=1ex] at (j12) {\( #4 \)};
\node[anchor=east] at (j23) {\( #5 \)};
\node[anchor=west, outer sep=1ex] at (j23b) {\( #5 \)};
\node[anchor=east] at (j1) {\( #1 \)};
\node[anchor=east] at (j2) {\( #2 \)};
\node[anchor=east] at (j3) {\( #3 \)};
\end{tikzpicture}
}}}
}

\newcommand{\intbubble}[4]{
\tikzsetnextfilename{intbubble#1#2#3#4}
\mathord{\vcenter{\hbox{
\footnotesize
\begin{tikzpicture}[vcenter=(c), x=1.8em,y=2em, line width=1.1pt, every node/.style={inner sep = 0pt, outer sep = 2pt}]
\coordinate (j') at (0,0);
\coordinate (j) at (4,0);
\coordinate (c) at (2,0);
\draw (j') -- (j);
\filldraw[color=white,draw=black] (c) circle (1.5em);
\node[anchor=east] at (j') {\( #1 \)};
\node[anchor=west] at (j) {\( #2 \)};
\node[anchor=south] at ($(c)+(0,1.5em)$) {\( #3 \)};
\node[anchor=north] at ($(c)-(0,1.5em)$) {\( #4 \)};
\end{tikzpicture}
}}}
}

\newcommand{\inthambubble}{
\tikzsetnextfilename{inthambubble}
\mathord{\vcenter{\hbox{
\footnotesize
\begin{tikzpicture}[vcenter=(c), x=1.85em,y=1.85em, line width=1.1pt, every node/.style={inner sep = 0pt, outer sep = 2pt}]
\coordinate (j1) at (0,2);
\coordinate (j6) at (0,0);
\coordinate (j5) at (2.2,1);
\coordinate (c)  at (1,1);
\coordinate (circ) at ($(j6)!0.5!(c)$);
\draw (j1) -- (c) -- (j6);
\draw (j5) -- (c);
\filldraw[color=white,draw=black] (circ) circle (0.55em);
\node[anchor=east] at (j1) {\( j_1 \)};
\node[anchor=east] at (j6) {\( j_6 \)};
\node[anchor=west] at (j5) {\( j_5 \)};
\node at (1.05,0.65) {\( j \)};
\node at (0.8,-0.15) {\( j_4 \)};
\node at (0,1) {\( j_2 \)};
\end{tikzpicture}
}}}
}

\newcommand{\inttet}{
\mathord{\vcenter{\hbox{
\tikzsetnextfilename{inttet}
\footnotesize
\begin{tikzpicture}[vcenter=(c), x=1.85em,y=1.85em, line width=1.1pt, every node/.style={inner sep = 0pt, outer sep = 2pt}]
\coordinate (j1) at (0,2);
\coordinate (j6) at (0,0);
\coordinate (j5) at (2.2,1);
\coordinate (c)  at (1,1);
\coordinate (circ) at ($(j6)!0.5!(c)$);
\draw (j5) -- (c);
\begin{scope}
\path[clip] (j1) -- (c) -- (j6);
\draw[circlemidarrow] (c) circle (1.4em);
\end{scope}
\draw (j1) -- (c) -- (j6);
\node[anchor=east] at (j1) {\( j_1 \)};
\node[anchor=east] at (j6) {\( j_6 \)};
\node[anchor=west] at (j5) {\( j_5 \)};
\node[anchor=south, inner sep =1ex] at (c) {\( j_3 \)};
\node[anchor=north, inner sep =1ex] at (c) {\( j_4 \)};
\node[anchor=east] at ($(c)-(1.6em,0)$) {\( j_2 \)};
\end{tikzpicture}
}}}
}

\tikzset{
midarrow/.style={postaction={decorate}, decoration={markings,mark=at position 0.5 with {\arrow[scale=0.6,xshift=4\pgflinewidth]{triangle 60}}}}}
\tikzset{
circlemidarrow/.style={postaction={decorate}, decoration={markings,mark=at position 0.5 with {\arrowreversed[scale=0.6,xshift=-3.5\pgflinewidth]{triangle 60}}}}}
\tikzset{
endarrowsmall/.style={postaction={decorate}, decoration={markings,mark=at position 1 with {\arrow[scale=0.5,xshift=2\pgflinewidth]{triangle 60}}}}}
\tikzset{
endarrow/.style={postaction={decorate}, decoration={markings,mark=at position 1 with {\arrow[scale=0.6,xshift=2\pgflinewidth]{triangle 60}}}}}

\title{Non-compact groups, tensor operators and applications to quantum gravity}
\author{Giuseppe Sellaroli}
\begin{document}
\frontmatter
\pagestyle{plain}
\newlength{\tpspace}
\setlength{\tpspace}{2em}
\colorlet{titlepage}{RedViolet}

\thispagestyle{empty}
\calccentering{\unitlength}
\begin{adjustwidth*}{\unitlength}{-\unitlength}
\begin{center}
\vspace*{0.1in}

\begin{Spacing}{1.5}
\makebox[\textwidth]{\color{titlepage} \Huge \lsstyle \uppercase{non-compact groups,}}
\end{Spacing}
\makebox[\textwidth]{\color{titlepage} \Huge \lsstyle \uppercase{tensor operators}}
\\[\baselineskip]
\makebox[\textwidth]{\huge \slshape and applications to}
\\[\baselineskip]
\makebox[\textwidth]{\color{titlepage} \Huge \lsstyle \uppercase{quantum gravity}}

\vspace{\tpspace}

{\Large \slshape by}

\vspace{\tpspace}

{\Large Giuseppe Sellaroli}

\vfill
{\huge \color{titlepage} \chapsymbol}
\vfill

\begin{Spacing}{1.1}
\Large \slshape A thesis\\
presented to the University of Waterloo\\
in fulfilment of the\\
thesis requirement for the degree of\\
Doctor of Philosophy\\
in\\
Applied Mathematics
\end{Spacing}

\vspace{3\tpspace}

{\Large \slshape Waterloo, Ontario, Canada, 2016}

\vspace{\tpspace}

{\Large © Giuseppe Sellaroli 2016}

\end{center}
\end{adjustwidth*}

\cleardoublepage
\chapter*{Author's declaration}
This thesis consists of material all of which I authored or co-authored: see statement of contributions included in the thesis. This is a true copy of the thesis, including any required final revisions, as
accepted by my examiners.
I understand that my thesis may be made electronically available to the public.

\openleft

\chapter*{Statement of contributions}

\begin{myquote}
\cref{chap:wigner-eckart} is based on the articles \cite{wigner_eckart} (published) and  \cite{Sellaroli:2015sca} (submitted for publication), of which I am the sole author.

\cref{sec:lorentzian_LQG} draws on \cite{Girelli:2015ija}, published work co-authored with my PhD supervisor
Florian Girelli.

\cref{chap:so*} contains new material which has never been published nor submitted for publication, and of which I am the sole author.
\end{myquote}

\openright

\chapter*{Abstract}

This work focuses on non-compact groups and their applications to quantum gravity, mainly through the use of tensor operators. Non-compact groups appear naturally if the space-time is of Lorentzian signature, but can also have an important role in the Euclidean case, as will be shown.

First, the mathematical theory of tensor operators for a Lie group is recast in a new way which is used to generalise the \WE\ theorem to non-compact groups. The result relies on the knowledge of the recoupling theory between finite-dimensional and infinite-dimensional irreducible representations of the group; here the previously unconsidered cases of the \( 3 \)D and \( 4 \)D Lorentz groups are investigated in detail. As an application, the \WE\ theorem is used to generalise the \JS\ representation of \( \SU(2) \) to both groups, for all representation classes.

Next, the results obtained for the \( 3 \)D Lorentz group are applied to \( (2+1) \) Lorentzian loop quantum gravity to develop an analogue of the well-known spinorial approach used in the Euclidean case. Tensor operators are used to construct observables and to generalise the Hamiltonian constraint introduced by Bonzom and Livine (2012) for \( 3 \)D gravity to the Lorentzian case. The Ponzano--Regge amplitude is shown to be a solution of this constraint by recovering the (opportunely generalised) Biedenharn--Elliott relations.

Finally, the focus is shifted on the intertwiner space based on \( \SU(2) \) representations, widely used in loop quantum gravity.
When  working in the spinorial formalism, it has been shown that the Hilbert space of \( n \)-valent intertwiners with fixed total area is a representation of \( \mathrm{U}(n) \).
Here it is shown that the full space of \emph{all} \( n \)-valent intertwiners forms an irreducible representation of the non-compact group \( \SO\Star(2n) \).
This fact is used to construct a new kind of  coherent intertwiner state (in the sense of Perelomov).
Although some of these states were known already, the majority of them was not until now; moreover, the underlying group structure was completely unknown.
Hints of how these coherent states can be interpreted in the semi-classical limit as convex polyhedra are provided.

\chapter*{Acknowledgements}

\begin{myquote}
I would like to thank my supervisor, Florian Girelli, for his continuous support, the endless discussions and in general for being the being the best supervisor I could hope for. I would also like to thank Aristide Baratin, Maïté Dupuis, Etera Livine and Simone Speziale for interesting discussions and comments.

I am grateful to my sister and my parents for their encouragement and help throughout the years; I could never have arrived to this point without them.

My greatest thanks go to Amanda, for the wonderful years we spent in Waterloo, her support during the composition of this thesis and for her help (together with Florian) in proofreading it and correcting my (many) mistakes.
\end{myquote}

\cleardoublepage

\begin{vplace}[0.5]
\centering
\itshape
to Harish-Chandra
\end{vplace}

\openleft
\clearpage

\tableofcontents

\clearpage

\renewcommand*{\listfigurename}{List of figures}
\listoffigures

\renewcommand*{\listtablename}{List of tables}
\listoftables

\openright

\cleardoublepage

\chapter{List of abbreviations}

\begin{tabularx}{\textwidth}{>{\hsize=.5\hsize}X>{\hsize=1.5\hsize}X}
LHS & left-hand side \\
RHS & right-hand side \\
LQG & loop quantum gravity
\end{tabularx}

{\let\clearpage\relax \chapter{List of symbols}}

\begin{tabularx}{\textwidth}{>{\hsize=.5\hsize}X>{\hsize=1.5\hsize}X}
\( \mathfrak g \) & Lie algebra associated to the Lie group \( G \)\\
\( \mathfrak{g}_\C \) & complexification of the Lie algebra \( g \) \\
\( \mathfrak{g}_\R \) & realification of the complex Lie algebra \( g \) \\
\( G_\C \) & complexification of the Lie group \( G \) \\
\( G_\R \) & realification of the complex Lie group \( G \) \\
\( M_n(\mathbb{F}) \) & space of \( n\times n \) matrices over the field \( \mathbb F \)\\
\( \GL(n,\mathbb{F}) \) &  group of invertible \( n\times n \) matrices over the field \( \mathbb F \)\\
\( \SL(n,\mathbb{F}) \) & subgroup of matrices in \( \GL(n,\mathbb{F}) \) with determinant \( 1 \)\\
\( \mathrm{U}(n) \) & group of \( n\times n \) unitary complex matrices\\
\( \1 \) & identity matrix or identity operator\\
\( M\Star \) & conjugate transpose of the complex matrix \( M \)\\
\( T\Dagger \) & adjoint of an operator in an inner product space\\
\( \ii \) & imaginary unit\\
\( \N \) & set of natural numbers (starting with \( 1 \))\\
\( \N_0 \) & set of natural numbers including \( 0 \)\\
\( x\mathbb{S} \) & set defined as \( \set{xs \setst s\in \mathbb{S}} \) (\( x\in \C \) and \( \mathbb S \subseteq \C \))\\
\( x+\mathbb{S} \) & set defined as \( \set{x+s \setst s\in \mathbb{S}} \) (\( x\in \C \) and \( \mathbb S \subseteq \C \))\\
\( f\circ g \) &composition of the functions \( f \) and \( g \)\\
\( \nabla \) & gradient operator\\
\( \wedge \) & wedge product

\end{tabularx}

\mainmatter
\pagestyle{ruled}
\counterwithout{figure}{chapter}

\chapter{Introduction}\label{chap:intro}

Lie groups are undoubtedly one of the most useful tools in mathematical and theoretical physics, especially in quantum theory. Although compact Lie groups play a more prominent role in physics, non-compact ones have important applications too: many of the \emph{dynamical groups} are non-compact\footcite{SinanoGlu1973}, and, as a matter of fact, the systematic study of the representation theory of non-compact Lie groups started in 1947 with \cite{Bargmann1947}, which was motivated by the importance of the Lorentz group in physics.

Non-compact groups are the main theme of this work, which focuses both on their mathematical properties and on their application to physics.
Part of this thesis---mostly \cref{chap:wigner-eckart}---is very mathematical in nature, as it deals with a rigorous construction of some important definitions and results in the representation theory of non-compact groups. The rest of the work focuses on two distinct but  related applications of non-compact groups to quantum gravity; the link between the two applications is the use---either explicitly or implicitly---of tensor operators. The three main topics covered in the following chapters are described in the sections below, together with a brief overview of loop quantum gravity.

\section{Tensor operators and \WE\ theorem for non-compact groups}

Among the many applications of Lie groups and Lie algebra to physics, tensor operators play a prominent role. Initially arising in the study of the quantum theory of angular momentum, these operators are a generalisation of the notion of classical tensors, in the sense that they  transform ``well'' under the action of a group (\( \SO(3) \) or its double cover \( \SU(2) \) for angular momentum); this statement can be formalised in terms of representation theory by requiring that tensor operators transform, under the adjoint action of the group, as vectors in one of its irreducible representations. Notable examples are the position and momentum operators \( \vec{q} \) and \( \vec{p} \) (vector operators) and their ``norms squared'' \( \vec{q}^2 \) and \( \vec{p}^2 \) (scalar operators): the latter are particularly important as, in general, scalar operators are exactly those which are invariant under the action of the group.

Tensor operators are extensively used in quantum mechanics, especially in atomic and nuclear physics\footcite[chap.~XIII]{messiah2}, essentially for two reasons:  two tensor operators can be combined to obtain another one (for example we can construct \( \vec q \times \vec p \) and \( \vec q \cdot \vec p \) from \( \vec q \) and \( \vec p \)), and in general the matrix elements of a tensor operator are easy to calculate, due to the result known as the \emph{\WE\ theorem}. The theorem states that\footcite[chap.~9]{barut}, when the group is compact (e.g., \( \SU(2) \)), the matrix elements of a tensor operator are proportional to the \emph{\CG\ coefficients}---quantities that appear in the study of the decomposition of a product of two representations into irreducible ones---with the proportionality constant independent of the specific component of the tensor operator and of the basis elements being considered: as a consequence, only one matrix element has to be explicitly calculated to know all of the others, of which, depending on the rank of the tensor and on the dimension of the vector spaces on which it acts, there can be quite a large number!

A generalisation to non-compact groups exists, although it is only for tensor operators transforming as unitary representations of the group\footcite{locallycompact}, which are necessarily either \( 1 \)-dimensional (trivial representation) or infinite-dimensional. In addition to the obvious drawback of having to work with infinitely many components, the latter have the disadvantage that, in general, they cannot be composed to obtain scalar operators\footnote{Mathematically, this is a consequence of the fact that the trivial representation does not appear in the decomposition of the product of two infinite\Hyphdash{}dimensional ones.}, which as noted before are the only ones invariant under the action of the group and thus, depending on the context, may be the only true observables of the theory.

In this thesis, a new generalisation of the Wigner\Endash{}Eckart theorem which allows tensor operators transforming as finite-dimensional (non-unitary) representation of non-compact groups is introduced. To do so, the theory of tensor operators will be revisited, introducing a basis-free definition which will make the proof of the theorem straightforward. As we will see, however, the theorem itself is quite useless without the explicit expression of the \CG\ coefficients, which for non-compact groups require the knowledge of the recoupling theory of finite-dimensional (non-unitary) and infinite-dimensional (unitary) representations, which is not known in general and has to be studied case by case. Here such a study will be presented for the particular cases of \( \Spin(2,1) \) and \( \Spin(3,1) \), the double covers of the \( 3 \)D and \( 4 \)D Lorentz groups, which are of great importance in physics.
In both cases, the recoupling theory between finite and infinite-dimensional representations was either only partially known\footnote{The \( 3 \)D case was considered only for representations in the \emph{discrete series} in \cite{haruo}, but even in these cases some results we are going to prove here are missing.} or completely unknown.
As we will see, despite these being amongst the simplest examples of non-compact groups, the study of their \CG\ decompositions is far from easy.

As an application, the \WE\ theorem will be used to obtain a generalisation of the \emph{\JS\ representation} of \( \SU(2) \) to infinite-dimensional representations of \( \Spin(2,1) \) and \( \Spin(2,1) \); in both cases this result was completely unknown for representations in the \emph{continuous series}, which for the \( 4 \)D Lorentz group contains \emph{all} the non-trivial unitary representations. The \JS\ representation is a way to construct the generators of infinitesimal rotations---the \( \su(2) \) generators, which can be seen as the components of a vector operator---in terms of smaller building blocks, namely a pair of uncoupled quantum harmonic oscillators, which are the components of two \emph{spinor operators}. The generalisation to the Lorentz groups exhibits similar features to the Euclidean counterpart, but in the case of continuous series representations the spinor operators can no longer be interpreted as harmonic oscillators, despite satisfying the same commutation relations.

\section{Applications to quantum gravity}

\subsection{Overview of loop quantum gravity}

An introductory overview of the most important aspects of loop quantum gravity is presented here, following \cite{RovelliVidotto}.
Loop quantum gravity (LQG) is a tentative approach to the quantisation of gravity whose main feature is that the quantisation is non-perturbative, i.e., the \emph{full metric} is quantised, not just the excitations of a fixed background metric.
LQG focuses on the quantum properties of geometrical quantities such as areas and volumes; the main result of the theory is that space is fundamentally discrete, in the sense that the spectra of the operators associated to the geometrical observables are discrete.

Loop quantum gravity attempts to quantise general relativity in the Palatini formalism: instead of the metric, the \emph{vielbein}\footnote{The equivalent of the \emph{tetrad} in arbitrary dimensions.} and the \emph{connection} are used as variables, without assuming that the latter is necessarily the Levi--Civita connection.
The quantisation is obtained in two steps: first the classical theory is discretised, then the resulting phase space is canonically quantised. The quantum theory obtained with this procedure is obviously a \emph{truncation} of the full theory; to recover the latter a \emph{continuum limit} has to be considered, where the discretisation is increasingly refined.

\subsubsection{Discretisation of classical boundary phase space}

\begin{figure}
\centering
\tikzsetnextfilename{dual_graph}
\begin{tikzpicture}[x=3em,y=3em,line width=1.1pt]
\coordinate (a) at (0,0);
\coordinate (b) at (1,2);
\coordinate (c) at (2,0.5);
\coordinate (d) at (3,1.5);
\fill (a) circle (2.2pt);
\fill (b) circle (2.2pt);
\fill (c) circle (2.2pt);
\fill (d) circle (2.2pt);
\draw[midarrow] (-1,1) -- (a);
\draw[midarrow] (a) -- (b);
\draw[midarrow] (a) -- (c);
\draw[midarrow] (b) -- (c);
\draw[midarrow] (d) -- (b);
\draw[midarrow] (d) -- (c);
\draw[midarrow] (d) -- ($(d)+(1,-0.5)$);
\end{tikzpicture}
\caption{Section of the dual graph associated to the boundary triangulation.}
\label{fig:dual graph}
\end{figure}
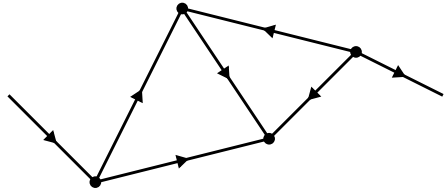

First, space-time is \emph{discretised} by introducing a triangulation, i.e., by approximating it with \emph{\( d \)-simplices}\footnote{The higher-dimensional analogue of triangles and tetrahedrons. Here \( d \) is the space-time dimension.}. The triangulation of a region of space-time induces a triangulation of its boundary, to which we associate its \emph{dual graph} (\cref{fig:dual graph}): to each node of the graph we associate a ``chunk of space''---a triangle for \( 3 \)D gravity or a tetrahedron for \( 4 \)D gravity--- and to each link coming out of the node we associate one of the \( (d-2) \)-simplices bounding it---lines in \( 3 \)D or triangles in \( 4 \)D (\cref{fig:tetra}). Two nodes are connected when their two associated tetrahedra/triangles are adjacent.

The connection and the vielbein are discretised by assigning an \( \SU(2) \) group element and an \( \su(2) \) algebra element to each link, known respectively as the \emph{holonomy} and \emph{flux}. This way the classical boundary phase space of the theory becomes \( \paren{\su(2)\times \SU(2)}^L \cong T^*\SU(2)^L \), where \( L \) is the number of links in the graph. Note that the group \( \SU(2) \) is only used in \( 3 \)D Euclidean gravity and \( 4 \)D gravity (both Euclidean and Lorentzian); as already mentioned, the non-compact group \( \Spin(2,1) \) is needed in the \( 3 \)D Lorentzian case.
\definecolor{tetra1}{HTML}{17385E}
\definecolor{tetra2}{HTML}{00428A}
\definecolor{tetra3}{HTML}{5A90CB}
\definecolor{tetra4}{HTML}{296BBB}
\tikzset{
hcenter/.style={
execute at end picture={
\path ([rotate around={180:#1}]perpendicular cs: horizontal line through={#1}, vertical line through={(current bounding box.east)})
([rotate around={180:#1}]perpendicular cs: horizontal line through={#1}, vertical line through={(current bounding box.west)});}}}
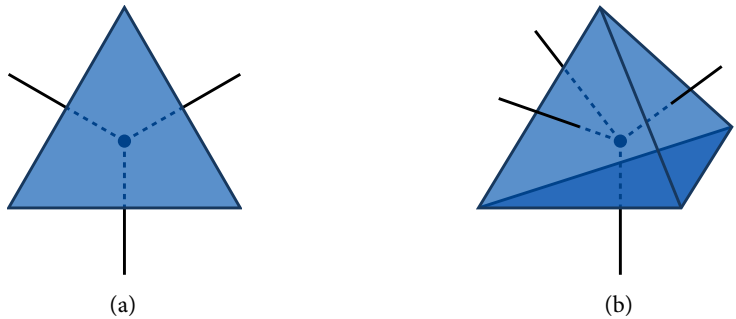
\begin{figure}
\tikzset{mydashed/.style={dash pattern=on 2pt off 2pt}}
\begin{minipage}{0.5\textwidth}
\centering
\tikzsetnextfilename{triangle}
\begin{tikzpicture}[x=4em, y=4em, line width=1.1pt]
\coordinate (A) at (0,0);
\coordinate (B) at (2,0);
\coordinate (C) at (1,{sqrt(3)});
\coordinate (c) at (barycentric cs:A=1,B=1,C=1);
\fill[color=tetra3] (A) -- (B) -- (C) -- cycle;
\fill[color=tetra2] (c) circle [radius=2.5pt];
\draw[color=tetra2,mydashed] (c) -- ($(A)!0.5!(C)$) coordinate[pos=2](p1);
\draw ($(A)!0.5!(C)$) to (p1);
\draw[color=tetra2,mydashed] (c) -- ($(B)!0.5!(C)$) coordinate[pos=2](p2);
\draw ($(B)!0.5!(C)$) to (p2);
\draw[color=tetra2,mydashed] (c) -- ($(B)!0.5!(A)$) coordinate[pos=2](p3);
\draw ($(A)!0.5!(B)$) to (p3);
\draw[color=tetra1] (A) -- (B) -- (C) -- cycle;
\end{tikzpicture}
\subcaption{\label{subcap:3d}}
\end{minipage}%
\begin{minipage}{0.5\textwidth}
\centering
\tikzsetnextfilename{tetrahedron}
\begin{tikzpicture}[x=3.5em, y=4em, line width=1.1pt, hcenter=(c)]
\coordinate (A) at (0,0);
\coordinate (B) at (2,0);
\coordinate (C) at (1.2,{sqrt(3)});
\coordinate (D) at (2.5,0.7);
\coordinate (c) at (1.4,{1/sqrt(3)});
\fill[color=tetra3] (A) -- (B) -- (C) -- cycle;
\fill[color=tetra3] (C) -- (B) -- (D) -- cycle;
\fill[color=tetra4] (A) -- (B) -- (D) -- cycle;
\fill[color=tetra2] (c) circle [radius=2.5pt];
\draw[color=tetra2,mydashed] (c) -- (1.4,0) coordinate[pos=2](p1);
\draw (1.4,0) to (p1);
\draw[color=tetra2,mydashed] (c) -- ($(A)!0.7!(C)$) coordinate[pos=1.5](p2);
\draw ($(A)!0.7!(C)$) to (p2);
\draw[color=tetra2,mydashed] (c) -- (1,0.7) coordinate[pos=3](p4);
\draw[color=tetra2,mydashed] (c) -- (1.9,0.9) coordinate[pos=2](p3);
\draw[color=tetra2,shorten <=0.5pt,shorten >=0.5pt] (A) -- (D);
\draw[color=tetra1] (A) -- (B) -- (D) -- (C) -- cycle;
\draw[color=tetra1,shorten <=0.5pt] (C) -- (B);
\draw (1.9,0.9) to (p3);
\draw (1,0.7) to (p4);
\end{tikzpicture}
\subcaption{\label{subcap:4d}}
\end{minipage}%
\caption{We associate to each node a triangle \subcaptionref{subcap:3d} in \( 3 \)D and a tetrahedron \subcaptionref{subcap:4d} in \( 4 \)D.}
\label{fig:tetra}
\end{figure}

\subsubsection{Hilbert space and spin networks}
Given a discretisation, or equivalently a boundary graph, the  Hilbert space describing the quantum states of the boundary geometry is given by canonically quantising \( T^*\SU(2)^L \) and taking into account the local \( \SU(2) \) invariance at each node of the graph. A basis for the Hilbert space is provided by \emph{spin networks}, i.e., graphs with irreducible \( \SU(2) \) representations (labelled by half-integer spins) attached to each link and \emph{intertwiners} attached to each node; the latter are mathematical objects needed to ensure that the sum of the \( \SU(2) \) generators (angular momenta) of the links connected to each node is zero---in other words they implement the local \( \SU(2) \) invariance.
The geometrical observables are constructed from the \( \SU(2) \) generators of each link. In particular, the operators associated to the area/length of the triangle/segment  dual to a link are proportional to the Casimir operator of its representation, and consequently have a discrete spectrum.

One should note that this Hilbert space is kinematical, i.e., the \emph{Hamiltonian constraint} has to be implemented to obtain the \emph{physical} Hilbert space.

\subsubsection{Open problems}

Loop quantum gravity still has some issues that need to be resolved. Of particular importance are the following:

\begin{itemize}

\item \emph{Hamiltonian constraint}: the \emph{dynamics} of the theory, i.e., the construction of the physical Hilbert space, is not fully understood yet, especially in the \( 4 \)D case. The difficulty lies in finding the solutions to the Hamiltonian constraint.

\item \emph{Semi-classical limit}: it is not yet known if loop quantum gravity has the right semi-classical limit, that is general relativity is recovered in the limit \( \hbar \rightarrow 0 \).
\end{itemize}

The results of this thesis are related to both these problems: in \cref{sec:lorentzian_LQG} a solvable Hamiltonian constraint is introduced for the \( 3 \)D \emph{Lorentzian} theory, while \cref{chap:so*} deals with a new kind of coherent states in loop quantum gravity, which could be used to achieve a better understanding of the semi-classical limit of the theory.

\subsection{Spinorial approach to \( 3 \)D Lorentzian loop quantum gravity}

\( 3 \)D quantum gravity is a useful ``theoretical laboratory'' to explore and test some  of the issues met in the \( 4 \)D theory\footcite{Carlip:1998uc}. For example, it is possible to solve the Hamiltonian constraint  and to relate loop quantum gravity (LQG) to the relevant spinfoam model: this was done in the Euclidean case, with either a vanishing or negative cosmological constant\footcite{Noui:2004ja, Bonzom:2011hm, Bonzom:2014bua}.
\( 3 \)D Euclidean LQG uses \( \SU(2) \) as a gauge groups, and its Hilbert space is spanned by \emph{spin networks}, graphs whose edges are labelled by irreducible representations of \( \SU(2) \) and whose vertices are associated to \emph{intertwiners} of the representations of the edges meeting at the vertex. These states diagonalise the operators describing geometrical quantities, such as lengths and angles, which are constructed out of the \( \su(2) \) algebra generators, and whose spectrum turns out to be discrete.

An important tool in the description of \( 3 \)D Euclidean LQG is given by what is known as the \emph{spinorial framework}, which uses the \JS\ representation to introduce a new family of \( \SU(2) \)-invariant observables, which can be used to construct all the usual geometrical observables and have the advantage of forming a closed algebra\footnote{In contrast, the length and angle operators do not form a closed algebra.}. Among the other things, these new observables can be used to construct a solvable Hamiltonian constraint\footcite{Bonzom:2011nv}. As mentioned above, the \JS\ representation can be recast in terms of tensor operators: this key realisation makes the generalisation of the spinorial framework to different gauge groups possible; for example, it was used to generalise it to the \emph{quantum group} \( \mathcal{U}_q(\su(2)) \) in \cite{Dupuis:2013lka}, in order to introduce a non-zero cosmological constant in the theory.

Having an equivalent of the \JS\ representation for \( \Spin(2,1) \) allows one to extend the spinorial formalism to the \( 3 \)D Lorentzian case, of which it is the gauge group\footcite{Freidel:2002hx}; this generalisation is one of the main topics of this thesis. As a first step, the classical LQG phase space is constructed by introducing classical tensors and, in particular, classical spinors. The spinors are used as fundamental building blocks, as they can be used to reconstruct both the flux and the holonomy variables of the phase space, similarly to the Euclidean case; moreover, following an approach similar to the one in \cite{Bonzom:2011nv}, the spinors and the group elements constructed with them are used to rewrite the \emph{flatness constraint} of the classical space in terms of a new set of variables, namely the classical equivalent of the observables constructed out of the spinor operators.

The quantisation of the classical phase space needs to be treated carefully, as there are subtleties involved due to the non-compacticity of the gauge group. For example, the quantum spinorial observables can take intertwiners between unitary representations to intertwiners involving some (infinite-dimensional) non-unitary representation. Nevertheless, these observables can be used to construct an Hamiltonian constraint as the quantisation of the spinorial flatness constraint. Focusing on a triangular face of a spin network, it is shown how the Lorentzian Ponzano--Regge amplitude, given by a \emph{Racah coefficient}, is a solution of the Hamiltonian constraint.

\vfil

\subsection{Intertwiner space as an \( \SO\Star(2n) \) representation}

This last topic deals with applications of non-compact groups to quantum gravity again, but it does so in the Euclidean setting.
When working with the spinorial formalism in loop quantum gravity with \( \SU(2) \) gauge group, an additional structure appears on the space of \( n \)-valent intertwiners with a \emph{fixed total area}: this space is finite dimensional, and it provides an irreducible unitary representation of the compact group \( \mathrm{U}(n) \), whose generators are constructed as \( \SU(2) \)-invariant quadratic polynomials in the \( 2n \) harmonic oscillators appearing in the \JS\ decomposition of each leg\footcite{freidel_fine_2010}. Here it will be shown how, even when working with a compact gauge group, the spinorial formalism naturally introduces a non-compact group in the theory: in fact, the \emph{full} space of \( n \)-valent intertwiners, with all possible areas, is shown to have the structure of an \( \SO\Star(2n) \) representation. This group, which is non-compact for all \( n>1 \), is a lesser-known real form of \( \SO(2n,\C) \); we will show how, using the fact that it is a subgroup of the symplectic group \( \Sp(4n,\R) \), it can be identified with the subgroup of \emph{Bogoliubov transformations} on the \( 2n \) harmonic oscillators that leaves the \( \SU(2) \) invariance of the intertwiners intact.

One of the most important consequences of this new result is that, as will be shown, the invariance of intertwiner space under \( \SO\Star(2n) \) can be used to construct a new kind of coherent intertwiner, just as \( \mathrm{U}(n) \) coherent intertwiners were introduced for fixed-area intertwiner space\footcite{FreidelLivine2011}. These are Gilmore--Perelomov coherent states, a generalisation of the well-known harmonic oscillator coherent states---living in a unitary representation of the \emph{Heisenberg group}---to arbitrary Lie groups. In this work the new kind of coherent states is introduced and analysed; in particular, the expectation values and variances of the physical observables measuring areas are calculated in these states, and in some specific cases it is shown how the full probability distribution can be calculated. Moreover, their semi-classical limit is investigated: when the areas involved are large, it is shown that the expectation values of the \( \SO\Star(2n) \) generators can be endowed with a Poisson algebra structure, which leads to the original space upon quantisation. This semi-classical limit can be related to a classical geometry by introducing \( n \) vectors that can be interpreted as the normals to the faces of a convex polyhedron in \( \R^3 \), although more works needs to be done to fully understand this process.

One should note that, although the understanding of the group structure underlying these coherent states is completely new, some of them have been considered before, for example in \cite{Freidel:2012ji}. Nevertheless, these form only a small subset of the full family of coherent states: in fact, as we will see, the \( \SO\Star(2n) \) coherent intertwiners are labelled by antisymmetric matrices \( \zeta \) satisfying some additional constraints; of these, only the ones with \( \rank(\zeta)=2 \) have been considered\footnote{In which case \( \zeta \) is sometimes said to satisfy the Plücker relations.}, as these are exactly the ones that can be obtained as a linear combination of the \( \mathrm{U}(n) \) coherent intertwiners.

\section{Organisation of the thesis}

The thesis is divided in three main chapters, based on the topics discussed above. \Cref{chap:wigner-eckart} starts with a brief review of the theory of \((\mathfrak{g},K)\)-modules, needed to rigorously treat infinite-dimensional group representations with algebraic methods, and a section dedicated to the study of tensor operators and the \WE\ theorem for arbitrary groups; subsequently, the results on recoupling theory of finite and infinite-dimensional representations and on the \JS\ representation are presented separately for \( \Spin(2,1) \) and \( \Spin(3,1) \). \Cref{sec:lorentzian_LQG} is roughly divided in two part: first the classical description of LQG is considered, then the focus is shifted to the quantum theory, with the description of the Lorentzian intertwiner space and the study of the quantum Hamiltonian constraint built out of the spinor operators.
\Cref{chap:so*} starts with an introduction of the Lie group \( \SO\Star(2n) \) and its Lie algebra, followed by a section describing their action on intertwiners space; the rest of the chapter is focused on the study of \( \SO\Star(2n) \) coherent states and their properties.

A number of appendices are included at the end of the thesis:
\cref{app:matrices} lists some useful properties of tridiagonal and antisymmetric matrices, \cref{app:CG} contains a table of the \CG\ coefficients used throughout the thesis and the proof of some of their properties, and \cref{app:bounded_symmetric_domains} provides some results on the groups \( \SO\Star(2n) \) and \( \Sp(2n,\R) \) and their action on \emph{bounded symmetric domains}, which are used to label the coherent states of \cref{chap:so*}.

\chapter[\WE\ theorem and \JS\ representation for the \texorpdfstring{\(3\)D}{3D} and \texorpdfstring{\(4\)D}{4D} Lorentz group][WE theorem and JS representation for the \texorpdfstring{\(3\)D}{3D} and \texorpdfstring{\(4\)D}{4D} Lorentz group]{\WE\ theorem and \JS\ representation for the \texorpdfstring{\(3\)D}{3D} and \texorpdfstring{\(4\)D}{4D} Lorentz group}
\label{chap:wigner-eckart}

The key question of this chapter is the following: does the \WE\ theorem admit a generalisation for non-compact groups? It is already known that this is possible if we use the theorem for tensor operators transforming as (infinite-dimensional) unitary representations\footcite{locallycompact}, so we will focus on tensor operators transforming as finite-dimensional (non-unitary) representations\footnote{As noted in \cref{chap:intro}, these are also more relevant from the physical point of view.}.
As is common in physics, we will work throughout the chapter with algebraic methods, i.e., we will consider everything from the Lie algebra perspective; in order to do this rigorously for infinite-dimensional representations, we will need the mathematical machinery of \((\mathfrak{g},K)\)-modules, so we will start by reviewing them in \cref{sec:gk}. We will then show in \cref{sec:tensor_operators} how tensor operators can be defined in a basis-independent way, and use this new definition to prove the \WE\ theorem for a generic Lie group.

As mentioned in \cref{chap:intro}, to actually use the theorem in the case on non-compact groups it is necessary to study the recoupling theory of the product of a finite-dimensional representation and an infinite-dimensional one. We will study in detail the cases of the \( 3 \)D and \( 4 \)D Lorentz groups, respectively in \cref{sec:3d-lorentz-group} and \cref{sec:4d-lorentz-group}. In both cases, as an application, we will use the \WE\ theorem to generalise the \JS\ representation, known for only some representation classes\footnote{Finite-dimensional and discrete series representations in the \( 3 \)D case, finite-dimensional only in the \( 4 \)D case.}, to all irreducible representations; the results for the \( 3 \)D case will be the basis for \cref{sec:lorentzian_LQG}.

The contents of this chapter are based on the results presented in the articles \cite{wigner_eckart} and  \cite{Sellaroli:2015sca}.

\section{Infinite-dimensional Lie group representations and \texorpdfstring{\( (\mathfrak{g},K) \)-modules}{(g,K)-modules}}\label{sec:gk}

When working with non-compact groups, as we are about to do in this thesis, we often have to deal with infinite-dimensional representations; we will see in this section how to rigorously treat them with algebraic methods, making use of the notion of \((\mathfrak{g},K)\)-modules. Recall that a continuous Lie group representation, which we will also refer to as a (topological) \( G \)-module, is a topological vector space\footnote{We will always assume that representations are on complex vector spaces.} \( V \) with a continuous action \( G\times V \rightarrow V \) such that
\begin{equation}
g(v+w)=gv+gw,\quad g\alpha v = \alpha g v,\quad \forall g\in G, \quad\forall v,w \in V,\quad\forall \alpha\in \C;
\end{equation}
as common when working with the module notation, we will denote the representation/module by its underlying vector space \( V \).
When working with finite-dimensional representation, we can obtain a \( \mathfrak g \)-module\footnote{i.e., a representation of the Lie algebra \( \mathfrak g \) of \( G \).} with the same vector space by defining
\begin{equation}\label{eq:g-module}
X v:= \eval{\der{}{t}}{t=0}e^{tX} v,\quad X\in\mathfrak g, \quad \forall v\in V,
\end{equation}
and we can often obtain all the information we need about the \( G \)-module by working on the corresponding Lie algebra representation: for example the \( G \)-module is irreducible if and only the associated \( \mathfrak g \)-module is.

These algebraic methods are extremely useful, and they are widely used in applications of representation theory to physics. However, when \( V \) is infinite-dimensional, the requirement that the group action be continuous makes things considerably more difficult; for example, an infinite-dimensional representation \( V \) is irreducible if there are no \emph{closed} invariant subspaces other than \( \set{0} \) and \( V \) itself, so that pure algebraic methods are a priori not enough to enstablish the irreducibility of \( V \). Moreover, working with the Lie algebra is not as straightforward as the finite-dimensional case:
one cannot always obtain a \( \mathfrak g \)-module from \( V \) as in \eqref{eq:g-module}, since the RHS may not be defined for a generic \( v \). In order to overcome these difficulties, we will work with \((\mathfrak{g},K)\) modules\footcite[See][chap.~3]{wallach}:
\begin{definition}
Let \( G \) be a real Lie group with Lie algebra \( \mathfrak g \) and maximal compact subgroup~\( K \). A \((\mathfrak{g},K)\)-module is a vector space \( V \) that is both a \( \g \)-module and a \( K \)-module, where we ignore the topology of \( K \), which satisfies the compatibility conditions
\begin{enumerate}
\item \(k\cdot X \cdot v=\operatorname{Ad}(k)X\cdot k \cdot v\) for all \(v\in V\), \(k\in K\), \(X\in\g\);
\item if \(v\in V\), \(Kv=\set{kv\setst k\in K}\) spans a finite-dimensional subspace of \(V\) on which the action of \(K\) is continuous;
\item if \(v \in V\) and \(Y\in \mathfrak{k}\) then \(\eval{\der{}{t}}{t=0}e^{tY}v=Yv\).
\end{enumerate}
The first condition is technical and is needed to extend the definition to disconnected groups; the second conditions is equivalent to saying that \( V \) is the algebraic direct sum\footnote{i.e., only sums of finitely many vectors are considered.} of finite-dimensional irreducible \( K\)-modules, while the third condition ensures that the infinitesimal action of \( K \) agrees with that of its Lie algebra \( \mathfrak k \subseteq \mathfrak g \).
\end{definition}
Although this definition may seem very technical, these objects have reasonable properties: they are essentially \( \mathfrak g \)-modules with some additional compatibility with the group. In order to understand how  \((\mathfrak{g},K)\)-modules provide the right tool to study infinite-dimensional representations with algebraic methods, we can take a look at some of the results from \cite{wallach}:

\begin{itemize}
\item every topological \( G \)-module \( H \) induces a \((\mathfrak{g},K)\)-action on the space of \emph{\( K \)-finite vectors}
\begin{equation*}
H_K:=\set{v\in V \setst \dim\Span\set{K v}<\infty},
\end{equation*}
which is also referred to as the \emph{underlying \((\mathfrak{g},K)\)-module} of \( H \); the action of \( \g \) on \( H_K \) is given by \eqref{eq:g-module} as expected.

\item An admissible \( G \)-module \( H \) is irreducible if and only if it is infinitesimally irreducible, i.e., \(H_K \) is irreducible\footnote{No need to check if the invariant subspaces are closed!}.

\item Two unitary representation on the Hilbert spaces \( H \), \( H' \) are unitarily equivalent if and only if they are infinitesimally equivalent, i.e., \( H_K\cong H\Prime_K \), both as a \( g \)-module and as a \( K\)-module.
\end{itemize}
Here a \((\mathfrak{g},K)\)-module \( V \) is  an \emph{admissible} if each irreducible representation of \( K \) appears only finitely many times in \( V \), while a \( G \)-module \( H \) is admissible if \( H_K \) is. Admissible representations are those that, in some sense, behave ``nicely'': for example, they include all irreducible unitary representations. We will implicitly assume that all representations we work with are admissible.

\section{Tensor operators and \WE\ theorem}\label{sec:tensor_operators}

Tensor operators are a class of operators that transform particularly well under the adjoint action of the group \( G \), namely they transform as vectors in a representation of \( G \). They are usually defined relative to a basis, i.e., a tensor operator is identified with the set of its components; here we will consider a basis-free definition instead. Moreover, we will distinguish between the concept of \emph{weak} and \emph{strong} tensor operators\footnote{This distinction is not found in the literature.} to allow for a rigorous treatment at the Lie algebra level for infinite-dimensional representations by using \((\mathfrak{g},K)\)-modules.
\begin{definition}[strong tensor operator]
Let \(V_0\), \(V\) and \(V\Prime\) be (topological) \(G\)-modules of a Lie group \(G\), with \(V_0\) finite-dimensional. A \emph{strong tensor operator} for \(G\) is an intertwiner between \(V_0\otimes V\) and \(V\Prime\), i.e., a \emph{continuous} linear map
\begin{equation*}
T:V_0\otimes V\rightarrow V\Prime
\end{equation*}
such that
\begin{equation*}
T\circ g = g \circ T,\quad \forall g\in G.
\end{equation*}
If \(V_0\) is irreducible, \(T\) is called an irreducible strong tensor operator.
\end{definition}

\begin{definition}[weak tensor operator]
Let \(V_0\), \(V\) and \(V\Prime\) be \( (\mathfrak{g},K) \)-modules of a Lie group \(G\), with \(V_0\) finite-dimensional. A \emph{weak tensor operator} for \(G\) is an intertwiner between \(V_0\otimes V\) and \(V\Prime\), i.e., a linear map\footnote{In this weaker definition \(T\) is not required to be continuous, as there is no topology specified on the \((\mathfrak{g},K)\)-modules.}
\begin{equation*}
T:V_0\otimes V\rightarrow V\Prime
\end{equation*}
such that
\begin{equation*}
T \circ X= X\circ T,\quad \forall X\in \mathfrak{g}
\quad \textnormal{and} \quad
T \circ k = k \circ T ,\quad \forall k \in K,
\end{equation*}
where \(\mathfrak{g}\) and \(K\) act on the product module as
\begin{align*}
X(v_0\otimes v)&=(X v_0)\otimes  v + v_0 \otimes (X v)\\
k(v_0\otimes v)&=(k v_0)\otimes (k v).
\end{align*}
If \(V_0\) is irreducible \(T\) is called an irreducible weak tensor operator.
\end{definition}
Note that this nomenclature is appropriate, as weak tensor operators are more general than strong ones; in fact
\begin{proposition}
An intertwiner \(T:V\rightarrow V\Prime\) between \(G\)-modules is also an intertwiner between the corresponding \((\mathfrak{g},K)\)-modules. As a consequence, a strong tensor operator is also a weak tensor operator.
\end{proposition}
\begin{proof}
Recall that the subspace \(V_K\subseteq V\) of \(K\)-finite vectors, i.e., the set of all vectors \(v\) such that \(\Span\{kv,k\in K\}\)  is finite-dimensional, is the \((\mathfrak{g},K)\)-module associated to \(V\), with
\begin{equation}
X v:=\eval{\frac{d}{dt}}{t=0}\exp(tX)v,\quad \forall X\in \mathfrak{g},\quad \forall v\in V_K.
\end{equation}
We have, since \(T\) commutes with the action of \(K\subseteq G\),
\begin{equation}
\begin{split}
\dim\Span\set*{k T v \setst k\in K }&=\dim\Span\set*{Tk v \setst k\in K }\\
&=
\dim T\paren{\Span\set*{k v \setst k\in K}}<\infty
\end{split}
\end{equation}
for each \(v\in V_K\), that is \(T(V_K)\subseteq V\Prime*_K\). Moreover, for each \(X\in\g\) and \(v\in V_K\),
\begin{equation}
XTv=\eval{\frac{d}{dt}}{t=0}\exp(tX)Tv=\eval{\frac{d}{dt}}{t=0}T\exp(tX)v=T\eval{\frac{d}{dt}}{t=0}\exp(tX)v=TXv,
\end{equation}
where the fact that \(T\) is continuous was used. It follows that \(\eval{T}{V_K}\) is an intertwiner between the \((\mathfrak{g},K)\)-modules \(V_K\) and \(V\Prime*_K\).
\end{proof}
As weak tensor operators are more general, in the following chapters we will refer to them simply as tensor operators, unless otherwise noted.
It is often preferable to have operators between \(V\) and \(V\Prime\): this can be achieved by defining the ``components'' of a tensor operator \(T\) in a basis \(\set{e_i}_{i\in I}\subseteq V_0\) as
\begin{equation}
T_i:v\in V \mapsto T (e_i\otimes v)\in V\Prime;
\end{equation}
the definitions of strong and weak tensor operators become respectively
\begin{equation}
\label{eq:intertwiner_def_basis1}
gT_i g^{-1}=\sum_{j\in I}\braket{e^j,ge_i}T_j,\quad \forall g \in G,
\end{equation}
and
\begin{equation}
\label{eq:intertwiner_def_basis2}
\begin{cases}
[X,T_i]=\sum_{j\in I}\braket{e^j,Xe_i}T_j, &\casestextn{}\forall X \in \mathfrak{g}
\\
kT_i k^{-1}=\sum_{j\in I}\braket{e^j,ke_i}T_j,&\casestextn{} \forall k \in K,
\end{cases}
\end{equation}
where \(\braket{\cdot,\cdot}\) is the dual pairing of \(V\Star*_0\) and \(V_0\) and \(\set{e^j}_{j\in I}\subseteq V\Star*_0\) is the dual basis\footnote{Here \( V\Star \) denotes the continuous dual space to \( V \), that is the space of continuous linear maps \( V \rightarrow \C\).} defined by
\begin{equation}
\braket{e^j,e_i}=e^j(e_i)=\tensor{\delta}{^j_i}.
\end{equation}

The definition of weak tensor operators can be simplified when \(K\) is connected, since one can simply require the the operator commutes with every element of \(\mathfrak{g}\). In fact we have\footcite[The proof is based on][]{stackexchange}
\begin{proposition}
If \(K\) is connected, a linear map \(T:V\rightarrow V\Prime\) is a \((\mathfrak{g},K)\)-module homomorphism, i.e., an intertwiner between \( V \) and \( V' \), if
\begin{equation*}
T \circ X= X\circ T,\quad \forall X\in \mathfrak{g}.
\end{equation*}
\end{proposition}
\begin{proof}
Let \(\mathfrak{k}\subseteq\mathfrak{g}\) be the Lie algebra of \(K\). For any \(X\in \mathfrak{k}\), \(v\in V\), \(\alpha\in V\Star\) one has
\begin{equation}
\eval{\frac{d}{dt}}{t=0}\braket{\alpha,\paren{ T\circ \exp(tX) - \exp(tX)\circ T }v}=\braket{\alpha,\paren{ T\circ X - X\circ T }v}=0.
\end{equation}
Since the derivative vanishes, it must be
\begin{equation}
\braket{\alpha,\paren{ T\circ \exp(X) - \exp(X)\circ T }v}=\braket{\alpha,\paren{ T\circ \exp(0) - \exp(0)\circ T }v}
\end{equation}
for each \(v\in V\), \(\alpha \in V\Star\), so that 
\begin{equation}
T\circ \exp(X)=\exp(X)\circ T,\quad \forall X\in \mathfrak{k};
\end{equation}
however, if \(K\) is connected, \(\exp(\mathfrak{k})\subseteq K\) generates it\footnote{In the sense that every \(k\in K\) can be obtained as a product of elements of \(\exp(\mathfrak{k})\). \cite[chap.~4]{kosmann2009}.}, hence
\begin{equation}
T\circ k = k \circ T,\quad \forall k \in K.
\end{equation} 
\end{proof}

One of the most useful properties of irreducible tensor operators is the Wigner\Endash{}Eckart theorem, originally proved for compact groups~\footcite{barut} and later extended to non-compact groups~\footcite{locallycompact} \emph{only} for the particular case of tensor operators transforming as (infinite-dimensional) \emph{unitary} representations, which we do not consider. Here we generalise it to tensor operators transforming as (possibly non-unitary) \emph{finite-dimensional} representations of arbitrary Lie groups. The theorem itself is trivial to prove, as it is essentially a corollary of Schur's lemma\footnote{We continue referring to it as a theorem solely for consistency with existing literature.}; in fact we have
\begin{lemma}
Let \(V_0\), \(V\), \(V\) be irreducible \((\mathfrak{g},K)\)-modules for a Lie group \(G\), with \(V_0\) finite-dimensional. If a decomposition into irreducible modules for \(V_0\otimes V\) exists, a non-zero intertwiner
\begin{equation*}
T:V_0\otimes V\rightarrow V\Prime
\end{equation*}
is possible if and only if \(V\Prime\) appears (at least once) in the decomposition.
If \(\mathcal{T}\) is the vector space of all such intertwiners, \(\dim \mathcal{T}\) equals the multiplicity of \(V\Prime\) in the decomposition, and a basis is provided by the projections in each of the submodules \(W_\alpha\subseteq V_0\otimes V\), \(W_\alpha\cong V\Prime\), with \(\alpha\) keeping track of the multiplicities.
\end{lemma}
\begin{proof}
Let \(T:V_0\otimes V\rightarrow V\Prime\) be a \((\mathfrak{g},K)\)-module homomorphism.
Schur's Lemma for irreducible \((\mathfrak{g},K)\)-modules\footcite[See][chap.~3]{wallach} guarantees that, if \(W\subseteq V_0\otimes V\) is a submodule,
\begin{equation}
\eval{T}{W}\propto
\begin{cases}
\1 & \casesif W\cong V\Prime\\
0 & \casestextn{otherwise}.
\end{cases}
\end{equation}
It is then trivial to see that any such \(T\) can be written as a linear combinations of the independent maps \(T^\alpha\) that project \(V_0\otimes V\) on each \(W_\alpha \cong V\Prime\).
\end{proof}
It trivially follows that
\begin{theorem*}[\WE]
Let \(T:V_0\otimes V \rightarrow V\Prime\) be an irreducible tensor operator, with \(V\), \(V\Prime\) irreducible. If a decomposition for \(V_0\otimes V\) exists, \(T\) is a linear combination of the projections \(T^\alpha:V_0\otimes V\rightarrow W_\alpha\) into each irreducible component \(W_\alpha\cong V\Prime\). If \(V\Prime\not\subseteq V_0\otimes V\) the tensor operator must necessarily vanish.
\end{theorem*}

The reason why this theorem is so useful is that it implies that a tensor operator \( T:V_0\otimes V\rightarrow V\Prime \) is \emph{fully} specified\footnote{Up to proportionality factors.} by the decomposition of the product module \( V_0\otimes V \). The non-trivial step is to study the decomposition of this product, when \( G \) is non-compact, for \( V_0 \) finite\Hyphdash{}dimensional and \( V \) infinite\Hyphdash{}dimensional. In the following sections we will  tackle all possible such representations for the specific case of the \( 3 \)D and \( 4 \)D Lorentz groups, which were previously unconsidered; in both cases the results will be used, with the aid of the \WE\ theorem, to generalise the \JS\ representation of \( \SU(2) \) to the respective non-compact group.

\section{\texorpdfstring{\(3\)D}{3D} Lorentz group}\label{sec:3d-lorentz-group}

This section focuses on the recoupling theory of a finite and an infinite-dimensional representation of the double cover of the \( 3 \)D Lorentz group, \( \Spin(2,1) \). First we will review the representation theory of the group in the language of \((\mathfrak{g},K)\)-modules, then we will study the \CG\ decomposition for all classes of infinite-dimensional representations. Finally the recoupling theory results are used, in conjunction with the \WE\ theorem to generalise the \JS\ representation to all representation classes of \( \Spin(2,1) \).

\subsection{Irreducible representations of \texorpdfstring{\(\Spin(2,1)\)}{Spin(2,1)}}

The \emph{proper orthochronous \( 3 \)D Lorentz group} \( \SO_0(2,1) \) is the identity component of the subgroup of \( \GL(3,\R) \) that preserves the indefinite quadratic form
\begin{equation}
S(x)=-(x_0)^2 + (x_1)^2 + (x_2)^2,\quad x=(x_0,x_1,x_2)\in\R^3.
\end{equation}
To allow for spin representations, we will work with its double cover \( \Spin(2,1) \), which is isomorphic to
\begin{equation}
\begin{split}
\SU(1,1)&=\set*{g\in \SL(2,\C)  \setstx g\Star
\begin{pmatrix}
1 & 0\\ 0 & -1
\end{pmatrix}
g=
\begin{pmatrix}
1 & 0\\ 0 & -1
\end{pmatrix}
}
\\
&=
\set*{
\begin{pmatrix}
\alpha & \beta\\
\conj\beta & \conj\alpha
\end{pmatrix}
\in M_2(\C) \setstx \abs{\alpha}^2-\abs{\beta}^2=1
}.
\end{split}
\end{equation}
Its maximal compact subgroup is given by
\begin{equation}
K=\set*{
\begin{pmatrix}
e^{\ii\theta} & 0\\
0 & e^{-\ii\theta}
\end{pmatrix}
\in M_2(\C) \setstx 0\leq \theta <2\pi
}
\cong \mathrm{U}(1).
\end{equation}
We will only consider complex representations, so we can work with the complexified Lie algebra \( \spin(2,1)_\C \), generated by
\begin{equation}
J_0=\tfrac{1}{2}
\begin{pmatrix}
1 & 0\\
0 & -1
\end{pmatrix},
\quad J_1=\tfrac{1}{2}
\begin{pmatrix}
0 & \ii \\
\ii & 0
\end{pmatrix},
\quad J_2=\tfrac{1}{2}
\begin{pmatrix}
0 & 1 \\
-1 & 0
\end{pmatrix},
\end{equation}
with commutation relations
\begin{equation}
[J_0,J_1]=\ii J_2,\quad [J_1,J_2]=-\ii J_0,\quad [J_2,J_0]=\ii J_1.
\end{equation}
The Casimir operator is given in this basis by\footnote{Note the resemblance with the quadratic form \( S(x) \).}
\begin{equation}
\label{eq:casimir}
Q=-(J_0)^2 +(J_1)^2 +(J_2)^2.
\end{equation}
It will prove useful to work with the \emph{ladder operators}
\begin{equation}
J_\pm:=J_1\pm\ii J_2
\end{equation}
satisfying
\begin{equation}
[J_0,J_\pm]=\pm J_\pm,\quad[J_+,J_-]=-2 J_0,
\end{equation}
so that the Casimir becomes
\begin{equation}\label{eq:SU(1,1)_casimir}
Q = -J_0(J_0+1)+J_-J_+\equiv-J_0(J_0-1)+J_+J_-.
\end{equation}
The \((\mathfrak{g},K)\)-modules induced by the irreducible admissible Hilbert space representations of \( \Spin(2,1) \) exhaust all the possible irreducible \((\mathfrak{g},K)\)-modules\footnote{See \cite[chap.~II]{knapp} for the explicit expression of the group representations.}, and the generators act on them as
\begin{equation}
\label{eq:reps}
\begin{cases}
J_0\ket{j,m}=m\ket{j,m}\\
J_\pm\ket{j,m}=\Gamma_\pm(j,m)\ket{j,m\pm 1}\\
Q\ket{j,m}=-j(j+1)\ket{j,m},
\end{cases}
\end{equation}
where
\begin{equation}
\Gamma_\pm(j,m):=\ii \sqrt{j \mp m}\sqrt{j \pm m + 1}.
\end{equation}
The vectors \(\ket{j,m}\) form an orthogonal basis\footnote{Unless the representation is unitary or finite-dimensional, it is not generally possible to renormalise the basis to \( 1 \) and keep the action \eqref{eq:reps} at the same time.} for the vector space of the representation, with \(j\) being a label for the representation and \(m\) enumerating the vectors; their possible values depend on the representation class, which can be one of the following\footnote{See \cite{wallach}, chap.~5, for the classification.}:
\begin{itemize}
\item \emph{Positive discrete series} \(D^+_j\): infinite-dimensional lowest weight (with a lower bound on \(m\)) modules, with
\begin{equation*}
j\in\set*{-\tfrac{1}{2},0,\tfrac{1}{2},1,\dotsc} \quad \mbox{and}\quad m\in\set*{ j+1,j+2,j+3,\dotsc}.
\end{equation*}
\item \emph{Negative discrete series} \(D^-_j\): infinite-dimensional highest weight (with an upper bound on \(m\)) modules, with
\begin{equation*}
j\in\set*{-\tfrac{1}{2},0,\tfrac{1}{2},1,\dotsc} \quad \mbox{and}\quad m\in\set*{-j-1,-j-2,-j-3,\dotsc}.
\end{equation*}
\item \emph{Continuous series} \(C_j^\varepsilon\): infinite-dimensional modules of \emph{parity} \(\varepsilon\in\set*{0,\tfrac{1}{2}}\), with
\begin{equation*}
m\in\varepsilon+\mathbb{Z}\quad\mbox{and}\quad j\in\mathbb{C};
\end{equation*}
when \(j\) is (half-)integer, there is the additional constraint
\begin{equation*}
j-\varepsilon\not\in\mathbb{Z}.
\end{equation*}
Moreover, the representations \(C^\varepsilon_j\) and \( C^\varepsilon_{-j-1}\) are isomorphic.
\item \emph{Finite-dimensional series} \(F_j\): isomorphic to the unitary \(\SU(2)\)-modules\footnote{As \( \spin(2,1)_C\cong \su(2)_C \) representations.}, with
\begin{equation*}
j\in\set*{0,\tfrac{1}{2},1,\dotsc}\quad\mbox{and}\quad m\in\set*{-j,-j+1,\dotsc,j-1,j}.
\end{equation*}
They are the only finite-dimensional modules, with dimension \(2j+1\).
\end{itemize}
Of these representations, the only unitary ones are the whole discrete (positive and negative) series, the continuous series with either
\begin{equation}
j\in\set*{-\tfrac{1}{2}+\ii s \setstx s\neq0 },\quad \varepsilon=0,\half,
\end{equation}
known as the \emph{principal series}, or
\begin{equation}
j\in \paren{-1,0},\quad \varepsilon =0,
\end{equation}
known as the \emph{complementary series}, and, among the finite-dimensional ones, only the \emph{trivial representation} \(F_0\). Of these, the only ones appearing in the \emph{Plancherel decomposition}\footnote{i.e., those with non-zero Plancherel measure.} are the principal series and the discrete series with \( j\geq 0 \). Note that the inner product on the Hilbert space of non-unitary representations, such as the \( F_j \), is \emph{not preserved} by the action of \( \spin(2,1) \).

\subsection{Product of finite and discrete modules}

Consider the coupling \(F_\gamma\otimes D^+_j\) of a finite-dimensional module and one from the discrete positive series, with \(\gamma\geq\frac{1}{2}\). The generators of \(\spin(2,1)\) act on this module as
\begin{equation}
J_0 \equiv J_0\otimes\1 + \1\otimes J_0,\quad {J}_\pm \equiv J_\pm\otimes\1 + \1\otimes J_\pm.
\end{equation}
\begin{remark}\label{rem:discrete+-}
The discrete negative module \(D^-_j\) is the \emph{dual module} to \(D^+_j\), i.e., they are related by the change
\begin{equation*}
J_0\rightarrow -J_0,\quad J_\pm \rightarrow -J_\mp,\quad \ket{j,m}\rightarrow (-1)^m\ket{j,-m}.
\end{equation*}
Conversely, \(F_\gamma\) is dual to itself, i.e., it remains unchanged under the same change.
For this reason, the results in this section will be proved for \(D^+_j\) only: the analogues for the negative module trivially follow by transforming operators and vectors for both the finite and the discrete series.
\end{remark}
Such a module is not generally irreducible. In order to find out if \(F_\gamma\otimes D^+_j\) can be decomposed in terms of irreducible modules of \(\spin(2,1)\)---a non-trivial task, since the module is not unitary---we will consider the algebraically equivalent problem of diagonalising (if possible) the Casimir \(Q\).
Solving the eigenvalue equation for generic \(\gamma\) is not easy; instead, the approach will be to explicitly find the eigenvectors and then show that, under certain conditions, they provide a basis for the product space.

To avoid confusion, the basis elements of the finite-dimensional series will be denoted by
\begin{equation}
\ket{\gamma,\mu},\quad \mu\in\set*{ -\gamma,\dotsc,\gamma}
\end{equation}
from now on. Since both \(F_\gamma\) and \(D^+_j\) are lowest weight modules, i.e., \(J_-\) annihilates one of their basis elements, their tensor product has to be as well. In fact, the vector
\begin{equation}
\ket{\psi_{(-\gamma)}}:=\ket{\gamma,-\gamma}\otimes\ket{j,j+1}
\end{equation}
satisfies
\begin{equation}
J_-\ket{\psi_{(-\gamma)}}=0;
\end{equation}
an element of \(F_\gamma\otimes D^+_j\) satisfying this property will be called a \emph{lowest weight vector}. \(\ket{\psi_{(-\gamma)}}\) is trivially a \(Q\)-eigenvector: from \eqref{eq:SU(1,1)_casimir} follows that
\begin{equation}
Q\ket{\psi_{(-\gamma)}}=-J_0(J_0 -1)\ket{\psi_{(-\gamma)}}=-(j-\gamma)(j-\gamma +1)\ket{\psi_{(-\gamma)}},
\end{equation}
since
\begin{equation}
J_0\ket{\gamma,\mu}\otimes\ket{j,m}=(m+\mu)\ket{\gamma,\mu}\otimes\ket{j,m}.
\end{equation}
This is not the only lowest weight vector; in fact, we have

\begin{proposition}\label{prop:lowest_weight}
For the coupling \(F_\gamma\otimes D^+_j\), the vectors
\begin{equation*}
\ket{\psi_{(\mu)}}=\sum_{\nu=-\gamma}^{\mu}(-1)^{\gamma+\nu}\prod_{\sigma=-\gamma}^{\nu-1}\frac{\Gamma_+(j,j+\mu-\sigma)}{\Gamma_+(\gamma,\sigma)}\ket{\gamma,\,\nu}\otimes\ket{j,\,j+1+\mu-\nu},
\end{equation*}
with \(\mu\in\set*{ -\gamma,\dotsc,\gamma}\) are lowest weight vectors and \(Q\)-eigenvectors, with respective eigenvalues
\begin{equation*}
q_{(\mu)}:=-(j+\mu)(j+\mu+1).
\end{equation*}
\end{proposition}
\begin{proof}
First notice that each \(\ket{\psi_{(\mu)}}\) is non-vanishing. Acting on it with \(J_-\), we get
\begin{multline}
\sum_{\nu=-\gamma}^{\mu}(-1)^{\gamma+\nu}\prod_{\sigma=-\gamma}^{\nu-1}\frac{\Gamma_+(j,j+\mu-\sigma)}{\Gamma_+(\gamma,\sigma)}\Gamma_+(\gamma,\nu-1)\ket{\gamma,\,\nu-1}\otimes\ket{j,\,j+1+\mu-\nu}+\\
\sum_{\nu=-\gamma}^{\mu}(-1)^{\gamma+\nu}\prod_{\sigma=-\gamma}^{\nu-1}\frac{\Gamma_+(j,j+\mu-\sigma)}{\Gamma_+(\gamma,\sigma)}\Gamma_+(j,\,j+\mu-\nu)\ket{\gamma,\,\nu}\otimes\ket{j,\,j+\mu-\nu},
\end{multline}
where the property
\begin{equation}\label{eq:C+-}
\Gamma_+(j,m-1)=\Gamma_-(j,m)
\end{equation}
was used. Relabelling the dummy index \(\nu\) in the first sum and noticing that the term \(\nu=\mu\) vanishes in the second one, we can rewrite this as
\begin{equation}
\sum_{\nu=-\gamma}^{\mu-1}\paren{-1+1}(-1)^{\gamma+\nu}\prod_{\sigma=-\gamma}^{\nu}\frac{\Gamma_+(j,j+\mu-\sigma)}{\Gamma_+(\gamma,\sigma)}
\Gamma_+(\gamma,\nu)\ket{\gamma,\,\nu}\otimes\ket{j,\,j+\mu-\nu}=0.
\end{equation}
Again, the action of the Casimir is trivially given by
\begin{equation}
Q\ket{\psi_{(\mu)}}=-J_0(J_0 -1)\ket{\psi_{(\mu)}}=-(j+\mu)(j+\mu +1)\ket{\psi_{(\mu)}}.
\end{equation}
\end{proof}

The fact that a finite number of eigenvectors exist does not mean \(Q\) is diagonalisable. Instead of working in an infinite-dimensional setting, however, we can take advantage of the tensor product basis vectors of \(F_\gamma\otimes D^+_j\) being \(J_{0}\)-eigenvectors: the space can be decomposed as\footnote{Here \(\oplus\) is the algebraic direct sum.}
\begin{equation}
F_\gamma\otimes D^+_j=\bigoplus_{M=j+1-\gamma}^\infty V_M,
\end{equation}
where the \(V_M\) are the orthogonal subspaces spanned by
\begin{equation}
\ket{(\mu)M}:=\ket{\gamma,\mu}\otimes\ket{j,M-\mu},\quad \mu\in\set*{-\gamma,\dotsc,\min(\gamma, M-j-1) }.
\end{equation}
Each \(V_M\) is finite-dimensional and, since \([Q,J_0]=0\), we can work with the restriction \(Q_M:=\eval{Q}{V_M}\), satisfying
\begin{equation}
Q_M(V_M)\subseteq V_M.
\end{equation}
The total Casimir \(Q\) will be diagonalisable if and only if each \(Q_M\) is.
In order to prove whether \(Q\) is diagonalisable or not and under which conditions, the following two lemmas will be needed.

\begin{lemma}\label{lem:repeated_J+}
If \(j>\gamma-1\), then the repeated action of \(J_+\) on a lowest weight vector never vanishes; that is, for every \(\mu\),
\begin{equation*}
(J_+)^n\ket{\psi_{(\mu)}}\neq0\quad\forall n\in\mathbb{N}.
\end{equation*}
\end{lemma}
\begin{proof}
Suppose the lemma is not true for an arbitrary \(\mu\), and let \(n\geq 1\) be the smallest integer such that
\begin{equation}
(J_+)^n\ket{\psi_{(\mu)}}=0;
\end{equation}
then we have \((J_+)^{n-1}\ket{\psi_{(\mu)}}\neq 0\) and, since \(Q\) and \(J_+\) commute,
\begin{equation}
Q(J_+)^{n-1}\ket{\psi_{(\mu)}}=(J_+)^{n-1}Q\ket{\psi_{(\mu)}}=q_{(\mu)}(J_+)^{n-1}\ket{\psi_{(\mu)}}.
\end{equation}
On the other hand
\begin{equation}
\begin{split}
Q(J_+)^{n-1}\ket{\psi_{(\mu)}}&=-J_0(J_0 + 1)(J_+)^{n-1}\ket{\psi_{(\mu)}}+\mathcal{J_-}(J_+)^{n}\ket{\psi_{(\mu)}}
\\
&=q_{(\mu+n)}(J_+)^{n-1}\ket{\psi_{(\mu)}},
\end{split}
\end{equation}
since
\begin{equation}
(J_+)^{n-1}\ket{\psi_{(\mu)}}\in V_{j+\mu+n}.
\end{equation}
This is only possible if \(q_{(\mu)}=q_{(\mu+n)}\), that is
\begin{equation}
(j+\mu)(j+\mu+1)=(j+\mu+n)(j+\mu+n+1),
\end{equation}
which is equivalent to
\begin{equation}
n(n+2j+2\mu+1)=0.
\end{equation}
However, since \(\mu\geq-\gamma\) and \(j>\gamma-1\), we have
\begin{equation}
\begin{cases}
n\geq 1\\
n+2j+2\mu+1> 1+2(\gamma - 1) -2\gamma +1= 0,
\end{cases}
\end{equation}
which leads to a contradiction.
\end{proof}

\begin{lemma}\label{lem:distinct_eigenvalues}
The values
\begin{equation*}
q_{(\mu)}=-(j+\mu)(j+\mu+1),\quad \mu\in\set*{ -\gamma,\dotsc,\gamma},\quad j\in\mathbb{C}
\end{equation*}
are all distinct if and only if
\begin{equation*}
j\not\in\mathbb{Z}/2\quad\mbox{or}\quad
\begin{cases}
j\in\mathbb{Z}/2\\
j\in (-\infty,-\gamma)\cup(\gamma-1,\infty).
\end{cases}
\end{equation*}
\end{lemma}
\begin{proof}
Consider arbitrary \(\mu\neq\nu\). One can easily check that
\begin{equation}
q_{(\mu)}=q_{(\nu)}\Leftrightarrow (\mu-\nu)(\mu+\nu +2j+1)=0.
\end{equation}
Since \(\mu\) and \(\nu\) are different, this is equivalent to solving
\begin{equation}
\mu+\nu = -2j-1.
\end{equation}
The LHS is an integer number, so if \(j\not\in\mathbb{Z}/2\) there is no solution, i.e., the \(q_{\mu}\)'s are all different.
Suppose now that \(j\in\mathbb{Z}/2\). The LHS is subject to the constraint (remember \(\mu\neq\nu\))
\begin{equation}
\abs{\mu+\nu}<2\gamma,
\end{equation}
so that a solution exists if and only if
\begin{equation}
\abs{2j+1}<2\gamma.
\end{equation}
Since \(j\) can only change by half-integer steps, it follows that coinciding \(q_{(\mu)}\)'s exist if and only if \(j\leq \gamma - 1\) and \(j\geq-\gamma\). Consequently, they are all different if and only if \(j>\gamma-1\) or \(j<-\gamma\).
\end{proof}

We can now prove, under certain conditions, the diagonalisability of \(Q\).
\begin{proposition}\label{prop:discrete_diagonalisable}
When \(j> \gamma -1\), the operator \(Q_M\) is diagonalisable, with distinct eigenvalues
\begin{equation*}
q_{(\mu)}=-(j+\mu)(j+\mu+1),\quad \mu\in\set*{ -\gamma,\dotsc,\min(\gamma,M-j-1)}.
\end{equation*}
It follows that \( Q \) is overall diagonalisable.
\end{proposition}
\begin{proof}
Define, up to a normalisation factor, the vectors
\begin{equation}
\ket{j+\mu,M}:=(J_+)^{M-j-1-\mu}\ket{\psi_{(\mu)}}\in V_M,\quad \mu\in\set*{ -\gamma,\dotsc,\min(\gamma,M-j-1)};
\end{equation}
owing to \cref{lem:repeated_J+}, they are all non-vanishing. 
Moreover, since \(Q\) commutes with \(J_+\), they are \(Q_M\)-eigenvectors, with eigenvalues \(q_{(\mu)}\). Finally, it follows from \cref{lem:distinct_eigenvalues} that the eigenvalues are all distinct: since the number of eigenvalues equals the dimension of \(V_M\), \(Q_M\) is diagonalisable.
\end{proof}
\begin{proposition}\label{prop:discrete_non-diagonalisable}
When \(j\leq \gamma -1\), the operator \(Q_{j+1+\gamma}\) is not diagonalisable. It follows that \( Q \) is overall not diagonalisable.
\end{proposition}
\begin{proof}
The proof is divided in two parts: first we show that the only possible eigenvalues of \(Q_{j+1+\gamma}\) are the \(q_{(\mu)}\)'s, then we use this fact to prove that \(Q_{j+1+\gamma}\) is not diagonalisable.
\begin{proofenumerate}
\item
Suppose there is a non-zero eigenvector \(\ket{\varphi}\in V_{j+1+\gamma}\), with eigenvalue
\begin{equation}
\varphi\neq q_{(\mu)},\quad \mu \in \set*{-\gamma,\dotsc,\gamma}.
\end{equation}
It must be
\begin{equation}
(J_-)^n\ket{\varphi}=0
\end{equation}
for some
\begin{equation}
n\in\set*{1,2,\dotsc,2\gamma+1},
\end{equation}
since there is only one vector in \(V_{j+1-\gamma}\) and it is annihilated by \(J_-\). Let \(N\) be the smallest such number; then \((J_-)^{N-1}\ket{\varphi}\neq 0\) and
\begin{equation}
Q(J_-)^{N-1}\ket{\varphi}=(J_-)^{N-1}Q\ket{\varphi}=\varphi(J_-)^{N-1}\ket{\varphi},
\end{equation}
while at the same time
\begin{equation}
Q(J_-)^{N-1}\ket{\varphi}=-J_0(J_0 -1)(J_-)^{N-1}\ket{\varphi}+ J_+(J_-)^N\ket{\varphi}=q_{(\gamma-N+1)}(J_-)^{N-1}\ket{\varphi}.
\end{equation}
It follows that \(\varphi\) equals one of the \(q_{(\mu)}\)'s, which is a contradiction.
\item
Notice that, since \(j\geq -\tfrac{1}{2}\), it is always \(j\geq -\gamma\). Then, since \(j\leq \gamma - 1\), it follows from \cref{lem:distinct_eigenvalues} that there are at most \(2\gamma\) distinct eigenvalues. However, by acting with \(Q_{j+1+\gamma}\) on the basis vectors
\begin{equation}
\ket{(\mu)j+1+\gamma}=\ket{\gamma,\mu}\otimes\ket{j,j+1+\gamma-\mu}\in V_{j+1+\gamma},
\end{equation}
we find that the matrix elements\footnote{Here \( \praket{e_i|A|e_j} \) denotes a matrix element of the operator \( A \) in the (possibly non-orthonormal) basis \( \set{e_i}_{i\in I} \). If the basis is orthonormal then \( \praket{e_i|A|e_j}\equiv \braket{e_i|A|e_j}. \)}
\begin{equation}
Q_{\mu\nu}:=\praket{(\mu)j+1+\gamma|Q|(\nu)j+1+\gamma}
\end{equation}
vanish unless
\begin{equation}
\mu=\nu\quad\mbox{or}\quad \mu=\nu\pm 1;
\end{equation}
in other words, \(Q_{\mu\nu}\) are the entries of a \emph{tridiagonal matrix} (see \cref{app:tridiagonal}).
In particular, since the superdiagonal entries
\begin{equation}
Q_{\mu,\mu+1}=\Gamma_+(\gamma,\mu)\Gamma_-(j,j+1+\gamma-\mu),\quad \mu\leq \gamma-1
\end{equation}
are all non-zero, it follows from \cref{prop:tridiagonal} that the eigenspaces of \( Q_{j+1+\gamma} \) are all \(1\)-dimensional. As a consequence, there are at most \(2\gamma\) eigenvectors, which means \(Q_{j+1+\gamma}\) is not diagonalisable, as \(\dim V_{j+1+\gamma}=2\gamma+1\).
Since \(Q_M\) is non-diagonalisable for at least one \(M\), \(Q\) will not be diagonalisable as well.
\end{proofenumerate}
\end{proof}
\subsubsection{Summary}
The coupling \(F_\gamma\otimes D^+_j\) can be decomposed in irreducible modules if and only if \(j>\gamma-1\).
An eigenbasis for \(Q\) can be constructed by defining recursively
\begin{equation}
\ket{J,M+1}=\frac{1}{\Gamma_+(J,M)}J_+\ket{J,M},\quad J\in\set*{ j-\gamma,\dotsc,j+\gamma},
\end{equation}
starting from\footnote{Up to a normalisation factor.}
\begin{equation}
\ket{J,J+1}\propto \ket{\psi_{J-j}},
\end{equation}
which satisfy
\begin{equation}
Q\ket{J,M}=-J(J+1)\ket{J,M},\quad J_0\ket{J,M}=M\ket{J,M}.
\end{equation}
\Cref{lem:repeated_J+} guarantees that these vectors are all non-zero, so that each \(Q\)-eigenspace behaves as the discrete positive module\footnote{Note that \( J_-\ket{J,J+1}=0 \) as \( \ket{J,J+1} \) is a lowest weight vector.} \(D^+_J\).
In terms of the the old basis elements, the change of basis must be of the form
\begin{equation}
\ket{j+\mu,M}=\sum_{\nu=-\gamma}^{\Omega_M}A^M_{\nu\mu}(j,\gamma)\ket{(\nu)M},\quad \mu\in\set*{-\gamma,\dotsc,\Omega_M},
\end{equation}
where
\begin{equation}
\Omega_M:=\min(\gamma, M-j-1),
\end{equation}
with the \(A^M_{\nu\mu}\)'s forming an invertible matrix; they will be called \emph{Clebsch--Gordan coefficients}, in analogy with \(\mathfrak{su}(2)\) representation theory.
More generally, we can write
\begin{equation}
\ket{J,M}=\sum_{\mu=-\gamma}^{\gamma}\sum_{m=j+1}^{\infty}A\paren{\gamma,\mu;j,m|J,M}\ket{\gamma,\mu}\otimes\ket{j,m},
\end{equation}
where\footnote{Note that \( \ket{\gamma,\mu}\otimes\ket{j,M-\mu} \) vanishes for \( \mu\geq \Omega_M \). }
\begin{equation}
A\paren{\gamma,\mu;j,m|J,M}:=A^M_{\mu,J-j}(j,\gamma)\delta_{m+\mu,M}
\end{equation}
will also be called \CG\ coefficients.

Analogous results are easily found for the coupling with \( D^-_j \). We can write the result in a compact form as
\begin{equation}
F_\gamma\otimes D^\pm_j = \bigboxplus_{J=j-\gamma}^{j+\gamma} D^\pm_J,\quad j>\gamma-1,
\end{equation}
where we use the symbol \( \boxplus \) to emphasise that this (algebraic) direct sum of modules is not an orthogonal direct sum\footnote{In fact, one can check that \( Q\) is not Hermitian, as \( F_\gamma \), \( \gamma\geq\half \) is non-unitary, so there is no reason to expect the Q-eigenbasis to be orthogonal.}.
\subsection{Product of finite and continuous modules}

Consider now the coupling \(F_\gamma\otimes C^\varepsilon_j\) of a finite-dimensional module and a generic one from the continuous series, not necessarily unitary. The technique used for the discrete series will not work here since the spectrum of \(J_0\) is unbounded, hence a different approach is needed.

We will work again individually on each \(J_0\)-eigenspace \(V_M\), \( M\in\varepsilon+\gamma+\Z \), with basis vectors
\begin{equation}
\ket{(\mu)M}=\ket{\gamma, \mu}\otimes \ket{j,M-\mu},\quad \mu\in\set*{-\gamma,\dotsc,\gamma },
\end{equation}
and try to diagonalise \(Q_M\). Explicitly, we are interested in finding a change of basis
\begin{equation}
\ket{J_{(\mu)},M}=\sum_{\nu=-\gamma}^\gamma A^M_{\nu\mu}(j,\gamma)\ket{(\nu)M},\quad \mu\in\set*{-\gamma,\dots,\gamma},
\end{equation}
with
\begin{equation}
Q\ket{J_{(\mu)},M}=-J_{(\mu)}\paren*{J_{(\mu)} +1}\ket{J_{(\mu)},M}.
\end{equation}
\begin{remark}
Since any non-trivial \(F_\gamma\) is not unitary, the total Casimir is not Hermitian; moreover, one can easily check that it is not a \emph{normal} operator either, i.e.
{\normalfont
\begin{equation*}
[Q\Dagger*_M,Q_M]\neq 0.
\end{equation*}}
As a consequence, not only the spectral theorem cannot be used to diagonalise it, but its eigenvectors will be non-orthogonal and the matrix \(A^M(j,\gamma)\) non-unitary.
\end{remark}
Solving the eigenvalue equation explicitly for arbitrary \(\gamma\) is too difficult; however, one can easily do it for the \(2\)-dimensional case \(\gamma=\frac{1}{2}\): each \(Q_M\) is diagonalisable if and only if \(j\neq-\frac{1}{2}\), with eigenvalues \(q_{(\pm\frac{1}{2})}\) (the corresponding Clebsch--Gordan coefficients are listed in \cref{tab:3d-1/2}). Using this information, we can prove by induction that, when \(j\not\in\mathbb{Z}/2\), \(Q\) is diagonalisable for all \(\gamma\geq\frac{1}{2}\). The case \(j\in\mathbb{Z}/2\) will be treated later with a different method.
\begin{proposition}\label{prop:continuous_non_Z/2}
When \(j\not\in\mathbb{Z}/2\), the eigenvalues of \(Q_M\) for the coupling \( F_\gamma\otimes C^\varepsilon_j \) are
\begin{equation*}
q_{(\mu)}=-(j+\mu)(j+\mu+1),\quad \mu\in\set*{ -\gamma,\dotsc,\gamma},
\end{equation*}
that is
\begin{equation*}
J_{(\mu)}=j+\mu.
\end{equation*}
These are all distinct and do not depend on \( M \), so \(Q\) is diagonalisable.
\end{proposition}
\begin{proof}
\begin{proofenumerate}
\item The proof proceeds by induction on half-integer \(\gamma\geq\frac{1}{2}\). As the statement is true for \(\gamma=\frac{1}{2}\), suppose that it is true for \(\gamma-\frac{1}{2}\) and consider the coupling \(F_\gamma\otimes C^\varepsilon_j\).
The finite-dimensional modules are isomorphic to the irreducible unitary modules of~\(\mathfrak{su}(2)\), seen as representations of the complexification \(\spin(2,1)_\mathbb{C}\cong \mathfrak{su}(2)_\mathbb{C}\). Consequently, the well-known result of \(\mathfrak{su}(2)\) recoupling theory\footcite{barut}
\begin{equation}
F_\gamma \subset  F_{\frac{1}{2}}\otimes F_{\gamma-\frac{1}{2}}\cong F_{\gamma-1}\oplus F_\gamma
\end{equation}
can be used; explicitly,
\begin{equation}
\ket{\gamma,\mu}\equiv \sum_{\sigma=-\frac{1}{2}}^\frac{1}{2}\sum_{\lambda=-\gamma+\frac{1}{2}}^{\gamma-\frac{1}{2}} \braket{  \tfrac{1}{2},\sigma ;\gamma-\tfrac{1}{2},\lambda | \gamma,\mu}  \ket{\tfrac{1}{2},\sigma}\otimes \ket{\gamma-\tfrac{1}{2},\lambda},
\end{equation}
where
\begin{equation}
\braket{  \tfrac{1}{2},\sigma ;\gamma-\tfrac{1}{2},\lambda | \gamma,\mu}
\end{equation}
are the \(\mathfrak{su}(2)\) Clebsch--Gordan coefficients. We can then write, since \(F_{\gamma-\frac{1}{2}}\otimes C^\varepsilon_j\) is decomposable by induction hypothesis,
\begin{equation}
\begin{split}
\ket{\gamma,\mu}\otimes\ket{j,M-\mu}=&\sum_{\sigma,\lambda} \braket{  \tfrac{1}{2},\sigma ;\gamma-\tfrac{1}{2},\lambda | \gamma,\mu} \ket{\tfrac{1}{2},\sigma} \otimes \paren[\bigg]{\ket{\gamma-\tfrac{1}{2},\lambda}\otimes\ket{j,M-\mu}}\\
=&\sum_{\sigma,\lambda} \braket{  \tfrac{1}{2},\sigma ;\gamma-\tfrac{1}{2},\lambda | \gamma,\mu} \ket{\tfrac{1}{2},\sigma} \otimes\sum_{\kappa=-\gamma+\frac{1}{2}}^{\gamma-\frac{1}{2}}B^{M-\sigma}_{\kappa\lambda}\paren[\big]{j,\gamma-\tfrac{1}{2}} \ket{j+\kappa,M-\sigma}
\end{split}
\end{equation}
where the \(B^M_{\kappa\lambda}\) are the \emph{inverse} Clebsch--Gordan coefficients, i.e., \(B^M\) is the inverse of the matrix \(A^M\).
In particular, when \(\mu=-\gamma\), the only non-zero \(\mathfrak{su}(2)\) coefficient is\footcite[Using the conventions from][]{CGtables}
\begin{equation}
\braket{ \tfrac{1}{2},-\tfrac{1}{2} ; \gamma-\tfrac{1}{2},-\gamma+\tfrac{1}{2} | \gamma,-\gamma}=1
\end{equation}
so that
\begin{multline}\label{eq:-gamma}
\ket{\gamma,-\gamma}\otimes\ket{j,M+\gamma}
=\sum_{\rho=-\frac{1}{2}}^\frac{1}{2}\sum_{\kappa=-\gamma+\frac{1}{2}}^{\gamma-\frac{1}{2}}B^{M+\frac{1}{2}}_{\kappa,-\gamma+\frac{1}{2}}(j,\gamma-\tfrac{1}{2})
\\ \times
B^{M}_{\rho,-\tfrac{1}{2}}\paren{j+\kappa,\tfrac{1}{2}}\ket{\paren{j+\kappa}j+\rho+\kappa,M},
\end{multline}
where the \((j+\kappa)\) label in the vector indicates it comes from the coupling
\begin{equation}
\ket{\tfrac{1}{2},-\tfrac{1}{2} }\otimes\ket{j+\kappa,M+\tfrac{1}{2}}.
\end{equation}
There are exactly \(4\gamma\) vectors on the RHS of~\eqref{eq:-gamma}: they are
\begin{equation}
\begin{cases}
\ket{\paren{j-\gamma+\tfrac{1}{2}}j-\gamma,M}
\\
\ket{\paren{j+\mu\pm\tfrac{1}{2}}j+\mu,M} \casestext{for} \mu\in\set*{-\gamma+1,\dotsc,\gamma-1}
\\
\ket{\paren{j+\gamma-\tfrac{1}{2}}j+\gamma,M};
\end{cases}
\end{equation}
their \(Q\)-eigenvalues are
\begin{equation}
q_{(\mu)}=-(j+\mu)(j+\mu+1),\quad \mu\in\set*{ -\gamma,\dotsc,\gamma},
\end{equation}
which are all distinct\footnote{It follows from \cref{lem:distinct_eigenvalues}, as \( j\not\in \Z/2 \).}, and they form a basis for the \(M\) eigenspace in \( V_\frac{1}{2}\otimes V_{\gamma-\frac{1}{2}} \otimes C_{j}^\varepsilon\), i.e., they are independent.

As shown in \cref{app:3D_CG}, Clebsch--Gordan coefficients satisfy the property
\begin{equation}
B_{\nu+1,-\gamma}^M(j,\gamma)=\alpha_\nu(j,\gamma)\frac{\sqrt{j+\nu-M +1}}{\sqrt{j+\nu+M+1}} B_{\nu,-\gamma}^M(j,\gamma),
\end{equation}
where \(\alpha_\nu\) is fixed by the normalisation convention and does not depend on \(M\). Using this formula and the fact that (see  \cref{tab:3d-1/2})
\begin{equation}
B^M_{\rho,-\frac{1}{2}}\paren{j+\kappa,\tfrac{1}{2}}=
\begin{cases}
-\frac{\sqrt{j+\kappa+M+\frac{1}{2}}}{\sqrt{2j+2\kappa+1}}&\casesif \rho=-\tfrac{1}{2}\\
\frac{\sqrt{j+\kappa-M+\frac{1}{2}}}{\sqrt{2j+2\kappa+1}}&\casesif \rho=\tfrac{1}{2},
\end{cases}
\end{equation}
we can write
\begin{equation}
\ket{(-\gamma)M}=\ket{\gamma,-\gamma}\otimes\ket{j,M+\gamma}=\sum_{\nu=-\gamma}^\gamma B^M_{\nu,-\gamma}(j,\gamma)\ket{j+\nu,M}
\end{equation}
for some coefficients \(B^M_{\nu,-\gamma}\), where the vectors on the RHS are defined up to a normalisation factor as
\begin{equation}
\label{eq:JM_def}
\ket{J,M}\propto
\begin{cases}
\ket{\paren{J+\tfrac{1}{2}}J,M} &\casesif J=j-\gamma
\\
\ket{\paren{J-\tfrac{1}{2}}J,M} &\casesif J=j+\gamma
\\
\frac{1}{\sqrt{2J}}\ket{\paren{J-\tfrac{1}{2}}J,M}-\frac{\beta(J)}{\sqrt{2J+2}}\ket{\paren{J+\tfrac{1}{2}}J,M}&\casestextn{otherwise},
\end{cases}
\end{equation}
with
\begin{equation}
\beta\paren{j+\kappa+\tfrac{1}{2}}=\alpha_\kappa\paren{j,\gamma-\tfrac{1}{2}}.
\end{equation}
Since these vectors live in different \(Q\)-eigenspaces, they are necessarily independent.

\item
Suppose now that the vectors \(\ket{J,M}\) defined in \eqref{eq:JM_def} belong to \(V_M\):
then they would be \(2\gamma+1\) independent eigenvectors in \(V_M\), i.e., an eigenbasis, which proves the proposition. It only remains to show that this is indeed true; it can be done by induction as well.
We can easily check that, for \(\mu<\gamma\),
\begin{equation}
Q\ket{(\mu)M}\in\mathrm{span}\set*{\ket{(\mu-1)M},\ket{(\mu)M},\ket{(\mu+1)M}},
\end{equation}
with
\begin{equation}
\praket{(\mu+1)M|Q|(\mu)M}=\Gamma_+(\gamma,\mu)\Gamma_-(j,M-\mu)\neq 0;
\end{equation}
consequently, it must be
\begin{equation}\label{eq:mu+1_span}
\ket{(\mu+1)M}\in\mathrm{span}\set{\ket{(\mu-1)M},\ket{(\mu)M},Q\ket{(\mu)M}}.
\end{equation}
Now suppose that
\begin{equation}
\ket{(\mu-1)M},\ket{(\mu)M}\in\mathrm{span}\set*{\ket{J,M}\setst J=j-\gamma,\dotsc,j+\gamma};
\end{equation}
then\footnote{Recall that \(Q\ket{J,M}\propto\ket{J,M}  \).}
\begin{equation}
Q\ket{(\mu)M}\in\mathrm{span}\set*{\ket{J,M}\setst J=j-\gamma,\dotsc,j+\gamma},
\end{equation}
so that, as a consequence of \eqref{eq:mu+1_span}, it must be
\begin{equation}
\ket{(\mu+1)M}\in\mathrm{span}\set*{\ket{J,M} \setst J=j-\gamma,\dotsc,j+\gamma}
\end{equation}
as well. 
Since when \( \mu=-\gamma \) the hypothesis is valid\footnote{Note that \(\ket{(-\gamma-1)M}\equiv0\).}, it follows by induction that every basis vector \(\ket{(\mu)M}\) can be written as a linear combination of the independent \(\ket{J,M}\) vectors. As their number match, the latter must form a basis for \(V_M\), so that they are, in fact, eigenvectors for \(Q_M\).
\end{proofenumerate}
\end{proof}
When \(j\in\mathbb{Z}/2\), \(Q_M\) is not always diagonalisable. In order to prove when it can be done, the following \namecref{lem:continuous_Z/2} is needed.
\begin{lemma}\label{lem:continuous_Z/2}
When \(j\in\mathbb{Z}/2\), the eigenvalues of \(Q_M\)  for the coupling \( F_\gamma\otimes C^\varepsilon_j \) are given by
\begin{equation*}
q_{(\mu)}=-(j+\mu)(j+\mu+1),\quad \mu \in \set*{ -\gamma,\dotsc,\gamma}.
\end{equation*}
\end{lemma}
\begin{proof}
The result follows by continuity from \cref{prop:continuous_non_Z/2}. First notice that the function\footnote{Here \( Q_M(j) \) is the matrix representing \( Q_M \) in the coupling \( F_\gamma \otimes C^\varepsilon_j \), where \( \gamma \) is fixed.}
\begin{equation}
d^M_\lambda:j\in \R \mapsto \det(Q_M(j)-\lambda\1)\in \C,
\end{equation}
is continuous since it is a product of continuous functions\footnote{One can easily check that the entries of \( Q_M(j) \) are continuous in \( j \).} of \(j\). Moreover, for \(j\not\in\mathbb{Z}/2\), it is given by
\begin{equation}
d_\lambda^M(j)=\prod_{\mu=-\gamma}^\gamma[-(j+\mu)(j+\mu+1)-\lambda],
\end{equation}
as a consequence of \cref{prop:continuous_non_Z/2}. Now let \(k\in\mathbb{Z}/2\); since \(d\) is continuous, it must be
\begin{equation}
d_\lambda^M(k)=\lim_{j\rightarrow k} d_\lambda^M(j)=\prod_{\mu=-\gamma}^\gamma[-(k+\mu)(k+\mu+1)-\lambda]
\end{equation}
so that the eigenvalues of \(Q_M\) are the \(q_{(\mu)}\)'s.
\end{proof}
It is now possible to prove that
\begin{proposition}\label{prop:continuous_Z/2}
When \(j\in\mathbb{Z}/2\), \(Q\) is diagonalisable  for the coupling \( F_\gamma\otimes C^\varepsilon_j \) if and only if \(j>\gamma-1\) or \(j<-\gamma\).
\end{proposition}
\begin{proof}
We know from \cref{lem:distinct_eigenvalues} that the eigenvalues of each \(Q_M\) (given by \cref{lem:continuous_Z/2}) are all distinct if and only if \(j>\gamma-1\) or \(j<-\gamma\). However, like in the discrete case (see proof of \cref{prop:discrete_non-diagonalisable}), \(Q_M\) is represented in the \(\ket{(\mu)M}\) basis by a tridiagonal matrix with non-zero superdiagonal entries. It follows from \cref{prop:tridiagonal} that the \(Q_M\) are diagonalisable if and only if the eigenvalues are all different, i.e., \(j>\gamma-1\) or \(j<-\gamma\), as required.
\end{proof}

\subsubsection{Summary}

The coupling \(F_\gamma\otimes C^\varepsilon_j\) can be decomposed in irreducible modules if and only if \(j\not\in\mathbb{Z}/2\) or, when \(j\) is (half-)integer, if \(j>\gamma-1\) or \(j<-\gamma\). One can check directly that each \(Q\)-eigenspace behaves as a continuous module: in fact, either
\begin{equation}
j\not\in \Z/2 \quad \Rightarrow \quad J\not\in \Z/2
\end{equation}
or
\begin{equation}
\begin{cases}
j\in \Z/2 \\
j-\varepsilon\not\in\Z
\end{cases}
 \Rightarrow \quad
\begin{cases}
J\in \Z/2 \\
J-E\not\in\Z,
\end{cases}
\end{equation}
where
\begin{equation}
E:=
\begin{cases}
0 &\casesif \varepsilon+\gamma\in \Z \\
\half & \casesif \varepsilon+\gamma\in \half +\Z;
\end{cases}
\end{equation}
in both cases
\begin{equation}
\Gamma_\pm(J,M)\neq 0,\quad \forall J\in\set{j-\gamma,\dotsc,j+\gamma},\quad\forall M\in E+\Z.
\end{equation}
The \CG\ coefficients can be found by solving the eigenvalue problem for the matrix representation of \( Q_M \), with \( M \) arbitrary; the coefficients for the specific cases of \( \gamma=\half,1 \) with arbitrary \( C^\varepsilon_j \) can be found respectively in \cref{tab:3d-1/2,tab:3d-1}.

As with the coupling with discrete representations, we can write the  result in the compact form
\begin{equation}
F_\gamma\otimes C^\varepsilon_j = \bigboxplus_{J=j-\gamma}^{j+\gamma} C^E_J,
\end{equation}
with the restriction that, if \( j\in\Z/2 \), \( j>\gamma-1 \) or \( j<-\gamma \).
\subsection{\JS\ representation}\label{sec:3D_JS}

An application of the Wigner--Eckart theorem to the non-compact group  \(\Spin(2,1)\) will be presented here. It is well known in the quantum theory of angular momentum, where the Lie group \(\SU(2)\) is used, that the generators of the algebra (physically corresponding to infinitesimal rotations) can be expressed in terms of a pair of uncoupled \emph{quantum harmonic oscillators}; this result is known as \emph{Jordan--Schwinger representation}\footcite{schwinger1952}. Explicitly, the \( \su(2)_\C \) generators\footnote{Unitary representations are those with \( K\Dagger*_z=K_z \), \( K\Dagger*_+=K_- \).} \(K_z\), \(K_+\) and \(K_-\) with commutation relations
\begin{equation}
[K_z,K_\pm]=\pm K_\pm,\quad [K_+,K_-]=2K_z
\end{equation}
can be expressed as
\begin{equation}
K_z=\tfrac{1}{2}\paren{a\Dagger a - b\Dagger b },\quad K_+=a\Dagger b,\quad K_-=b\Dagger a,
\end{equation}
where \(a\) and \(b\) are quantum harmonic oscillators, i.e., satisfy
\begin{equation}
[a,a\Dagger]=[b,b\Dagger]=\1,
\end{equation}
and all the other commutators vanish. More generally \(a\), \(a\Dagger\), \(b\), \(b\Dagger\) and \(\1\) form a complex unitary representation of the \(5\)-dimensional Heisenberg algebra \(\mathfrak{h}_2(\mathbb{R})\).

One may ask if a similar result holds for the \( (\mathfrak{g},K) \)-modules of \( \Spin(2,1) \): the answer is positive for the discrete and finite-dimensional series, but an analogous construction for the continuous series is not easily guessed and, in fact, was not available until recently\footcite{wigner_eckart}. It will be shown here how the Wigner--Eckart theorem can be used to find an analogue of the Jordan--Schwinger representation for \( \Spin(2,1) \), which covers all representation classes.

First notice that the \WE\ theorem for a \( \Spin(2,1) \) tensor operator\footnote{Here  \( V_j \) is an irreducible \((\mathfrak{g},K)\)-module of \( \Spin(2,1) \), on which the Casimir acts as \( Q\equiv -j(j+1)\1 \). Moreover, we assume the module \( F_\gamma\otimes V_j \) is decomposable.}
\begin{equation}
\tau:F_\gamma\otimes V_{j}\rightarrow V_{j\Prime},
\end{equation}
with components
\begin{equation}
\tau_\mu:\ket{j,m}\in V_j \rightarrow \tau\paren[\big]{\ket{\gamma,\mu}\otimes \ket{j,m}}\in V_{j\Prime},
\end{equation}
takes the form
\begin{equation}
\praket{j\Prime,m\Prime|\tau_\mu|j,m}=\praket{j\Prime\Vert\tau\Vert j} B\paren{j\Prime,m\Prime|\gamma,\mu;j,m},
\end{equation}
where \( \praket{j\Prime\Vert\tau\Vert j} \) (usually called the \emph{reduced matrix element}) does not depend on \( m \), \( m\Prime \) or \( \mu \) and \( B\paren{j\Prime,m\Prime|\gamma,\mu;j,m} \) are the inverse \CG\ coefficients\footnote{If \( V_j \) is finite dimensional these are the \( \SU(2) \) \CG\ coefficients. It is implicitly assumed that the coefficients vanish if \( V_{j\Prime} \) does not belong to the decomposition.}
\begin{equation}
B\paren{j\Prime,m\Prime|\gamma,\mu;j,m}:=B^{m\Prime}_{j\Prime-j,\mu}(j,\gamma)\delta_{m+\mu,m\Prime}.
\end{equation}
Now note that a tensor operator
\begin{equation}
V:F_1\otimes V_j \rightarrow V_j
\end{equation}
can be constructed out of the algebra generators, with components
\begin{equation}\label{eq:vector_operator}
V_{\pm 1}= \mp \ii J_\pm,\quad V_0=-\sqrt{2}J_0;
\end{equation}
in fact
\begin{equation}
[J_0,V_\mu]=\mu V_\mu,\quad [J_\pm,V_\mu]=\Gamma_\pm(1,\mu).
\end{equation}
An alternative way to look at the Jordan--Schwinger construction is to look for two tensor operators
\begin{equation}
T:F_{\shalf}\otimes V_j \rightarrow V_{j-\shalf},\quad \widetilde T:F_{\shalf}\otimes V_j \rightarrow V_{j+\shalf}
\end{equation}
that can be combined to obtain \(V\). Explicitly, we make the ansatz
\begin{equation}\label{eq:ansatz1}
V_\mu=\sum_{\mu_1=-\frac{1}{2}}^\frac{1}{2} \sum_{\mu_2=-\frac{1}{2}}^\frac{1}{2} \braket{\tfrac{1}{2},\mu_1;\tfrac{1}{2},\mu_2|1,\mu}
\, T_{\mu_1}\widetilde{T}_{\mu_2}.
\end{equation}
It can be shown\footcite[chap.~9]{barut} that the RHS \eqref{eq:ansatz1} is indeed a tensor operator \( F_1\otimes V_j \rightarrow V_j\); the \CG\ coefficients appearing in it are\footcite{CGtables}
\begin{equation}
\braket{\tfrac{1}{2},\mu_1;\tfrac{1}{2},\mu_2|1,\mu}=\frac{\sqrt{(1-\mu)!}\sqrt{(1+\mu)!}}{\sqrt{2}}
\end{equation}
so that, in terms of the generators,
\begin{equation}
J_\pm=\pm \ii T_\pm \widetilde{T}_\pm,\quad J_0=-\tfrac{1}{2} \paren{ T_- \widetilde{T}_+ + T_+ \widetilde{T}_- },
\end{equation}
with the shorthand notation
\begin{equation}
T_\pm:=T_{\pm\frac{1}{2}}.
\end{equation}
We know from the \WE\ theorem that the matrix elements of \( T \) and \( \widetilde T \) are
\begin{subequations}
\begin{align}
\praket{j-\half,m\Prime|T_\mu|j,m}&=f(j) B\paren*{j-\half,m\Prime|\half,\mu;j,m}
\\
\praket{j+\half,m\Prime|\widetilde T_\mu|j,m}&=\widetilde f(j) B\paren*{j+\half,m\Prime|\half,\mu;j,m},
\end{align}
\end{subequations}
where \( f \) and \( \widetilde f \) are arbitrary functions.
The matrix elements of the generators are known: using the ansatz, we get
\begin{equation}
\Gamma_+(j,m)=\praket{j,m+1|J_+|j,m}=\ii\praket{j,m+1|T_+|j+\half,m+\half}\praket{j+\half,m+\half|\widetilde{T}_+|j,m}.
\end{equation}
The RHS can be evaluated, assuming the decomposition
exists of \( F_{\shalf}\otimes V_j \) exists, with the \CG\ coefficients from \cref{tab:3d-1/2}, which give
\begin{equation}
\ii\frac{f\paren{j+\half}\widetilde{f}(j)}{\sqrt{2j+2}\sqrt{2j+1}}\sqrt{j-m}\sqrt{j+m+1}=\frac{f\paren{j+\half}\widetilde{f}(j)}{\sqrt{2j+2}\sqrt{2j+1}}\, \Gamma_+(j,m),
\end{equation}
so that it must be
\begin{equation}\label{eq:ansatz_condition}
\frac{f(j-\tfrac{1}{2})\widetilde{f}(j)}{\sqrt{2j}\sqrt{2j+1}} = 1.
\end{equation}
Similarly
\begin{equation}
\praket{j,m-1|J_-|j,m}=\frac{f\paren{j+\half}\widetilde{f}(j)}{\sqrt{2j+2}\sqrt{2j+1}}\,\Gamma_-(j,m)
\end{equation}
and
\begin{equation}
\praket{j,m|J_0|j,m}=\frac{f\paren{j+\half}\widetilde{f}(j)}{\sqrt{2j+2}\sqrt{2j+1}}\,m,
\end{equation}
which means the ansatz is true whenever \eqref{eq:ansatz_condition} holds. The choice
\begin{equation}
f(j)=\widetilde{f}(j)=\sqrt{2j+1}
\end{equation}
will be used here.
The action of \(T\) and \(\widetilde T\) is thus
\begin{subequations}
\begin{align}
T_-\ket{j,m}&=-\sqrt{j+m}\,\ket{j-\half,m-\half}\\
T_+\ket{j,m}&=\sqrt{j-m}\,\ket{j-\half,m+\half}\\
\widetilde{T}_-\ket{j,m}&=\sqrt{j-m+1}\,\ket{j+\half,m-\half}\\
\widetilde{T}_+\ket{j,m}&=\sqrt{j+m+1}\,\ket{j+\half,m+\half},
\end{align}
\end{subequations}
from which it follows that
\begin{equation}\label{eq:heisenberg_cr}
[T_+,\widetilde{T}_-]=[\widetilde{T}_+,T_-]=\1,
\end{equation}
with all other commutators vanishing.

These commutation relations closely resemble those of the harmonic oscillator and, in fact, generalise them. For example, when the representation considered is \(F_j\) we find by inspection
\begin{equation}
\widetilde T_\pm=\mp T\Dagger*_\mp.
\end{equation}
Renaming
\begin{equation}
T_-=-a,\quad T_+=b
\end{equation}
we get
\begin{equation}
J_+=\ii a\Dagger b,\quad J_-=\ii b\Dagger a,\quad J_0=\tfrac{1}{2}\paren{a\Dagger a - b\Dagger b },
\end{equation}
with \(a\) and \(b\) satisfying the harmonic oscillator commutation relation.
Analogously, for the discrete series \(D^\pm_j\) with \(j\geq 0\) we have
\begin{equation}
\widetilde T_\pm=
\begin{cases}
-{T}\Dagger*_\mp &\quad \mbox{for }D^+_j\\
{T}\Dagger*_\mp &\quad \mbox{for }D^-_j;
\end{cases}
\end{equation}
with the choice
\begin{equation}
\begin{cases}
{T}_-=a,\quad {T}_+=\ii b\Dagger &\quad \mbox{for }D^+_j\\
{T}_-=a\Dagger,\quad {T}_+=\ii b &\quad \mbox{for }D^-_j
\end{cases}
\end{equation}
we get
\begin{equation}
\begin{cases}
J_+=a\Dagger b\Dagger,\quad J_-=a b,\quad J_0=
\tfrac{1}{2}\paren{a\Dagger a + b\Dagger b +1} &\casestext{for} D^+_j
\\
J_+=-ab,\quad J_-=-a\Dagger b\Dagger,\quad J_0=-\half\paren{a\Dagger a + b\Dagger b +1} &\casestext{for} D^-_j.
\end{cases}
\end{equation}
Note that, despite the fact that the \CG\ decomposition of \( F_{\shalf}\otimes D^\pm_{-\shalf} \) does not exist\footnote{Recall that it must be \( j>\gamma-1 \).}, the \JS\ representation in terms of harmonic oscillators also works for \( D^\pm_{-\shalf} \); in fact the action of \( \widetilde T \) is well defined even when \( j=-\half \), while \( T \) only acts on \( D^\pm_{\shalf}\), on which it is defined.

The continuous series generators cannot be rewritten in terms of harmonic oscillators because, while
\begin{equation}
\praket{j+\tfrac{1}{2},m\pm \tfrac{1}{2}|\widetilde T_\pm|j,m}=\mp\praket{j,m|{T}_\mp|j+\tfrac{1}{2},m\pm \tfrac{1}{2}},
\end{equation}
these matrix elements are never always real or imaginary, as that depends on the value of \(m\). This is to be expected, as if the generators could be written in terms of harmonic oscillators, the Casimir element \(Q\) would be expressible in terms of the \emph{number operators}
\begin{equation}
N_a=a\Dagger a,\quad N_b=b\Dagger b,
\end{equation}
which have discrete spectrum\footcite{messiah1}: this contradicts the fact that the eigenvalues of \(Q\) are continuous. 
Nevertheless, thanks to the Wigner--Eckart theorem, we constructed an an analogue of the Jordan--Schwinger representation that works even in this case. This, together with the \CG\ decomposition of \( F_\gamma \otimes C^\varepsilon_j \), is the most important result of this chapter, as it will allow us to generalise the spinorial formalism of loop quantum gravity to \( 3 \)D Lorentzian space-time in such a way that continuous representations can be considered, as we will see in \cref{sec:lorentzian_LQG}.
One should note that,despite the fact that the components of \( T \) and \( \widetilde T \) are not harmonic oscillators, the commutation relations \eqref{eq:heisenberg_cr} are still those of a Heisenberg algebra representation, where one of the generators acts as the identity, so that for each \( j\), \( \varepsilon \) the space\footnote{Although \( F_{\shalf} \otimes C^\varepsilon_{j} \) is not decomposable when \( j=-\half \) and \( \varepsilon=0 \), the Jordan\Endash{}Schwinger representation works even in this case, as with the discrete series.}
\begin{equation}
\bigboxplus_{k\in \Z} C^{\varepsilon_k}_{j+k/2},
\end{equation} 
where \( \varepsilon_{k} \) changes parity every time \( k \) increases by \( 1 \) and \( \varepsilon_0\equiv\varepsilon\), carries the structure of a (non-unitary) \( \mathfrak{h}_2(\R)_\C\)-module.

\section{\texorpdfstring{\(4\)D}{4D} Lorentz group}\label{sec:4d-lorentz-group}

In this section we are going to study the recoupling theory of finite and infinite\Hyphdash{}dimensional representations of the double cover of the \( 4 \)D Lorentz group, \( \Spin(3,1) \), using the results of the \( 3 \)D case as a guideline. The results we obtain are very similar to the \( \Spin(2,1) \) case, but their proofs are more elaborate due to the more sophisticated nature of the representations. 
We will first review the \( \Spin(3,1) \) representation theory, then study the product of a finite dimensional representation and an infinite-dimensional one. As for the \( 3 \)D case, we are then going to use the \WE\ theorem to generalise the \JS\ representation, known only for the finite-dimensional modules, to infinite-dimensional representations.

\subsection{Irreducible representations of \texorpdfstring{\(\Spin(3,1)\)}{Spin(3,1)}}
\label{sec:4dlorentz_representation_theory}

The \emph{proper orthochronous Lorentz group} \(\SO_0(3,1)\), henceforth simply referred to as the Lorentz group, is the identity component of the subgroup of \(\GL(4,\R)\) that preserves the quadratic form
\begin{equation}
Q(x)=-(x_0)^2 + (x_1)^2 + (x_2)^2 + (x_3)^2,\quad x=(x_0,x_1,x_2,x_3)\in \R^4.
\end{equation}
To allow for spin representations, the double cover \(\Spin(3,1)\cong \SL(2,\C)_\R\) of \(\SO_0(3,1)\) will be used here; moreover, only complex representations will be considered, so that one may work with  a complexified Lie algebra.

The Lie algebra \(\spin(3,1)_\C\) has 6 generators
\begin{equation}
\vec{J}=(J_0,J_1,J_2),\quad \vec{K}=(K_0,K_1,K_2),
\end{equation}
with commutation relations\footnote{Here \( \tensor{\varepsilon}{_{ab}^c} \) is the Levi\Endash{}Civita tensor, and we use the Einstein convention of summation over repeated indices.}
\begin{equation}\label{eq:spin(3,1)_commutation}
[J_a,J_b]=\ii\tensor{\varepsilon}{_{ab}^c}J_c,\quad [J_a,K_b]=\ii\tensor{\varepsilon}{_{ab}^c}K_c,\quad[K_a,K_b]=-\ii\tensor{\varepsilon}{_{ab}^c}J_c.
\end{equation}
The \(J\)'s generate the subalgebra \(\spin(3)\cong\su(2)\) (i.e., spatial rotations), while the \(K\)'s are the generators of boosts.
The algebra has two Casimirs
\begin{equation}
\cC_1=\vec{J}\cdot\vec{K},\quad \cC_2=J^2-K^2
\end{equation}
which, introducing the operators
\begin{equation}
J_\pm:=J_1\pm i J_2,\quad K_\pm:=K_1\pm i K_2
\end{equation}
and making use of~\eqref{eq:spin(3,1)_commutation},
can be rewritten as
\begin{equation}
\cC_1=J_0K_0+\tfrac{1}{2}(J_-K_+ + J_+K_-),\quad \cC_2=J^2-(J_0 + K_0^2+K_+K_-).
\end{equation}

As in \cref{sec:3d-lorentz-group}, we will work with the \((\mathfrak{g},K)\)-modules induced by irreducible admissible Hilbert space representations\footnote{The maximal compact subgroup is \( \SU(2) \).},  with \(\mathfrak{g}=\spin(3,1)\) and \(K=SU(2)\);
for \(\Spin(3,1)\) these exhaust all the possible irreducible \((\mathfrak{g},K)\)-modules\footnote{See \cite[chap.~II]{knapp} for the explicit expression of the group representations.}.
The general irreducible admissible \((\mathfrak{g},K)\)-module\footcite{gelfand}, labelled by a pair \((\lambda,\rho)\in\Z/2\times\C\), is the \emph{algebraic} direct sum
\begin{equation}
V_{\lambda,\rho} =\bigoplus_{j=\abs{\lambda}}^{j_\textnormal{max}} V^j_{\lambda,\rho}
\end{equation}
of unitary irreducible \(\SU(2)\)-modules \(V^j_{\lambda,\rho}\), where the sum is in integer steps and, depending on the values of \(\lambda\) and \(\rho\), it is either \(j_\textnormal{max}\in \abs{\lambda}+\N_0\) or \(j_\textnormal{max}=\infty\) (see later discussion). The (complex) vector space \(V_{\lambda,\rho}^j\) is spanned by the basis
\begin{equation}
\label{eq:basis}
\ket{(\lambda,\rho)j,m}, \quad m\in\cM_j:=\set{-j,-j+1,\dotsc,j-1,j},
\end{equation}
on which the \(\su(2)_\C\) generators act as\footnote{\( J^2=\vec{J}\cdot\vec{J} \) is the \( \su(2) \) Casimir.}
\begin{equation}
\begin{cases}
J_0\ket{(\lambda,\rho)j,m}=m\ket{(\lambda,\rho)j,m}\\
J_\pm \ket{(\lambda,\rho)j,m}=C_\pm(j,m)\ket{(\lambda,\rho)j,m\pm 1}\\
J^2\ket{(\lambda,\rho)j,m}=j(j+1)\ket{(\lambda,\rho)j,m},
\end{cases}
\end{equation}
with
\begin{equation}
C_\pm(j,m):=\sqrt{j\mp m}\sqrt{j \pm m+1},
\end{equation}
i.e., they are eigenvectors for \(J_0\) and \(J^2\); since \(\SU(2)\) is simply connected, its action on \(V_{\lambda,\rho}^j\) is completely determined by the corresponding Lie algebra action.
The \((\mathfrak{g},K)\)-module will be given an inner product by requiring the \(\SU(2)\)-modules to be orthogonal to each other and the vectors in~\eqref{eq:basis} to be orthogonal to each other\footnote{As in \cref{sec:3d-lorentz-group}, we cannot guarantee they are of norm \( 1 \) unless the module is unitary or finite-dimensional.}.

The possible matrix elements\footnote{All the other matrix elements necessarily vanish. In fact, the boost generators \( K_0 \), \( K_\pm \) are (proportional to) the components of an \( \SU(2) \) tensor operator, so that all other possibilities are excluded by the \WE\ theorem.} of the boost generators are
\begin{subequations}
\label{eq:K_actions}
\begin{align}
\praket{j+1,m\pm1|K_\pm|j,m}&=\mp P^+_{\lambda,\rho}(j)\sqrt{j\pm m +1}\sqrt{j\pm m +2}\\
\praket{j+1,m|K_0|j,m}&= P^+_{\lambda,\rho}(j)\sqrt{j+ m +1}\sqrt{j- m +1}
\\\addlinespace
\praket{j,m\pm1|K_\pm|j,m}&= P_{\lambda,\rho}(j)\,C_\pm(j,m)\\
\praket{j,m|K_0|j,m}&= P_{\lambda,\rho}(j)\,m
\\\addlinespace
\praket{j-1,m\pm1|K_\pm|j,m}&=\pm P^-_{\lambda,\rho}(j)\sqrt{j\mp m}\sqrt{j\mp m -1}\\
\praket{j-1,m|K_0|j,m}&= P^-_{\lambda,\rho}(j)\sqrt{j+ m}\sqrt{j- m},
\end{align}
\end{subequations}
where
\begin{equation}
P^-_{\lambda,\rho}(j)=\frac{\sqrt{j+\lambda}\sqrt{j-\lambda}\sqrt{j+\rho}\sqrt{j-\rho}}{j\sqrt{2j+1}\sqrt{2j-1}},\quad P_{\lambda,\rho}^+(j)=P_{\lambda,\rho}^-(j+1)
\end{equation}
and
\begin{equation}
P_{\lambda,\rho}(j)=
\begin{cases}
\frac{\ii\lambda\rho}{j(j+1)}&\casesif j\neq 0\\
0&\casesif j=0.
\end{cases}
\end{equation}
The Casimirs act on \(V_{\lambda,\rho}\) as
\begin{equation}
\cC_1=\ii\lambda\rho\,\1,\quad \cC_2=(\lambda^2+\rho^2-1)\1.
\end{equation}
The values \(j\) can take have an upper bound \(j_\textnormal{max}\in \abs{\lambda}+\N_0\) if and only if
\begin{equation}
P_{\lambda,\rho}^+(j_\textnormal{max})=0\quad\textnormal{and}\quad P_{\lambda,\rho}^+(j)\neq 0 \quad\forall j<j_\textnormal{max},
\end{equation}
i.e.\footnote{Recall that \( j\geq\abs{\lambda} \).},
\begin{equation}
\rho=\pm(j_\textnormal{max}+1).
\end{equation}
It follows that \(V_{\lambda,\rho}\) is finite-dimensional when \(\rho\in \pm (\abs{\lambda}+\N)\) and it is infinite-dimensional in all other cases.
\begin{remark}[isomorphic modules]
The values of the Casimirs and of \(P_{\lambda,\rho}(j)\), \(P^+_{\lambda,\rho}(j)\) and \(P^-_{\lambda,\rho}(j)\) are invariant under the change \((\lambda,\rho)\rightarrow (-\lambda,-\rho)\);
moreover, whether the module is finite-dimensional and the value of \(j_\textnormal{max}\) are unaffected by the change as well. It follows that the modules \(V_{\lambda,\rho}\) and \(V_{-\lambda,-\rho}\) are isomorphic.
\end{remark}

Unitary modules are those for which
\begin{equation}
K\Dagger*_0=K_0,\quad K\Dagger*_+=K_-,
\end{equation}
with respect to the inner product on \(V_{\lambda,\rho}\).
Explicitly, it must be
\begin{equation}
P_{\lambda,\rho}(j)\in\R,\quad P^-_{\lambda,\rho}(j)\in\R,
\end{equation}
which is satisfied by three possible classes of modules:
\begin{itemize}
\item \emph{principal series}: \(\lambda\in\Z/2\) and \(\rho \in \ii\R\);
\item \emph{complementary series}: \(\lambda=0\) and \(\rho\in(-1,0)\cup(0,1)\);
\item \emph{trivial representation}: \(\lambda=0\) and \(\rho=\pm1\).
\end{itemize}

\subsubsection*{Finite-dimensional modules}\label{par:finite-dimensional}
It was shown that the \((\mathfrak{g},K)\)-module with \(\rho=B(\omega+1)\), \(\omega\in \abs{\lambda}+\N_0\), \(B=\pm 1\) is finite-dimensional. We will assume, for finite-dimensional modules (and for those only), that \(\lambda\geq 0\). It is then easy to check that
\begin{equation}
\dim V_{\lambda,\rho} =
\sum_{j=\lambda}^\omega (2j+1)=(\omega-\lambda+1)(\omega+\lambda+1).
\end{equation}
Finite-dimensional modules can be given an alternative construction using the fact that
\begin{equation}
\spin(3,1)_\C\cong \su(2)_\C \oplus \su(2)_\C,
\end{equation}
i.e., by changing to the basis
\begin{equation}
\vec{M}^A:=\half(\vec{J}-\ii A \vec{K}),\quad A=\pm 1,
\end{equation}
with commutation relations
\begin{equation}
[M^A_a,M^B_b]=\ii \tensor{\varepsilon}{_{ab}^c}M^A_c\delta_{AB};
\end{equation}
one can easily show that, for finite-dimensional modules,
\begin{equation}
K\Dagger*_0=-K_0,\quad K\Dagger*_+ = - K_-,
\end{equation}
so that each \(M_a^A\) is self-adjoint, i.e., each \( \su(2)_\C \) subalgebra acts as a unitary \( \su(2) \) representation.

From \(\su(2)\) representation theory we know that, if \(V_j\) is the \((2j+1)\)-dimensional unitary irreducible \(\su(2)\)-module,
\begin{equation}
\bigoplus_{j=\lambda}^\omega V_j \cong V_{\half[\omega+\lambda]}\otimes V_{\half[\omega-\lambda]}\cong V_{\half[\omega-\lambda]}\otimes V_{\half[\omega+\lambda]}.
\end{equation}
Since \(\vec{J}=\vec{M}^{(-)} + \vec{M}^{(+)}\), we can then change the basis to
\begin{equation}
\ket{j_1,m_1}\otimes \ket{j_2,m_2}=\sum_{j=\lambda}^\omega \sum_{m=-j}^j \braket{j,m|j_1,m_1;j_2,m_2}\ket{(\lambda,\rho)j,m},
\end{equation}
where \(\braket{j,m|j_1,m_1;j_2,m_2}\) are the \(\su(2)\) \CG\ coefficients\footcite{CGtables} and
\begin{equation}
\begin{cases}
j_1=\frac{\omega+\lambda}{2}\\
j_2=\frac{\omega-\lambda}{2}
\end{cases}
\textnormal{or}\quad
\begin{cases}
j_1=\frac{\omega-\lambda}{2}\\
j_2=\frac{\omega+\lambda}{2};
\end{cases}
\end{equation}
it is assumed that \(\vec{M}^{(-)}\) and \(\vec{M}^{(+)}\) only act respectively on \(\ket{j_1,m_1}\) and \(\ket{j_2,m_2}\). The dimension of the new basis is
\begin{equation}
(2j_1+1)(2j_2+1)=(\omega+\lambda+1)(\omega-\lambda+1),
\end{equation}
as expected.
The choice of \(j_1\), \(j_2\) depends on the sign of \(\rho\): in fact, we have
\begin{equation}
\label{eq:finite-dim_consistency}
\cC_1\ket{j_1,m_1}\otimes \ket{j_2,m_2}=\ii B \lambda(\omega+1)\ket{j_1,m_1}\otimes \ket{j_2,m_2},
\end{equation}
but also
\begin{equation}
\cC_1=\sum_A \ii A \paren{M^A}^2,
\end{equation}
so that~\eqref{eq:finite-dim_consistency} is consistent if and only if
\begin{equation}
j_1=\frac{\omega-B\lambda}{2},\quad j_2=\frac{\omega+B\lambda}{2}.
\end{equation}

Conversely, one can show that every product of \(\su(2)\)-modules
\begin{equation}
V_{j_1}\otimes V_{j_2},\quad j_1,j_2\in\N_0/2
\end{equation}
gives rise to a Lorentz group \((\mathfrak{g},K)\)-module with
\begin{equation}
\lambda=\abs{j_1-j_2},\quad \rho=
\begin{cases}
(j_1+j_2+1)&\casesif j_1 < j_2\\
-(j_1+j_2+1) & \casesif j_1\geq j_2.
\end{cases}
\end{equation}
As a consequence, every finite-dimensional irreducible \((\mathfrak{g},K)\)-module can be specified by a pair \((j_1,j_2)\in \N_0/2 \times \N_0/2\); it is customary to use the pair to denote the module itself.

Examples of finite-dimensional modules are
\begin{itemize}
\item \((0,0)\): the \emph{scalar} module (trivial representation);
\item \((\half,0)\) and \((0,\half)\): respectively the left and right \emph{Weyl spinor} modules;
\item \((\half,\half)\): the (complexified) \emph{vector} module;
\item \((\half,0)\oplus(0,\half)\): the \emph{Dirac spinor} module (not irreducible).
\end{itemize}
It is not difficult to infer the decomposition of the product of two finite-dimensional modules from \(\su(2)\) results; we have
\begin{equation}
(j_1,k_1)\otimes(j_2,k_2)\cong \bigoplus_{j=\abs{j_1-j_2}}^{j_1+j_2}\bigoplus_{k=\abs{k_1-k_2}}^{k_1+k_2}(j,k),
\end{equation}
and, in particular,
\begin{equation}
\label{eq:finite_recoup}
(j_1,j_2)\equiv (j_1,0)\otimes(0,j_2).
\end{equation}
We will refer to modules of the kind \((j,0)\) and \((0,j)\) respectively as \emph{left} and \emph{right} modules; it follows from~\eqref{eq:finite_recoup} that any other irreducible module can be constructed from the product of a left and right one. To allow to easily specify if a module is left or right, the notation
\begin{equation}
F^A_j:=
\begin{cases}
(j,0) & \casesif A=-1\\
(0,j) &\casesif A=1
\end{cases}
\end{equation}
will be used in the following sections. A basis for \(F^A_j\) is given by
\begin{equation}
\ket{j_A,\mu},\quad \mu\in\cM_j,
\end{equation}
with
\begin{equation}
\begin{cases}
J_0\ket{j_A,\mu}=\mu \ket{j_A,\mu}\\
J_\pm \ket{j_A,\mu}=C_\pm(j,\mu)\ket{j_A,\mu\pm 1}
\end{cases}
\textnormal{and}\quad
\begin{cases}
K_0\ket{j_A,\mu}=\ii A \mu \ket{j_A,\mu}\\
K_\pm \ket{j_A,\mu}=\ii A C_\pm(j,\mu)\ket{j_A,\mu\pm 1}.
\end{cases}
\end{equation}

\subsection{Product of finite and infinite-dimensional modules}

In order to use the \WE\ theorem with infinite-dimensional modules, we need to study the \CG\ decomposition of the tensor product of a non-trivial finite-dimensional module (necessarily non-unitary) and an infinite-dimensional one (either unitary or non-unitary), which was previously unconsidered.
In light of the consequences of~\eqref{eq:finite_recoup} mentioned above, we will start by considering couplings of the kind \(F^A_\gamma \otimes V_{\lambda,\rho}\), where \(\gamma\geq \half\) and \(\lambda\), \(\rho\) are such that 
\begin{equation}
P^+_{\lambda,\rho}(j)\neq 0,\quad \forall j \in \abs{\lambda}+\N_0.
\end{equation}
A \CG\ decomposition exists if and only if it is possible to \emph{simultaneously} diagonalise the two Casimirs in the product module\footnote{One can check explicitly that the Casimirs acting on the product space are neither self-adjoint nor normal operators, i.e., they do not commute with their adjoint, so that the spectral theorem cannot be used.}, where the generators act as
\begin{equation}
\vec{J}\equiv\vec{J}\otimes \1 + \1\otimes \vec{J},\quad \vec{K}\equiv\vec{K}\otimes \1 + \1\otimes \vec{K}
\end{equation}
on the basis elements
\begin{equation}
\label{eq:product_basis}
\ket{\gamma_A,\mu}\otimes\ket{(\lambda,\rho)j,m},\quad j\in \abs{\lambda}+\N_0,\quad m \in \cM_j,\quad \mu \in \cM_\gamma.
\end{equation}

Instead of working with an infinite dimensional vector space, we can decompose the product space into a sum of finite-dimensional spaces by diagonalising \(J_0\) and \(J^2\) first. Using \(\su(2)\) recoupling theory, we find that the vectors
\begin{equation}
\begin{split}
\ket{(j)J,M}:=\,&\sum_{\mu\in\cM_\gamma}\sum_{m\in\cM_j} \braket{\gamma,\mu;j,m|J,M}\ket{\gamma_A,\mu}\otimes\ket{(\lambda,\rho)j,m},
\\
&j\in \abs{\lambda}+\N_0,\quad J\in\set{ \abs{j-\gamma},\dotsc,j+\gamma},\quad M\in\cM_J
\end{split}
\end{equation}
provide an orthogonal basis of \((J_0,J^2)\)-eigenvectors for the product space.
\begin{proposition}
The set of possible values \(J\) can take for the vectors \(\ket{(j)J,M}\) is
\[
\cJ(\lambda,\gamma)=\max(\varepsilon,\abs{\lambda}-\gamma)+\N_0,\quad \varepsilon=
\begin{cases}
0 &\casesif \lambda+\gamma\in \Z\\
\half  &\casesif \lambda+\gamma\in \half+\Z.
\end{cases}
\]
\end{proposition}
\begin{proof}
\begin{proofenumerate}
\item First consider the case \(\abs{\lambda}\geq\gamma\). As \(j\geq\abs{\lambda}\geq \gamma\), the possible values \(J\) can take for fixed \(j\) are
\begin{equation}
j-\gamma,\dotsc,j+\gamma,
\end{equation}
so that
\begin{equation}
\cJ(\lambda,\gamma)\equiv \bigcup_{j\in\abs{\lambda}+\N_0}\set{j-\gamma,\dotsc,j+\gamma}=\abs{\lambda}-\gamma+\N_0\equiv \max(\varepsilon,\abs{\lambda}-\gamma)+\N_0,
\end{equation}
as \(|\lambda|-\gamma\geq \varepsilon\).
\item
Now let \(\gamma>\abs{\lambda}\); in this case, we have
\begin{equation}
J\in
\begin{cases}
\set{j-\gamma,\dotsc,j+\gamma} &\casesif{j\geq \gamma}\\
\set{\gamma-j,\dotsc,\gamma+j} &\casesif{j< \gamma}.
\end{cases}
\end{equation}
It follows that, with \(\varepsilon\) defined as in the statement,
\begin{equation}
\begin{split}
\cJ(\lambda,\gamma) &=\bigcup_{j=\abs{\lambda}}^{\gamma+\varepsilon-1} \set{\gamma-j,\dotsc,\gamma+j} \cup \bigcup_{j=\gamma+\varepsilon}^{\infty} \set{j-\gamma,\dotsc,j+\gamma}
\\
&=\bigcup_{j=\abs{\lambda}}^{\gamma+\varepsilon-1} \set{\gamma-j,\dotsc,\gamma+j} \cup \paren{\varepsilon +\N_0}
\\
&=\varepsilon+\N_0 \equiv \max(\varepsilon,\abs{\lambda}-\gamma)+\N_0,
\end{split}
\end{equation}
as \(\bigcup_{j=\abs{\lambda}}^{\gamma+\varepsilon-1} \set{\gamma-j,\dotsc,\gamma+j}\) is necessarily\footnote{Note that
\( \gamma-j\geq 1-\varepsilon\geq\varepsilon\) when \(\abs{\lambda}\leq j\leq\gamma+\varepsilon-1\).} a subset of \(\varepsilon+\N_0\).
\end{proofenumerate}
\end{proof}
The eigenspace \(V^J_M\), defined by
\begin{equation}
J_0\ket{\psi}=M\ket{\psi},\quad J^2\ket{\psi}=J(J+1)\ket{\psi},\quad \forall\ket{\psi}\in V^J_M,
\end{equation}
is spanned by the basis vectors\footnote{The form of \( \Omega_J(\lambda,\gamma) \) can be evinced from \cref{tab:j_values}.}
\begin{equation}
\ket{(j)J,M},\quad j\in\Omega_J(\lambda,\gamma):=
\begin{cases}
\set{ \max(\abs{\lambda},J-\gamma),\dotsc,J+\gamma} &\casesif J\geq \gamma-\abs{\lambda}\\
\set{\gamma-J,\dotsc,\gamma+J} & \casesif J< \gamma-\abs{\lambda},
\end{cases}
\end{equation}
so that
\begin{equation}
\dim V^J_M =
\begin{cases}
\min(J+\gamma-\abs{\lambda}+1,2\gamma+1) &\casesif J\geq \gamma-\abs{\lambda}\\
2J+1&\casesif J< \gamma-\abs{\lambda};
\end{cases}
\end{equation}
note that, when \(\abs{\lambda}\geq \gamma\), it is always true that \(J\geq 0\geq \gamma-\abs{\lambda}\), so that the case \(J< \gamma-\abs{\lambda}\) need only be considered when \(\abs{\lambda}<\gamma\).
\begin{table}
\footnotesize
\centering
\begin{minipage}{\textwidth}
\subcaption{\(\gamma=\abs{\lambda}+\shalf+n,\quad n\in\N_0\)}
\begin{tabularx}{0.36\textwidth}{lX}
\toprule
\(j\) & possible values of \(J\) \\
\midrule
\(\abs{\lambda}\) & \(n+ \half,\dotsc,2\gamma-n-\half\)\\
\addlinespace[0.5\defaultaddspace]
\(\abs{\lambda}+1\) & \(n-\frac{1}{2},\dotsc,2\gamma-n+\half\)\\
\addlinespace[0.5\defaultaddspace]
\multicolumn{2}{c}{\(\vdots\)}\\
\addlinespace[0.5\defaultaddspace]
\(\abs{\lambda}+n-1\) & \(\frac{3}{2},\dotsc,2\gamma-\frac{3}{2}\)\\
\addlinespace[0.5\defaultaddspace]
\(\abs{\lambda}+n\) & \(\frac{1}{2},\dotsc,2\gamma-\half\)\\
\addlinespace[0.5\defaultaddspace]
\(\abs{\lambda}+n+1\) & \(\frac{1}{2},\dotsc,2\gamma+\half\)\\
\addlinespace[0.5\defaultaddspace]
\(\abs{\lambda}+n+2\) & \(\frac{3}{2},\dotsc,2\gamma+\frac{3}{2}\)\\
\addlinespace[0.5\defaultaddspace]
\multicolumn{2}{c}{\(\vdots\)}\\
\bottomrule
\end{tabularx}
\hfill
\begin{tabularx}{0.63\textwidth}{Xll}
\toprule
\(J\) & possible values of \(j\)\\
\midrule
\(\frac{1}{2}\) & \(\abs{\lambda}+n,\abs{\lambda}+n+1\) & \((\gamma-J,\dotsc,\gamma+J)\)\\
\addlinespace[0.5\defaultaddspace]
\(\frac{3}{2}\) & \(\abs{\lambda}+n-1,\dotsc,\abs{\lambda}+n+2\) & \((\gamma-J,\dotsc,\gamma+J)\)\\
\addlinespace[0.5\defaultaddspace]
\multicolumn{3}{c}{\(\vdots\)}\\
\addlinespace[0.5\defaultaddspace]
\(n+\frac{1}{2}\) & \(\abs{\lambda},\dotsc,\abs{\lambda}+2n+1\) & \((\abs{\lambda},\dotsc,J+\gamma)\)\\
\addlinespace[0.5\defaultaddspace]
\(n+\frac{3}{2}\) & \(\abs{\lambda},\dotsc,\abs{\lambda}+2n+2\) & \((\abs{\lambda},\dotsc,J+\gamma)\)\\
\addlinespace[0.5\defaultaddspace]
\multicolumn{3}{c}{\(\vdots\)}\\
\addlinespace[0.5\defaultaddspace]
\(2\gamma-n-\frac{1}{2}\) & \(\abs{\lambda},\dotsc,\abs{\lambda}+2\gamma\) & \((\abs{\lambda},\dotsc,J+\gamma)\)\\
\addlinespace[0.5\defaultaddspace]
\(2\gamma-n-\frac{1}{2}+k\) & \(\abs{\lambda}+k,\dotsc,\abs{\lambda}+2\gamma+k\) & \((J-\gamma,\dotsc,J+\gamma)\)\\
\bottomrule
\end{tabularx}
\end{minipage}
\\[2\defaultaddspace]
\begin{minipage}{\textwidth}
\subcaption{\(\gamma=\abs{\lambda}+1+n,\quad n\in\N_0\)}
\begin{tabularx}{0.36\textwidth}{lX}
\toprule
\(j\) & possible values of \(J\) \\
\midrule
\(\abs{\lambda}\) & \(n+ 1,\dotsc,2\gamma-n-1\)\\
\addlinespace[0.5\defaultaddspace]
\(\abs{\lambda}+1\) & \(n,\dotsc,2\gamma-n\)\\
\addlinespace[0.5\defaultaddspace]
\multicolumn{2}{c}{\(\vdots\)}\\
\addlinespace[0.5\defaultaddspace]
\(\abs{\lambda}+n-1\) & \(2,\dotsc,2\gamma-2\)\\
\addlinespace[0.5\defaultaddspace]
\(\abs{\lambda}+n\) & \(1,\dotsc,2\gamma-1\)\\
\addlinespace[0.5\defaultaddspace]
\(\abs{\lambda}+n+1\) & \(0,\dotsc,2\gamma\)\\
\addlinespace[0.5\defaultaddspace]
\(\abs{\lambda}+n+2\) & \(1,\dotsc,2\gamma+1\)\\
\addlinespace[0.5\defaultaddspace]
\multicolumn{2}{c}{\(\vdots\)}\\
\bottomrule
\end{tabularx}
\hfill
\begin{tabularx}{0.63\textwidth}{Xll}
\toprule
\(J\) & possible values of \(j\)\\
\midrule
\(0\) & \(\abs{\lambda}+n+1\) & \((\gamma-J,\dotsc,\gamma+J)\)\\
\addlinespace[0.5\defaultaddspace]
\(1\) & \(\abs{\lambda}+n,\dotsc,\abs{\lambda}+n+2\) & \((\gamma-J,\dotsc,\gamma+J)\)\\
\addlinespace[0.5\defaultaddspace]
\multicolumn{3}{c}{\(\vdots\)}\\
\addlinespace[0.5\defaultaddspace]
\(n+1\) & \(\abs{\lambda},\dotsc,\abs{\lambda}+2n+1\) & \((\abs{\lambda},\dotsc,J+\gamma)\)\\
\addlinespace[0.5\defaultaddspace]
\(n+2\) & \(\abs{\lambda},\dotsc,\abs{\lambda}+2n+2\) & \((\abs{\lambda},\dotsc,J+\gamma)\)\\
\addlinespace[0.5\defaultaddspace]
\multicolumn{3}{c}{\(\vdots\)}\\
\addlinespace[0.5\defaultaddspace]
\(2\gamma-n-1\) & \(\abs{\lambda},\dotsc,\abs{\lambda}+2\gamma\) & \((\abs{\lambda},\dotsc,J+\gamma)\)\\
\addlinespace[0.5\defaultaddspace]
\(2\gamma-n-1+k\) & \(\abs{\lambda}+k,\dotsc,\abs{\lambda}+2\gamma+k\) & \((J-\gamma,\dotsc,J+\gamma)\)\\
\bottomrule
\end{tabularx}
\end{minipage}
\caption{Possible values of \(J\) given \(j\) (left) and of \(j\) given \(J\) (right) when \(\gamma>\abs{\lambda}\).}
\label{tab:j_values}
\end{table}
%

Since the Casimirs commute with both \(J_0\) and \(J^2\), we can work with their restriction on the finite-dimensional subspaces \(V^J_M\) and diagonalise those; moreover, it suffices to consider the restrictions to \(V_J:=V^J_J\) thanks to the following
\begin{proposition}\label{prop:JM}
Let \(J\in\mathcal{J}(\lambda,\gamma)\). The eigenvalues of the Casimirs \(\cC_1\) and \(\cC_2\) are the same on each \(V^J_M\), \(M\in\cM_J\).
\end{proposition}
\begin{proof}
The basis vectors of \(V^J_M\) satisfy
\begin{equation}
J_\pm\ket{(j)J,M}=C_\pm(J,M)\ket{(j)J,M\pm 1},
\end{equation}
so that
\begin{equation}
J_\pm (V^J_M) \subseteq V^J_{M\pm 1},\quad
\begin{cases}
\ker J_+|_{V^J_M}=\set{0}, \quad &\forall M<J\\
\ker J_-|_{V^J_M}=\set{0}, \quad &\forall M>-J.
\end{cases}
\end{equation}
Since \(J_\pm\) commutes with the Casimirs, given a \(\cC_a\)-eigenvector \(\ket{\alpha_a}\in V^J_M\) with eigenvalue \(\alpha_a\in \C\) we have
\begin{equation}
0\neq J_\pm\ket{\alpha_a}\in V^J_{M\pm 1},\quad \cC_a J_\pm \ket{\alpha_a}=J_\pm \cC_a\ket{\alpha_a}=\alpha_a J_\pm\ket{\alpha_a}
\end{equation}
whenever \(V^J_{M\pm 1}\) is defined\footnote{Respectively when \( M<J \) and \( M>-J \).}, so that each \(V^J_M\) has the same eigenvalues. 
\end{proof}
The action of the Casimirs on the basis vectors of \(V_J\)
\begin{equation}
\ket{(j)J}:=\ket{(j)J,J},\quad j\in\Omega_J(\lambda,\gamma)
\end{equation}
is given by
\begin{subequations}
\label{eq:casimir_actions}
\begin{align}
\begin{split}
\cC_1\ket{(j)J}=&\,\bracks[\Big]{J(J+1)\half[\ii A + P_{\lambda,\rho}(j)] - \paren[\big]{j(j+1)-\gamma(\gamma+1)}{\half[\ii A - P_{\lambda,\rho}(j)]}}\ket{(j)J}\\
&+\half[P^+_{\lambda,\rho}(j)]\sqrt{J+j+\gamma+2}\sqrt{j+\gamma-J+1}\sqrt{J+j-\gamma+1}\sqrt{J-j+\gamma}\ket{(j+1)J}\\
&+\half[P^-_{\lambda,\rho}(j)]\sqrt{J+j+\gamma+1}\sqrt{j+\gamma-J}\sqrt{J+j-\gamma}\sqrt{J-j+\gamma+1}\ket{(j-1)J}
\end{split}
\\
\begin{split}
\cC_2\ket{(j)J}=&\,\bracks[\big]{\paren[\big]{J(J+1)-j(j+1)}\paren[\big]{1 -\ii A P_{\lambda,\rho}(j)}
\\
&+ \gamma(\gamma+1) \paren[\big]{1 +\ii A P_{\lambda,\rho}(j)}+\lambda^2 +\rho^2 -1}\ket{(j)J}\\
&-\ii A P^+_{\lambda,\rho}(j)\sqrt{J+j+\gamma+2}\sqrt{j+\gamma-J+1}\sqrt{J+j-\gamma+1}\sqrt{J-j+\gamma}\ket{(j+1)J}\\
&-\ii A P^-_{\lambda,\rho}(j)\sqrt{J+j+\gamma+1}\sqrt{j+\gamma-J}\sqrt{J+j-\gamma}\sqrt{J-j+\gamma+1}\ket{(j-1)J},
\end{split}
\end{align}
\end{subequations}
where it is implicitly assumed that \(\ket{(j)J}=0\) if \(j\not\in \Omega_J(\lambda,\gamma)\).
Note that the matrix form of each \(\cC_a\) is \emph{tridiagonal} (see \cref{app:tridiagonal}) and that, for the subdiagonal entries,
\begin{equation}
\praket{(j+1)J|\cC_a|(j)J}=0\quad\Leftrightarrow\quad j=J+\gamma=\max\Omega_J(\lambda,\gamma),
\end{equation}
i.e., they are all non-zero\footnote{Note that when \( j=J+\gamma \) the vector \( \ket{(j+1)J}\) vanishes.};
it then follows from \cref{prop:tridiagonal} that the eigenspaces of \(\cC_a\) are all \(1\)-dimensional, so that it is diagonalisable if and only if it has \(\dim V_J\) \emph{distinct} eigenvalues.
Explicitly, the Casimirs are simultaneously diagonalisable on \(V_J\) if and only if there is a basis
\begin{equation}
\ket{(\Lambda,\Rho)J}=\sum_{j\in\Omega_J}\mathrm{A}\lbrace\gamma_A;(\lambda,\rho)j|(\Lambda,\Rho)J\rbrace\ket{(j)J},\quad (\Lambda,\Rho)\in\cC^A_J(\lambda,\rho,\gamma)\subseteq \C^2,
\end{equation}
with
\begin{equation}
\card{\cC^A_J(\lambda,\rho,\gamma)}=\dim V_J,
\end{equation}
such that
\begin{subequations}
\label{eq:lambda_rho_eigenvectors}
\begin{align}
\cC_1\ket{(\Lambda,\Rho)J}&=\ii\Lambda\Rho\ket{(\Lambda,\Rho)J}\\
\cC_2\ket{(\Lambda,\Rho)J}&=(\Lambda^2+\Rho^2-1)\ket{(\Lambda,\Rho)J}
\end{align}
\end{subequations}
and for every \((\Lambda,\Rho)\), \((\Lambda\Prime,\Rho\Prime)\in \cC_J^A(\lambda,\rho,\gamma)\)
\begin{equation}
\begin{cases}
\Lambda\Rho=\Lambda\Prime\Rho\Prime\\
\Lambda^2+\Rho^2=(\Lambda\Prime)^2+(\Rho\Prime)^2
\end{cases}
\quad \Leftrightarrow \quad
(\Lambda\Prime,\Rho\Prime)=(\Lambda,\Rho);
\end{equation}
note that at this stage \(\Lambda\) is allowed to be any complex number, to ensure that any pair of eigenvalues of the Casimirs can be written as in~\eqref{eq:lambda_rho_eigenvectors}.
The coefficients of the change of basis \(\mathrm{A}\lbrace\gamma_A;(\lambda,\rho)j|(\Lambda,\Rho)J\rbrace\) will be called as usual \emph{\CG\ coefficients}.
Conversely, the inverse change of basis is
\begin{equation}
\ket{(j)J}=\sum_{(\Lambda,\Rho)\in\cC_J}\mathrm{B}\lbrace (\Lambda,\Rho)J|\gamma_A;(\lambda,\rho)j \rbrace\ket{(\Lambda,\Rho)J},\quad j\in\Omega_J(\lambda,\rho),
\end{equation}
where the \(\mathrm{B}\lbrace (\Lambda,\Rho)J|\gamma_A;(\lambda,\rho)j \rbrace\) are the \emph{inverse \CG\ coefficients}.
As a consequence of \cref{prop:JM}, the eigenvectors in \(V^J_M\), \(M<J\) will be
\begin{equation}
\ket{(\Lambda,\Rho)J,M}:=\sum_{j\in\Omega_J}\mathrm{A}\lbrace\gamma_A;(\lambda,\rho)j|(\Lambda,\Rho)J\rbrace\ket{(j)J,M}
\end{equation}
so that, more generally,
\begin{multline}
\ket{(\Lambda,\Rho)J,M}=\sum_{j\in\Omega_J}\sum_{\mu\in\cM_\gamma}\sum_{m\in\cM_j}
\mathrm{A}\lbrace\gamma_A;(\lambda,\rho)j|(\Lambda,\Rho)J\rbrace
\\[-0.7em]
\times\braket{\gamma,\mu;j,m|J,M}\ket{\gamma_A,\mu}\otimes\ket{(\lambda,\rho)j,m}.
\end{multline}

Solving the eigenvalue equations for arbitrary \(\gamma\) is not an easy task: instead, we will solve explicitly the case \(\gamma=\half\) and proceed by induction for the other cases.
When \(\gamma=\half\), we have
\begin{equation}
\mathcal{J}(\lambda,\half)=
\begin{cases}
\half+\N_0&\casesif\lambda=0\\
\abs{\lambda}-\half+\N_0&\casesif\lambda\neq0
\end{cases}
\quad\textnormal{and}\quad \dim V_J=
\begin{cases}
1&\casesif J=\abs{\lambda}-\half\\
2&\casesif J\geq\abs{\lambda}+\half,
\end{cases}
\end{equation}
and it can be explicitly checked by solving the eigenvalue problem for \eqref{eq:casimir_actions} that, when \(\lambda\neq-A\rho\),
\begin{equation}\label{eq:4d-1/2_eigenvalues}
\cC_J^A(\lambda,\rho,\half)=
\begin{cases}
\set*{(\lambda-\half,\rho-\half[A]),(\lambda+\half,\rho+\half[A])} \subseteq\Z/2 \times \C &\casesif J\geq\abs{\lambda}+\half\\
\set*{(\lambda-\half\sgn(\lambda),\rho-\half[A]\sgn(\lambda))} \subseteq\Z/2 \times \C  &\casesif J=\abs{\lambda}-\half;
\end{cases}
\end{equation}
the corresponding \CG\ coefficients can be found in \cref{tab:4d-1/2} (\cref{app:CG}). When \(\rho=-A\lambda\) the eigenvalues for \(J\geq \abs{\lambda}+\half\) coincide, so that, as pointed out earlier, the Casimirs cannot be diagonalised.

As the eigenvalues do not depend on \(J\) and
\begin{equation}
\cC_J^A(\lambda,\rho,\half)\subseteq\cC_{J+1}^A(\lambda,\rho,\half),\quad \forall J\in\cJ(\lambda,\gamma),
\end{equation}
the eigenvectors can be extended to an eigenbasis 
\begin{equation}
\ket{(\Lambda,\Rho)J,M},\quad (\Lambda,\Rho)=(\lambda\pm\half,\rho\pm\half[A]),\quad J\in\abs{\Lambda}+\N_0,\quad M\in \cM_J,
\end{equation}
as it follows from \eqref{eq:4d-1/2_eigenvalues} that when \( \lambda\geq\half \)
\begin{equation}
\begin{cases}
\Lambda=\lambda-\half &\quad \Rightarrow \quad J\geq \abs{\lambda}-\half = \abs{\Lambda}
\\
\Lambda=\lambda+\half &\quad \Rightarrow \quad J\geq \abs{\lambda}+\half = \abs{\Lambda}
\end{cases}
\end{equation}
and when \( \lambda\leq-\half \)
\begin{equation}
\begin{cases}
\Lambda=\lambda-\half &\quad \Rightarrow \quad J\geq \abs{\lambda}+\half = \abs{\Lambda}
\\
\Lambda=\lambda+\half &\quad \Rightarrow \quad J\geq \abs{\lambda}-\half = \abs{\Lambda}.
\end{cases}
\end{equation}
One can check, for all eigenvalue pairs\footnote{Assuming \( \rho \not \in (\abs{\lambda}+\N)\).}, that
\begin{equation}
\Rho\not\in\pm(\abs{\Lambda}+\N),
\end{equation}
so that \(F^A_{\shalf}\otimes V_{\lambda,\rho}\) splits in two infinite-dimensional irreducible modules
\begin{equation}
V_{\Lambda,\Rho},\quad (\Lambda,\Rho)=(\lambda\pm\half,\rho\pm\half[A]).
\end{equation}
These modules are \emph{never} both unitary: a list of the possible pairs \((\lambda,\rho)\) such that there  is \emph{one} unitary module in the decomposition can be found in \cref{tab:LR_unitary}. Notice that there are, up to isomorphisms, only two unitary modules that coupled with \(F^A_{\shalf}\) have a unitary one in the decomposition.
\begin{table}[!ht]
\centering
\begin{tabularx}{\textwidth}{Xccc}
\toprule
& \(\lambda\) & \(\rho\) & \(V_{\lambda,\rho}\) unitary if  \\
\midrule
principal series & any & \(\pm \half + \ii\R\) & \((\lambda,\rho)=(0,\pm\half)\)\\
\addlinespace
complementary series & \(\pm\half\) & \(\sgn(\lambda)\half[A]+(-1,0)\cup(0,1)\) & \((\lambda,\rho)=(\pm\half,0)\) \\
\bottomrule
\end{tabularx}
\caption{The possible pairs \((\lambda,\rho)\) such that one \(V_{\Lambda,\Rho}\) is unitary (principal or complementary series).}
\label{tab:LR_unitary}
\end{table}

A generalisation of the case \(\gamma=\half\) to arbitrary \(\gamma\) is provided by the following Propositions:
\begin{lemma}
\label{lem:no_double_eigenvalues}
Let \(\gamma\in\N/2\), \(A=\pm1\) and \((\lambda,\rho)\in\Z/2\times\C\), with \(\rho\not\in\pm(\abs{\lambda}+\N)\). Then
\begin{equation*}
\begin{cases}
\ii(\lambda+\mu)(\rho+A\mu)\neq \ii(\lambda+\nu)(\rho+A\nu)\\[0.5em]
(\lambda+\mu)^2+(\rho+A\mu)^2-1\neq (\lambda+\nu)^2+(\rho+A\nu)^2-1
\end{cases}
\quad \forall \mu\neq\nu\in\cM_\gamma
\end{equation*}
if and only if \(\rho+A\lambda\not\in(-2\gamma,2\gamma)\cap\Z\).
\end{lemma}
\begin{proof}
Let \(\mu\neq\nu\in\cM_\gamma\). The statement reduces to
\begin{equation}
\mu^2+\mu(\lambda+A\rho)\neq \nu^2+\nu(\lambda+A\rho),
\end{equation}
which is equivalent to\footnote{Recall that \( \mu\neq\nu \).}
\begin{equation}
\label{eq:no_double_eigenvalues}
(\mu-\nu)(\lambda+A\rho+\mu+\nu)\neq 0\quad \Leftrightarrow \quad \lambda+A\rho+\mu+\nu\neq0.
\end{equation}
The possible values the sum \(\mu+\nu\) can take are
\begin{equation}
\set{\mu+\nu \setst \mu\neq\nu\in \cM_\gamma }\equiv \set*{-2\gamma+1,-2\gamma+2,\dotsc,2\gamma-2,2\gamma-1 },
\end{equation}
so that~\eqref{eq:no_double_eigenvalues} is true if and only if \(\rho+A\lambda\not\in(-2\gamma,2\gamma)\cap \Z\).
\end{proof}
\begin{proposition}
\label{prop:coupling}
Consider the product \(F^A_\gamma\otimes V_{\lambda,\rho}\), with \(\gamma\geq\half\) and \(V_{\lambda,\rho}\) infinite\Hyphdash{}dimensional. When \(\rho+A\lambda\not\in(-2\gamma,2\gamma)\cap \Z\) the Casimirs are simultanously diagonalisable, with
\begin{equation*}
(\Lambda,\Rho)\in\set {(\lambda+\nu,\rho+A\nu)\setst \nu\in\cM_\gamma}.
\end{equation*}
\end{proposition}
\begin{proof}
The proof proceeds by induction on \(\gamma\in\N/2\). Assume that the statement is true for \(\gamma-\half\), and consider the product \(F^A_\gamma\otimes V_{\lambda,\rho}\), \(\gamma>\half\).
It is known from \(\su(2)\) representation theory that
\begin{equation}
F^A_{\shalf}\otimes F^A_{\gamma-\shalf}=F^A_{\gamma-1}\oplus F^A_\gamma,
\end{equation}
so that
\begin{equation}
\ket{\gamma_A,\mu}\equiv \sum_{\sigma\in\cM_{\shalf}}\sum_{\tau\in\cM_{\gamma-\shalf}}\braket{\half,\sigma;\gamma-\half,\tau|\gamma,\mu}\ket{\half_A,\sigma}\otimes\ket{(\gamma-\half)_A,\tau};
\end{equation}
in particular
\begin{equation}
\label{eq:1/2+gamma-1/2}
\ket{\gamma_A,\gamma}=\ket{\half_A,\half}\otimes\ket{(\gamma-\half)_A,\gamma-\half}.
\end{equation}
Consider now the \(J^2\)-eigenspace \(V_J\), \(J\geq \abs{\lambda}+\gamma\),
so that \(J-\gamma\in\Omega_J(\lambda,\gamma)\) and the vector
\begin{equation}
\ket{(J-\gamma)J}=\ket{\gamma_A,\gamma}\otimes\ket{(\lambda,\rho) J-\gamma, J-\gamma}
\end{equation}
exists. Using~\eqref{eq:1/2+gamma-1/2}, \(\ket{(J-\gamma)J}\) can be rewritten as
\begin{equation}
\begin{split}
\ket{(J-\gamma)J}&=\ket{\half_A,\half}\otimes\Big( \ket{(\gamma-\half)_A,\gamma-\half}\otimes\ket{(\lambda,\rho) J-\gamma, J-\gamma}\Big)\\
&=
\begin{multlined}[t][0.7\textwidth]
\sum_{\tau\in\cM_{\gamma-\shalf}}
\mathrm{B}\set {(\lambda+\tau,\rho+A\tau)J-\half | (\gamma-\half)_A;(\lambda,\rho)J-\gamma}
\\[-0.7em]
\times\ket{\half_A,\half}\otimes\ket{(\lambda+\tau,\rho+A\tau)J-\half},
\end{multlined}
\end{split}
\end{equation}
where we used the inductive hypothesis and the fact that
\begin{equation}
\rho+A\lambda\not\in(-2\gamma,2\gamma)\cap \Z\quad \Rightarrow \quad \rho+A\lambda\not\in(-2\gamma+1,2\gamma-1)\cap \Z.
\end{equation}
Since in particular \(\rho\neq-A\lambda\), the results of the case \(\gamma=\half\) can be used, so that
\begin{multline}
\label{eq:decomposition 1}
\ket{(J-\gamma)J}=\sum_{\sigma\in\cM_{\shalf}}
\sum_{\tau\in\cM_{\gamma-\shalf}}
\mathrm{B}\lbrace (\lambda+\tau+\sigma,\rho+A\tau+A\sigma)J | \half_A;(\lambda+\tau,\rho+A\tau)J-\half \rbrace\\
\times\mathrm{B}\lbrace (\lambda+\tau,\rho+A\tau)J-\half | (\gamma-\half)_A;(\lambda,\rho)J-\gamma \rbrace \ket{[\lambda+\tau](\lambda+\tau+\sigma,\rho+A\tau+A\sigma)J},
\end{multline}
where \([\sigma]\) keeps track of the fact that \((\lambda+\tau+\sigma,\rho+A\tau+A\sigma)\) comes from \((\lambda+\tau,\rho+A\tau)\). There are 
exactly \(4\gamma\) (independent) vectors on the RHS of~\eqref{eq:decomposition 1}, namely
\begin{equation}
\begin{cases}
\ket{[+\half](\lambda+\gamma,\rho+A\gamma)J}\\
\ket{[-\half](\lambda+\nu,\rho+A\nu)J}\quad\textnormal{and}\quad\ket{[+\half](\lambda+\nu,\rho+A\nu)J},\quad \nu \in \cM_{\gamma-1}\\
\ket{[-\half](\lambda-\gamma,\rho-A\gamma)J},
\end{cases}
\end{equation}
with \(2\gamma+1\equiv \dim V_J\) distinct eigenvalue pairs (see \cref{lem:no_double_eigenvalues}).

As shown in \cref{prop:4d_CG_property}, when \(J\geq \abs{\lambda}+\gamma\) the \CG\ coefficients satisfy
\begin{equation}
\frac{\mathrm{B}\lbrace(\Lambda+1,\Rho+A)J-\half|(\gamma-\half)_A;(\lambda,\rho)J-\gamma\rbrace}{\mathrm{B}\lbrace(\Lambda,\Rho)J-\half|(\gamma-\half)_A;(\lambda,\rho)J-\gamma\rbrace}=
\alpha(\Lambda,\Rho)\frac{\sqrt{J+\Lambda+\half}\sqrt{J+A\Rho+\half}}{\sqrt{J-\Lambda-\half}\sqrt{J-A\Rho-\half}},
\end{equation}
where \(\alpha\) is fixed by the normalisation convention and is independent of \(J\). Using this formula and the \(\gamma=\half\) \CG\ coefficients from \cref{tab:4d-1/2}
\begin{equation}
\mathrm{B}\lbrace (\Lambda+\sigma,\Rho+A\sigma)J | \half_A;(\Lambda,\Rho)J-\half \rbrace=
\begin{cases}
\ii A\frac{\sqrt{J-\Lambda+\shalf}\sqrt{J-A\Rho+\shalf}}{\sqrt{2J+1}\sqrt{\Lambda+A\Rho}} & \casesif \sigma=-\half
\\\addlinespace
\frac{\sqrt{J+\Lambda+\shalf}\sqrt{J+A\Rho+\shalf}}{\sqrt{2J+1}\sqrt{\Lambda+A\Rho}} & \casesif \sigma=+\half,
\end{cases}
\end{equation}
it is possible to write
\begin{equation}
\ket{(J-\gamma)J}=\sum_{\nu\in\cM_\gamma}\mathrm{B}\lbrace (\lambda+\nu,\rho+A\nu)J | \gamma_A; (\lambda,\rho)J-\gamma \rbrace \ket{(\lambda+\nu,\rho+A\nu)J},
\end{equation}
where the vectors on the RHS are defined (up to a normalisation factor) as
\begin{equation}
\ket{(\Lambda,\Rho)J}\propto\ket{[\pm\half](\Lambda,\Rho)J}\casesif (\Lambda,\Rho)=(\lambda\pm\gamma,\rho\pm A\gamma)
\end{equation}
and
\begin{equation}
\ket{(\Lambda,\Rho)J}\propto\tfrac{1}{\sqrt{\Lambda+A\Rho-1}}\ket{[+\half](\Lambda,\Rho)J}+\ii A \tfrac{\alpha(\Lambda-\shalf,\Rho-\shalf[A])}{\sqrt{\Lambda+A\Rho+1}} \ket{[-\half](\Lambda,\Rho)J}
\end{equation}
otherwise.
As these vectors live in different \((\cC_1,\cC_2)\)-eigenspaces, they are necessarily independent. Moreover, they form a basis of \(V_J\): in fact, we know from~\eqref{eq:casimir_actions} that
\begin{equation}
\cC_1\ket{(j) J}\in\Span\set{ \ket{(j-1) J},\ket{(j) J},\ket{(j+1) J}} ,
\end{equation}
with
\begin{equation}
\praket{(j-1) J|\cC_1|(j) J}=0\quad\Leftrightarrow\quad j=J+\gamma,
\end{equation}
so that
\begin{equation}
\ket{(j+1) J}\in\Span\set{\ket{(j-1) J},\ket{(j) J},\cC_1\ket{(j) J}}.
\end{equation}
Since \(\ket{(J-\gamma) J}\) is a linear combination of the \(\ket{(\Lambda,\Rho) J}\) vectors, it follows recursively that
\begin{equation}
\ket{(j) J}\in\Span \set{\ket{(\lambda+\nu,\rho+A\nu) J}\setst 
\nu\in\cM_\gamma},\quad \forall j\in\Omega_J(\lambda,\gamma).
\end{equation}

It follows from \cref{prop:JM} that all the results obtained for \(V_J\) hold for each \(V^J_M\), \(M\in\cM_J\). One can then extend them to every \(J\in \cJ(\lambda,\gamma)\) by defining recursively (once the \CG\ coefficients  and the vectors have been appropriately normalised, so that the generators act on the \((\cC_1,\cC_2)\)-eigenvectors as~\eqref{eq:K_actions})
\begin{multline}
\ket{(\Lambda,\Rho)J-1,J-1}\propto K_-\ket{(\Lambda,\Rho)J,J}-P_{\Lambda,\Rho}(J)\sqrt{2J}\ket{(\Lambda,\Rho)J,J-1}
\\
-P^+_{\Lambda,\Rho}(J)\sqrt{2}\ket{(\Lambda,\Rho)J+1,J-1},
\end{multline}
for \(J\leq\abs{\lambda}-\gamma\), which are trivially still eigenvectors. Since a basis of eigenvectors has been constructed for the whole product space, it follows that the Casimirs are diagonalisable.
\end{proof}

The results of \cref{prop:coupling} does not apply when \(\rho+A\lambda\in(-2\gamma,2\gamma)\cap \Z\); in this case we have
\begin{proposition}
\label{prop:continuity}
Consider the product \(F^A_\gamma\otimes V_{\lambda,\rho}\), with \(\gamma\geq\half\) and \(V_{\lambda,\rho}\) infinite\Hyphdash{}dimensional.
When \(\rho+A\lambda \in(-2\gamma,2\gamma)\cap \Z\) the Casimirs are not diagonalisable on the product module.
\end{proposition}
\begin{proof}
Consider the \(J^2\)-eigenspace \(V_J\), \(J\geq\abs{\lambda}+\gamma\). The function
\begin{equation}
d^\lambda_k:\rho\in\R \mapsto \det(\cC_1|_{V_J}-k\1)\in \C,\quad k\in\C
\end{equation}
is continuous, as it is a product of continuous functions of \(\rho\) (see eq.~\eqref{eq:casimir_actions}). From \cref{prop:coupling} we have that, for \(\rho+A\lambda\not\in(-2\gamma,2\gamma)\cap \Z\),
\begin{equation}
d_k^\lambda(\rho)=\prod_{\nu\in\cM_\gamma}\big[\ii(\lambda+\nu)(\rho+A\nu)-k\big];
\end{equation}
it follows from continuity that, for each fixed \(\lambda\in \Z/2\), \(n\in \{-2\gamma+1,\dotsc,2\gamma-1\}\),
\begin{equation}
d_k^\lambda(-A\lambda+n)=\lim_{\rho\rightarrow -A\lambda+n}d_k^\lambda(\rho)= \prod_{\nu\in\cM_\gamma}\big[\ii(\lambda+\nu)(-A\lambda + n+A\nu)-k\big].
\end{equation}
From \cref{lem:no_double_eigenvalues} we know that there are  \emph{at most} \(2\gamma\) distinct eigenvalues in this case, while \(\dim V_J=2\gamma+1\). As pointed out earlier, the matrix form of \(\cC_a|_{V_J}\) satisfies the assumptions of \cref{prop:tridiagonal}, so that it has at most \(2\gamma\) eigenvectors, i.e., it is not diagonalisable on \(V_J\) (and hence on the whole product space). 
\end{proof}

Finally, the result for left and right modules can be generalised to arbitrary finite-dimensional ones with the following
\begin{corollary}
The Casimirs are simultaneously diagonalisable in \((\gamma_1,\gamma_2)\otimes V_{\lambda,\rho}\), with \(\gamma_1,\gamma_2\geq \shalf\) and \(V_{\lambda,\rho}\) infinite-dimensional, if and only if \(\rho-\lambda\not\in(-2\gamma_1,2\gamma_1)\cap\Z\) and \(\rho+\lambda\not\in(-2\gamma_2,2\gamma_2)\cap\Z\), with
\begin{equation*}
(\Lambda,\Rho)\in\set{(\lambda+\nu_1+\nu_2,\rho-\nu_1+\nu_2) \setst \nu_1\in \cM_{\gamma_1}, \nu_2\in \cM_{\gamma_2}};
\end{equation*}
the eigenvalue pairs are not necessarily distinct.
\end{corollary}
\begin{proof}
As already noted, one has \((\gamma_1,\gamma_2)\cong F^{-}_{\gamma_1}\otimes F^{+}_{\gamma_2}\), so that we may diagonalise the Casimirs in \(F^{+}_{\gamma_2}\otimes V_{\lambda,\rho}\) first, and then, for each resulting eigenspace \(V_{\Lambda,\Rho}\), in \(F^{-}_{\gamma_1}\otimes V_{\Lambda,\Rho}\). We can distinguish \(3\) cases:
\begin{proofenum}
\item if \(\rho+\lambda\not\in(-2\gamma_2,2\gamma_2)\cap\Z\) the product \(F^{+}_{\gamma_2}\otimes V_{\lambda,\rho}\) admits a decomposition. The second decomposition exists if and only if, for each \(\nu\in\cM_{\gamma_2}\),
\begin{equation}
\rho+\nu-(\lambda+\nu)=\rho-\lambda\not\in(-2\gamma_1,2\gamma_1)\cap\Z.
\end{equation}
\item If \(\rho+\lambda\in(-2\gamma_2,2\gamma_2)\cap\Z\) but \(\rho-\lambda\not\in(-2\gamma_1,2\gamma_1)\cap\Z\)  the product \(F^{+}_{\gamma_2}\otimes V_{\lambda,\rho}\) is not decomposable, but we can  use the fact that \(F^{-}_{\gamma_1}\otimes F^{+}_{\gamma_2}\cong F^{+}_{\gamma_2}\otimes F^{-}_{\gamma_1}\) and decompose the product \(F^{-}_{\gamma_1}\otimes V_{\lambda,\rho}\) first. Following the same reasoning of the previous case, the product of each resulting submodule with \(F^+_{\gamma_2}\) will not be decomposable, as for each \(\nu\in\cM_{\gamma_1}\)
\begin{equation}
\rho+\nu+(\lambda-\nu)=\rho+\lambda\in(-2\gamma_2,2\gamma_2)\cap\Z.
\end{equation}
\item Finally, if \(\rho-\lambda\in(-2\gamma_1,2\gamma_1)\cap\Z\) and \(\rho+\lambda\in(-2\gamma_2,2\gamma_2)\cap\Z\), both \(F^{-}_{\gamma_1}\otimes V_{\lambda,\rho}\) and  \(F^{+}_{\gamma_2}\otimes V_{\lambda,\rho}\) are non-decomposable. The only results we have are on the product of \(F^A_\gamma\) with \emph{irreducible} modules, so we are not in a position to say anything in this case.
However, as we saw in the proof of \cref{prop:continuity}, the eigenvalues of \( \cC_1 \) and \( \cC_2 \) on \(F^{+}_{\gamma_2}\otimes V_{\lambda,\rho}\) are still respectively
\begin{equation}
\ii(\lambda+\nu)(\rho+\nu)\quad\mbox{and}\quad (\lambda+\nu)^2 + (\rho+\nu)^2 -1,\quad \nu\in\cM_\gamma,
\end{equation}
although they are not all distinct. Moreover, as each \( \cC_1,\cC_2 \)-eigenspace is \( 1 \)-dimensional, in each \((J_0,J^2) \)-eigenspace  \( V^J_M \) there is exactly one vector \( \ket{(\lambda+\gamma_2,\rho+\gamma_2)J,M} \) such that
\begin{subequations}
\begin{align}
\cC_1 \ket{(\lambda+\gamma_2,\rho+\gamma_2)J,M} &= \ii (\lambda+\gamma_2)(\rho+\gamma_2)\ket{(\lambda+\gamma_2,\rho+\gamma_2)J,M}
\\
\cC_2 \ket{(\lambda+\gamma_2,\rho+\gamma_2)J,M} &= \bracks*{ (\lambda+\gamma_2)^2 + (\rho+\gamma_2)^2 -1}\ket{(\lambda+\gamma_2,\rho+\gamma_2)J,M}.
\end{align}
\end{subequations}
It is easy to see that\footnote{We are assuming for simplicity, and without loss of generality, that \( \lambda\geq 0 \).}
\begin{equation}
\Span\set{\ket{(\lambda+\gamma_2,\rho+\gamma_2)J,M} \setst J\in \lambda+\gamma_2+\N_0, M\in \cM_J}
\end{equation}
behaves as a Lorentz group \((\mathfrak{g},K)\)-module under the action of \( \vec{J} \) and \( \vec{K} \), that is \(F^{+}_{\gamma_2}\otimes V_{\lambda,\rho}\), although not completely reducible, has at least one submodule \(V_{\lambda+\gamma_2,\rho+\gamma_2}\). Since the product \(F^-_{\gamma_1}\otimes V_{\lambda+\gamma_2,\rho+\gamma_2}\) is not decomposable\footnote{As \( (\lambda + \gamma_2)-(\rho+\gamma_2)=\lambda-\rho \).}, \((\gamma_1,\gamma_2)\otimes V_{\lambda,\rho}\supseteq F^-_{\gamma_1}\otimes V_{\lambda+\gamma_2,\rho+\gamma_2}\) will be indecomposable as well.
\end{proofenum}

The values of the Casimirs follow from \cref{prop:coupling}, and it can be checked explicitly that they need not be all distinct: for example, when \(\gamma_1=\gamma_2=1\), two possible pairs are \((\lambda,\rho)\) and \((\lambda-2,\rho)\), which are equivalent if \((\lambda,\rho)=(1,0)\).
\end{proof}

\subsubsection{Summary}

A \CG\ decomposition of the product \(F^A_\gamma\otimes V_{\lambda,\rho}\), with \( V_{\lambda,\rho} \) infinite\Hyphdash{}dimensional, is possible if and only if \(\rho+A\lambda\not\in(-2\gamma,2\gamma)\cap \Z\), with the modules in the decomposition having
\begin{equation}
(\Lambda,\Rho)\in\set{(\lambda+\nu,\rho+A\nu) \setst \nu\in\cM_\gamma};
\end{equation}
we can write this result in the compact form
\begin{equation}
F^A_\gamma \otimes V_{\lambda,\rho}=\bigboxplus_{\nu\in\cM_\gamma} V_{\lambda+\nu,\rho+A\nu}. 
\end{equation}
Likewise, the product \( (\gamma_1,\gamma_2)\otimes V_{\lambda,\rho} \) is decomposable if and only if \(\rho-\lambda\not\in(-2\gamma_1,2\gamma_1)\cap\Z\) and \(\rho+\lambda\not\in(-2\gamma_2,2\gamma_2)\cap\Z\), in which case
\begin{equation}
(\gamma_1,\gamma_2)\otimes V_{\lambda,\rho}=\bigboxplus_{\substack{\nu_1\in\cM_{\gamma_1}\\\nu_2\in\cM_{\gamma_2}}} V_{\lambda+\nu_1+\nu_2,\rho-\nu_1+\nu_2}.
\end{equation}

\subsection{\JS\ representation}\label{sec:4D_JS}

Just as in the \( 3 \)D case, we can use the \WE\ theorem to generalise the \( \SU(2) \) \JS\ representation to infinite-dimensional \(\Spin(3,1)\) representations. Recall that the \( \su(2)_\C \) generators can be rewritten, when acting on unitary irreducible \( \SU(2) \) representations, as
\begin{equation}
J_0=\tfrac{1}{2}\paren{a\Dagger a - b\Dagger b },\quad J_+=a\Dagger b,\quad J_-=b\Dagger a;
\end{equation}
the extension of this result to finite-dimensional \(\Spin(3,1)\) representations trivially follows from the fact that \(\spin(3,1)_\C\cong \su(2)_\C \oplus \su(2)_\C\) (see \cref{sec:4dlorentz_representation_theory}). A generalisation to infinite-dimensional \((\mathfrak{g},K)\)-modules can be obtained by making use of tensor operators as follows.
\begin{proposition}
Let \(\vec{M}^A=\half(\vec{J}-\ii A \vec{K})\), \(A=\pm1\) be the generators of \(\su(2)_\C\oplus\su(2)_\C\cong\spin(3,1)_\C\). There exist four tensor operators
\[
T^A:F^A_{\shalf}\otimes V_{\lambda,\rho}\rightarrow V_{\lambda-\shalf,\rho-\shalf[A]},
\quad
\widetilde{T}^A:F^A_{\shalf}\otimes V_{\lambda,\rho}\rightarrow V_{\lambda+\shalf,\rho+\shalf[A]},
\quad A=\pm 1,
\]
where \(V_{\lambda,\rho}\) is an arbitrary infinite-dimensional \((\mathfrak{g},K)\)-module with
\begin{equation}
\rho\neq\pm\lambda,\quad \rho\neq\pm(\lambda+1),
\end{equation}
such that
\[
M_0^A=-\half \big(T^A_-\widetilde{T}^A_+ + T^A_+\widetilde{T}^A_-\big),\quad M_\pm^A=\pm T^A_\pm \widetilde{T}^A_\pm,
\]
when acting on \( V_{\lambda,\rho} \), where
\[
T^A_\pm\ket{(\lambda,\rho)j,m}:=T^A\ket{\half_A,\pm\half}\otimes\ket{(\lambda,\rho)j,m}.
\]
Their action on \( V_{\lambda,\rho} \) is
\begin{subequations}
\label{eq:JS_matrix}
\begin{align*}
T^A_\pm\ket{(\lambda,\rho)j,m} =& \pm\frac{\sqrt{j\mp m}\sqrt{j+\lambda}\sqrt{j+A\rho}}{\sqrt{2j}\sqrt{2j+1}}\ket{(\lambda-\half,\rho-\half[A])j-\tfrac{1}{2},m\pm\tfrac{1}{2}}
\\
&+ \ii A\frac{\sqrt{j\pm m+1}\sqrt{j-\lambda+1}\sqrt{j-A\rho+1}}{\sqrt{2j+1}\sqrt{2j+2}} \ket{(\lambda-\half,\rho-\half[A])j+\tfrac{1}{2},m\pm\tfrac{1}{2}}
\\\addlinespace
\widetilde{T}^A_\pm \ket{(\lambda,\rho)j,m} =& \mp\ii A\frac{\sqrt{j\mp m}\sqrt{j-\lambda}\sqrt{j-A\rho}}{\sqrt{2j}\sqrt{2j+1}} \ket{(\lambda+\half,\rho+\half[A])j-\tfrac{1}{2},m\pm\tfrac{1}{2}}
\\
&+ \frac{\sqrt{j\pm m+1}\sqrt{j+\lambda+1}\sqrt{j+A\rho+1}}{\sqrt{2j+1}\sqrt{2j+2}} \ket{(\lambda+\half,\rho+\half[A])j+\tfrac{1}{2},m\pm\tfrac{1}{2}},
\end{align*}
\end{subequations}
and they satisfy the commutation relations
\begin{equation*}
\label{eq:JS_commutation}
[T^A_+,\widetilde{T}^B_-]=[\widetilde{T}^A_+,T^B_-]=\delta^{AB},\quad [T^A_\mu,T^B_\nu]=[\widetilde{T}^A_\mu,\widetilde{T}^B_\nu]=0.
\end{equation*}
\end{proposition}
\begin{proof}
Consider the tensor operators \(T^A\), \(\widetilde{T}^A\) described above. As a consequence of the \WE\ theorem, it must be
\begin{subequations}
\begin{align}
\begin{multlined}[c][0.8\textwidth]
T^A_\mu\ket{(\lambda,\rho)j,m} =  t^A(\lambda,\rho)\sum_{J=j-\shalf}^{j+\shalf} \mathrm{B}\{(\lambda-\half,\rho-\half[A])J|\half_A;(\lambda,\rho) j \}
\\
\times \braket{J,M|\half,\mu;j,m} \ket{(\lambda-\half,\rho-\half[A])J,M}
\end{multlined}
\\\addlinespace
\begin{multlined}[c][0.8\textwidth]
\widetilde{T}^A_\mu\ket{(\lambda,\rho)j,m} =  \widetilde{t}^A(\lambda,\rho) \sum_{J=j-\shalf}^{j+\shalf} \mathrm{B}\{(\lambda+\half,\rho+\half[A])J|\half_A;(\lambda,\rho) j \}
\\
\times \braket{J,M|\half,\mu;j,m} \ket{(\lambda+\half,\rho+\half[A])J,M},
\end{multlined}
\end{align}
\end{subequations}
with \(t^A\), \(\widetilde{t}^A\) arbitrary functions of \(\lambda\) and \(\rho\).
Let now
\begin{equation}
V^A_0:=-\sqrt{2}M^A_0,\quad V^A_{\pm 1}:=\pm M^A_\pm;
\end{equation}
one can check that they are the components in the basis \(\ket{1_A,\mu}\) of a tensor operator \(V^A:F^A_1\otimes V_{\lambda,\rho}\rightarrow V_{\lambda,\rho}\); in fact
\begin{equation}
\commutator{M_0^B,V^A_\mu}=\mu\delta_{AB}V^A_\mu,\quad \commutator{M_\pm^B,V^A_\mu}=C_\pm(1,\mu)\delta_{AB}V^A_{\mu\pm 1}.
\end{equation}
Suppose, as an ansatz, that
\begin{equation}
\label{eq:ansatz}
V_\mu^A=\sum_{\mu_1\in\cM_{\shalf}}\sum_{\mu_2\in\cM_{\shalf}}\braket{\half,\mu_1;\half,\mu_2|1,\mu}T^A_{\mu_1}\widetilde T^A_{\mu_2};
\end{equation}
it is a standard result for \(\SU(2)\) tensor operators\footcite[chap.~9]{barut}\footnote{Remember that \(F^A_\gamma\) is also an \(\SU(2)\) representation.} that the RHS is indeed the \(\mu\) component of a tensor operator transforming like \(F^A_1\), so that the ansatz is consistent. Evaluating the \CG\ coefficients, one can rewrite~\eqref{eq:ansatz} as
\begin{equation}
M_0^A=-\half \big(T^A_-\widetilde{T}^A_+ + T^A_+\widetilde{T}^A_-\big),\quad M_\pm^A=\pm T^A_\pm \widetilde{T}^A_\pm.
\end{equation}

Comparing the possible matrix elements of both sides of~\eqref{eq:ansatz} one can explicitly check that they agree, i.e., the ansatz is verified, if and only if\footnote{Notice that the RHS is non-zero if and only if \( \rho\neq \pm\lambda\) and \(\rho\neq\pm(\lambda+1) \).}
\begin{equation}
t^A(\lambda+\half,\rho+\half[A])\,\widetilde{t}^A(\lambda,\rho)=\sqrt{\lambda+A\rho}\sqrt{\lambda+A\rho+1}\neq 0.
\end{equation}
We choose here the particular solution
\begin{equation}
t^A(\lambda,\rho)=\widetilde{t}^A(\lambda,\rho)=\sqrt{\lambda+A\rho};
\end{equation}
with this choice we recover the required matrix elements\footnote{Using the \CG\ coefficients from \cref{tab:4d-1/2}.} and, after some  tedious but simple calculations, the required commutation relations.
\end{proof}
Note that the matrix elements of the \( T^A \), \( \widetilde T^A \) operators satisfy
\begin{equation}
\praket{(\lambda-\half,\rho-\half[A])j-\tfrac{1}{2},m\pm\tfrac{1}{2}|T^A_\pm|(\lambda,\rho)j,m} =\pm \praket{(\lambda,\rho)j,m|\widetilde T^A_\pm|(\lambda-\half,\rho-\half[A])j-\tfrac{1}{2},m\pm\tfrac{1}{2}}
\end{equation}
and
\begin{equation}
\praket{(\lambda-\half,\rho-\half[A])j+\tfrac{1}{2},m\pm\tfrac{1}{2}|T^A_\pm|(\lambda,\rho)j,m} =\mp \praket{(\lambda,\rho)j,m|\widetilde T^A_\pm|(\lambda-\half,\rho-\half[A])j+\tfrac{1}{2},m\pm\tfrac{1}{2}},
\end{equation}
and, like in the \( 3 \)D case for continuous representations, they are never always real or always imaginary if \( V_{\lambda,\rho} \) is infinite-dimensional, so the components of the tensor operators are not harmonic oscillators. Nevertheless, the commutation relations from \cref{eq:JS_commutation} are still those of the Lie algebra \(\mathfrak{h}_2(\R)_\C\oplus \mathfrak{h}_2(\R)_\C\), the same as the finite-dimensional case; in the infinite-dimensional case, however, since \(\vec{M}^A\) does \emph{not} act on \(V_{\lambda,\rho}\) as a unitary \(\su(2)\) representation, the \(T^A\), \(\widetilde{T}^A\) (with \(A\) fixed) operators will not act unitarily as a Heisenberg algebra either. This result is analogous to the one for the continuous series in \cref{sec:3D_JS}, and was similarly unknown until now. Let us emphasise that, unlike the case of \( \Spin(2,1) \), in the \( 4 \)D case there is no discrete series, so that this is the first version of the \JS\ representation that works for unitary representations.

\section{Concluding remarks}

We have seen in this chapter how the well-known \WE\ theorem admits a simple generalisation to arbitrary Lie groups, possibly non-compact. Despite the simplicity of the proof, it is still important to remember that, to actually gain any useful information from the theorem, it is necessary to know which representations appear in the \CG\ decomposition of the product of the representation the tensor operator transforms as and the representation it acts on, as well as the \CG\ coefficients themselves for the explicit values of the matrix elements. When the representation acted on is infinite-dimensional---as it happens when the group in non-compact---not only are these results not known in general, but as we have seen they are not easy to obtain. We have studied the particular cases of \( \Spin(2,1) \) and \( \Spin(3,1) \) for their potential applications to physics, an example of which we will see in \cref{sec:lorentzian_LQG}, and in the hope that the techniques we used will prove useful to investigate more complicated cases.

Regarding the Jordan-Schwinger representation, we have discovered the new result that, even when the representation is in the continuous principal series, it is possible to express the algebra generators in terms of two spinor operators, which generalise the harmonic oscillators. Although for continuous representations the spinors are not harmonic oscillators anymore, it is interesting that they still have the commutation relations of a Heisenberg algebra.

\newcommand{\Hom}{\operatorname{Hom}}
\newcommand{\racah}[6]{
{\arraycolsep=0.7\arraycolsep\ensuremath
\begin{bmatrix}
#1 & #2 & #3 \\
#4 & #5 & #6
\end{bmatrix}
}}

\chapter{Spinor operators in \texorpdfstring{\(3\)D}{3D} Lorentzian loop quantum gravity}\label{sec:lorentzian_LQG}

With this chapter we start investigating the applications of non-compact groups to quantum gravity; in particular, we will construct a model of \( 3 \)D Lorentzian loop quantum gravity, and make use of the results of \cref{sec:3d-lorentz-group}---particularly the \JS\ representation---to implement the Lorentzian version of the spinorial framework used in the Euclidean case. Our main goal is to generalise to Lorentzian signature the results of \cite{Bonzom:2011nv}, where the spinorial framework is used in the \( 3 \)D Euclidean case (with \( \SU(2) \) as gauge group) to construct a solvable Hamiltonian constraint. In the Lorentzian case the gauge group is given by \( \Spin(2,1) \); as we should expect by now, the treatment will be considerably more complicated than the compact case.

We will first work at the classical level. In \cref{sec:classical-tensors} we will define classical tensors, which are essentially the equivalent of tensor operators for Poisson algebras; in particular we will use  classical spinors to obtain a classical analogue of the \JS\ representation. In \cref{sec:classical_LQG} we will then use the spinors to construct a set of observables, which we will use to express the classical Hamiltonian constraint. The rest of the chapter is dedicated to the study of the quantised model. We start by constructing the space of quantum states in \cref{sec:relativistic-spin-networks}; in \cref{sec:lorentzian-ponzano--regge-model} we will then quantise the classical Hamiltonian constraint, and we will show that the Lorentzian Ponzano--Regge amplitude\footcite{Davids:2000kz,Freidel:2000uq}, given by the Racah coefficient, is in its kernel.
Finally, we will see in \cref{sec:relationship-with-su2-theory} that our formalism is general enough that it can be used to cover the Euclidean case as well.

The content of this chapter is based on the results presented in the article \cite{Girelli:2015ija}.

\section{Classical tensors and tensor operators for \texorpdfstring{\(\SU(1,1)\)}{SU(1,1)}}\label{sec:classical-tensors}

This section focuses on the notion of \emph{classical tensors}. We will first define these quantities, then show how, upon quantisation, they become the tensor operators we defined in \cref{sec:tensor_operators}.

\subsection{Classical tensors} \label{sec:classical_tensors}

Classical \( \Spin(2,1) \) tensors are the Poisson analogue of the tensor operators we defined in \cref{sec:tensor_operators}. Explicitly, a tensor is a set of functions \( \tau^\gamma_\mu \) that transform as the vectors in a \( \Spin(2,1) \) representation, where the infinitesimal \( \Spin(2,1) \) action is implemented as a Poisson bracket\footnote{We follow here a ``canonical dequantisation'' procedure, i.e., we replace \( [\cdot,\cdot] \) with \( \ii\braces{\cdot,\cdot} \).}, i.e.,
\begin{equation}
\label{eq:tensor}
\braces{x_0, \tau^\gamma_\mu} = -\ii \mu \tau^\gamma_\mu, \quad \braces{x_\pm, \tau^{\gamma}_\mu} = -\ii \Gamma_\pm(\gamma,\mu) \tau^\gamma_{\mu\pm 1}.
\end{equation}
The Poisson structure on \( \R^3 \) analogous to the \( \spin(2,1) \) commutation relations is given by
\begin{equation}
\braces{x_0,x_\pm} = \pm \ii x_\pm, \quad \braces{x_+,x_-}=2\ii x_0,
\end{equation}
where the algebra is parametrised by \( x_0 \in \R\), \( x_\pm \in \C \) with \( x_- = \conj x_+ \).
We will only consider tensors transforming like finite-dimensional representations: as we will see they are the only ones that can be contracted together to obtain \( \Spin(2,1) \)-invariant quantities. We will call respectively \emph{vectors}, \emph{spinors} and \emph{scalars} the tensors transforming like \( F_1 \), \( F_{\shalf} \) and \( F_0 \).
We will also define \emph{contravariant tensors}, i.e., tensors \( \tau^{\gamma\Star} \) transforming as the \emph{dual representation} \( F_\gamma\Star \). Recall that the Lie algebra acts on the dual space \( F_\gamma\Star \) as
\begin{equation}\label{eq:dual1}
X \bra{\gamma,\mu} := -\bra{\gamma,\mu} X,\quad X\in\spin(2,1)_\C.
\end{equation}
One can easily show that \( F_\gamma\Star \cong F_\gamma \) as representations, with the isomorphism given by
\begin{equation}\label{eq:dual2}
\varphi_\gamma:\bra{\gamma,\mu}\in F_\gamma\Star \mapsto (-1)^{\gamma-\mu} \ket{\gamma,-\mu} \in F_\gamma;
\end{equation}
consequently, we define the components of the tensor dual to \( \tau^\gamma \) as
\begin{equation}\label{eq:contravariant_tensor}
\tau^{\gamma\Star}_\mu := (-1)^{\gamma-\mu}\tau^{\gamma}_{-\mu},
\end{equation}
which satisfy
\begin{equation}
\braces{x_0, \tau^{\gamma\Star}_\mu} = \ii \mu \tau^{\gamma\Star}_\mu, \quad \braces{x_\pm, \tau^{\gamma\Star}_\mu} = \ii \Gamma_\mp(\gamma,\mu) \tau^{\gamma\Star}_{\mu\mp 1}.
\end{equation}

Analogously to tensor operators, tensors can be composed to obtain new tensors  using the Clebsch--Gordan coefficients; in fact\footcite[chap.~9]{barut},
\begin{equation}\label{eq:concatenation}
\tau^\gamma_\mu=\sum_{\mu_1,\mu_2}A(\gamma_1,\mu_1;\gamma_2,\mu_2|\gamma,\mu)\,\tau^{\gamma_1}_{\mu_1}\tau^{\gamma_2}_{\mu_2}
\end{equation}
is the \( \mu \) component of a tensor transforming as \( F_\gamma \), as long as \( F_\gamma\subseteq F_{\gamma_1} \otimes F_{\gamma_1} \).
We can use this fact to construct \( x_0 \), \( x_\pm \) out of two spinors, retracing our steps from \cref{sec:3D_JS} (\JS\ representation),
i.e., we consider two spinors
\(
\widetilde{\tau}=
\begin{smallpmatrix}
\widetilde{\tau}_-\\\widetilde{\tau}_+
\end{smallpmatrix}
\), 
\(
\tau=
\begin{smallpmatrix}
\tau_-\\\tau_+
\end{smallpmatrix}\) such that
\begin{equation}
x_0=-\half (\tau_-\widetilde\tau_+ + \tau_+\widetilde\tau_-),\quad x_\pm=\pm\ii \tau_\pm \widetilde\tau_\pm,
\end{equation}
with
\begin{equation}
\braces{\tau_+,\widetilde\tau_-}=\braces{\widetilde\tau_+,\tau_-}=-\ii
\end{equation}
and all other Poisson brackets vanishing.
At this stage we are working with a symplectic structure on \( \C^4 \), which we have to reduce by imposing the reality constraints \( \conj x_0=x_0 \), \( \conj x_+=x_- \); two natural choices to implement these constraints are
\begin{equation}\label{constraints}
{\widetilde{\tau}}_\pm = \conj\tau_\mp \quad\mbox{and}\quad  {\widetilde{\tau}}_\pm = -\conj\tau_\mp,
\end{equation}
which reduce \( \C^4 \) to \( \C^2 \) equipped with the canonical symplectic form.

We can concatenate the  spinors to form scalars using \eqref{eq:concatenation}: using the \CG\ coefficients
\begin{equation}
A(\half,\mu_1;\half,\mu_2|0,0)=  \frac{(-1)^{\shalf-\mu_1}}{\sqrt{2}} \delta_{\mu_1+\mu_2,0},
\end{equation}
we define a bilinear form
\begin{equation}\label{eq:bilinear_form}
\mathcal{B}(\sigma,\tau):=-\sqrt{2}\sum_{\mu_1,\mu_2}A(\half,\mu_1;\half,\mu_2|0,0)\sigma_{\mu_1} \tau_{\mu_2}= \sigma_- \tau_+ - \sigma_+ \tau_-,
\end{equation}
which assigns a scalar to two spinors \( \sigma \), \( \tau \).

It will be useful to introduce a bra--ket notation for the spinors. We define
\begin{equation}
\ket{\tau}:= \tau,\quad \ketb{\tau}:= \widetilde\tau 
\end{equation}
and
\begin{equation}
\bra{\tau} := \mathcal{B}(\widetilde\tau,\cdot) = (-\widetilde\tau_+,\widetilde\tau_-), \quad \brab{\tau} := \mathcal{B}(\tau,\cdot)= (-\tau_+,\tau_-);
\end{equation}
note that \( \bra{\tau} \) and \( \brab{\tau} \) are respectively the dual spinors of \( \ketb{\tau} \) and \( \ket{\tau} \). With this notation we can write the \( \spin(2,1) \) generators in the compact form
\begin{equation}\label{flux}
x_a=\half\braket{\tau|\sigma_a|\tau},
\end{equation}
where
\begin{equation}
x_1:= \frac{x_+ + x_-}{2},\quad x_2:= \frac{x_+ - x_-}{2\ii}
\end{equation}
and
\begin{equation}
\sigma_0 =
\begin{pmatrix}
1 & 0 \\ 0 & -1
\end{pmatrix},\quad
\sigma_1 =
\begin{pmatrix}
0 & -\ii \\ -\ii & 0
\end{pmatrix},\quad
\sigma_2 =
\begin{pmatrix}
0 & -1 \\ -1 & 0
\end{pmatrix}
\end{equation}
are the equivalent of the Pauli matrices.

An interesting feature of the spinors is the possibility to use them to construct \( \SU(1,1)\cong \Spin(2,1) \) group elements
\begin{equation}
g=
\begin{pmatrix}
\alpha&\beta\\ \conj{\beta}&\conj{\alpha}
\end{pmatrix}
\quad |\alpha|^2-|\beta|^2=1.
\end{equation}
using tensor products of spinors and contravariant spinors; explicitly, we introduce another pair of spinors \( w \), \( \widetilde w \) and define\footnote{One should note that this definition differs from the one presented in \cite{Girelli:2015ija}, as an inconsistency was later discovered in the old definition. The rest of the content in the chapter has also been adapted to fit the change.}
\begin{equation}\label{eq:holonomy}
g= \frac{\ket{w} \bra{\tau}- \ketb{w} \brab{\tau}}{\sqrt{\braket{\tau|\tau}\braket{w|w}}} = \frac{1}{\sqrt{\braket{\tau|\tau}\braket{w|w}}}
\begin{pmatrix}
\widetilde{w}_-{\tau}_+ - w_- \widetilde\tau_+& - \widetilde w_-\tau_- + { w}_-\widetilde{\tau}_-\\ 
\widetilde{w}_+{\tau}_+ - {w}_+ \widetilde\tau_+ & - \widetilde{w}_+\tau_- + {w}_+\widetilde{\tau}_-
\end{pmatrix},
\end{equation}
with the normalisation factor ensuring that \( \det(g)=1 \). We also require that \( g_{11}= \conj g_{22} \) and \( g_{12}= \conj g_{21} \); one can easily check that these conditions are satisfied when using one of the constraints from \eqref{constraints} for the spinors.
The inverse group element is given by 
\begin{equation}
g^{-1} = \frac{\ket{\tau}\bra{w} - \ketb{\tau}\brab{w}}{\sqrt{\braket{\tau|\tau}\braket{w|w}}}. 
\end{equation}
Note that \( g \) acts on spinors by interchanging \( \tau \) with \( w \); in fact, introducing the \emph{matching constraint}\footcite[Following][]{Bonzom:2011nv}
\begin{equation}\label{eq:matching}
 \braket{ w|w} = \braket{\tau|\tau},
\end{equation}
we have 
\begin{equation}
\label{eq:g_action}
\begin{cases}
g \ket{\tau} = \ket{w}, & g \ketb{\tau}=\ketb{w},\\
\bra{w}g=\bra{\tau}, & \brab{w}g=\brab{\tau}
\end{cases}
\quad \mbox{and} \quad
\begin{cases}
g^{-1} \ket{w} = \ket{\tau}, & g^{-1}\ketb{w}=\ketb{\tau},\\
\bra{\tau}g^{-1}=\bra{w}, & \brab{\tau}g^{-1}=\brab{w}.
\end{cases}
\end{equation}
We will see in \cref{sec:classical_LQG} that we can use this fact to interpret the group element \( g \) as the \emph{parallel transport} between the spinors on the edge of a graph.

\subsection{Quantisation of classical tensors}\label{sec:holo}

Let us now consider the quantisation of the phase space we constructed. The \( \spin(2,1) \) generators become the operators \( J_0 \), \( J_\pm \) acting on an irreducible representation; the reality constraints \( \conj x_0 = x_0 \) and \( \conj x_+ = x_- \)
are quantised as
\begin{equation}
J\Dagger_0 = J_0,\quad J\Dagger*_+=J_-,
\end{equation}
i.e., the representation is unitary.
If we quantise the spinors as
\begin{equation}
\tau_\pm \rightarrow T_\pm,\quad \widetilde{\tau}_\pm \rightarrow \widetilde{T}_\pm
\end{equation}
and the Poisson brackets as \( \braces{\cdot,\cdot}\rightarrow -\ii\bracks{\cdot,\cdot} \), we obtain the tensor operators we defined when we treated the \JS\ representation in \cref{sec:3D_JS}, satisfying
\begin{equation}
[T_+,\widetilde{T}_-]=[\widetilde T_+,T_-]=\1
\end{equation}
and such that
\begin{equation}
J_\pm=\pm\ii T_\pm \widetilde T_\pm,\quad J_0=\half(T_- \widetilde T_+ + T_+ \widetilde T_-);
\end{equation}
we will thus henceforth refer to them as \emph{spinor operators}.

Recall that we have a constraint on the spinors imposed by the reality conditions. A priori, we have a choice: we can first implement the reality constraints, then quantise, or alternatively first quantise and then implement a quantum version of the reality constraints. The quantisation of the two natural reality constraints \eqref{constraints}
\begin{equation}\label{natural q constraint}
\widetilde\tau_\pm=
\begin{cases}
-\conj \tau_\mp\\
\conj \tau_\mp 
\end{cases}  \rightarrow\quad  T_\pm=
\begin{cases}
-\widetilde{T}\Dagger*_\mp \\
\widetilde{T}\Dagger*_\mp.
\end{cases}
\end{equation}
leads to the spinor operators acting on the discrete series \( D^\pm_j \) as harmonic oscillators, as we saw in \cref{sec:3D_JS}. However, there is no natural reality condition at the classical level that upon quantisation leads to an action on the continuous series, as in this case
\begin{equation}
\praket{j+\tfrac{1}{2},m\pm \tfrac{1}{2}|\widetilde T_\pm|j,m}=\mp\praket{j,m|{T}_\mp|j+\tfrac{1}{2},m\pm \tfrac{1}{2}},
\end{equation}
i.e., in some sense\footnote{The notion of \emph{transpose} here is basis dependent: there is no guarantee that the matrix elements in a different basis satisfy the same conditions.}, \( \widetilde T_\pm = \mp T\Transpose*_\mp \), which does not have a classical analogue.
Despite this, the quantum constraints \( J_0
\Dagger = J_0 \) and \( J\Dagger*_+=J_- \) are still satisfied. For this reason, we will adopt the second quantisation scheme (first quantise, then implement the reality constraints), which allows us to have continuous representations at the quantum level; in other words, we will treat \( \tau \) and \( \widetilde \tau \) as independent variables until we quantise them.

The quantisation of the spinors \( \ket{w} \) and \( \ketb{w} \) appearing in the spinorial description of the \( \SU(1,1) \) group elements is analogous, i.e.,
\begin{equation}
w_\pm\rightarrow T\pm,\quad \widetilde w_\pm \rightarrow \widetilde T_\pm,
\end{equation}
with the difference that we will have them act on covectors (bras).
It should be noted that there is an ambiguity in the quantisation of \( \braket{\tau|\tau} \) and \( \braket{w|w} \), as the operators appearing in their naïve quantisation do not commute; we will choose the \emph{symmetric} quantisation
\begin{equation}
\widetilde \tau_- \tau_+ - \widetilde \tau_+ \tau_- \rightarrow \half \paren{ \widetilde T_- T_+ - \widetilde T_+ T_- +  T_+ \widetilde T_- - T_- \widetilde T_+}\equiv \widetilde T_- T_+ - \widetilde T_+ T_- + \1=:E,
\end{equation}
which has the double advantage of regularising the denominator of the (quantised) group element when \( j=0 \) and, as we will see later, closing the Lie algebra of scalar operators we can build from the spinor ones.
Note that with this choice the quantised versions of \( \braket{\tau|\tau} \) and \( \braket{w|w} \) satisfy respectively\footnote{Using the action of the spinor operators from \cref{sec:3D_JS}.}
\begin{equation}
E\ket{j,m}=(2j+1)\ket{j,m}\quad\mbox{and}\quad E\bra{j,m}=(2j+1)\bra{j,m},
\end{equation}
which is the quantum version of the matching constraint \eqref{eq:matching}.

\section{Classical description of Lorentzian \texorpdfstring{\(3\)D}{3D} loop quantum gravity}\label{sec:classical_LQG}

\begin{figure}[h]
\centering
\tikzsetnextfilename{flux}
\begin{tikzpicture}[x=4em, line width=1.1pt,inner sep=0.5em]
\coordinate (c1) at (0,0);
\coordinate (c2) at (2,0);
\fill[] (c1) circle[radius=2.2pt];
\fill[] (c2) circle[radius=2.2pt];
\draw[midarrow] (c1) -- (c2);
\node[anchor=south] at ($(c1)!0.5!(c2)$) {\( g \)};
\coordinate (v1) at ($(c1)+(-0.15,-0.9)$);
\coordinate (v2) at ($(c2)+(0.15,-0.9)$);
\draw[endarrowsmall,line width=0.8pt] (c1) -- (v1);
\draw[endarrowsmall,line width=0.8pt] (c2) -- (v2);
\node[anchor=west] at ($(c1)!0.5!(v1)$) {\( \vec{X} \)};
\node[anchor=east] at ($(c2)!0.5!(v2)$) {\( \widetilde{\vec{X}} \)};
\draw[endarrowsmall,line width=0.8pt] (2.5,0) -- (3,0);
\coordinate (c3) at (3.5,0);
\coordinate (c4) at (5.5,0);
\fill[] (c3) circle[radius=2.2pt];
\fill[] (c4) circle[radius=2.2pt];
\draw[midarrow] (c3) -- (c4);
\node[anchor=south] at ($(c3)!0.5!(c4)$) {\( g \)};
\node[anchor=south west,inner sep=0pt,yshift=0.5em] at (c3) {\( \ket{\tau} \)};
\node[anchor=north west,inner sep=0pt,yshift=-0.5em] at (c3) {\( \ketb{\tau} \)};
\node[anchor=south east,inner sep=0pt,yshift=0.5em] at (c4) {\( \ket{w} \)};
\node[anchor=north east,inner sep=0pt,yshift=-0.5em] at (c4) {\( \ketb{w} \)};
\path ($(current bounding box.south west)+(-1ex,-0.5ex)$) rectangle ($(current bounding box.north east)+(1ex,0)$);

\end{tikzpicture}
\caption{The information about fluxes is now encoded by a pair of spinors.}\label{fig:flux}
\end{figure}
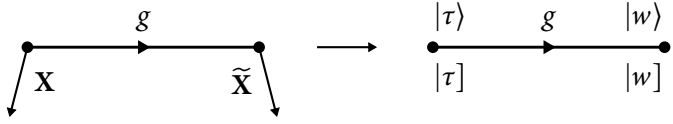

We recall now the standard construction of the loop quantum gravity phase space\footcite[See for example][]{Bonzom:2011nv}, specialising it to the \(\Spin(2,1)\) case. The triad and connection \((e,\omega)\) are discretised into the holonomy and flux variables \((g,\vec{X})\in T\Star\Spin(2,1)\). More precisely, we consider a graph \(\Gamma\) and to  each edge \(e\) we associate two fluxes \( \vec{X} \), \( \widetilde{\vec{X}} \) and a group element \( g \); the fluxes sit respectively at the \emph{source} and \emph{target} vertex, and the group element parallel transports  \( \vec{X} \) to \(\widetilde{\vec{X}} \) (\cref{fig:flux}). 
The idea behind the spinorial framework is to  replace the fluxes and holonomies  by attaching a pair of spinors \(\ket{\tau}\), \(\ketb{\tau}\) at each vertex. For each edge the two pairs of spinors provide the full information about \(T\Star\Spin(2,1)\), since we can reconstruct from them both the flux and the holonomy\footnote{see \eqref{flux} and \eqref{eq:holonomy}.}.

The dynamics of gravity is encoded by two constraints, the Gauß constraint and the flatness constraint. The Gauß constraint is discretised at the vertices of \(\Gamma\), and corresponds to an (infinitesimal) \(\Spin(2,1)\) invariance at the vertex; due to the proportionality between the fluxes and the \(\su(1,1)\) generators, this invariance can be interpreted as the requirement that the total flux at each vertex is zero\footnote{One should be aware that this is merely a coincidence: when dealing with quantum groups, the invariance under the quantum group cannot be interpreted as the requirement that the equivalent of the fluxes sum to zero. See \cite{Bonzom:2014bua}.}, i.e.,
\begin{equation}
\sum_i  \vec{X}_i = 0.
\end{equation}
Given a vertex \(v\), we can construct a set of functions which  commute with the Gauß constraint, and as such they will be called observables. They are defined in terms of  the spinors living on different legs of the vertex in such a way that they are \(\Spin(2,1)\) invariant; these functions are
\begin{equation}
\label{eq:classical_observables}
\begin{aligned}
f_{ab} &:= \mathcal{B}(\tau_a,\tau_b) = \braketlb{\tau_a|\tau_b},
&\widetilde{f}_{ab} &:= \mathcal{B}(\widetilde{\tau}_a,\widetilde{\tau}_b) = \braketrb{ \tau_a|\tau_b},\\
e_{ab} &:= \mathcal{B}(\widetilde{\tau}_a,\tau_b) = \braket{ \tau_a|\tau_b},
\quad &\widetilde{e}_{ab} &:= \mathcal{B}(\tau_a,\widetilde{\tau}_b) = \braketbb{ \tau_a|\tau_b}=-e_{ba}.
\end{aligned}
\end{equation}  
The observables \(f_{ab}\) and \(\widetilde{f}_{ab}\)  are not all independent when reality conditions are implemented:  for example, if we use either \(\widetilde{\tau}_{\pm}=-\conj{\tau}_\mp\) or \(\widetilde{\tau}_{\pm}=\conj{\tau}_\mp\) on both of the legs \(a\) and \(b\), we get that \(\widetilde{f}_{ab}= \conj f_{\!ab}\). If instead we use \(\widetilde{\tau}_{\pm}=-\conj{\tau}_\mp\) on leg \(a\) and  \(\widetilde{\tau}_{\pm}=\conj{\tau}_\mp\) on leg \(b\) (or vice versa), we get \(\widetilde{f}_{ab}= -\conj f_{\!ab}\).
The functions \(e\), \(f\) and \(\widetilde{f}\) satisfy the closed Poisson relations
\begin{subequations}
\begin{align}
\braces{e_{ab}, e_{cd}} &= -\ii(\delta_{cb} e_{ad} -\delta_{ad} e_{cb})
\\
\braces{e_{ab}, f_{cd}} &= -\ii (\delta_{ad} f_{bc} - \delta_{ac} f_{bd})
\\
\braces{e_{ab}, \widetilde{f}_{cd}} &=- \ii (\delta_{bc} \widetilde{f}_{ad}-\delta_{bd} \widetilde{f}_{ac})
\\
\braces{f_{ab}, \widetilde{f}_{cd}}&= -\ii \paren{\delta_{db}e_{ca}+\delta_{ca}e_{db}-\delta_{cb}e_{da}-\delta_{da}e_{cb} }
\\
\braces{f_{ab}, f_{cd}}&= \braces{\widetilde{f}_{ab}, \widetilde{f}_{cd}} =0.
\end{align}
\end{subequations}
These quantities are equivalent to the spinorial observables of the Euclidean case\footnote{These will be analysed in detail in \cref{chap:so*}, where we will work in the Euclidean regime.}, and in fact satisfy the same Poisson algebra: the only difference is in the choice of real structure, i.e., which variables are conjugate to each other.
As it is well-known in the Euclidean case, we can use these observables to generate all the standard  LQG observables.

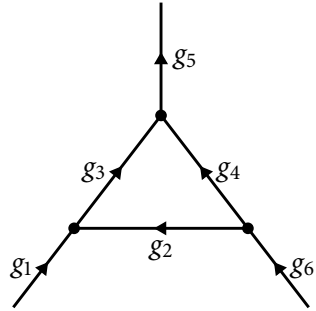
\begin{figure}[h]
\centering
\tikzsetnextfilename{flatness}
\begin{tikzpicture}[x=3em,y=3em,line width=1.1pt]
\coordinate (v123) at (0,0);
\coordinate (v246) at (2,0);
\coordinate (v345) at (1,1.3);
\coordinate (v1) at (-0.7,-0.91);
\coordinate (v6) at (2.7,-0.91);
\coordinate (v5) at (1,2.6);
\fill (v123) circle[radius=2.2pt];
\fill (v246) circle[radius=2.2pt];
\fill (v345) circle[radius=2.2pt];
\draw[midarrow] (v246) -- (v123);
\draw[midarrow]  (v246) -- (v345);
\draw[midarrow]  (v123) -- (v345);
\draw[midarrow]  (v345) -- (v5);
\draw[midarrow]  (v1) -- (v123);
\draw[midarrow]  (v6) -- (v246);
\node[anchor=east] at ($(v123)!0.5!(v1)$) {\( g_1 \)};
\node[anchor=north] at ($(v123)!0.5!(v246)$) {\( g_2 \)};
\node[anchor=east] at ($(v123)!0.5!(v345)$) {\( g_3 \)};
\node[anchor=west] at ($(v246)!0.5!(v345)$) {\( g_4 \)};
\node[anchor=west] at ($(v5)!0.5!(v345)$) {\( g_5 \)};
\node[anchor=west] at ($(v6)!0.5!(v246)$) {\( g_6 \)};
\end{tikzpicture}
\caption{The flatness constraint on the triangular face is \(g_2 g_4^{-1} g_3=\1\).}\label{fig:flatness} 
\end{figure}

The flatness constraint is discretised by requiring that the product of the holonomies around each face \(f\) of the graph is the identity, i.e.,
\begin{equation}\label{c1}
\prod_{e\in f} g_e = \1.
\end{equation}
In the Euclidean case it was discovered that this constraint can be recast in a natural constraint involving either the fluxes\footcite{Bonzom:2011hm} or the spinors\footcite{Bonzom:2011nv}, according to the initial choice of variables on the graph. In the spinorial framework, one essentially projects the flatness constraints on the basis provided by the spinors, to obtain a set of scalar constraints. The physical interpretation is that the scalar product of two spinors at a vertex is left invariant when parallel transporting the spinors  along the edges around the relevant face.
To generalise this result to the Lorentzian case, we will focus on a triangular face of the graph, such as in \cref{fig:flatness}, following Bonzom and Livine\footnote{One should be aware that this construction should be generalised to any face. This was done for vector constraints (using the fluxes) by Bonzom, in an unpublished work.}. 
Sitting at the vertex between \(g_2\) and \(g_3\) and proceeding clockwise (i.e. along the cycle \(\braket{342}\)), the constraints are given by
\begin{subequations}
\begin{align}
H^{\braket{}}_{342}&:=
\braket{ w_2| (\1 - g_2 g_4^{-1} g_3)|\tau_3} \braketbb{ w_2|\tau_3}
\\
H^{\braketbb{}}_{342}&:=
\braketbb{ w_2| (\1 - g_2 g_4^{-1} g_3)|\tau_3} \braket{ w_2|\tau_3}
\\
H^{\braketrb{ }}_{342}&:=
\braketrb{ w_2| (\1 - g_2 g_4^{-1} g_3)|\tau_3}\braketlb{ w_2|\tau_3}
\\
H^{\braketlb{}}_{342}&:=
\braketlb{ w_2| (\1 - g_2 g_4^{-1} g_3)|\tau_3}\braketrb{ w_2|\tau_3},
\end{align}
\label{list}
\end{subequations}
where the factors on the right, e.g., \( \braketbb{w_2|\tau_3} \), are introduced for convenience, as they will be important when we quantise these contraints.

The constraint  \eqref{c1} is actually a set of 3 real scalar constraints: in fact \(g_2 g_4^{-1} g_3\), as an \(\SU(1,1)\cong\Spin(2,1)\) group element, is parametrised by 3 real parameters, so that the constraint 
\begin{equation}
\label{eq:342_constraint}
\1-g_2 g_4^{-1} g_3=0
\end{equation}
has 3 (real) degrees of freedom. The four complex constraints in \eqref{list}, being proportional to the matrix elements of \eqref{eq:342_constraint}, are equivalent to it and thus carry the same degrees of freedom.
Using the parallel transport of the spinors we can simplify the expression of the previous Hamiltonian constraints, namely we can express them in terms of the vertex observables \(e_{ab}, f_{ab}, \widetilde f_{ab}\). For example, using \eqref{eq:g_action} for \(\bra{w_2}g_2\) and \(g_3\ketb{\tau_3}\), we have that 
\begin{equation}
\begin{split}
H^{\braketrb{ }}_{342}&=
\braketrb{ w_2| (\1 -  g_2 g_4^{-1} g_3 )|\tau_3} \braketlb{ w_2|\tau_3}\\
&=
\braketrb{ w_2| \tau_3}\braketlb{ w_2|\tau_3} - \braketrb{ \tau_2| g_4^{-1} | w_3} \braketlb{ w_2|\tau_3} \\
&=
\braketrb{ w_2| \tau_3}\braketlb{ w_2|\tau_3} -\bra{\tau_2}  \paren[\bigg]{\frac{\ket{\tau_4} \bra {w_4}- \ketb{\tau_4}\brab{w_4}}{\sqrt{\braket{\tau_4|\tau_4}\braket{  w_4| w_4}}}} \ketb{w_3} \braketlb{ w_2|\tau_3} \\
&=
\widetilde f_{23} {f}_{23} -\paren{e_{24} \widetilde f_{24}  -\widetilde{f}_{43}\widetilde{e}_{43}} e_4^{-1} {f}_{23},
\label{hamiltonian constraint}
\end{split}
\end{equation}
where we define that \(e_4:=e_{44}=\braket{ \tau_4|\tau_4}= \braket{  w_4| w_4}\).

Different sets of constraints can be obtained by considering the other possible cycles; the general expression for them is
\begin{subequations}
\label{eq:general_hamiltonian_constraint}
\begin{align}
H_{abc}^{\braket{}}=& e_{ca}\widetilde{e}_{ca} - \paren{ e_{cb} {e}_{ba} - \widetilde f_{cb} f_{ba}} e_b^{-1} \widetilde e_{ca}
\\
H_{abc}^{\braketbb{}}=& \widetilde{e}_{ca}e_{ca} - \paren{ {f}_{cb} \widetilde f_{ba} -\widetilde{e}_{cb}\widetilde{e}_{ba}} e_b^{-1} {e}_{ca}
\\
H_{abc}^{\braketrb{}}=& \widetilde{f}_{ca}f_{ca} - \paren{ e_{cb}\widetilde{f}_{ba} - \widetilde f_{cb} \widetilde e_{ba} } e_b^{-1} f_{ca}
\\
H_{abc}^{\braketlb{}}=& f_{ca}\widetilde{f}_{ca} - \paren{ {f}_{cb} e_{ba} - \widetilde{e}_{cb} {f}_{ba} } e_b^{-1} \widetilde{f}_{ca},
\end{align}
\end{subequations}
where \( \braket{abc}\) is any permutation of \( \braket{342} \). Note that it suffices to consider even permutations only\footnote{In other words, the counter\Hyphdash{}clockwise cycles provide the same constraints as the clockwise ones.}, as it is easy to check that
\begin{equation}
H_{cba}^{\braket{}}\equiv H_{abc}^{\braketbb{}},\quad H_{cba}^{\braketrb{}}\equiv H_{abc}^{\braketrb{}},\quad H_{cba}^{\braketlb{}}\equiv H_{abc}^{\braketlb{}}.
\end{equation}
One can check by direct computation that
\begin{equation}
H_{abc}^{\braketrb{}} + H_{abc}^{\braketlb{}}-H_{abc}^{\braket{}}-H_{abc}^{\braketbb{}}= e_a e_c  \tr(\1 - g_a g_b g_c),
\end{equation}
where the trace is calculated using the fact that, for two column vectors \( x \) and \( y \),
\begin{equation}
\tr(x\otimes y\Transpose)\equiv y\Transpose x,
\end{equation}
so that for example
\begin{equation}
\tr\paren[\big]{\ket{\tau_a}\bra{\tau_b}}=\braket{\tau_b|\tau_a}=e_{ba},
\end{equation}
and the identity
\begin{equation}
\begin{split}
f_{ca} \widetilde f_{ca} - e_{ca} \widetilde e_{ca} &=
\paren{\tau^c_- \tau^a_+ - \tau^c_+ \tau^a_-}\paren{\widetilde \tau^c_- \widetilde \tau^a_+ - \widetilde \tau^c_+ \widetilde \tau^a_-} - \paren{\widetilde \tau^c_- \tau^a_+ - \widetilde \tau^c_+ \tau^a_-} \paren{\tau^c_- \widetilde \tau^a_+ - \tau^c_+ \widetilde \tau^a_-}
\\
&= \paren{\widetilde \tau^a_- \tau^a_+ - \widetilde \tau^a_+ \tau^a_-} \paren{\widetilde \tau^c_- \tau^c_+ - \widetilde \tau^c_+ \tau^c_-}
\\
&= e_a e_c
\end{split}
\end{equation}
is used.

\section{Relativistic spin networks}\label{sec:relativistic-spin-networks}

Our goal in this section is to quantise the classical LQG space we constructed in \ref{sec:classical_LQG}. We will first recall some notions of \( \Spin(2,1) \) recoupling theory that we will need to proceed, then construct the space of \( \Spin(2,1) \) intertwiners, which we will use as building blocks of our quantum theory. It should be noted that, due to the non-compactness of the group, \emph{closed} spin networks, which are proportional to the intertwiner which maps the trivial representation to itself, are generally divergent; spin networks for non-compact groups have been studied in detail in \cite{non-compact}, where it was shown how to deal with these divergencies. We will mostly focus on \( 3 \)-valent intertwiners, but we will also consider \( 4 \)-valent ones to introduce the notion of Racah coefficients. In the last subsection we will introduce an inner product in the space of intertwiners.

%

\subsection{\texorpdfstring{\( \Spin(2,1) \)}{Spin(2,1)} recoupling theory}

Some results of \( \Spin(2,1) \) representation theory, namely the known recouplings between irreducible \((\mathfrak{g},K)\)-modules are reviewed here. The ones we discovered in \cref{sec:3d-lorentz-group} are recalled as well.

\subsubsection{Coupling of finite-dimensional representations}
The finite-dimensional representations of \( \Spin(2,1) \) coincide with those of \( \mathrm{SU}(2) \). In particular, their recoupling will have the same \emph{\CG\ decomposition}, i.e.
\begin{equation}
F_{j}\otimes F_{j\Prime}=\bigoplus_{J=|j-j\Prime|}^{j+j\Prime}F_J.
\end{equation}

\subsubsection{Coupling of unitary representations}
The known recouplings for unitary representations are\footcite{mukunda}
\begin{subequations}
\label{eq:unitary_coupling}
\begin{gather}
D^\pm_{j}\otimes D^\pm_{j\Prime}=\bigoplus_{J=j+j\Prime+1}^\infty D^\pm_J
\\
D^\pm_j\otimes D^\mp_{j\Prime}=\bigoplus_{J=J_\text{min}}^{j-j\Prime-1}D^\pm_J \oplus \bigoplus_{J=J_\text{min}}^{j\Prime-j-1}D^\mp_J\oplus \int_{\mathbb{R}_+}^\oplus C^\varepsilon_{-\frac{1}{2}+\ii S}\,\eder S,\quad J_\text{min}=\varepsilon=\varsigma(j+j\Prime)
\\
D^\pm_j\otimes C^\varepsilon_{-\frac{1}{2}+\ii s}=\bigoplus_{J=J_\text{min}}^\infty D^\pm_J \oplus \int_{\mathbb{R}_+}^\oplus C^E_{-\frac{1}{2}+\ii S}\,\eder S,\quad J_\text{min}=E=\varsigma(j+\varepsilon)
\\
\label{eq:recoupling_CxC}
C^\varepsilon_{-\frac{1}{2}+\ii s} \otimes C^{\varepsilon\Prime}_{-\frac{1}{2}+\ii s\Prime} = \bigoplus_{J=J_\text{min}}^\infty D^+_J \oplus \bigoplus_{J=J_\text{min}}^\infty D^-_J \oplus 2  \int_{\mathbb{R}_+}^\oplus C^E_{-\frac{1}{2}+\ii S}\,\eder S,\quad J_\text{min}=E=\varsigma(\varepsilon+\varepsilon\Prime),
\end{gather}
\end{subequations}
where \( j,j\Prime\geq -\tfrac{1}{2} \) and \( s,s\Prime>0 \), the function \( \varsigma \) is defined by
\begin{equation}
\varsigma(x)=
\begin{cases}
0\quad&\mbox{if }x\in \mathbb{Z}\\
\frac{1}{2} & \mbox{if }x\in \frac{1}{2}+\mathbb{Z}
\end{cases}
\end{equation}
and it is to be understood that \( \bigoplus_{J=a}^b \) vanishes if \( b<a \). Notice in particular that only representations in the Plancherel decomposition appear in the Clebsch--Gordan decomposition, even when we consider couplings involving discrete representations with \( j=-\frac{1}{2} \). Moreover, the trivial representation \( F_0 \) does not appear in any of the representations. The factor \( 2 \) in \eqref{eq:recoupling_CxC} denotes that each continuous representation appears twice in that decomposition.

\subsubsection{Coupling of finite and infinite-dimensional representations}

Recall from \cref{sec:3d-lorentz-group} that for the coupling of a finite-dimensional representation and one from the discrete or continuous series we have
\begin{equation}
F_\gamma\otimes D^\pm_j=\bigboxplus_{J=j-\gamma}^{j+\gamma}{D}^\pm_J,
\end{equation}
with the restriction \( j>\gamma-1 \) and
\begin{equation}
F_\gamma \otimes C^\varepsilon_j=\bigboxplus_{J=j-\gamma}^{j+\gamma} {C}^E_{J},\quad E=\varsigma(\gamma+\varepsilon),
\end{equation}
with the restriction that, if $j\in\mathbb{Z}/2$, $j>\gamma-1$ or $j<-\gamma$.

\subsubsection{Clebsch--Gordan coefficients and label notation} So far  our results of recoupling theory are very heterogeneous. In order to have a uniform notation across different cases, we will introduce a new convention: the quantum number \( j\in\mathbb{C} \) will become a label, i.e., we will, with abuse of notation, continue to call \( j \) the pair \( (j,\alpha) \), where
\begin{equation}
\alpha\in\set{D^+,D^-,C^0,C^\frac{1}{2},F}
\end{equation}
is a symbol denoting the representation class. The label \( j \) now completely determines the module, which we denote by \( V_j \), spanned by the standard basis \( \ket{j,m} \). The set of possible \( m \) values will be denoted by \( \mathcal{M}_j \).

Consider now a generic coupling \(  V_j\otimes V_{j\Prime} \). If a decomposition exists, we are going to denote by \( \mathcal{D}(j,j\Prime) \) the set containing the labels of all representations appearing in it. We then have
\begin{equation}\label{eq:label_notation_CG}
\ket{J,M}=\sum_{m,m\Prime}A(j,m;j\Prime,m\Prime|J,M)\ket{j,m}\otimes\ket{j\Prime,m\Prime},\quad J\in\mathcal{D}(j,j\Prime),\quad M\in\mathcal{M}_j,
\end{equation}
where the \( A(j,m;j\Prime,m\Prime|J,M) \)'s are the Clebsch--Gordan coefficients of the decomposition. To account for the case \( F_\gamma\otimes  V_j \), in which this map is generally \emph{not} unitary, we will write
\begin{equation}
\ket{j,m}\otimes\ket{j\Prime,m\Prime}=\int_{\mathcal{D}(j,j\Prime)} \eder\xi(J) \sum_{M\in\mathcal{M}_J}B(J,M|j,m;j\Prime,m\Prime)\ket{J,M},
\end{equation}
where the \( B(J,M|j,m;j\Prime,m\Prime) \)'s are the components of \( A^{-1} \), i.e., the inverse \CG\ coefficients. The integral is taken with respect to a measure \( \xi \) defined as follows: if \( \mathcal{D}_\alpha(j,j\Prime)\subseteq\mathcal{D}(j,j\Prime) \) is the subset of labels with representation class \( \alpha \), then\footnote{Here \( \abs{S} \) denotes the \emph{cardinality} of a set \( S \).}
\begin{equation}
\label{eq:measure}
\xi|_{\mathcal{D}_\alpha}:=
\begin{cases}
\lambda &\casesif\abs{\mathcal{D}_\alpha}=\abs{\mathbb{R}}\\
\sum_{J\in\mathcal{D}_\alpha}\delta_J&\casesif\abs{\mathcal{D}_\alpha}=\abs{\mathbb{N}},
\end{cases}
\end{equation}
where \( \lambda \) is the Lebesgue measure on \( \C \) and \( \delta_J \) is the Dirac measure defined by
\begin{equation}
\delta_J(A)=
\begin{cases}
1 & \casesif J\in A\\
0 & \casesif J\not\in A.
\end{cases}
\end{equation}
Clebsch--Gordan coefficients possess many interesting properties. It follows from their definition that they satisfy the orthogonality relations
\begin{subequations}
\begin{gather}
\label{eq:CG_orth1}
\int\eder\xi(J)\sum_M A(j,m;j\Prime,m\Prime|J,M)B(J,M|j,n;j\Prime,n\Prime)=\delta_{m,n}\,\delta_{m\Prime,n\Prime}
\\
\label{eq:CG_orth2}
\sum_{m,m\Prime} B(J,M|j,m;j\Prime,m\Prime) A(j,m;j\Prime,m\Prime|J\Prime,M\Prime)=\delta(J,J\Prime)\,\delta_{M,M\Prime},
\end{gather}
\end{subequations}
where 
\begin{equation}
\label{eq:delta}
\delta|_{\mathcal{D}_\alpha \times \mathcal{D}_\beta}
\begin{cases}
\mbox{is a Dirac delta}&\casestextn{if \( \alpha=\beta \) and \( |\mathcal{D}_\alpha|=|\mathbb{R}| \)}\\
\mbox{is a Kronecker delta}&\casestextn{if \( \alpha=\beta \) and \( |\mathcal{D}_\alpha|=|\mathbb{N}| \)}\\
\mbox{identically vanishes}&\casestextn{if \( \alpha\neq\beta \)}.
\end{cases}
\end{equation}
Moreover, they can be normalised so that
\begin{equation}
A(j,m;j\Prime,m\Prime|J,M)\equiv B(J,M|j,m;j\Prime,m\Prime),
\end{equation}
so that we may refer to both of them as Clebsch--Gordan coefficients.  With this normalisation, they satisfy the recursion relations
\begin{multline}
\Gamma_\pm(J,M)A(j,m;j\Prime,m\Prime|J,M\pm 1)=
\Gamma_\pm(j, m\mp 1)A(j,m \mp 1;j\Prime,m\Prime|J,M)\\
+\Gamma_\pm(j\Prime,m\Prime\mp 1)A(j,m;j\Prime,m\Prime\mp 1|J,M),
\end{multline}
which easily follow by acting with \( J_\pm \) on both sides of \eqref{eq:label_notation_CG}.

\subsection{Intertwiners}
Recall that an intertwiner between \((\mathfrak{g},K)\)-modules for \( \Spin(2,1) \), \( V \) and \( W \), is a linear map \( \psi:V\rightarrow W \) satisfying
\begin{equation}\label{eq:intertwiner_conditions}
\psi\circ X = X\circ \psi,\quad \forall X\in\spin(2,1).
\end{equation}
The set of all possible intertwiners from \(V\) to \(W\) forms a vector space, which will be denoted by \(\Hom(V,W)\).
We will only work with intertwiners between representations that are either irreducible or a tensor product of irreducible ones\footnote{When speaking of products we assume that none of the representations involved is the trivial one, for obvious reasons.}. An intertwiner
\begin{equation}
\psi:\bigotimes_{a=1}^k V_{j_a}\rightarrow \bigotimes_{b=r+1}^n V_{j_b}
\end{equation}
will be called \(n\)-valent and, for reasons that will become clear shortly, we will say it has \(k\) incoming legs and \((n-k)\) outgoing ones.

Of particular interest are the \(3\)-valent intertwiners. If the decomposition of  \( V_{j_1}\otimes  V_{j_2}\) exists, the vector space \(\Hom( V_{j_1}\otimes  V_{j_2}, V_{j_3})\) is completely specified by it, as a non-vanishing intertwiner only exists if \( V_{j_3}\) appears in the decomposition; the number of independent intertwiners equals the multiplicity of \( V_{j_3}\) in the decomposition (\( 1 \) or \( 2 \) for the known decompositions). These basis elements will be denoted by
\begin{equation}
\trivalentlr{j_1}{j_2}{j_3}:\ket{j_1,m_1}\otimes\ket{j_2,m_2}\mapsto \sum_{m_3\in \mathcal{M}_{j_3}}B(j_3,m_3|j_1,m_1;j_2,m_2)\ket{j_3,m_3},
\end{equation}
where we assume \(j_3\) also includes an appropriate label for multiplicities, when necessary. On the LHS we used a graphical notation for the map, which will turn out to be very useful. It is to be read this way: incoming representations (legs) are on the left, while outgoing ones are on the right; an arrow will be used to make the direction clear if needed.

Analogously, the basis elements for \(\Hom( V_{j_3}, V_{j_1}\otimes  V_{j_2})\) are given by the intertwiners
\begin{equation}
\trivalentrl{j_1}{j_2}{j_3}:\ket{j_3,m_3}\mapsto \sum_{\substack{m_1\in\mathcal{M}_{j_1} \\ m_2 \in \mathcal{M}_{j_2}}}A\paren{j_1,m_1;j_2,m_2|j_3,m_3}\ket{j_1,m_1}\otimes\ket{j_2,m_2}.
\end{equation}
Moreover the unique intertwiner in the \(1\)-dimensional space \(\Hom( V_j, V_j)\) will be denoted by
\begin{equation}
\intidA{j}{j}\equiv \1_{V_j}.
\end{equation}

The two kinds of \(3\)-valent intertwiners can be used as building blocks of all the others, provided that the necessary Clebsch--Gordan decomposition exists: this can be achieved by composing intertwiners, to obtain maps on bigger representations; with our graphical notation, this amounts to ``glueing'' them together.
We will call any such composition of intertwiners a spin network.
Note that, when working with unitary representations, there is no way to obtain a \emph{closed} spin network\footnote{i.e., an element of the space \(\Hom(F_0,F_0)\).} with this glueing procedure, as the trivial representation \(F_0\) does not appear in any recoupling of infinite-dimensional representations. Closing a spin network by tracing, which graphically amounts to connecting an incoming leg with an incoming one of the same intertwiner, leads to divergencies, which will have to be dealt with; here however we are only interested in the nodes \emph{inside} a spin network, which would be unaffected by any regularisation procedure.

\subsection{Racah coefficients}\label{sec:racah}

Consider a \(4\)-valent intertwiner \(\psi\) with \( 3 \) incoming legs \( V_{j_1}\otimes  V_{j_2}\otimes  V_{j_3}\) and a single outgoing one \( V_{j}\): it will generally not be unique, unless one of the representations involved is the trivial one. Two possible \emph{bases} of intertwiners, whose linear combinations can be used to construct any \(4\)-valent one of this type can be obtained by exploiting the associativity of tensor products, i.e.
\begin{equation}
V_{j_1}\otimes V_{j_2}\otimes V_{j_3}\cong (V_{j_1}\otimes V_{j_2})\otimes V_{j_3} \cong V_{j_1}\otimes (V_{j_2}\otimes V_{j_3}).
\end{equation}
Assuming the decomposition in irreducible representations of \(V_{j_1}\otimes V_{j_2}\) exists, the generic \(\psi\) can be written as a linear combinations of the intertwiners
\begin{equation}\label{eq:4valent_basis1}
\racahintA{j_1}{j_2}{j_3}{j}{j_{12}},\quad j_{12}\in\mathcal{D}(j_1,j_2);
\end{equation}
analogously, if \(V_{j_2}\otimes V_{j_3}\) is decomposable, the intertwiners
\begin{equation}\label{eq:4valent_basis2}
\racahintB{j_1}{j_2}{j_3}{j}{j_{23}},\quad j_{23}\in\mathcal{D}(j_2,j_3)
\end{equation}
form a basis as well. We will now study how the two bases are related to each other.
\pagebreak

First notice that\footnote{To simplify notation, the range of the \(j\)'s in the summation is omitted: it is implied to only assume the values for which a non-vanishing intertwiner exists. Moreover, when a subset of labels \(j\) appears continuously in a decomposition, over that subset the sum is to be considered an integration.} 
\begin{equation}
\sum_{j_{12},j}\intspiderA{j_1}{j_2}{j_3}{j}{j_{12}}=\sum_{j_{23},j}\intspiderB{j_1}{j_2}{j_3}{j}{j_{23}}
=
\intidB{j_1}{j_2}{j_3},
\end{equation}
as can be checked explicitly using the properties of Clebsch--Gordan coefficients. This equation can be ``glued'' to the basis elements \eqref{eq:4valent_basis1} to obtain
\begin{equation}
\racahintA{j_1}{j_2}{j_3}{j}{j_{12}}=\sum_{j_{23},j'}
\intracahsquare{j_1}{j_2}{j_3}{j_{12}}{j_{23}}{j}{j'}.
\end{equation}
The intertwiner
\begin{equation}
\intsquareA{j_1}{j_2}{j_3}{j_{12}}{j_{23}}{j}{j'},
\end{equation}
having only one incoming and outgoing representation, must \emph{necessarily} be proportional to the unique intertwiner between \(j\Prime\) and \(j\). Since the latter vanishes when \(j\neq j\Prime\), it must be
\begin{equation}\label{eq:racah1}
\intsquareA{j_1}{j_2}{j_3}{j_{12}}{j_{23}}{j}{j'}\propto
\delta(j,j\Prime)\intidA{j}{j},
\end{equation}
where the \(\delta\) is to be considered a Dirac delta over continuous subsets in both \(\mathcal{D}(j_{12},j_3)\) and \(\mathcal{D}(j_1,j_{23})\), and a Kronecker delta otherwise. The proportionality factor in \eqref{eq:racah1}, which we will call \emph{Racah coefficient}, is given by
\begin{multline}\label{eq:racah2}
\racah{j_1}{j_2}{j_{12}}{j_3}{j}{j_{23}}=
\sum_{\substack{m_1,m_2,m_3\\m_{12},m_{23}}}
A(j_1,m_1;j_{23},m_{23}|j,m)A(j_2,m_2;j_3,m_3|j_{23},m_{23})
\\ \times
B(j_{12},m_{12}|j_1,m_1;j_2,m_2)B(j,m|j_{12},m_{12};j_3,m_3),
\end{multline}
with \(m\in\mathcal{M}_j\); one can check using the Clebsch--Gordan recursion relations that the result does not depend on which \(m\) is chosen. We finally get that
\begin{equation}
\label{eq:racah_coeff1}
\racahintA{j_1}{j_2}{j_3}{j}{j_{12}}
=\sum_{j_{23}}
\racah{j_1}{j_2}{j_{12}}{j_3}{j}{j_{23}}
\racahintB{j_1}{j_2}{j_3}{j}{j_{23}},
\end{equation}
i.e., the Racah coefficients are the components of the elements of one basis in terms of the other. An analogous argument can be made for the basis elements \eqref{eq:4valent_basis2}. With our convention for the Clebsch--Gordan coefficients, we can check that
\begin{equation}
\intsquareB{j_1}{j_2}{j_3}{j_{12}}{j_{23}}{j}{j'}=\intsquareA{j_1}{j_2}{j_3}{j_{12}}{j_{23}}{j}{j'}
\end{equation}
so that
\begin{equation}
\label{eq:racah_coeff2}
\racahintB{j_1}{j_2}{j_3}{j}{j_{23}}
=\sum_{j_{12}}
\racah{j_1}{j_2}{j_{12}}{j_3}{j}{j_{23}}
\racahintA{j_1}{j_2}{j_3}{j}{j_{12}}.
\end{equation}

\begin{remark}
Note how there was \emph{no} mention of unitary representations in the discussion of Racah coefficients: what was presented is well-defined any time the appropriate Clebsch--Gordan decomposition exists. This means in particular that we can consider Racah coefficients \emph{involving both unitary and non-unitary representations}, which be relevant when discussing the quantum version of the observables \(e\), \(f\) and \(\widetilde f\).
\end{remark}

\subsection{Inner product space structure}

We already saw that \(\Hom(V,W)\) is a vector space; we will now see how an inner product can be defined naturally on it. This time we will only use unitary irreducibly representations from the Plancherel decomposition.

The space of intertwiners can inherit an inner product by requiring that\footnote{Note that it is always true that the LHS can be split in the sum of independent subspaces on the right: what we are really requiring is for these subspaces to be orthogonal.}
\begin{equation}
\Hom(\textstyle \bigotimes_a V_{j_a}\oplus\bigotimes_b  V_{j_b},\bigotimes_c  V_{j_c})\equiv
\Hom(\textstyle \bigotimes_a V_{j_a},\bigotimes_c  V_{j_c}) \oplus
\Hom(\textstyle \bigotimes_b  V_{j_b},\bigotimes_c  V_{j_c}).
\end{equation}
It is then easy to convince ourselves that the composition of two intertwiners belongs to (a space isomorphic to) the tensor product of their respective intertwiner spaces, so that, for example,
\begin{equation}
\Hom( V_{j_1}\otimes V_{j_2}, V_{j_3}\otimes  V_{j_4})=
\int^\oplus \eder\xi(j)\Hom( V_{j_1}\otimes V_{j_2}, V_j)\otimes\Hom( V_{j}, V_{j_3}\otimes  V_{j_4})
\end{equation}
or
\begin{equation}
\Hom( V_{j_1}\otimes V_{j_2}\otimes V_{j_3}, V_{j})=
\int^\oplus \eder\xi(j_{12})\Hom( V_{j_1}\otimes V_{j_2}, V_{j_{12}})\otimes\Hom( V_{j_{12}}\otimes V_{j_3}, V_j).
\end{equation}
We can repeat this process until we only have sums of products of \(3\)-valent spaces, so that it only remains to define the inner product on the latter. This is easily achieved:
\begin{itemize}
\item when the space is one dimensional there is only one basis vector which we may normalise to \(1\);
\item when the space is two dimensional, i.e., there is multiplicity, we choose the two basis elements\footnote{Note that there is a natural choice for the basis elements once a convention has been chosen for the \CG\ coefficients appearing in \eqref{eq:unitary_coupling}.} to be orthonormal.
\end{itemize}
One can check explicitly that this is consistent with the possibility of using different decompositions for the same space, e.g.
\begin{equation}
\Hom( V_{j_1}\otimes V_{j_2}\otimes V_{j_3}, V_{j})=
\int^\oplus \eder\xi(j_{23})\Hom( V_{j_2}\otimes V_{j_3}, V_{j_{23}})\otimes\Hom( V_{j_1}\otimes  V_{j_{23}}, V_{j}),
\end{equation}
so that the procedure is well defined.

Restricting ourselves to representations in the Plancherel decomposition makes our construction possible, as it guarantees that the direct sums in the Clebsch--Gordan decomposition are orthogonal. Note, however, that if we only use finite-dimensional representations the same would be true, and the inner product would still be well defined: in fact, the total Casimir acting on a product of finite-dimensional representations is self-adjoint, so that the modules appearing in the decomposition are orthogonal to each other.

\section[\texorpdfstring{\(3\)D}{3D} Lorentzian LQG and Lorentzian Ponzano--Regge model]{\texorpdfstring{\(3\)D}{3D} Lorentzian loop quantum gravity and Lorentzian Ponzano--Regge model}\label{sec:lorentzian-ponzano--regge-model}

In this final section we will construct a quantum version of the spinorial observables, and determine their action on intertwiners. We will then discuss some properties of the Racah coefficients, which are defined even when both unitary and non-unitary representations are coupled; in particular, we will show that the Biedenharn--Elliott relation---also known as pentagon identity---holds, even when one of the representations involved are finite-dimensional. Finally, we will quantise the Hamiltonian constraints defined in \cref{sec:classical_LQG}, and show that the intertwiner associated to the Racah coefficients, which generates the Lorentzian Ponzano--Regge model, is in their kernel; the proof consists in showing that the action of each constraint is implemented as a recursion relation (the Biedenharn--Elliott relation), whose solution is the Racah coefficient.

\subsection{Intertwiner observables} \label{sec:int_obs}\label{lqg-obs}
We want observables in loop quantum gravity to be invariant under the action of the gauge group \(\Spin(2,1)\): this is exactly what \emph{scalar operators} (tensor operators transforming as \( F_0 \)) are.
The usual observables we consider are those built from the algebra generators, which are essentially the components of the vector operator \(V\) defined in \eqref{eq:vector_operator}. When acting on a product of representations \(\bigotimes_a  V_{j_a}\) they are defined as
\begin{equation}
Q_{ab}:=\frac{\sqrt{3}}{2}\sum_{\mu}A(1,\mu;1,-\mu|0,0) V^a_\mu V^b_{-\mu}=-J^a_0J^b_0 +\frac{1}{2}\left(J^a_- J^b_+ + J^a_+ J^b_-\right),
\end{equation}
where \(V^a\) denotes the operator acting only on representation \(a\); equivalently, we can write them in the suggestive form
\begin{equation}
Q_{ab}=\eta^{ij}J^a_i J^b_j,\quad \eta=\mathrm{diag}(-1,1,1),
\end{equation}
where \(i\) and \(j\) are space-time indices.

When working in the spinorial setting, we can construct  scalar operators by combining the two spinor operators \(T\) and \(\widetilde T\). The four  kinds of operators we can get are
\begin{subequations}
\begin{align}
E_{ab}&=-\sqrt{2}\sum_{\mu}A(\tfrac{1}{2},\mu;\tfrac{1}{2},-\mu|0,0)\half \paren[\big]{\widetilde T^a_\mu {T}^b_{-\mu} + T^a_\mu \widetilde{T}^b_{-\mu}}= \widetilde T^a_- {T}^b_+ - \widetilde T^a_+ {T}^b_- + \delta_{ab}\1,
\\
F_{ab}&=-\sqrt{2}\sum_{\mu}A(\tfrac{1}{2},\mu;\tfrac{1}{2},-\mu|0,0) T^a_\mu {T}^b_{-\mu}= T^a_- {T}^b_+ - T^a_+ {T}^b_- ,
\\
\widetilde{F}_{ab}&=-\sqrt{2}\sum_{\mu}A(\tfrac{1}{2},\mu;\tfrac{1}{2},-\mu|0,0) \widetilde{T}^a_\mu \widetilde{T}^b_{-\mu}= \widetilde{T}^a_- \widetilde{T}^b_+ - \widetilde{T}^a_+ \widetilde{T}^b_-,
\end{align}
\end{subequations}
which are the quantum analogues of the classical observables \eqref{eq:classical_observables}. Note that an ordering factor was introduced in the quantisation of \( e_{ab}\); this particular ordering was chosen to ensure that these operators form a closed Lie algebra. The commutation relations of these operators are
\begin{subequations}
\begin{align}
[E_{ab},E_{cd}]		&= \delta_{cb}E_{ad} - \delta_{ad}E_{cb} \\
[E_{ab},\widetilde F_{cd}]	&= \delta_{bc}\widetilde F_{ad} - \delta_{bd}\widetilde F_{ac} \\
[E_{ab},F_{cd}]		&= \delta_{ad}F_{bc} - \delta_{ac}F_{bd} \\
[F_{ab},\widetilde F_{cd}]	&= \delta_{db}E_{ca} + \delta_{ca}E_{db}  - \delta_{cb}E_{da} -\delta_{da}E_{cb} \\
[F_{ab},F_{cd}]		&= [\widetilde F_{ab},\widetilde F_{cd}] = 0.
\end{align}
\end{subequations}
Note that, when acting on the continuous class, the operators \(E,F,\widetilde F\)  take unitary representations (in the Plancherel decomposition) to \emph{non-unitary} ones\footnote{The \(T\) and \( \widetilde T \) operators would send the module \(V_{-\shalf+\ii s}\) respectively to \(V_{-1+\ii s}\) and \(V_{\ii s}\), neither of which would be unitary anymore.}.
As such, they are not proper observables when acting on continuous representations; however, one can choose quadratic functions of these observables such that the representation is sent to itself. 

Due to the relation between \(T\) and \(\widetilde{T}\), the operators we defined are not all independent. One has, in general\footnote{Here the transpose is defined with respect to the matrix elements in the standard \( \ket{j,m} \) basis.},
\begin{equation}
F\Transpose*_{ab}=\widetilde{F}_{ab},\quad E\Transpose*_{ab}={E}_{ba}.
\end{equation}
In particular cases the transposes can be converted to adjoints; for example, if \(a\) and \(b\) both denote a representation in the discrete positive (negative) class \(F\Transpose*_{ab}=F\Dagger*_{ab}\), while if one is them is discrete positive and the other discrete negative \(F\Transpose*_{ab}=-F\Dagger*_{ab}\).
 
The operators we defined act on representations; their action can be extended to intertwiners as follows. Let
\begin{equation}
\psi:\bigotimes_{a=1}^k\ket{j_a,m_a}\rightarrow \sum_{m_{r+1}}\dotsm\sum_{m_n}\alpha(m_1,\dotsc,m_n)\bigotimes_{b=k+1}^n\ket{j_b,m_b}
\end{equation}
be a generic \(n\)-valent intertwiner\footnote{The operators we define will always act on a single \emph{node}, i.e., \(n\)-valent intertwiner, inside a generic spin network. }, where \(\alpha\) is a function depending on Clebsch--Gordan coefficients; this intertwiner can be also expressed in the form
\begin{equation}\label{eq:dualised_intertwiner}
\psi^\star :=\sum_{m_1}\dotsm\sum_{m_n}\alpha(m_1,\dotsc,m_n)\bigotimes_{b=k+1}^n\ket{j_b,m_b}\otimes \bigotimes_{a=1}^k\bra{j_a,m_a},
\end{equation}
which does indeed return the same values when acting on \(\bigotimes_a V_{j_a}\). However, \eqref{eq:dualised_intertwiner} is not necessarily an element of the space \(\bigotimes_b V_{J_b}\otimes \bigotimes_a V\Star*_{j_a}\), since it does not generally have finite norm for infinite-dimensional representations\footnote{This can be easily checked in the case of an intertwiner with one incoming and one outgoing leg, which is necessarily proportional to the identity.}: it is only to be regarded as a formal expression, similarly to the usual way of representing the identity of a separable Hilbert space as
\begin{equation}
\sum_{i\in I}\ket{i}\bra{i},
\end{equation}
where \(\{\ket{i}\}_{i\in I}\) is an orthonormal basis.

One can easily check, using \eqref{eq:intertwiner_conditions}, that \(\psi\) is in intertwiner if and only if
\begin{equation}
J_0\psi^\star=0,\quad J_\pm\psi^\star=0,
\end{equation}
where the generators act on dual vectors as the dual representation (see \cref{eq:dual1,eq:dual2}), i.e.
\begin{equation}
J_0\bra{j,m}=-m\bra{j,m},\quad J_\pm\bra{j,m}=-C_\mp(j,m)\bra{j,m\mp 1}.
\end{equation}
The action of an operator of the form
\begin{equation}\label{eq:int_operator}
T:\bigotimes_{b=k+1}^n V_{j_b}\otimes \bigotimes_{a=1}^k V\Star*_{j_a}\rightarrow \bigotimes_{b=k+1}^n V_{j\Prime*_b}\otimes \bigotimes_{a=1}^k V\Star*_{j\Prime*_a}
\end{equation}
is then defined by inverting transformation \eqref{eq:dualised_intertwiner} for \(T\psi^\star\). One can check that  the resulting map is an intertwiner if and only if \(T\) is a scalar operator.

The \(E\), \(F\) and \(\widetilde{F}\) operators can be expressed as a sum of operators of the form \eqref{eq:int_operator} by having \(T^a\) and \(\widetilde{T}^a\) act as the identity on anything but the \(a\)-th leg (incoming or outgoing) and by extending their action to dual vectors as
\begin{equation}
T_\pm\bra{j,m}:=\bra{j,m} T_\pm, \quad \widetilde T_\pm\bra{j,m}:=\bra{j,m} \widetilde T_\pm,
\end{equation}
that is
\begin{subequations}
\begin{align}
T_-\bra{j,m} &= -\sqrt{j+m+1}\,\bra{j+\half,m+\half}\\
T_+\bra{j,m} &= \sqrt{j-m+1}\,\bra{j+\half,m-\half}\\
\widetilde T_-\bra{j,m} &= \sqrt{j-m}\,\bra{j-\half,m+\half}\\
\widetilde T_+\bra{j,m} &= \sqrt{j+m}\,\bra{j-\half,m-\half}.
\end{align}
\end{subequations}
The actions of the scalar operators on some \(3\)-valent intertwiners of interest are listed here, where the notation
\begin{equation}
D(j):=\sqrt{2j+1}
\end{equation}
is used; these actions are\footnote{Note that we have \(F_{ba}=-F_{ab}\) and \(\widetilde{F}_{ba}=-\widetilde{F}_{ab}\) by definition.}
\begin{subequations}\label{eq:intertwiner_actions}
\begin{align}
E_{12}\trivalentlr{j_1}{j_2}{j_3}&=D(k_1)D(j_2)
\racah{k_1}{\shalf}{j_1}{j_2}{j_3}{k_2}
\delta_{k_1,j_1-\shalf}\delta_{k_2,j_2+\shalf}
\trivalentlr{k_1}{k_2}{j_3}
\\ \addlinespace
E_{21}\trivalentlr{j_1}{j_2}{j_3}&=-D(k_1)D(j_2)
\racah{k_1}{\shalf}{j_1}{j_2}{j_3}{k_2}
\delta_{k_1,j_1+\shalf}\delta_{k_2,j_2-\shalf}
\trivalentlr{j_1}{j_2}{j_3}
\\ \addlinespace
F_{12}\trivalentlr{j_1}{j_2}{j_3}&=-D(k_1)D(j_2)
\racah{k_1}{\shalf}{j_1}{j_2}{j_3}{k_2}
\delta_{k_1,j_1+\shalf}\delta_{k_2,j_2+\shalf}
\trivalentlr{k_1}{k_2}{j_3}
\\ \addlinespace
\widetilde{F}_{12}\trivalentlr{j_1}{j_2}{j_3}&=-D(k_1)D(j_2)
\racah{k_1}{\shalf}{j_1}{j_2}{j_3}{k_2}
\delta_{k_1,j_1-\shalf}\delta_{k_2,j_2-\shalf}
\trivalentlr{k_1}{k_2}{j_3}
\end{align}
\begin{align}
E_{12}\trivalentrl{j_1}{j_2}{j_3}&=-D(k_1)D(j_2)
\racah{k_1}{\shalf}{j_1}{j_2}{j_3}{k_2}
\delta_{k_1,j_1+\shalf}\delta_{k_2,j_2-\shalf}
\trivalentrl{k_1}{k_2}{j_3}
\\\addlinespace
E_{21}\trivalentrl{j_1}{j_2}{j_3}&=+D(k_1)D(j_2)
\racah{k_1}{\shalf}{j_1}{j_2}{j_3}{k_2}
\delta_{k_1,j_1-\shalf}\delta_{k_2,j_2+\shalf}
\trivalentrl{k_1}{k_2}{j_3}
\\\addlinespace
F_{12}\trivalentrl{j_1}{j_2}{j_3}&=-D(k_1)D(j_2)
\racah{k_1}{\shalf}{j_1}{j_2}{j_3}{k_2}
\delta_{k_1,j_1-\shalf}\delta_{k_2,j_2-\shalf}
\trivalentrl{k_1}{k_2}{j_3}
\\\addlinespace
\widetilde{F}_{12}\trivalentrl{j_1}{j_2}{j_3}&=-D(k_1)D(j_2)
\racah{k_1}{\shalf}{j_1}{j_2}{j_3}{k_2}
\delta_{k_1,j_1+\shalf}\delta_{k_2,j_2+\shalf}
\trivalentrl{k_1}{k_2}{j_3}
\end{align}
\begin{align}
E_{23}\trivalentlr{j_1}{j_2}{j_3}&=D(j_2)D(j_3)
\racah{j_1}{j_2}{j_3}{\shalf}{k_3}{k_2}
\delta_{k_2,j_2-\shalf}\delta_{k_3,j_3-\shalf}
\trivalentlr{j_1}{k_2}{k_3}
\\\addlinespace
E_{32}\trivalentlr{j_1}{j_2}{j_3}&=D(j_2)D(j_3)
\racah{j_1}{j_2}{j_3}{\shalf}{k_3}{k_2}
\delta_{k_2,j_2+\shalf}\delta_{k_3,j_3+\shalf}
\trivalentlr{j_1}{k_2}{k_3}
\\\addlinespace
F_{23}\trivalentlr{j_1}{j_2}{j_3}&=D(j_2)D(j_3)
\racah{j_1}{j_2}{j_3}{\shalf}{k_3}{k_2}
\delta_{k_2,j_2+\shalf}\delta_{k_3,j_3-\shalf}
\trivalentlr{j_1}{k_2}{k_3}
\\\addlinespace
\widetilde{F}_{23}\trivalentlr{j_1}{j_2}{j_3}&=- D(j_2)D(j_3)
\racah{j_1}{j_2}{j_3}{\shalf}{k_3}{k_2}
\delta_{k_2,j_2-\shalf}\delta_{k_3,j_3+\shalf}
\trivalentlr{j_1}{k_2}{k_3}.
\end{align}
\end{subequations}
The Racah coefficients we have used may involve both unitary and non-unitary representations; as discussed in \cref{sec:racah}, they are still  defined in this case.
\draftnote{show how to calculate these actions}

\subsection[Biedenharn--Elliott relations and symmetries]{Biedenharn--Elliott relations and symmetries of the Racah coefficients}
Some useful properties of the Racah coefficients are presented here, namely some symmetries and the Biedenharn\Endash{}Elliott relations, essential to our goal.

\subsubsection{Symmetries}

When at least one of \( V_{j_1}\), \( V_{j_2}\) and \( V_{j_3}\) is the finite-dimensional representation \(F_\frac{1}{2}\) the Racah coefficients possess the symmetries\footnote{This is not by any means the only case in which some symmetries arise, but explicit knowledge of the Clebsch\Endash{}Gordan coefficients is needed in order to prove them.}
\begin{subequations}
\begin{align}
\racah{j_1}{\tfrac{1}{2}}{k_1}{j_2}{J}{k_2}
&= (-1)^{j_1+j_2-k_1-k_2}\frac{D(k_1)D(k_2)}{D(j_1)D(j_2)}
\racah{k_1}{\tfrac{1}{2}}{j_1}{k_2}{J}{j_2}
\\\addlinespace
\racah{\tfrac{1}{2}}{j_1}{k_1}{J}{j_2}{k_2}
&= (-1)^{j_1+j_2-k_1-k_2}\frac{D(k_1)D(j_2)}{D(j_1)D(k_2)}
\racah{\tfrac{1}{2}}{k_1}{j_1}{J}{k_2}{j_2}
\\\addlinespace
\racah{J}{j_1}{j_2}{\tfrac{1}{2}}{k_2}{k_1}
&= (-1)^{j_1+k_2-k_1-j_2}\frac{D(k_1)D(k_2)}{D(j_1)D(j_2)}
\racah{J}{k_1}{k_2}{\tfrac{1}{2}}{j_2}{j_1};
\end{align}
\end{subequations}
note that the numbers on the exponents are always in \(\mathbb{Z}/2\): for example, in the first equation, it must be \(k_i\in\mathcal{D}(\frac{1}{2},j_i)\) so that \(j_i-k_i=\pm\frac{1}{2}\).
The proof of these symmetries is straightforward, and can be checked by inserting the explicit values of the Clebsch--Gordan coefficients from \cref{tab:3d-1/2} in \eqref{eq:racah2}.

\subsubsection{Biedenharn--Elliott}

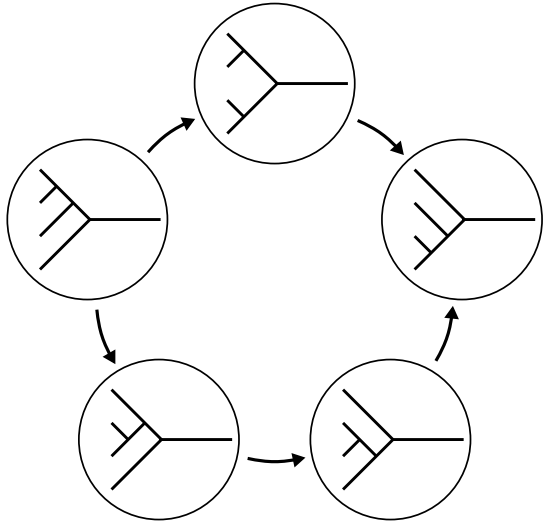
\begin{figure}
\centering
\tikzsetnextfilename{pentagon}
\begin{tikzpicture}
\foreach \x in {1,2,...,5} {
\node[circle, draw,inner sep=-6pt,line width=0.65pt,outer sep=4pt] (p\x) at ({-54+72*\x}:6.8em) {
\tikzset{external/export next=false}
\begin{tikzpicture}[line width=1.1, x=1.15em,y=1.15em]
\coordinate (v1) at (0,3);
\coordinate (v2) at (0,2);
\coordinate (v3) at (0,1);
\coordinate (v4) at (0,0);
\coordinate (c) at (1.5,1.5);
\coordinate (v) at ($(c)+({1.5*sqrt(2)},0)$);
\path (c) circle ({1.5*sqrt(2)});
\draw (v1)--(c) --(v4);
\draw (v)--(c);
\ifnum\x=1
\draw (v2)--($(v4)!2/3!(c)$);
\draw (v3)--($(v4)!1/3!(c)$);
\fi
\ifnum\x=2\draw (v2)--($(v1)!1/3!(c)$);
\draw (v3)--($(v4)!1/3!(c)$);
\fi
\ifnum\x=3
\draw (v2)--($(v1)!1/3!(c)$);
\draw (v3)--($(v1)!2/3!(c)$);
\fi
\ifnum\x=4
\draw (v2)--({1.5/3},1.5);
\draw (v3)--($(v1)!2/3!(c)$);
\fi
\ifnum\x=5
\draw (v2)--($(v4)!2/3!(c)$);
\draw (v3)--({1.5/3},1.5);
\fi
\end{tikzpicture}};
};
\draw[bend left=12, line width=1.2pt, endarrow]  (p2) to (p1);
\draw[bend left=12, line width=1.2pt, endarrow]  (p3) to (p2);
\draw[bend right=12, line width=1.2pt, endarrow]  (p3) to (p4);
\draw[bend right=12, line width=1.2pt, endarrow]  (p4) to (p5);
\draw[bend right=12, line width=1.2pt, endarrow]  (p5) to (p1);
\end{tikzpicture}
\caption{Graphical representation of the pentagon identity.}\label{fig:pentagon}
\end{figure}
Racah coefficients, \textit{regardless} of the representation classes involved, satisfy the \emph{Biedenharn\Endash{}Elliott relations} or \emph{pentagon identity}, which can be represented graphically as in \cref{fig:pentagon}: one can go from the leftmost intertwiner to the rightmost one by repeated Racah transformations in two possible ways; equating the Racah coefficients appearing in the two transformations we get that
\begin{subequations}
\begin{gather}
\sum_{j_{23}}
\racah{j_1}{j_2}{j_{12}}{j_3}{j_{123}}{j_{23}}
\racah{j_1}{j_{23}}{j_{123}}{j_4}{j}{j_{234}}
\racah{j_2}{j_3}{j_{23}}{j_4}{j_{234}}{j_{34}}
=
\racah{j_1}{j_2}{j_{12}}{j_{34}}{j}{j_{234}}
\racah{j_{12}}{j_3}{j_{123}}{j_4}{j}{j_{34}}.
\intertext{Analogously, we can repeat the process starting from one of the other \( 4 \) intertwiners, to get the remaining identities}
\sum_{j_{123}}
\racah{j_{12}}{j_3}{j_{123}}{j_4}{j}{j_{34}}
\racah{j_1}{j_2}{j_{12}}{j_3}{j_{123}}{j_{23}}
\racah{j_1}{j_{23}}{j_{123}}{j_4}{j}{j_{234}}
=
\racah{j_2}{j_3}{j_{23}}{j_4}{j_{234}}{j_{34}}
\racah{j_1}{j_2}{j_{12}}{j_{34}}{j}{j_{234}},
\\\addlinespace
\sum_{j_{12}}
\racah{j_1}{j_2}{j_{12}}{j_{34}}{j}{j_{234}}
\racah{j_{12}}{j_3}{j_{123}}{j_4}{j}{j_{34}}
\racah{j_1}{j_2}{j_{12}}{j_3}{j_{123}}{j_{23}}
=
\racah{j_1}{j_{23}}{j_{123}}{j_4}{j}{j_{234}}
\racah{j_2}{j_3}{j_{23}}{j_4}{j_{234}}{j_{34}},
\\\addlinespace
\label{eq:BE}
\sum_{j_{34}}
\racah{j_2}{j_3}{j_{23}}{j_4}{j_{234}}{j_{34}}
\racah{j_1}{j_2}{j_{12}}{j_{34}}{j}{j_{234}}
\racah{j_{12}}{j_3}{j_{123}}{j_4}{j}{j_{34}}
=
\racah{j_1}{j_2}{j_{12}}{j_3}{j_{123}}{j_{23}}
\racah{j_1}{j_{23}}{j_{123}}{j_4}{j}{j_{234}},
\\\addlinespace
\sum_{j_{234}}
\racah{j_1}{j_{23}}{j_{123}}{j_4}{j}{j_{234}}
\racah{j_2}{j_3}{j_{23}}{j_4}{j_{234}}{j_{34}}
\racah{j_1}{j_2}{j_{12}}{j_{34}}{j}{j_{234}}
=
\racah{j_{12}}{j_3}{j_{123}}{j_4}{j}{j_{34}}
\racah{j_1}{j_2}{j_{12}}{j_3}{j_{123}}{j_{23}}.
\end{gather}
\end{subequations}
One can equivalently obtain all relations from the first one by repeatedly applying the Racah coefficients orthogonality relations
\begin{subequations}
\begin{gather}
\sum_{j_{12}}
\racah{j_1}{j_2}{j_{12}}{j_3}{j}{j_{23}}
\racah{j_1}{j_2}{j_{12}}{j_3}{j}{j_{23}\Prime}
= \delta(j_{23},j_{23}\Prime)
\mathcal{D}(j_1,j_2|j_{12})\mathcal{D}(j_2,j_3|j_{23})
\\\addlinespace
\sum_{j_{23}}
\racah{j_1}{j_2}{j_{12}}{j_3}{j}{j_{23}}
\racah{j_1}{j_2}{j_{12}\Prime}{j_3}{j}{j_{23}}
= \delta(j_{12},j_{12}\Prime)
\mathcal{D}(j_1,j_2|j_{12})\mathcal{D}(j_2,j_3|j_{23}),
\end{gather}
\end{subequations}
where
\begin{equation}
\mathcal{D}(j_1,j_2|j_{12}):=
\begin{cases}
1\quad&\mbox{if } j_{12}\in \mathcal{D}(j_1,j_2)\\
0 &\mbox{if } j_{12}\not\in \mathcal{D}(j_1,j_2).
\end{cases}
\end{equation}

\subsection[Recovering the Lorentzian Ponzano--Regge model]{Recovering the Lorentzian Ponzano--Regge model from the Hamiltonian constraint}\label{recover}

We have now all the tools to discuss the quantum Hamiltonian constraint and the Lorentzian Ponzano--Regge model\footcite{Freidel:2000uq, Davids:2000kz}. The classical Hamiltonians given in \eqref{eq:general_hamiltonian_constraint} can be quantised using the quantum observables \(E\), \(F\) and \(\widetilde F\); we will choose the ordering exactly as it appears in the classical equations. The quantum Hamiltonians are given by
\begin{align}
\hat{H}_{abc}^{\braket{}}=& {E}_{ca} \widetilde{E}_{ca} -\paren*{ {E}_{cb} {E}_{ba} - \widetilde{F}_{cb} {F}_{ba} } \frac{\widetilde{E}_{ca}}{E_b}
\\
\hat{H}_{abc}^{\braketbb{}}=& \widetilde{E}_{ca} {E}_{ca} - \paren*{ {F}_{cb} \widetilde{F}_{ba} - \widetilde{E}_{cb} \widetilde{E}_{ba} } \frac{{E}_{ca}}{E_b}
\\
\hat{H}_{abc}^{\braketrb{}}=& \widetilde{F}_{ca} {F}_{ca} - \paren*{ {E}_{cb} \widetilde{F}_{ba} - \widetilde{F}_{cb} \widetilde{E}_{ba} } \frac{{F}_{ca}}{{E_b}}
\\
\hat{H}_{abc}^{\braketlb{}}=& {F}_{ca}\widetilde {F}_{ca} - \paren*{ {F}_{cb} {E}_{ba} - \widetilde{E}_{cb} {F}_{ba} } \frac{\widetilde{F}_{ca}}{E_b}.
\end{align}
Note that there is no ordering ambiguity in the fractional term, as \( E_b \) and, say, \( \widetilde E_{ca} \) act on different nodes. On the other hand, there is an ordering ambiguity between \( E_b^{-1} \) and the other terms where one of the indices is \( b \): this particular ordering was chosen to ensure that the Lorentzian Ponzano--Regge amplitude, given by the Lorentzian Racah coefficient, is a solution of these constraints. To prove this we restrict ourselves to a triangular subgraph, given by the spin network\footnote{This is the intertwiner equivalent of the triangular face of \cref{fig:flatness}.}
\begin{equation}
\psi(j_2,j_3,j_4):=\inttet;
\end{equation}
we made explicit only the dependence on \(j_2\), \(j_3\) and \(j_4\) as these are the only legs that can be changed by \(\hat H_{abc}\), when \( \braket{abc} \) is a permutation of \( \braket{342} \).

Let us consider the particular quantum Hamiltonian constraint given by 
\begin{equation}
\hat{H}_{342}^{\braketrb{}}= \widetilde{F}_{23} {F}_{23} - \paren*{ {E}_{24} \widetilde{F}_{43} - \widetilde{F}_{24} \widetilde{E}_{43} } \frac{{F}_{23}}{{E_4}}
\equiv
\widetilde{F}_{23} {F}_{23} + \paren*{ {E}_{24} \widetilde{F}_{34} - \widetilde{F}_{24} {E}_{34} } \frac{{F}_{23}}{{E_4}};
\end{equation}
all the other cases can be treated in the same way.
The proof  that \(\psi\) it is annihilated by the operator \(\hat{H}^{\langle]}_{342}\) consists in showing that the action of \(\hat{H}^{\langle]}_{342}\) on \(\psi\) provides a recursion relation for the Racah coefficient, essentially the Biedenharn--Elliott relation. This was already discussed at length, for both the  (undeformed and deformed) vector case \footcite{Bonzom:2011hm, Bonzom:2014bua} and the spinor case \footcite{Bonzom:2011nv} in the Euclidean framework; we will see here that this is also happening in the Lorentzian case\footnote{\cite{Girelli:2015ija}.}. Note that this new result relies on the knowledge of recoupling theory between finite and infinite\Hyphdash{}dimensional representations investigated in \cref{sec:3d-lorentz-group}: without it the Racah coefficients appearing in \eqref{eq:intertwiner_actions} would not be defined.

Making use of equations \eqref{eq:intertwiner_actions} we can compute the explicit action of \( \hat H^{\braketrb{}}_{342} \) on \( \psi \). For the first part of \( \hat H^{\braketrb{}}_{342} \) we get
\begin{multline}
\widetilde{F}_{23} F_{32} \psi{(j_2,j_3,j_4)} = -D(j_2) D(j_3) D(j_2+\half) D(j_3-\half)
\\
\times \racah{j_1}{j_2}{j_3}{\shalf}{j_3-\frac{1}{2}}{j_2+\frac{1}{2}} \racah{j_1}{j_2+\half}{j_3-\half}{\shalf}{j_3}{j_2}
\psi(j_2,j_3,j_4),
\end{multline}
while for the other 2 parts we have\footnote{Note that each operator is acting on a different node.}
\begin{multline}
{E}_{24} \widetilde{F}_{34} \frac{{F}_{23}}{{E_4}} \psi{(j_2,j_3,j_4)}=D(j_2) D(j_3) D(j_2+\half) D(j_3-\half)
\racah{j_1}{j_2}{j_3}{\frac{1}{2}}{j_3-\frac{1}{2}}{j_2+\frac{1}{2}}
\\
\times
\racah{j_3-\half}{\half}{j_3}{j_4}{j_5}{j_4-\half}
\racah{j_2+\half}{\half}{j_2}{j_4}{j_6}{j_4-\half}
\psi(j_2+\half,j_3-\half,j_4-\half)
\end{multline}
and
\begin{multline}
\widetilde{F}_{24} {E}_{34} \frac{{F}_{23}}{{E_4}} \psi{(j_2,j_3,j_4)}=-D(j_2) D(j_3) D(j_2+\half) D(j_3-\half)
\racah{j_1}{j_2}{j_3}{\frac{1}{2}}{j_3-\frac{1}{2}}{j_2+\frac{1}{2}}
\\
\times
\racah{j_3-\half}{\half}{j_3}{j_4}{j_5}{j_4+\half}
\racah{j_2+\half}{\half}{j_2}{j_4}{j_6}{j_4+\half}
\psi(j_2+\half,j_3-\half,j_4+\half).
\end{multline}
Using the definitions of the Racah coefficients \eqref{eq:racah_coeff1} and the fact that, as a consequence of \eqref{eq:CG_orth2}, when \(j,j\Prime\in\mathcal{D}(j_1,j_2)\)
\begin{equation}
\intbubble{j'}{j}{j_1}{j_2}=\delta(j,j\Prime)\intidA{j}{j}\equiv \delta(j,j\Prime)\intidA{j'}{j'},
\end{equation}
we see that
\begin{equation}
\psi(j_2,j_3,j_4)=\sum_j\racah{j_1}{j_2}{j_3}{j_4}{j_5}{j}\inthambubble=\racah{j_1}{j_2}{j_3}{j_4}{j_5}{j_6}\trivalentlr{j_1}{j_6}{j_5};
\end{equation}
moreover, we can adapt \eqref{eq:BE} to our situation as\footnote{Note that \( J_4\in\mathcal{D}(\half,j_4) \).}
\begin{equation}
\racah{j_1}{J_2}{J_3}{\half}{j_3}{j_2}
\racah{j_1}{j_2}{j_3}{j_4}{j_5}{j_6}
=\sum_{J_4=j_4-\half}^{j_4+\half}
\racah{J_2}{\half}{j_2}{j_4}{j_6}{J_4}
\racah{j_1}{J_2}{J_3}{J_4}{j_5}{j_6}
\racah{J_3}{\half}{j_3}{j_4}{j_5}{J_4}.
\end{equation}
Substituting these results in the action of the Hamiltonian, it follows that
\begin{equation}
\hat{H}^{\langle]}_{342} \psi(j_2,j_3,j_4)=0.
\end{equation}

\section{Relationship with \texorpdfstring{\(\SU(2)\)}{SU(2)} theory}\label{sec:relationship-with-su2-theory}

The framework we have constructed automatically describes the Euclidean case as well. Mathematically, this is a consequence of the fact that \(\SU(2)\) and \(\SU(1,1)\) are two real forms of the complex Lie group \(\SL(2,\mathbb{C})\), i.e.
\begin{equation}
\SU(2)_\mathbb{C}\cong \SU(1,1)_\mathbb{C} \cong \SL(2,\mathbb{C}).
\end{equation}
As a consequence, the complex representations of the two groups coincide; in particular, \(\SU(1,1)\) representation theory contains as a subcase all the finite-dimensional representations of \(\SU(2)\) used in Euclidean LQG.
In our description, every notion at the representation theory level (spinor operators, Racah coefficients, etc.) has \emph{by design} not been restricted to unitary representations, instead allowing for any irreducible one. The only exception is the definition of the Hilbert space structure, which however, as we noted, is still valid if we restrict to finite-dimensional representations alone; consequently, everything at the quantum level can be used to described the Euclidean case as well, by using intertwiners between finite-dimensional representations.

The same is true at the classical level. Recall that for finite-dimensional representations the spinor operators satisfy \(\widetilde T_\pm=\mp {T}\Dagger*_\mp\); using the equivalent reality condition \(\widetilde{\tau}_\pm = \mp \conj{\tau}_\mp\) for classical spinors, the group element \eqref{eq:holonomy} becomes
\begin{equation}
g = \frac{1}{\sqrt{\braket{\tau|\tau}\braket{w|w}}}
\begin{pmatrix}
\conj{w}_+ {\tau}_+ + {w}_- \conj\tau_- & {w}_- \conj\tau_+ - \conj{w}_+ {\tau}_-\\
{w}_+ \conj{\tau}_- - \conj{w}_- \tau_+ & \conj{w}_- \tau_- + {w}_+ \conj{\tau}_+
\end{pmatrix}.
\end{equation}
Since
\begin{equation}
{\braket{\tau|\tau}}=\abs{\tau_-}^2 + \abs{\tau_+}^2 \geq 0,
\end{equation}
we have
\begin{equation}
g_{22}=\conj{g}_{11},\quad g_{21}=-\conj{g}_{12},
\end{equation}
which makes \(g\) an element\footnote{Recall that \( g \) was normalised so that \( \det(g)=1 \).} of \(\SU(2)\). The \(\su(2)\) Poisson brackets are recovered by letting \(x_\pm\rightarrow -\ii x_\pm\); the same transformation, at the quantum level, makes the finite-dimensional representations unitary (as \(\SU(2)\) representations).

\section{Concluding remarks}

We have seen in this chapter that, thanks to our results from \cref{sec:3d-lorentz-group}, it was possible to extend the spinorial formalism of loop quantum gravity to the \( 3 \)D Lorentzian case; moreover, we were able to reproduce the results of \cite{Bonzom:2011nv}, namely we constructed an Hamiltonian constraint using the spinor operators, and we showed, using an opportunely generalised Biedenharn--Elliott relation, that it is solved by the Ponzano--Regge amplitude.
The Racah coefficients we defined in \cref{sec:racah} were essential in obtaining this result, as they appear in the action of the spinorial observables and in the Biedenharn--Elliott relation. It is important to note that, when working with continuous representations, the Racah coefficients we use may involve non-unitary infinite-dimensional representations, which are not in the inner product space we defined: this should not be seen as an issue\footnote{In fact, even though the action of the Hamiltonian takes us outside of the inner product space, it maps the Ponzano--Regge amplitude to the zero vector.}, but rather as a consequence of the parametrisation of the Hamiltonian constraint using complex variables.
One should note that, following Bonzom and Livine, we only considered the case of a triangular face; for a complete treatment more general cases should be investigated.

\newcommand{\jtot}{J_\text{tot}}

\chapter{\texorpdfstring{\(\SU(2)\)}{SU(2)} intertwiners from \texorpdfstring{\(\SO\Star(2n)\)}{SO*(2n)} representations}
\label{chap:so*}

In this last chapter we will consider a second application of non-compact groups to quantum gravity. Unlike \cref{sec:lorentzian_LQG}, we will work with Euclidean loop quantum gravity, with \( 
\SU(2) \) as gauge group; despite the fact that the gauge group is compact, we will show that a non-compact group appears naturally when working in the spinorial formalism, i.e., when we rewrite the \( \SU(2) \) generators using the \JS\ representation. Two main results are presented in the chapter: we will first show how the non-compact group \( \SO\Star(2n) \), whose properties are reviewed in \cref{sec:so*_def}, has a natural action on the space of all \( n \)-valent intertwiners, a generalisation of the known fact that the space of \( n \)-valent intertwiners with fixed area provides a \( \mathrm{U}(n)\subset\SO\Star(2n) \) representation\footcite{freidel_fine_2010}; this result, together with a review of the \( \mathrm{U}(n) \) one, is the topic of \cref{sec:action-on -int-space}. The second result is an application of the first one: in \cref{sec:coherent-intertwiners} we will use the \( \SO\Star(2n) \) structure to construct a new kind of coherent intertwiners, following the Gilmore--Perelomov construction of coherent states for arbitrary Lie groups. We will then study the properties of these coherent states, in particular the matrix elements and expectation values of the algebra generators, and the semi-classical limit. Finally we will see how these states are connected to the symplectic group and to Bogoliubov transformations, which will allow us to give a physical interpretation to \( \SO\Star(2n) \) as the subgroup of Bogoliubov transformation of the \JS\ harmonic oscillators which are compatible with the \( \SU(2) \) invariance.

\section{The Lie group \texorpdfstring{\(\SO\Star(2n)\)}{SO*(2n)} and its Lie algebra}\label{sec:so*_def}

The non-compact Lie group \(G=\SO\Star(2n)\) is the subgroup of \(\SU(n,n)\)  consisting of matrices that preserve the symmetric form
\begin{equation}
x_1 y_{n+1}+y_1 x_{n+1}+{x}_2 y_{n+2}+y_2 {x}_{n+2} + \dotsb + {x}_n y_{2n}+y_{n}{x}_{2n},\quad x,y\in \C^{2n},
\end{equation}
that is
\begin{equation}
\SO\Star(2n) = \set*{ g \in \SU(n,n) \setstx g\Transpose
\begin{pmatrix}
0  & \1_n \\ \1_n & 0
\end{pmatrix}
g = 
\begin{pmatrix}
0  & \1_n \\ \1_n & 0
\end{pmatrix}
}.
\end{equation}
Recall that \(\SU(n,n)\) is the group of complex matrices with determinant \(1\) preserving the indefinite Hermitian form
\begin{equation}
\conj{x}_1 y_1 + \conj{x}_2 y_2 + \dotsb + \conj{x}_n y_n - \conj{x}_{n+1} y_{n+1}-\dotsb - \conj{x}_{2n} y_{2n},\quad x,y\in\C^{2n},
\end{equation}
i.e.,
\begin{equation}
\SU(n,n)=\set*{g\in \SL(2n,\C)  \setstx g\Star
\begin{pmatrix}
\1_n & 0\\ 0 & -\1_n
\end{pmatrix}
g=
\begin{pmatrix}
\1_n & 0\\ 0 & -\1_n
\end{pmatrix}
}.
\end{equation}
Elements of \( \SO\Star(2n) \) can be parametrised\footnote{Details can be found in \cref{app:bounded_symmetric_domains}.} as \( 2\times 2 \) block matrices
\begin{equation}
g=\begin{pmatrix}
A & B \\ -\conj B & \conj A
\end{pmatrix},\quad A,B\in M_n(\C)
\end{equation}
with \( \det(A)\neq 0 \) and
\begin{equation}
\label{eq:so*_conditions}
\begin{aligned}
AA\Star-BB\Star&=\1,\\
A\Star A-B\Transpose\conj B &=\1,
\end{aligned}
\qquad\quad
\begin{aligned}
A\Star B &=- B\Transpose \conj A,\\
BA\Transpose &= - A B\Transpose,
\end{aligned}
\end{equation}
with inverse
\begin{equation}
g^{-1}=
\begin{pmatrix}
A\Star & B\Transpose\\
-B\Star & A\Transpose
\end{pmatrix}.
\end{equation}
The maximal compact subgroup \(K\subseteq G\) is isomorphic to \(\mathrm{U}(n)\), and is given by the elements of the form
\begin{equation}
\begin{pmatrix}
U & 0 \\
0 & \conj{U}
\end{pmatrix},
\quad U\in \mathrm{U}(n).
\end{equation}
The group is non-compact for all \( n\geq 2 \), while \( \SO\Star(2)\cong U(1) \).

The Lie algebra of \( \SO\Star(2n) \) is
\begin{equation}
\so\Star(2n) = \set*{V \in \su(n,n) \setstx V\Transpose
\begin{pmatrix}
0  & \1_n \\ \1_n & 0
\end{pmatrix}
= -
\begin{pmatrix}
0  & \1_n \\ \1_n & 0
\end{pmatrix}
V
},
\end{equation}
where
\begin{equation}
\su(n,n)=\set*{V\in \mathfrak{sl}(2n,\C)  \setstx V\Star
\begin{pmatrix}
\1_n & 0\\ 0 & -\1_n
\end{pmatrix}
= -
\begin{pmatrix}
\1_n & 0\\ 0 & -\1_n
\end{pmatrix}
V
}
\end{equation}
and
\begin{equation}
\mathfrak{sl}(2n,\C) =\set*{V\in M_n(\C)  \setstx \tr V=0 };
\end{equation}
its elements are parametrised by \( 2\times 2 \) block matrices
\begin{equation}
V=
\begin{pmatrix}
X & Y \\ - \conj Y & \conj X
\end{pmatrix},\quad X,Y\in M_n(\C)
\end{equation}
satisfying
\begin{equation}
X\Star=-X,\quad Y\Transpose =- Y,
\end{equation}
so that \( \dim \so\Star(2n)=n(2n-1) \).
A basis for \( \so\Star(2n)_\C \cong \so(2n,\C) \) is given by the matrices
\begin{equation}
E_{ab}=
\begin{pmatrix}
\Delta_{ab} & 0 \\ 0 & -\Delta_{ba}
\end{pmatrix},\quad
F_{ab}=
\begin{pmatrix}
0 & 0 \\ \Delta_{ab} -\Delta_{ba} & 0
\end{pmatrix},\quad
\widetilde F_{ab}=
\begin{pmatrix}
0 & \Delta_{ab} -\Delta_{ba} \\ 0 & 0
\end{pmatrix},\quad
\end{equation}
where \( a,b=1,\dotsc,n \) and \( \Delta_{ab}\in M_n(\C) \) is the matrix with entries
\begin{equation}
(\Delta_{ab})_{cd}=\delta_{ac}\delta_{bd};
\end{equation}
the \( E_{ab} \) matrices span the complexification of the subalgebra \( \mathfrak{u}(n) \).
The commutation relations of the \( \so\Star(2n) \) complexified generators are
\begin{subequations}
\label{eq:so*_comm}
\begin{align}
[E_{ab},E_{cd}]		&= \delta_{cb}E_{ad} - \delta_{ad}E_{cb} \\
[E_{ab},\widetilde F_{cd}]	&= \delta_{bc}\widetilde F_{ad} - \delta_{bd}\widetilde F_{ac} \\
[E_{ab},F_{cd}]		&= \delta_{ad}F_{bc} - \delta_{ac}F_{bd} \\
[F_{ab},\widetilde F_{cd}]	&= \delta_{db}E_{ca} + \delta_{ca}E_{db}  - \delta_{cb}E_{da} -\delta_{da}E_{cb} \\
[F_{ab},F_{cd}]		&= [\widetilde F_{ab},\widetilde F_{cd}] = 0,
\end{align}
\end{subequations}
and unitary representations are those for which
\begin{equation}
E\Dagger*_{ab}=E_{ba},\quad F\Dagger*_{ab}=\widetilde F_{ab}.
\end{equation}
It will prove useful to also introduce the notation
\begin{equation}
E_\alpha:=\alpha^{ab}E_{ab},\quad \widetilde F_z := z^{ab}\widetilde F_{ab}, \quad F_z := \conj z^{ab} F_{ab},\quad \alpha,z \in M_n(\C),
\end{equation}
where we use the complex conjugate of \( z \) in \( F_z \) to ensure that when the representation is unitary \( (F_z)\Dagger = \widetilde F_z \); these \( \so\Star(2n)_\C \) elements satisfy the commutation relations
\begin{subequations}
\begin{align}
[E_\alpha,E_\beta] &= E_{[\alpha,\beta]}
\\
[E_\alpha,\widetilde F_z] &= \widetilde F_{\alpha z + z \alpha\Transpose}
\\
[E_\alpha, F_z] &= - F_{\alpha\Star z + z \conj\alpha}
\\
[F_w,\widetilde F_z] &= E_{(z-z\Transpose)(w-w\Transpose)\Star}.
\end{align}
\end{subequations}

\section{\texorpdfstring{\( \SO\Star(2n) \)}{SO*(2n)} action on intertwiner space}\label{sec:action-on -int-space}

We will now see how, when working in the spinorial setting\footnote{i.e., making use of the \JS\ representation.}, the space of all \( \SU(2) \) intertwiners with \( n \) legs possesses a natural \( \SO\Star(2n) \) action, and is in fact an irreducible representation of the latter. Recall that the \JS\ representation for \( \SU(2) \) takes the form
\begin{equation}
J_z = \half(A\Dagger A - B\Dagger B),\quad J_+ = A\Dagger B,\quad J_- = B\Dagger A
\end{equation}
where
\begin{equation}
[A,A\Dagger]=[B,B\Dagger]=\1
\end{equation}
are two decoupled harmonic oscillators\footnote{i.e., all \( A \)'s and \( B \)'s commute.}. They act as a Heisenberg group representation on the orthonormal basis\footcite[chap.~XII]{messiah1}
\begin{equation}
\ket{n_A,n_B}_\text{HO}\equiv \ket{n_A}_\text{HO} \otimes \ket{n_B}_\text{HO},\quad n_A,n_b \in \N_0,
\end{equation}
where 
\begin{equation}
\begin{aligned}
A\ket{n_A}_\text{HO}=\sqrt{n_A}\ket{n_A}_\text{HO},
\\\addlinespace 
B\ket{n_B}_\text{HO}=\sqrt{n_B}\ket{n_B}_\text{HO},
\end{aligned}
\quad
\begin{aligned}
A\Dagger\ket{n_A-1}_\text{HO}=\sqrt{n_A+1}\ket{n_A+1}_\text{HO},
\\\addlinespace 
B\Dagger\ket{n_B-1}_\text{HO}=\sqrt{n_B+1}\ket{n_B+1}_\text{HO};
\end{aligned}
\end{equation}
the numbers \( n_A \) and \( n_B \) are the eigenvalues of the number operators
\begin{equation}
N_A:= A\Dagger A,\quad N_B := B\Dagger B.
\end{equation}
The standard \( \SU(2) \) basis for the representation \( F_j \) can be rewritten in the harmonic oscillator basis as
\begin{equation}
\ket{j,m}=\ket{j+m,j-m}_\text{HO},\quad m\in\cM_j.
\end{equation}
One can easily check that
\begin{equation}
J^2 = \tfrac14 (E-\1)(E+\1),\quad E:=A\Dagger A + B\Dagger B + \1,
\end{equation}
with
\begin{equation}
E\ket{j,m}=(2j+1)\ket{j,m},
\end{equation}
that is, in some sense, \( E \) provides (almost) a square root of the Casimir.

We can now extend this construction to the intertwiner space as follows. We denote by \( \operatorname{Inv}_{\SU(2)}(F_{j_1}\otimes\dotsb \otimes F_{j_n}) \) the set of \( \SU(2) \) invariant vectors in the tensor product of \( n \) \( \SU(2) \) irreducible unitary representations, that is those that are annihilated by the \emph{total angular momentum}\footnote{Here the components of \( \vec{J}^{(a)} \) are the the generators acting on the representation \( F_{j_a} \).}
\begin{equation}
\vec{J}:=\sum_{a=1}^n \vec{J}^{(a)},
\end{equation}
which we can identify with \( n \)-legged intertwiners. We then introduce the \JS\ representation for each leg, i.e., we use \( 2n \) harmonic oscillators\footnote{It is implicitly assumed that the operators with subscript \( a \) only act on \( F_{j_a} \).}
\begin{equation}
[A_a,A\Dagger*_b]=[B_a,B\Dagger*_b]=\delta_{ab}\1
\end{equation}
to write
\begin{equation}
J^{(a)}_z=\half \paren{ A\Dagger*_a A_a - B\Dagger*_a B_a}, \quad J^{(a)}_+ = A\Dagger*_a B_a,\quad J^{(a)}_- = B\Dagger*_a B_a.
\end{equation}
The \( E \) operator we defined for a single leg generalises to the \( 2n \) operators\footnote{These are the Euclidean equivalent of the observables we defined in \cref{sec:lorentzian_LQG} from the spinors operators \( T \) and \( \widetilde T \), and are in fact scalar operators for \( \SU(2) \).}
\begin{equation}\label{eq:ho_E}
E_{ab} = A\Dagger*_a A_b + B\Dagger*_a B_b +\delta_{ab}\1
\end{equation}
satisfying the commutation relations
\begin{equation}
[E_{ab},E_{cd}] = \delta_{cb} E_{ad} - \delta_{ad} E_{cb},
\end{equation}
which are those of a \( \mathfrak{u}(n)_\C \) algebra (see \eqref{eq:so*_comm}). These operators can be used to construct all the usual LQG observables, namely
\begin{equation}
\vec{J}^{(a)}\cdot \vec{J}^{(b)} \equiv 2 \mathcal{A}_{ab}\mathcal{A}_{ba} - \mathcal{A}_a \mathcal{A}_b - (1-2\delta_{ab})\mathcal{A}_a,
\end{equation}
where
\begin{equation}
\mathcal{A}_{ab}:= \half (E_{ab}-\delta_{ab}\1),\quad \mathcal{A}_a:=\mathcal{A}_{aa},
\end{equation}
as it is easy to show. We are going to interpret the eigenvalues of the operator \( \mathcal{A}_a \)
\begin{equation}
\mathcal{A}_a\ket{j_a,m_a}= j_a \ket{j_a,m_a}
\end{equation}
as the \emph{area} associated to the leg \( a \), hence we will refer to  the \( \mathcal{A}_a \)'s as \emph{area operators}; the operator \( \mathcal{A}:= \sum_a \mathcal{A}_a \) gives us the total area of the intertwiner.

It was shown in \cite{freidel_fine_2010} that the space of intertwiners with a fixed total area\footnote{the fact that the total area must be an integer follows from the selection rules of the addition of angular momenta.} \( J \in \N_0\)
\begin{equation}
\mathcal{H}^J_n = \bigoplus_{\sum_{a}j_a=J} \operatorname{Inv}_{\SU(2)}(V_{j_1}\otimes\dotsb \otimes V_{j_n})
\end{equation}
has the structure of an irreducible unitary representation of \( U(n) \), whose infinitesimal action is given by the \( E_{ab} \) operators we defined\footnote{Our notation differs from that of the article, namely our \( E_{ab} \) have an additional \( \delta_{ab} \) term, which as we will see is essential to construct the \( \SO\Star(2n) \) representation.}. Explicitly,
\begin{equation}
\mathcal{H}^J_n \equiv [J+1,J+1,1,\dotsc,1],
\end{equation}
where the \( [\lambda_1,\lambda_2,\dotsc,\lambda_n] \), with
\begin{equation}
\lambda_1\geq \lambda_2 \geq\dotsb \geq \lambda_n\geq 0,
\end{equation}
denotes the \( U(n) \) representation with highest weight vector \( \ket{\lambda} \), for which
\begin{equation}
E_{aa}\ket{\lambda}=\lambda_a\ket{\lambda} \quad\mbox{and}\quad E_{ab}\ket{\lambda}=0,\quad \forall a<b;
\end{equation}
this particular choice of \( \lambda \)'s is required to for \( \SU(2) \) invariance. The dimension of \( U(n) \) representations can be computed with the \emph{hook-length formula}\footcite[chap.~4]{Iachello_2015}
\begin{equation}
\dim [\lambda_1,\dotsc,\lambda_n]= \prod_{a<b}\frac{\lambda_a - \lambda_b + b-a}{b-a},
\end{equation}
which in our specific case gives
\begin{equation}
\dim [\lambda_1,\lambda_2,1,\dotsc,1]= \frac{\lambda_1 - \lambda_2 +1}{\lambda_1} \binom{\lambda_1+n-2}{\lambda_1-1} \binom{\lambda_2+n-3}{\lambda_2-1},
\end{equation}
so that
\begin{equation}
\dim \mathcal{H}^J_n = \frac{1}{J+1} \binom{J+n-1}{J}\binom{J+n-2}{J};
\end{equation}
one can check\footnote{Refer to the aforementioned paper.}
that this is indeed the dimension of the space of \( n \)-legged intertwiners with fixed total area.

We will now show how, in addition to the action of \( U(n) \) on each \( \mathcal{H}^J_n \), there is an action of \( \SO\Star(2n) \)the full space on \( n \)-legged intertwiners
\begin{equation}
\mathcal{H}_n:=\bigoplus_{J=0}^\infty \mathcal{H}^J_n
\end{equation}
which is a new result. To do so, we introduce the operators\footnote{These operators were already known: what is new is the fact that, thanks to the \( \delta_{ab} \) we introduced in \eqref{eq:ho_E}, they form a closed algebra together with the \( E_{ab} \)'s.}
\begin{subequations}
\begin{align}
F_{ab}		& = B_a A_b - A_a B_b\\
\widetilde F_{ab}	& = B\Dagger*_a A\Dagger*_b - A\Dagger*_a B\Dagger*_b
\end{align}
\end{subequations}
which act respectively as \emph{ladder operators} for the total area, i.e.,
\begin{equation}
[\mathcal{A} , \widetilde F_{ab}] = \widetilde F_{ab},\quad [\mathcal{A} , F_{ab}] =  F_{ab};
\end{equation}
together with the \( E_{ab} \) operators we defined in \eqref{eq:ho_E}, they satisfy the commutation relations \eqref{eq:so*_comm}, which are those of an \( \so\Star(2n)_\C \) algebra. Since
\begin{equation}
E\Dagger*_{ab} = E_{ba}, \quad F\Dagger*_{ab} = \widetilde F_{ab},
\end{equation}
we can see \( \mathcal{H}_n \) as a unitary representation\footnote{Technically speaking it is a unitary \((\mathfrak{g},K)\)-module, but we will refer to it as an \( \SO\Star(2n) \) representation for simplicity.} of \( \SO\Star(2n) \), which is irreducible since, as it easy to see, \( \ker \widetilde F_{ab} =\set{0} \), for \( a\neq b \). The fact that the repeated action of the \( E \), \( F \), and \( \widetilde F \) operators on \( \ket{0} \) is still an intertwiner follows from the fact that they satisfy
\begin{equation}
[\vec{J},E_{ab}]= [\vec{J},F_{ab}] = [\vec{J},\widetilde F_{ab}]=0,
\end{equation}
i.e., they are all scalar operators.

Finally, as an aside, note that the \( \widetilde F_{ab} \) operators we introduced can be used to obtain an explicit expression for the highest weight vectors in the \( \mathrm{U}(n) \) representation \( \mathcal{H}^J_n \); in fact
\begin{proposition}
The highest weight vector for the \( \mathrm{U}(n) \) representation \( \mathcal{H}^J_n \) is defined up to a phase factor as
\begin{equation*}
\ket{\psi_J} := \mathcal{N}_J (\widetilde F_{12})^J \ket{0}, \quad \mathcal{N}_J = \frac{1}{\sqrt{J! (J+1)!}}.
\end{equation*}
\end{proposition}
\begin{proof}
First note that, as a consequence of the commutation relations \eqref{eq:so*_comm}, we have
\begin{equation}
[E_{ab},\widetilde F_{12}] = \delta_{b1} \widetilde F_{a2} + \delta_{b2} \widetilde F_{a1} =
\begin{cases}
\widetilde F_{12} & \casesif a=b\leq 2
\\
0 & \casesif a=b>2
\\
0 & \casesif a<b.
\end{cases}
\end{equation}
If we assume that \( \ket{\psi_J} \) is a highest weight vector,
using the fact that
\begin{equation}
\ket{\psi_{J+1}} = \frac{\mathcal{N}_{J+1}}{\mathcal{N}_{J}} \widetilde F_{12} \ket{\psi,J}
\end{equation}
we get
\begin{equation}
E_{ab}\ket{\psi_{J+1}} \propto \paren*{ \widetilde F_{12} E_{ab} + [E_{ab},\widetilde F_{12}] } \ket{\psi_J} = 
\begin{cases}
(J+2)\ket{\psi_{J+1}} & \casesif a=b\leq 2
\\
\ket{\psi_{J+1}} & \casesif a=b> 2
\\
0 & \casesif a<b;
\end{cases}
\end{equation}
since the result is true for \( J =0 \), as
\begin{equation}
E_{ab}\ket{\psi_0}= E_{ab}\ket{0}=\delta_{ab}\ket{0},
\end{equation}
it follows by induction that \( \ket{\psi_J} \) is a highest weight vector for all \( J \in \N_0 \). The normalisation factor is chosen so that \( \braket{\psi_J|\psi_J} = 1\); we will see how to calculate this inner product later in the chapter, with \cref{prop:probability}.
\end{proof}

\section{Coherent intertwiners}\label{sec:coherent-intertwiners}

In this section we will consider an application of the fact that the space of all \( n \)-valent \( \SU(2) \) intertwiners forms an irreducible \( \SO\Star(2n) \) representation. Following Perelomov, we will introduce a set of coherent states for \( \SO\Star(2n) \) which, being based on the intertwiner representations, provide a new kind of coherent intertwiners. We are first going to review the construction of Gilmore--Perelomov coherent states, then apply it to the specific case we are interested in. We are then going to analyse the properties of these states, specifically the matrix elements and expectation values of the \( \so\Star(2n) \) generators and the semi-classical limit. Finally, we are going to investigate the connection of the coherent states with the symplectic group \( \Sp(4n,\R) \) and with Bogoliubov transformations.

\subsection{\texorpdfstring{\( \SO\Star(2n) \)}{SO*(2n)} coherent states}

The full understanding of the group structure underlying the \( E_{ab} \), \( F_{ab} \) and \( \widetilde{F}_{ab} \) operators allows us to construct a new kind of coherent states in the intertwiner space, namely the \emph{Gilmore--Perelomov generalised coherent states\footcite{perelomov_article}} for \( \SO\Star(2n) \). This construction generalises and complements the coherent intertwiners presented in \cite{FreidelLivine2011}, which make use of the \( \mathrm{U}(n) \) structure  and live in the space of intertwiners with a fixed area.

Recall that generalised coherent states for a unitary irreducible module \( V \) of a generic Lie Group \( G \) are defined as
\begin{equation}
\ket{g}:=g\ket{\psi_0},\quad g\in G,
\end{equation}
where \( \ket{\psi_0}\in V \) is a fixed state of norm \( 1 \). Note that, at this stage, there is no guarantee that two coherent states labelled by different group elements indeed describe physically different states (i.e., they are not the same vector up to a phase\footnote{Note that since the representation is unitary and \( \ket{\psi_0} \) has norm \( 1 \), so does every \( \ket{g} \).}). In fact, let  \( H\subseteq G \) be the maximal subgroup that leaves \( \ket{\psi_0} \) invariant up to a phase, that is
\begin{equation}
h\ket{\psi_0}=e^{\ii\theta(h)}\ket{\psi_0},\quad \forall h\in H,
\end{equation}
which will be called the \emph{isotropy subgroup} for \( \ket{\psi_0} \): it is obvious that if \( g_2\in g_1 H \) then
\begin{equation}
\ket{g_2}=e^{\ii\theta} \ket{g_1},
\end{equation}
i.e., the two states are equivalent. The inequivalent coherent states are labelled by elements of the \emph{left coset space}
\begin{equation}
G/H:=\set{gH\setst g\in G},
\end{equation}
and are given by
\begin{equation}
\ket{x}:=\ket{g_x}=g_x\ket{\psi_0}, \quad \forall x\in G/H,
\end{equation}
where \( g_x\in x \) is a representative of the equivalence class \( x \).

For the particular case of the intertwiner representation of \( \SO\Star(2n) \), we will choose the harmonic oscillator vacuum \( \ket 0 \) as our fixed state. It is easy to see that the isotropy subgroup for \( \ket{0} \) is the maximal compact subgroup \( K=\mathrm{U}(n)\subset \SO\Star(2n) \); the coset space \( \SO\Star(2n)/\mathrm{U}(n) \) can be identified with one of the \emph{bounded symmetric domains} classified by Cartan\footnote{More information on this bounded symmetric domain, as well as some the proofs of some of the statements presented in the following can be found in \cref{app:bounded_symmetric_domains}.}, namely
\begin{equation}
\SO\Star(2n)/\mathrm{U}(n)\cong\Omega_n:=\set{\zeta\in M_n(\C)\setst \zeta\Transpose=-\zeta \mbox{ and } \zeta\Star\zeta<\1},
\end{equation}
on which \(  \SO\Star(2n)  \) acts holomorphically and transitively as
\begin{equation}\label{eq:so*_action_bounded}
g(\zeta)\equiv
\begin{pmatrix}
A & B \\ C & D
\end{pmatrix}
(\zeta):=\paren{A\zeta+B}\paren{C\zeta+D}^{-1}.
\end{equation}
The isotropy subgroup\footnote{Here we mean the subgroup of all \( g\in G \) such that \( g(0)=0 \).} at \( \zeta=0 \) is given by \( K \), and the correspondence between \( \Omega_n \) and \( \SO\Star(2n)/\mathrm{U}(n) \) is given by
\begin{equation}
\zeta\in\Omega_n \mapsto \set{g\in G \setst g(0)=\zeta}\equiv g_\zeta K  \in \SO\Star(2n)/\mathrm{U}(n),
\end{equation}
where\footnote{Here \( \sqrt{M} \) denotes the \emph{unique} positive semi-definite square root of a positive semi-definite matrix \( M \) . Recall that, since the square root is unique, we have \((\sqrt{A})\Transpose\equiv \sqrt{A\Transpose}\) and analogous expressions for \(\conj A\), and~\(A\Star\).}
\begin{equation}
g_\zeta:=
\begin{pmatrix}
X_\zeta & \zeta \conj X_\zeta\\
\zeta\Star X_\zeta & \conj X_\zeta
\end{pmatrix},
\quad X_\zeta:=\sqrt{(\1-\zeta\zeta\Star)^{-1}};
\end{equation}
the coherent intertwiner states are then given by
\begin{equation}
\ket{\zeta}:=g_\zeta\ket{0}, \quad \zeta\in\Omega_n.
\end{equation}
Note how
\begin{equation}
\ket{\zeta}\equiv \ket{g_\zeta(0)}
\end{equation}
and, indeed,
\begin{equation}
g\ket{\zeta}=e^{\ii\theta(g,\zeta)}\ket{g(\zeta)},\quad \forall g\in G,\forall \zeta\in\Omega_n.
\end{equation}
A more explicit expression for these states can be obtained using the following \namecref{lem:UDL}.
\begin{lemma}[Block \(UDL\) decomposition]\label{lem:UDL}
Any element of \(\SO\Star(2n)\) can be decomposed as
\begin{equation*}
\begin{split}
\begin{pmatrix}
A & B\\-\conj B& \conj A
\end{pmatrix}
&=
\begin{pmatrix}
\1 & B\conj A^{-1}\\0&\1
\end{pmatrix}
\begin{pmatrix}
(A\Star)^{-1} &0\\0&\conj A
\end{pmatrix}
\begin{pmatrix}
\1 &0\\ -\conj A^{-1}\conj B &\1
\end{pmatrix}\\
&=\exp\paren*{\half \widetilde F_{B\conj A^{-1}}}
\exp\paren{E_L}
\exp\paren*{-\half F_{ A^{-1} B}}
\end{split}
\end{equation*}
where \(L\) is such that \(e^{-L}=A\Star\). Note that, unless \( B=0 \), the factors do not belong to \(\SO\Star(2n)\) anymore, but to its \emph{complexification} \( \SO(2n,\C) \) instead.
\end{lemma}
\begin{proof}
\begin{proofenumerate}
\item Since \(A\) must be invertible, the matrices appearing in the decomposition are well-defined. One can check explicitly that the LHS equals the RHS, making use of the fact that
\begin{equation}
\begin{split}
(A\Star)^{-1}-B\conj A^{-1}\conj B &= (A\Star)^{-1} + (A\Star)^{-1}B\Transpose \conj B\\&= (A\Star)^{-1}-(A\Star)^{-1}(A\Star A - \1)\\
&=A.
\end{split}
\end{equation}
\item
To see how the exponentials arise, notice that both \(B\conj A^{-1}\) and \(\conj A^{-1}\conj B\) are antisymmetric, as a consequence of \eqref{eq:so*_conditions}. For any antisymmetric matrix \(T\) we have
\begin{equation}
\half  \widetilde F_T =
\half \sum_{a,b} T_{ab}
\begin{pmatrix}
0 & \Delta_{ab}-\Delta_{ba}\\ 0 & 0
\end{pmatrix}
=
\sum_{a,b}
\begin{pmatrix}
0 & T_{ab}\Delta_{ab}\\ 0 & 0
\end{pmatrix}
=
\begin{pmatrix}
0 & T\\ 0 & 0
\end{pmatrix}
\end{equation}
so that
\begin{equation}
\exp\paren*{ \half \widetilde F_{ B\conj A^{-1}}}=
\begin{pmatrix}
\1 &  B\conj A^{-1}\\ 0 & \1
\end{pmatrix};
\end{equation}
similarly
\begin{equation}
\exp\paren*{-\half F_{ A^{-1} B}}=
\begin{pmatrix}
\1 & 0\\ - \conj A^{-1}\conj B & \1
\end{pmatrix}.
\end{equation}
For the middle matrix in the factorisation, recall that any invertible complex matrix admits a (non-unique) logarithm. Since \(A\Star\) is invertible, there is \(L\) such that \(e^{L}=(A\Star)^{-1}\); moreover, it follows from the properties of the matrix exponential that \(\conj A=e^{-L\Transpose}\). Then
\begin{equation}
E_L=
\sum_{a,b}L_{ab}
\begin{pmatrix}
\Delta_{ab} & 0 \\
0 & - \Delta_{ba}
\end{pmatrix}
=
\begin{pmatrix}
L & 0\\ 0 & -L\Transpose
\end{pmatrix}
\end{equation}
so that
\begin{equation}
\exp(E_L)=
\begin{pmatrix}
e^L & 0\\ 0 & e^{-L\Transpose}
\end{pmatrix}
=
\begin{pmatrix}
(A\Star)^{-1} &0\\0&\conj A
\end{pmatrix},
\end{equation}
which concludes the proof.
\end{proofenumerate}
\end{proof}
As a consequence of \cref{lem:UDL} we can rewrite \( g_\zeta \) as
\begin{equation}
g_\zeta=\exp\paren*{\half \widetilde{F}_\zeta}\exp(E_L)\exp\paren[\Big]{-\half F_{X_\zeta^{-1} \zeta \conj X_\zeta}}
\end{equation}
where \( L \) is such that
\begin{equation}
e^L=\sqrt{\1-\zeta\zeta\Star}.
\end{equation}
Since \( \ket{0} \) is annihilated by every \( F_{ab} \) and
\begin{equation}
e^{ E_L}\ket{0}=e^{\tr L}\ket{0}=\det(e^L)\ket{0}=\det(1-\zeta\Star\zeta)^{\shalf}\ket{0}
\end{equation}
we can eventually write the coherent states as
\begin{equation}
\ket{\zeta}=\mathcal{N}(\zeta)\exp\paren*{\half\widetilde F_\zeta}\ket{0},\quad \mathcal{N}(\zeta)=\det(1-\zeta\Star\zeta)^{\shalf}.
\end{equation}
Using the fact that the representation is unitary, we can write the inner product between two coherent states as
\begin{equation}
\braket{\omega|\zeta}=\braket{0|g_\omega^{-1} g_\zeta|0},
\end{equation}
with
\begin{equation}
g_\omega^{-1} g_\zeta=
\begin{pmatrix}
X_\omega(\1-\omega \zeta\Star)X_\zeta & X_\omega (\zeta-\omega)\conj X_\zeta\\
\conj X_\omega (\zeta\Star-\omega\Star) X_\zeta & \conj X_\omega(\1-\omega\Star \zeta)\conj X_\zeta
\end{pmatrix}
\end{equation}
which automatically ensures
\begin{equation}
\det(\1-\omega\Star\zeta)\neq 0,
\end{equation}
as \(\conj X_\omega(\1-\omega\Star \zeta)\conj X_\zeta\) must be invertible. We know from \Cref{lem:UDL} that the group element can be written as
\begin{equation}
g_\omega^{-1} g_\zeta=
\exp\paren{ \widetilde F_\alpha}
\exp\paren{ E_\Lambda}
\exp\paren{ F_\beta}
\end{equation}
for some \(\alpha\) and \(\beta\), with \( \Lambda \) such that
\begin{equation}
e^\Lambda=X_\omega^{-1}(\1-\zeta\omega\Star)^{-1}X_{\omega}^{-1}=\sqrt{\1-\zeta\zeta\Star}(\1-\zeta\omega\Star)^{-1}\sqrt{\1-\omega\omega\Star},
\end{equation}
so that
\begin{equation}
\braket{\omega|\zeta}=\det(e^\Lambda)\braket{0|0}=\frac{\det(\1-\zeta\Star\zeta)^{\shalf}\det(\1-\omega\Star\omega)^{\shalf} }{\det(\1-\omega\Star\zeta)};
\end{equation}
the Cauchy--Schwarz inequality ensures that
\begin{equation}
\abs{\braket{\omega|\zeta}}^2\leq 1
\end{equation}
where the equality only holds when \(\omega=\zeta\), as by definition states labelled by different cosets are not proportional to each other.

\subsubsection{Summary}

To summarise, we have constructed a set of coherent intertwiners
\begin{equation}
\ket{\zeta}=\det(1-\zeta\Star\zeta)^{\shalf}(\zeta)\exp\paren*{\half\widetilde F_\zeta}\ket{0},\quad \zeta\in\Omega_n,
\end{equation}
where \( \Omega_n \) is the set of anti-symmetric matrices \( \zeta\in M_n(\C) \) satisfying \( \zeta\Star\zeta<\1 \). They are all independent from each other, and their inner product is given by
\begin{equation}
\braket{\omega|\zeta}=\frac{\det(\1-\zeta\Star\zeta)^{\shalf}\det(\1-\omega\Star\omega)^{\shalf} }{\det(\1-\omega\Star\zeta)} \leq 1,
\end{equation}
with the equality holding only when \( \omega=\zeta \). They are Gilmore--Perelomov coherent states, as they satisfy
\begin{equation}
g\ket{\zeta}=e^{\ii\theta(g,\zeta)}\ket{g(\zeta)}, \quad g\in\SO\Star(2n),
\end{equation}
where the action of \( g \) on \( \zeta \) is given by \eqref{eq:so*_action_bounded}, i.e., up to a phase factor, the action of the group goes through the coherent states.

It is important to mention that, although this construction is new, some of these states have been considered before in \cite{Freidel:2012ji}, \cite{Freidel:2013fia}, and \cite{Bonzom:2012bn}, although the underlying group structure was not known. Nevertheless, the coherent states presented in those articles are only those such that \( \rank(\zeta)=2 \)---which are exactly those that can be seen as a linear combination of the \( \mathrm{U}(n) \) coherent states for all possible areas---so that the vast majority of the states we constructed in this section are indeed new.
\subsection{Matrix elements of the \texorpdfstring{\(\so\Star(2n)\)}{so*(2n)} generators}
The easiest way to compute the matrix elements of the \( \so\Star(2n) \) generators \( E_{ab} \), \( F_{ab} \) and \( \widetilde{F}_{ab} \) in the coherent state basis is to make use of the \( 2n \) harmonic oscillator operators \( A_a \), \( B_a \); in particular, we are going to project the states \( \ket{\zeta} \) on the well-known harmonic oscillator coherent states.
Recall that\footcite[chap.~3]{perelomov_book} coherent states for the representation of the Heisenberg group \( \mathrm{H}_{2n} \) with generators acting as
\begin{equation}
\commutator{A_a,A\Dagger*_b}=\commutator{B_a,B\Dagger*_b}=\delta_{ab}\1,
\end{equation}
on the vector space spanned by the vectors\footnote{Here we use the \emph{multi-index notation}, that is we have  \( \paren{A\Dagger}^{\mu}:= \paren{A\Dagger*_1}^{\mu_1}\dotsb\paren{A\Dagger*_n}^{\mu_n} \) and \( \mu!:=\mu_1!\dotsc\mu_n! \), with \( \mu\in \N_0^n \).}
\begin{equation}
\ket{\mu,\nu}=\frac{\paren{A\Dagger}^{\mu}}{\sqrt{\mu!}}\frac{\paren{B\Dagger}^{\nu}}{\sqrt{\nu!}}\ket{0},\quad \mu,\nu\in \N_0^n,
\end{equation}
where \( \ket{0}\equiv \ket{0,0} \) is the harmonic oscillator vacuum
\begin{equation}
A_a\ket{0}=B_a\ket{0}=0,
\end{equation}
are the vectors
\begin{equation}
\ket{\alpha,\beta}:=e^{-\shalf \paren{\alpha\Star\alpha +\beta\Star\beta} }\sum_{\mu,\nu\in\N_0^n}\frac{\alpha^\mu}{\sqrt{\mu!}}\frac{\beta^\nu}{\sqrt{\nu!}}\ket{\mu,\nu},\quad \alpha,\beta\in\C^n
\end{equation}
satisfying
\begin{equation}
A_a\ket{\alpha,\beta}=\alpha_a\ket{\alpha,\beta},\quad B_a\ket{\alpha,\beta}=\beta_a\ket{\alpha,\beta}.
\end{equation}
The resolution of the identity in terms of these coherent states is given by
\begin{equation}
\int_{\C^{2n}}\eder\mu(\alpha,\beta)\ket{\alpha,\beta}\bra{\alpha,\beta}=\1,
\end{equation}
where the measure of integration is\footnote{Here \( \Re \) and \( \Im \) denote respectively the real and imaginary part of a complex number.}
\begin{equation}
\eder\mu(\alpha,\beta)= \frac{1}{\pi^{2n}}\eder^n\Re(\alpha)\,\eder^n\,\Im(\alpha) \, \eder^n\Re(\beta)\,\eder^n\,\Im(\beta).
\end{equation}

We can now use the fact that
\begin{equation}
\begin{split}
\braket{\alpha,\beta|\zeta}&=\mathcal{N}(\zeta)\braket{\alpha,\beta|\exp\paren*{\half \widetilde{F}_\zeta}|0}
\\
&=\mathcal{N}(\zeta)\braket{\alpha,\beta|0}e^{\beta\Star\zeta\conj{\alpha}}
\\
&=\mathcal{N}(\zeta)e^{\beta\Star\zeta\conj{\alpha}-\shalf\paren{\alpha\Star\alpha+\beta\Star\beta}}
\end{split}
\end{equation}
to write
\begin{equation}\label{eq:gaussian_int1}
\begin{split}
\braket{\omega|\zeta}&=\int_{\C^{2n}} \eder\mu(\alpha,\beta)\braket{\omega|\alpha,\beta}\braket{\alpha,\beta|\zeta}
\\
&= \mathcal{N}(\omega)\mathcal{N}(\zeta)
\int_{\C^{2n}} \eder\mu(\alpha,\beta)e^{\beta\Star\zeta\conj\alpha+\beta\Transpose \conj\omega\alpha - \alpha\Star\alpha-\beta\Star\beta}
\\
&=\mathcal{N}(\omega)\mathcal{N}(\zeta)
\int_{\C^{2n}}\eder\mu(\alpha,\beta)\,\exp \bracks*{
-\half
\begin{smallpmatrixDiv}{*{4}{c}} 
\alpha\Transpose & \beta\Transpose & \conj \alpha\Transpose & \conj \beta\Transpose
\end{smallpmatrixDiv}
\begin{smallpmatrixDiv}{cc|cc}
0 & \conj\omega & \1 & 0\\
-\conj\omega & 0 & 0 & \1\\\hline
\1 & 0 & 0 & \zeta \\
0 & \1 & -\zeta & 0
\end{smallpmatrixDiv}
\begin{smallpmatrix}
\alpha \\ \beta \\ \conj \alpha \\ \conj\beta
\end{smallpmatrix}
},
\end{split}
\end{equation}
which is a Gaussian integral\footnote{In fact, we could also have calculated \( \braket{\omega|\zeta} \) by evaluating this integral.}. Although we already know the value of \( \braket{\omega|\zeta} \), we can use this expression to calculate the matrix elements of any operator built as a polynomial in the harmonic oscillator operators thanks to the following \namecref{prop:gaussian_integral}.
\begin{proposition}\label{prop:gaussian_integral}
Let
\begin{equation*}
S=\int e^{-\shalf x\Transpose A x} \,\eder^n x,
\end{equation*}
with \(A\in M_{n}(\C)\) symmetric and invertible, be a convergent Gaussian integral\footnote{It is assumed that all the requirements on $A$ such that the integral converges are satisfied.}. Then
\begin{equation*}
\int x_{a_1}x_{a_2}\dotsb x_{a_k} e^{-\shalf x\Transpose A x} \,\eder^n x
=
S
\eval{\pder{}{J_{a_1}}\pder{}{J_{a_2}}\dotsb \pder{}{J_{a_k}}}{J=0} e^{\shalf J\Transpose A^{-1} J}
\end{equation*}
for any \(k\in \N\); in particular, the integral vanishes whenever \(k\) is odd.
\end{proposition}
\begin{proof}
First note that
\begin{equation}
\int x_{a_1}x_{a_2}\dotsb x_{a_k} e^{-\shalf x\Transpose A x} \,\eder^n x
=
\int
\eval{\pder{}{J_{a_1}}\pder{}{J_{a_2}}\dotsb \pder{}{J_{a_k}}
}{J=0} e^{-\shalf x\Transpose A x + J\Transpose x} \,\eder^n x.
\end{equation}
With the change of variable \(x\rightarrow x+ A^{-1} J\) one has
\begin{equation}
\begin{split}
-\half x\Transpose A x + J\Transpose x \rightarrow&\, -\half x\Transpose A x - \half J\Transpose x - \half x\Transpose J - \half J\Transpose A^{-1} J+ J\Transpose x + J\Transpose A^{-1}J\\
&=-\half x\Transpose A x + \half J\Transpose A^{-1}J
\end{split}
\end{equation}
so that
\begin{equation}
\int x_{a_1}x_{a_2}\dotsb x_{a_k} e^{-\shalf x\Transpose A x} \,\eder^n x = \eval{\pder{}{J_{a_1}}\pder{}{J_{a_2}}\dotsb \pder{}{J_{a_k}}
}{J=0} 
e^{\shalf J\Transpose A^{-1}J}
\int e^{-\shalf x\Transpose A x } \,\eder^n x
\end{equation}
as required. Note that, since \(J\Transpose A^{-1} J\) is a quadratic polynomial in \(J_1,\dotsc,J_n\), if \(k\) is odd there is a leftover factor of \(J\) after \(k\) derivatives, which makes the whole integral vanish when evaluated at \(J=0\). 
\end{proof}

\Cref{prop:gaussian_integral} can be used to find matrix elements by starting with \eqref{eq:gaussian_int1} and setting
\begin{equation}
J:=
\begin{pmatrix}
X \\ Y \\ \conj X \\ \conj Y
\end{pmatrix}
,\quad
A:=
\begin{pmatrixDiv}{cc|cc}
0 & \conj\omega & \1 & 0\\
-\conj\omega & 0 & 0 & \1\\\hline
\1 & 0 & 0 & \zeta \\
0 & \1 & -\zeta & 0
\end{pmatrixDiv};
\end{equation}
one can easily check that
\begin{equation}
A^{-1}=
\begin{smallpmatrixDiv}{cc|cc}
0 & -\zeta(\1-\omega\Star\zeta)^{-1} & (\1-\zeta\omega\Star)^{-1} & 0 \\
\zeta(\1-\omega\Star\zeta)^{-1} & 0 & 0 & (\1-\zeta\omega\Star)^{-1}\\
\hline
(\1-\omega\Star\zeta)^{-1} & 0 & 0 & -(\1-\omega\Star\zeta)^{-1}\conj\omega\\
0 & (\1-\omega\Star\zeta)^{-1} & (\1-\omega\Star\zeta)^{-1}\conj\omega & 0
\end{smallpmatrixDiv},
\end{equation}
so that
\begin{equation}\label{eq:gaussian_source}
\begin{split}
S(\conj\omega,\zeta):=\half J\Transpose A^{-1} J =&\, Y\Transpose \zeta \paren{\1-\omega\Star\zeta}^{-1} X + \conj Y\Transpose \paren{\1-\omega\Star\zeta}^{-1}\conj\omega\conj X
\\
&+ \conj X\Transpose \paren{\1-\omega\Star\zeta}^{-1} X + \conj Y\Transpose \paren{\1-\omega\Star\zeta}^{-1} Y.
\end{split}
\end{equation}
Then, for any operator of the form\footnote{Here we use the multi-index notation again.}
\begin{equation}
p(A,B,A\Dagger,B\Dagger)=
A^{k_1} B^{k_2} \paren{A\Dagger}^{k_3} \paren{B\Dagger}^{k_4},
\quad k_1,k_2,k_3,k_4\in \N_0^n,
\end{equation}
where it is important that all the raising operators are on the right (anti-normal ordering)\footnote{If they are not, they can always be rewritten in this form up to some summands proportional to the identity, for which it is trivial to compute matrix elements.}, we have, as a consequence of \cref{prop:gaussian_integral},
\begin{equation}
\begin{split}
\braket{\omega|p(A,B,A\Dagger,B\Dagger)|\zeta} &= \int_{\C^{2n}}\eder\mu(\alpha,\beta)p(\alpha,\beta,\conj\alpha,\conj\beta)\braket{\omega|\alpha,\beta}\braket{\alpha,\beta|\zeta}
\\
&=\braket{\omega|\zeta}\eval{p\paren[\big]{\nabla_X,\nabla_Y,\nabla_{\conj X},\nabla_{\conj Y}}}{X=Y=0} e^{S(\conj\omega,\zeta)},
\end{split}
\end{equation}
where
\begin{equation}
(\nabla_X)^k=\pder{^{k_1}}{X_1^{k_1}}\pder{^{k_2}}{X_2^{k_2}}\dotsb \pder{^{k_n}}{X_n^{k_n}},\quad k\in\N_0^n.
\end{equation}
In particular, we have:
\begin{proposition}\label{prop:coherent_matrix_elements}
The matrix elements of the \( \so\Star(2n) \) generators in the coherent state basis are given by
\begin{align*}
\braket{\omega|E_{ab}|\zeta}&=\braket{\omega|\zeta}\big[ \1 + 2 \omega\Star\zeta (\1-\omega\Star\zeta)^{-1}\big]_{ab}\\
\braket{\omega|F_{ab}|\zeta}&=\braket{\omega|\zeta}\big[2 \zeta(\1-\omega\Star\zeta)^{-1}\big]_{ab}\\
\braket{\omega|\widetilde F_{ab}|\zeta}&=\braket{\omega|\zeta}\big[2 (\1-\omega\Star\zeta)^{-1}\conj \omega\big]_{ab}
\end{align*}
\end{proposition}
\begin{proof}
\begin{proofenumerate}
\item First let us rewrite \( E_{ab} \) as
\begin{equation}
E_{ab}=A_b A\Dagger*_a + B_b B\Dagger*_a - \delta_{ab}
\end{equation}
using the commutation relations of the harmonic oscillators. Then we can insert the resolution of the identity for the \( \mathrm{H}_{2n} \) coherent states to obtain
\begin{equation}
\begin{split}
\braket{\omega|A_b A\Dagger*_a|\zeta} &= \int_{\C^{2n}}\eder\mu(\alpha,\beta)\braket{\omega|A_b|\alpha,\beta}\braket{\alpha,\beta|A\Dagger*_a|\zeta}
\\
&= \int_{\C^{2n}}\eder\mu(\alpha,\beta)\,\conj{\alpha_a}\alpha_b\braket{\omega|\alpha,\beta}\braket{\alpha,\beta|\zeta}
\\
&= \mathcal{N}(\omega)\mathcal{N}(\zeta) \int_{\C^{2n}}\eder\mu(\alpha,\beta)\,\conj{\alpha_a}\alpha_b \,e^{\beta\Star\zeta\conj\alpha+\beta\Transpose \conj\omega\alpha - \alpha\Star\alpha-\beta\Star\beta};
\end{split}
\end{equation}
applying \cref{prop:gaussian_integral} together with \cref{eq:gaussian_int1,eq:gaussian_source} we obtain
\begin{equation}
\begin{split}
\braket{\omega|A_b A\Dagger*_a|\zeta} &= \braket{\omega|\zeta}\eval{\pder{}{\conj X_a}\pder{}{X_b}}{X=Y=0} e^{\conj X\Transpose \paren{\1-\omega\Star\zeta}^{-1} X + \dotsb}
\\
&= \braket{\omega|\zeta}\bracks[\big]{\paren{\1-\omega\Star\zeta}^{-1}}_{ab}.
\end{split}
\end{equation}
Similarly
\begin{equation}
\braket{\omega|B_b B\Dagger*_a|\zeta} = \braket{\omega|\zeta}\bracks[\big]{\paren{\1-\omega\Star\zeta}^{-1}}_{ab}
\end{equation}
so that
\begin{equation}
\braket{\omega|E_{ab}|\zeta}=\braket{\omega|\zeta}\bracks[\big]{2\paren{\1-\omega\Star\zeta}^{-1} -\1}_{ab}=\braket{\omega|\zeta}\bracks[\big]{\1 + 2\omega\Star\zeta\paren{\1-\omega\Star\zeta -\1}^{-1}}_{ab}
\end{equation}
as
\begin{equation}
\paren{\1-X}^{-1}=\1+ X\paren{\1-X}^{-1}.
\end{equation}
\item To obtain the matrix elements of \( F_{ab} \) we insert the resolution of the identity again, which gives
\begin{equation}
\begin{split}
\braket{\omega|B_a A_b|\zeta} &= \int_{\C^{2n}}\eder\mu(\alpha,\beta)\braket{\omega|B_a A_b|\alpha,\beta}\braket{\alpha,\beta|\zeta}
\\
&= \int_{\C^{2n}}\eder\mu(\alpha,\beta)\,\beta_a\alpha_b\braket{\omega|\alpha,\beta}\braket{\alpha,\beta|\zeta}
\\
&= \braket{\omega|\zeta} \eval{\pder{}{Y_a}\pder{}{X_b}}{X=Y=0} e^{Y\Transpose \zeta\paren{\1-\omega\Star\zeta}^{-1} X + \dotsb}
\\
&= \braket{\omega|\zeta} \bracks[\big]{\zeta\paren{\1-\omega\Star\zeta}^{-1}}_{ab},
\end{split}
\end{equation}
leading to
\begin{equation}
\braket{\omega|F_{ab}|\zeta}=\braket{\omega|\zeta} \bracks[\big]{2\zeta\paren{\1-\omega\Star\zeta}^{-1}}_{ab}
\end{equation}
as
\begin{equation}
\bracks[\big]{\zeta\paren{\1-\omega\Star\zeta}^{-1}}\Transpose=-\paren{\1-\zeta\omega\Star}^{-1}\zeta=-\zeta\paren{\1-\omega\Star\zeta}^{-1}.
\end{equation}
\item The matrix elements of \( \widetilde{F}_{ab} \) are easily obtained from the \( F_{ab} \) ones as
\begin{equation}
\begin{split}
\braket{\omega|\widetilde{F}_{ab}|\zeta} &= \conj{\braket{\zeta|F_{ab}|\omega}}
\\
&= \conj{\braket{\zeta|\omega}} \bracks[\big]{2\conj\omega\paren{\1-\zeta\omega\Star}^{-1}}_{ab}
\\
&= {\braket{\omega}|\zeta} \bracks[\big]{2\paren{\1-\omega\Star\zeta}^{-1}\conj\omega}_{ab}.
\end{split}
\end{equation}
\end{proofenumerate}
\end{proof}

\begin{proposition}[Expectation values of areas]\label{prop:expected_values}
The expectation values of the area operators in a particular coherent state \( \ket{\zeta} \) are
\begin{equation*}
\braket{\mathcal{A}_a}=\bracks[\big]{\zeta\Star\zeta\paren{\1-\zeta\Star\zeta}^{-1}}_{aa},
\quad
\braket{\mathcal{A}}=\tr\bracks*{\zeta\Star\zeta\paren{\1-\zeta\Star\zeta}^{-1}}
\end{equation*}
and their variance is
\begin{equation*}
\Var(\mathcal{A}_a)=\half\braket{\mathcal{A}_a}\paren[\big]{\braket{\mathcal{A}_a}+1},
\quad
\Var(\mathcal{A})=\sum_{a,b}\braket{\mathcal{A}_{ab}}\paren[\big]{\braket{\mathcal{A}_{ab}}+\delta_{ab}}.
\end{equation*}
Moreover, when the non-zero eigenvalues of \( \zeta\Star\zeta \) approach \( 1 \), although \(\Var(\mathcal{A})\) grows without bound, the \emph{coefficient of variation} \( \frac{\sqrt{\Var(\mathcal{A})}}{\braket{\mathcal{A}}} \) approaches a value in \( (0,1] \).
\end{proposition}
\begin{proof}
\begin{proofenumerate}
\item
The form of the expected values follows directly from \cref{prop:coherent_matrix_elements}. In order to calculate the variances, we will need the covariance\footnote{Note that the \( \mathcal{A}_a \) all commute, so there is no ordering ambiguity.}
\begin{equation}
\Cov(\mathcal{A}_a,\mathcal{A}_b):=\braket{\mathcal{A}_a\mathcal{A}_b}-\braket{\mathcal{A}_a}\braket{\mathcal{A}_b}.
\end{equation}
First note that
\begin{equation}
\begin{split}
4\mathcal{A}_a\mathcal{A}_b =&\, \paren{A_a A\Dagger*_a + B_a B\Dagger*_a-2}\paren{A_b A\Dagger*_b + B_b B\Dagger*_b-2}
\\
=&\, A_a A\Dagger*_a A_b A\Dagger*_b + B_a B\Dagger*_a B_b B\Dagger*_b + A_a A\Dagger*_a B_b B\Dagger*_b + B_a B\Dagger*_a A_b A\Dagger*_b
\\
&-4\mathcal{A}_a -4\mathcal{A}_b -4
\\
=&\, A_a A_b A\Dagger*_a A\Dagger*_b + B_a B_b B\Dagger*_a B\Dagger*_b + A_a B_b A\Dagger*_a B\Dagger*_b + A_b B_a A\Dagger*_b B\Dagger*_a
\\
&-4\mathcal{A}_a -4\mathcal{A}_b -2\delta_{ab}\mathcal{A}_a - 4 -2\delta_{ab}.
\end{split}
\end{equation}
Making use of the resolution of the identity for the \( \mathrm{H}_{2n} \) coherent states we get\footnote{To simplify notation we define \( \sigma := \paren{\1-\zeta\Star\zeta}^{-1} \).}
\begin{equation}
\begin{split}
\braket{A_a A_b A\Dagger*_a A\Dagger*_b} &= \int_{\C^{2n}}\eder\mu(\alpha,\beta)\braket{\zeta|A_a A_b|\alpha,\beta}\braket{\alpha,\beta| A\Dagger*_a A\Dagger*_b|\zeta}
\\
&= \int_{\C^{2n}}\eder\mu(\alpha,\beta)\,\alpha_a \alpha_b \conj\alpha_a \conj\alpha_b \braket{\zeta|\alpha,\beta}\braket{\alpha,\beta|\zeta}
\\
&= \eval{\pder{}{X_a}\pder{}{X_b}\pder{}{\conj X_a}\pder{}{\conj X_b}}{X=Y=0}e^{\conj{X}\Transpose\sigma X+\dotsb}
\\
&= \pder{}{X_a}\pder{}{X_b} \paren[\big]{(\sigma X)_a + (\sigma X)_b}
\\
&=\sigma_{aa}\sigma_{bb}+\sigma_{ab}\sigma_{ba}
\end{split}
\end{equation}
and similarly
\begin{equation}
\braket{B_a B_b B\Dagger*_a B\Dagger*_b} = \sigma_{aa}\sigma_{bb}+\sigma_{ab}\sigma_{ba},
\end{equation}
while for the term with both harmonic oscillators we have
\begin{equation}
\begin{split}
\braket{A_a B_b A\Dagger*_a B\Dagger*_b} &= \int_{\C^{2n}}\eder\mu(\alpha,\beta)\braket{\zeta|A_a B_b|\alpha,\beta}\braket{\alpha,\beta| A\Dagger*_a B\Dagger*_b|\zeta}
\\
&= \int_{\C^{2n}}\eder\mu(\alpha,\beta)\,\alpha_a \beta_b \conj\alpha_a \conj\beta_b \braket{\zeta|\alpha,\beta}\braket{\alpha,\beta|\zeta}
\\
&= \eval{\pder{}{X_a}\pder{}{Y_b}\pder{}{\conj X_a}\pder{}{\conj Y_b}}{X=Y=0}e^{Y\Transpose \zeta \sigma X + \conj Y\Transpose \sigma \conj\zeta \conj X +\conj{X}\Transpose\sigma X+ \conj Y \sigma Y}
\\
&= \eval{\pder{}{X_a}\pder{}{Y_b}\pder{}{\conj X_a}}{X=Y=0} \paren[\big]{(\sigma Y)_b + \paren[\big]{\sigma \conj\zeta \conj X}_b}
\, e^{Y\Transpose \zeta \sigma X +\conj{X}\Transpose\sigma X+ \dotsb}
\\
&= \eval{\pder{}{X_a}\pder{}{Y_b}}{X=Y=0} \paren[\big]{(\sigma X)_a (\sigma Y)_b + \paren[\big]{\sigma \conj\zeta}_{ba}}
\, e^{Y\Transpose \zeta \sigma X + \dotsb}
\\
&= \sigma_{aa}\sigma_{bb}+\paren[\big]{\sigma \conj\zeta}_{ba} \paren{\zeta\sigma}_{ba}
\\
&= \sigma_{aa}\sigma_{bb} + \paren{\sigma\zeta\Star}_{ab}\paren{\zeta\sigma}_{ba}
\\
&= \sigma_{aa}\sigma_{bb} + \paren{\sigma\zeta\Star}_{ba}\paren{\zeta\sigma}_{ab}
\end{split}
\end{equation}
Eventually we can compute the covariance as\footnote{Recall that \( \zeta\Star\zeta\sigma=\sigma-\1 \), so that \( \braket{\mathcal{A}_a}=\sigma_{aa}-1 \).}
\begin{equation}
\begin{split}
\Cov(\mathcal{A}_a,\mathcal{A}_b) =& \sigma_{aa}\sigma_{bb}+\half\sigma_{ab}\sigma_{ba} + \half \paren{\sigma\zeta\Star}_{ab}\paren{\zeta\sigma}_{ba} -\sigma_{aa} - \sigma_{bb} 
\\
& - \half \delta_{ab}\sigma_{ab} + 1 - \sigma_{aa}\sigma_{bb}+\sigma_{aa}+\sigma_{bb}-1
\\
=& \half\sigma_{ab}\sigma_{ba} + \half \paren{\sigma\zeta\Star}_{ab}\paren{\zeta\sigma}_{ba} - \half \delta_{ab}\sigma_{ab},
\end{split}
\end{equation}
which leads to\footnote{Note that \( \zeta\sigma \) is anti\Hyphdash{}symmetric.}
\begin{equation}
\Var(\mathcal{A}_a) := \Cov(\mathcal{A}_a,\mathcal{A}_a)=\half\sigma_{aa}(\sigma_{aa}-1)\equiv
\half\braket{\mathcal{A}_a}\paren[\big]{\braket{\mathcal{A}_a}+1}
\end{equation}
and
\begin{equation}
\Var(\mathcal{A}) :=\sum_{a,b}\Cov(\mathcal{A}_a,\mathcal{A}_b)=\tr(\sigma^2-\sigma) \equiv
\sum_{a,b}\braket{\mathcal{A}_{ab}}\paren[\big]{\braket{\mathcal{A}_{ab}}+\delta_{ab}}.
\end{equation}
\item
The coefficient of variation for the total area is then given by
\begin{equation}
\frac{\sqrt{\Var{\mathcal{A}}}}{\braket{\mathcal{A}}}=\frac{\sqrt{\tr\bracks[\big]{\sigma(\sigma-\1)}}}{\tr(\sigma-\1)}\geq 0;
\end{equation}
making use of the fact that, as both \( \sigma \) and \( \sigma-\1 \) are positive semi-definite\footnote{Note that as \( (\1-\zeta\Star\zeta) \leq \1\), it must be \( \sigma\geq \1 \).}\footnote{Recall that, if \( A,B\in M_n{\C} \) are positive semi-definite matrices, \( \tr(AB)\leq \tr(A)\tr(B) \).},
\begin{equation}
\tr\bracks[\big]{\sigma(\sigma-1)}\leq\tr(\sigma)\tr(\sigma-\1),
\end{equation}
we obtain an upper bound for the coefficient of variation,
\begin{equation}
\frac{\sqrt{\Var{\mathcal{A}}}}{\braket{\mathcal{A}}} \leq \paren*{\frac{\tr(\sigma)}{\tr(\sigma)-n}}^{\shalf}.
\end{equation}
When the non-zero eigenvalues of \( \zeta\Star\zeta\) approach \( \1 \) we have \( \tr(\sigma)\rightarrow\infty \), so that
\begin{equation}
\frac{\sqrt{\Var{\mathcal{A}}}}{\braket{\mathcal{A}}} \lesssim 1 \quad \mbox{when} \quad \tr(\sigma)\rightarrow \infty,
\end{equation}
as expected.
\end{proofenumerate}
\end{proof}
Let us spend a few words on the last result of \cref{prop:expected_values}, regarding the coefficient of variation. This coefficient measures the \emph{relative} standard deviation, i.e., the amount of dispersion compared to the value of the mean. In our particular case, the result is telling us that, even though the dispersion gets bigger as the total area increases, the relative standard deviation is bounded by a value that approaches \( 1 \) for sufficiently large area. Note that the coefficient of variation does not provide any useful information when the area is very small, as\footnote{Using the fact that, as \( \sigma \geq 0  \), \(\tr(\1) \tr(\sigma^2)\geq \tr(\sigma)^2 \).}
\begin{equation}
\frac{\sqrt{\Var{\mathcal{A}}}}{\braket{\mathcal{A}}}=\frac{\sqrt{\tr\bracks[\big]{\sigma(\sigma-\1)}}}{\tr(\sigma-\1)}\geq
\frac{\sqrt{\frac{1}{n}\tr(\sigma)\tr(\sigma-\1)}}{\tr(\sigma-\1)}\rightarrow \infty
\end{equation}
when \( \braket{\mathcal{A}} \rightarrow 0 \), i.e., \( \sigma \rightarrow \1\).

\begin{figure}
\centering
\caption{Distribution of total area for different values of $\tr(\zeta\Star\zeta)$ when $\zeta$ is of rank $2$.}\label{fig:distribution_plots}
\setlength{\tabcolsep}{12pt}
\begin{tabular}{cc}
\tikzsetnextfilename{distribution_plot1}
\begin{tikzpicture}[trim axis left, trim axis right]
\begin{axis}[font=\tiny,width=0.55\textwidth,samples=21,domain=0:20,xmin=0,ymin=0,xmax=20,title={$\tr(\zeta\Star\zeta)=0.8$}]
\addplot[color=NavyBlue,mark=*, mark size=1pt,only marks] gnuplot {((0.6)**2)*((0.4)**x)*(x+1)}; \end{axis} 
\end{tikzpicture}
&
\tikzsetnextfilename{distribution_plot2}
\begin{tikzpicture}[trim axis left, trim axis right]
\begin{axis}[font=\tiny,width=0.55\textwidth,samples=51,domain=0:50,xmin=0,ymin=0,xmax=50,title={$\tr(\zeta\Star\zeta)=1.6$}]
\addplot[color=NavyBlue,mark=*, mark size=1pt,only marks] gnuplot {((0.2)**2)*((0.8)**x)*(x+1) }; \end{axis}
\end{tikzpicture}
\\ \addlinespace
\tikzsetnextfilename{distribution_plot3}
\begin{tikzpicture}[trim axis left, trim axis right]
\begin{axis}[font=\tiny,width=0.55\textwidth,samples=81,domain=0:80,xmin=0,ymin=0,xmax=80,title={$\tr(\zeta\Star\zeta)=1.8$}]
\addplot[color=NavyBlue,mark=*, mark size=1pt,only marks] gnuplot {((0.1)**2)*((0.9)**x)*(x+1) }; \end{axis}
\end{tikzpicture}
&
\tikzsetnextfilename{distribution_plot4}
\begin{tikzpicture}[trim axis left, trim axis right]
\begin{axis}[font=\tiny,width=0.55\textwidth,samples=301,domain=0:300,xmin=0,ymin=0,xmax=300,title={$\tr(\zeta\Star\zeta)=1.96$}]
\addplot[color=NavyBlue,mark=*, mark size=1pt,only marks] gnuplot {((0.02)**2)*((0.98)**x)*(x+1) }; \end{axis}
\end{tikzpicture}
\end{tabular}
\end{figure}
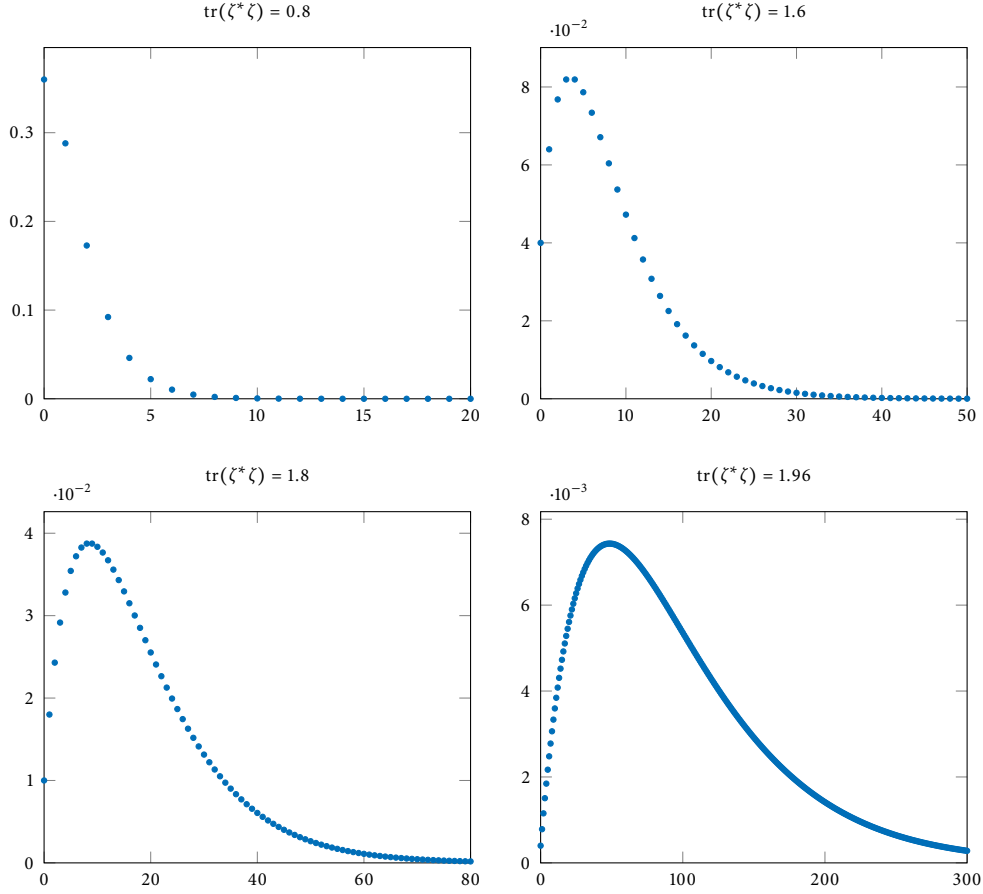

In the specific case when \( \rank(\zeta)=2 \) we can do much more than computing expectation values and variances: in fact, we can produce the complete probability distribution of the total area as follows\footnote{When \( \rank(\zeta)>2 \) an important simplifying assumption is missing, namely, as we will see, the fact that \( \zeta\zeta\Star\zeta  \) is proportional to \( \zeta \).}.
\begin{proposition}[Probability distribution of total area]\label{prop:probability}
When \(\zeta\) is of rank \(2\) the probability distribution for the total area in the state \(\ket{\zeta}\) is
\[
P_\zeta(J)=\det(\1-\zeta\Star\zeta)\paren*{\half\tr(\zeta\Star\zeta)}^J (J+1),\quad J\in \N_0.
\]
\end{proposition}
\begin{proof}
\begin{proofenumerate}
\item
Let
\begin{equation}
\ket{J,\zeta}:=\left(\half \widetilde F_\zeta\right)^J\ket{0},\quad J\in \N_0;
\end{equation}
these are eigenvectors of $\mathcal{A}$, with
\begin{equation}
\mathcal{A}\ket{J,\zeta}=J\ket{J,\zeta},
\end{equation}
as $\widetilde F_\zeta$ adds one quantum of area each time\footnote{In fact \( [\mathcal{A},\widetilde F_{\zeta}]=\widetilde F_{\zeta}\).}. 
The $\SO\Star(2n)$ coherent states can then be written as
\begin{equation}
\begin{split}
\ket{\zeta}&=\det(\1-\zeta\Star\zeta)^{\shalf} \exp\paren*{\half \widetilde F_\zeta} \ket{0}
\\
&=\det(\1-\zeta\Star\zeta)^{\shalf} \sum_{J=0}^\infty\frac{1}{J!}\ket{J,\zeta}.
\end{split}
\end{equation}
Since the $\ket{J,\zeta}$ states are mutually orthogonal\footnote{As they are eigenvectors of a self-adjoint operator, with different eigenvalues.}, the probability that $\ket{\zeta}$ is measured with total area $J$ is given by
\begin{equation}\label{eq:probability_from_projection}
P_\zeta(J)\equiv\frac{\abs{\braket{J,\zeta|\zeta}}^2}{\braket{J,\zeta|J,\zeta}}
=\frac{\det(\1-\zeta\Star\zeta)}{(J!)^2}\braket{J,\zeta|J,\zeta};
\end{equation}
it remains to calculate the norm squared of the state $\ket{J,\zeta}$.
\item
Recall that
\begin{equation}
\commutator[\big]{\half F_\zeta,\half \widetilde  F_\zeta}=E_{\frac{1}{4}(\zeta-\zeta\Transpose)(\zeta-\zeta\Transpose)\Star}= E_{\zeta\zeta\Star}
\end{equation}
and
\begin{equation}
\commutator[\big]{E_{\zeta\zeta\Star},\half \widetilde  F_\zeta}=\half \widetilde F_{\zeta\zeta\Star\zeta+\zeta\conj{\zeta}\zeta\Transpose}=\widetilde F_{\zeta\zeta\Star\zeta};
\end{equation}
moreover, since $\zeta$ is of rank $2$, one has (see \cref{app:antisymmetric})
\begin{equation}
\zeta\zeta\Star\zeta=\half \tr(\zeta\zeta\Star)\zeta,
\end{equation}
so that
\begin{equation}
\begin{split}
\commutator[\big]{E_{\zeta\zeta\Star},\paren*{\half \widetilde  F_\zeta}^k}&=\sum_{\ell=1}^{k}\paren*{\half \widetilde  F_\zeta}^{\ell-1}
\,\commutator[\big]{E_{\zeta\zeta\Star},\half \widetilde  F_\zeta}\,
\paren*{\half \widetilde  F_\zeta}^{k-\ell}
\\
&=\sum_{\ell=1}^{k}\paren*{\half \widetilde  F_\zeta}^{\ell-1}
\,\widetilde F_{\zeta\zeta\Star\zeta}\,
\paren*{\half \widetilde  F_\zeta}^{k-\ell}
\\
&=k\tr(\zeta\zeta\Star)\paren*{\half \widetilde  F_\zeta}^{k}.
\end{split}
\end{equation}
It follows that\footnote{Recall that $F_\alpha \ket{0}=0$ and that $E_{\alpha}\ket{0}=\tr(\alpha)\ket{0}$.}
\begin{equation}
\begin{split}
\half F_\zeta\,\paren*{\half \widetilde  F_\zeta}^{J}\ket{0}&=\commutator[\big]{\half  F_\zeta,\paren*{\half \widetilde  F_\zeta}^{J}}\ket{0}
\\
&=\sum_{k=1}^J \paren*{\half \widetilde  F_\zeta}^{k-1}
\,\commutator[\big]{\half F_\zeta,\half \widetilde  F_\zeta}\,
\paren*{\half \widetilde  F_\zeta}^{J-k}\ket{0}
\\
&=\sum_{k=1}^J \paren*{\half \widetilde  F_\zeta}^{k-1}
\,E_{\zeta\zeta\Star}\,
\paren*{\half \widetilde  F_\zeta}^{J-k}\ket{0}
\\
&=J\tr(\zeta\zeta\Star)\paren*{\half \widetilde  F_\zeta}^{J-1}\ket{0}+
\sum_{k=1}^J \paren*{\half \widetilde  F_\zeta}^{k-1}
\,\commutator[\big]{E_{\zeta\zeta\Star},\paren*{\half \widetilde  F_\zeta}^{J-k}}\ket{0}
\\
&=J(J+1)\half\tr(\zeta\Star\zeta)\paren*{\half \widetilde  F_\zeta}^{J-1}\ket{0}
\end{split}
\end{equation}
and in particular
\begin{equation}
\begin{split}
\braket{J,\zeta|J,\zeta}&=\braket{0|\paren*{\half  F_\zeta}^{J}\paren*{\half \widetilde  F_\zeta}^{J}|0}\\
&=J(J+1)\tr(\zeta\Star\zeta)\braket{0|\paren*{\half  F_\zeta}^{J-1}\paren*{\half \widetilde  F_\zeta}^{J-1}|0}
\\
&=J(J+1)\half\tr(\zeta\Star\zeta)\braket{J-1,\zeta|J-1,\zeta}.
\end{split}
\end{equation}
Solving the recurrence relation with $\braket{0,\zeta|0,\zeta}=\braket{0|0}=1$ we obtain
\begin{equation}
\braket{J,\zeta|J,\zeta}=J!(J+1)!\paren*{\half\tr(\zeta\Star\zeta)}^J
\end{equation}
which, plugged in \eqref{eq:probability_from_projection}, gives
\begin{equation}
P_\zeta(J)=\det(\1-\zeta\Star\zeta)\paren*{\half\tr(\zeta\Star\zeta)}^J (J+1)
\end{equation}
as expected.
\end{proofenumerate}
\end{proof}
Plots for the probability distribution can be found in \cref{fig:distribution_plots}.
Note how, as  the non-zero eigenvalues of \( \zeta\Star\zeta \) approach \( 1 \) (equivalently \( \tr(\zeta\Star\zeta)\rightarrow 2 \)), the relative shape of the distribution remains the same, as a consequence of \cref{prop:expected_values}.

\subsection{Semi-classical limit}
Let us now consider the semi-classical limit of our coherent intertwiners. Our goal is to obtain out of the expectation values of the algebra generators a set of variables that, endowed with the appropriate Poisson structure, we can interpret as a classical geometry (similarly to the classical space of \cref{sec:lorentzian_LQG}). In particular, we want to be able to construct a set of vectors that sum to zero, and as such can be regarded as the normals to a convex polyhedron\footcite{return_spinor}.

In order to investigate the semi-classical limit , it will prove useful to  rewrite the expected values of the \( \so\Star(2n) \) generators in a different way; note the similarity with the bra-ket notation we introduced in \cref{sec:classical_tensors} when working with classical spinors.
\begin{proposition}\label{prop:semiclassical}
The expected values of the \(\so\Star(2n)\) generators can be written in the form
\begin{equation*}
\braket{\zeta|F_{ab}|\zeta}=\sum_{\alpha=1}^{k}\frac{1}{\lambda_\alpha}\braketlb{z^\alpha_a|z^\alpha_b}
,\quad
\braket{\zeta|\widetilde{F}_{ab}|\zeta}=\sum_{\alpha=1}^{k}\frac{1}{\lambda_\alpha}\braketrb{z^\alpha_a|z^\alpha_b}
,\quad
\braket{\zeta|E_{ab}|\zeta}=\delta_{ab}+\sum_{\alpha=1}^{k} \braket{z^\alpha_a|z^\alpha_b},
\end{equation*}
where \( k=\shalf\rank(\zeta) \),  \(\lambda_\alpha^2\) is a non-zero eigenvalue of \( \zeta\Star\zeta \) and
\begin{equation*}
\ket{z^\alpha_a}=
\begin{pmatrix}
x^\alpha_a \\ y^\alpha_a
\end{pmatrix},
\quad
\ketb{z^\alpha_a}=
\begin{pmatrix}
\conj{y}^\alpha_a \\ -\conj{x}^\alpha_a
\end{pmatrix},
\quad
\bra{z^\alpha_a}=
\begin{pmatrix}
\conj{x}^\alpha_a & \conj{y}^\alpha_a
\end{pmatrix},
\quad
\brab{z^\alpha_a}=
\begin{pmatrix}
y^\alpha_a & -x^\alpha_a
\end{pmatrix}
\end{equation*}
satisfy
\begin{equation*}
\sum_{a=1}^{n}\ket{z^\alpha_a}\bra{z^\beta_a}=\delta_{\alpha\beta} \sum_{a=1}^n \half \braket{z^\alpha_a|z^\alpha_a}\1_2.
\end{equation*}
\end{proposition}
\begin{proof}
From \cref{lem:antisymmetric_decomposition} we know that $\zeta = U M U\Transpose$, where $U$ is unitary and
\begin{equation}
M=\bigoplus_{\alpha=1}^k \lambda_\alpha 
\begin{pmatrix}
0 & -1 \\ 1 & 0
\end{pmatrix}
\oplus \0_{n-2k},\quad \lambda_\alpha>0;
\end{equation}
then
\begin{equation}
M\Star M=\bigoplus_{\alpha=1}^k \lambda_\alpha^2
\1_2 \oplus \0_{n-2k}
\end{equation}
and
\begin{equation}
(\1-M\Star M)^{-1}=\bigoplus_{\alpha=1}^k (1-\lambda_\alpha^2)^{-1}\,
\1_2 \oplus \1_{n-2k}.
\end{equation}
\vfil
It follows that
\begin{equation}
\begin{split}
\braket{\zeta|F_{ab}|\zeta} &= \left[ 2\zeta(\1-\zeta\Star\zeta)^{-1}\right]_{ab}
\\
&= \left[ 2 U M (\1-M\Star M)^{-1}U\Transpose \right]_{ab}
\\
&= \sum_{\alpha=1}^{k}\sum_{c,d=1}^n \frac{2\lambda_\alpha}{1-\lambda_\alpha^2}
U_{ac}\paren{\delta_{c,2\alpha}\delta_{d,2\alpha-1}-\delta_{c,2\alpha-1}\delta_{d,2\alpha}}U_{bd}
\\
&= \sum_{\alpha=1}^{k} \frac{2\lambda_\alpha}{1-\lambda_\alpha^2} \paren{U_{a,2\alpha}U_{b,2\alpha-1}-U_{a,2\alpha-1}U_{b,2\alpha}}
\end{split}
\end{equation}
and
\begin{equation}
\begin{split}
\braket{\zeta|E_{ab}|\zeta} -\delta_{ab} &= \left[ 2\zeta\Star\zeta(\1-\zeta\Star\zeta)^{-1}\right]_{ab}
\\
&= \left[ 2 \conj{U} M\Star M (\1-M\Star M)^{-1}U\Transpose \right]_{ab}
\\
&= \sum_{\alpha=1}^{k}\sum_{c,d=1}^n \frac{2\lambda_\alpha^2}{1-\lambda_\alpha^2}
\conj{U}_{ac}\paren{\delta_{c,2\alpha-1}\delta_{d,2\alpha-1}+\delta_{c,2\alpha}\delta_{d,2\alpha}}U_{bd}
\\
&= \sum_{\alpha=1}^{k} \frac{2\lambda_\alpha^2}{1-\lambda_\alpha^2} \paren{\conj{U}_{a,2\alpha-1}U_{b,2\alpha-1} + \conj{U}_{a,2\alpha}U_{b,2\alpha}}.
\end{split}
\end{equation}
Choosing
\begin{equation}
\ket{z^\alpha_a}= \paren*{\frac{2\lambda^2_\alpha}{1-\lambda_\alpha^2}}^{\shalf}
\begin{pmatrix*}[l]
U_{a,2\alpha-1} \\ U_{a,2\alpha}
\end{pmatrix*}
\quad\Rightarrow\quad
\ketb{z^\alpha_a}= \paren*{\frac{2\lambda^2_\alpha}{1-\lambda_\alpha^2}}^{\shalf}
\begin{pmatrix*}[c]
\conj{U}_{a,2\alpha} \\ -\conj{U}_{a,2\alpha-1}
\end{pmatrix*}
\end{equation}
we find
\begin{equation}
\braket{\zeta|F_{ab}|\zeta}=\sum_{\alpha=1}^{k}\frac{1}{\lambda_\alpha}\braketlb{z^\alpha_a|z^\beta_b},\quad \braket{\zeta|E_{ab}|\zeta}=\delta_{ab} +\sum_{\alpha=1}^{n}  \braket{z^\alpha_a|z^\alpha_b};
\end{equation}
moreover,
\begin{equation}
\begin{split}
\sum_{a=1}^{n}\ket{z^\alpha_a}\bra{z^\beta_a} &=
\paren*{\frac{2\lambda^2_\alpha}{1-\lambda_\alpha^2}}^{\shalf}
\paren*{\frac{2\lambda^2_\beta}{1-\lambda_\beta^2}}^{\shalf}
\sum_{a=1}^n
\begin{pmatrix}
U_{a,2\alpha-1}\conj{U}_{a,2\beta-1} & U_{a,2\alpha}\conj{U}_{a,2\beta-1}
\\ \addlinespace
U_{a,2\alpha-1}\conj{U}_{a,2\beta} & U_{a,2\alpha}\conj{U}_{a,2\beta}
\end{pmatrix}
\\
&= \delta_{\alpha\beta} \frac{2\lambda^2_\alpha}{1-\lambda_\alpha^2} \1_2
\\
&= \delta_{\alpha\beta} \sum_{a=1}^n \half \braket{z^\alpha_a|z^\alpha_a}\1_2
\end{split}
\end{equation}
as expected.
\end{proof}
As consequence of this fact, in the limit
\begin{equation}
\lambda_\alpha\rightarrow 1,\quad \alpha=1,\dotsc,k
\end{equation}
where the expected value of the total area
\begin{equation}
\braket{\mathcal{A}}=\sum_{\alpha=1}^{k}\frac{\lambda_\alpha^2}{1-\lambda_\alpha^2}\rightarrow \infty,
\end{equation}
we have
\begin{equation}
\braket{\zeta|F_{ab}|\zeta} \sim \sum_{\alpha=1}^{k} \braketlb{z^\alpha_a|z^\alpha_b},
\quad
\braket{\zeta|E_{ab}|\zeta} = \delta_{ab} + \sum_{\alpha=1}^{k}\braket{z^\alpha_a|z^\alpha_b}.
\end{equation}
We can interpret the semi-classical limit as a classical geometry by introducing the canonical symplectic structure on \( \C^{2kn} \)
\begin{equation}
\omega=\ii\sum_{a=1}^{n}\sum_{\alpha=1}^{k}\paren*{\eder x^\alpha_a \wedge \eder\conj{x}^\alpha_a +  \eder y^\alpha_a \wedge \eder\conj{y}^\alpha_a},
\end{equation}
with Poisson brackets
\begin{equation}
\braces{f,g}=-\ii \sum_{a=1}^{n}\sum_{\alpha=1}^{k}\paren*{\pder{f}{x^\alpha_a}\pder{g}{\conj{x}^\alpha_a} - \pder{f}{\conj{x}^\alpha_a}\pder{g}{x^\alpha_a} + \pder{f}{y^\alpha_a}\pder{g}{\conj{y}^\alpha_a} - \pder{f}{\conj{y}^\alpha_a}\pder{g}{y^\alpha_a} },
\end{equation}
so that
\begin{equation}\label{eq:poisson_ho}
\braces{x^\alpha_a,\conj{x}^\beta_b}=\braces{y^\alpha_a,\conj{y}^\beta_b}= -\ii \delta^{\alpha\beta}\delta_{ab}
\end{equation}
with all other brackets vanishing. With this symplectic structure, the functions
\begin{equation}
e_{ab}:=\sum_{\alpha=1}^{k}\braket{z^\alpha_a|z^\alpha_b},\quad f_{ab}:=\sum_{\alpha=1}^{k}\braketlb{z^\alpha_a|z^\alpha_b}
\end{equation}
satisfy
\begin{subequations}
\begin{align}
\braces{e_{ab},e_{cd}} &= -\ii \paren{\delta_{cb}e_{ad}-\delta_{ad}e_{cb}}
\\
\braces{e_{ab},f_{cd}} &= -\ii \paren{\delta_{ad}f_{bc}-\delta_{ac}f_{bd} }
\\
\braces{e_{ab},\conj{f}_{\!cd}} &= -\ii \paren{\delta_{bc}\conj{f}_{\!ad}-\delta_{bd}\conj{f}_{\!ac} }
\\
\braces{f_{ab},\conj{f}_{\!cd}} &= -\ii \paren{\delta_{db}e_{ca}+\delta_{ca}e_{db}-\delta_{cb}e_{da}-\delta_{da}e_{cb} }
\\
\braces{f_{ab},f_{cd}}&=\braces{\conj{f}_{\!ab},\conj{f}_{\!cd}}=0,
\end{align}
\end{subequations}
which are the classical analogue of the \( \so\Star(2n) \) commutation relations \eqref{eq:so*_comm}; in fact, upon quantisation we have
\begin{equation}
x^\alpha_a\rightarrow A^\alpha_a,\quad \conj x^\alpha_a\rightarrow {A^\alpha_a}\Dagger,\quad y^\alpha_a\rightarrow B^\alpha_a,\quad \conj y^\alpha_a\rightarrow {B^\alpha_a}\Dagger,
\end{equation}
which satisfy the commutation relations of \( 2kn \) decoupled harmonic oscillators when \( \braces{\cdot,\cdot}\rightarrow -\ii\bracks{\cdot,\cdot} \), so that\footnote{The \( \delta_{ab} \) term in the definition of \( E_{ab} \) comes from the choice of ordering of the harmonic oscillators, and is needed to ensure that the \( E\), \( F \) and \( \widetilde F \) operators form a closed algebra.}
\begin{subequations}
\begin{align}
e_{ab} & \rightarrow E_{ab}:= \sum_{\alpha=1}^{k}\paren[\big]{{A^\alpha_a}\Dagger A^\alpha_b + {B^\alpha_a}\Dagger B^\alpha_b +\delta_{ab}}
\\
f_{ab} & \rightarrow F_{ab}:= \sum_{\alpha=1}^{k}\paren[\big]{B^\alpha_a A^\alpha_b + A^\alpha_a B^\alpha_b}
\\
\conj{f}_{\!ab} &\rightarrow \widetilde{F}_{ab} \:= F\Dagger*_{ab}
\end{align}
\end{subequations}
which satisfy \eqref{eq:so*_comm}.

To recover a classical geometry, we construct \( n \) (\( 3 \)-dimensional) vectors out of the \( 2kn \) \emph{spinors} \( \ket{z^\alpha_a} \), namely
\begin{equation}
\vec{V}^{(a)}(z):=\sum_{\alpha=1}^{k}\half \braket{z^\alpha_a|\vecsymbol{\sigma}|z^\alpha_a},\quad a=1,\dotsc,n,
\end{equation}
where \( \vecsymbol{\sigma} \) is the \emph{Pauli vector}
\begin{equation}
\vecsymbol\sigma=\sigma_x \hat{\vec{x}} + \sigma_y \hat{\vec{y}} + \sigma_z \hat{\vec{z}}
,\quad
\sigma_x=
\begin{pmatrix}
0 & 1 \\ 1 & 0
\end{pmatrix},
\quad
\sigma_y=
\begin{pmatrix}
0 & -\ii \\ \ii & 0
\end{pmatrix},
\quad
\sigma_y=
\begin{pmatrix}
1 & 0 \\ 0 & -1
\end{pmatrix},
\end{equation}
which in components read
\begin{subequations}
\begin{align}
V^{(a)}_x&=\sum_{\alpha=1}^{k}\half \paren{\conj{x}^\alpha_a y^\alpha_a + \conj{y}^\alpha_a x^\alpha_a}
\\
V^{(a)}_y&=\sum_{\alpha=1}^{k}\tfrac{1}{2\ii} \paren{\conj{x}^\alpha_a y^\alpha_a - \conj{y}^\alpha_a x^\alpha_a}
\\
V^{(a)}_z&=\sum_{\alpha=1}^{k}\half \paren{\conj{x}^\alpha_a x^\alpha_a - \conj{y}^\alpha_a y^\alpha_a}.
\end{align}
\end{subequations}
It follows from \cref{prop:semiclassical} that the spinors satisfy the \emph{closure contraints}
\begin{equation}
\sum_{a=1}^n\sum_{\alpha=1}^k \ket{z^\alpha_a}\bra{z^\alpha_a}=
\sum_{a=1}^n\sum_{\alpha=1}^k 
\begin{pmatrix}
\conj x^\alpha_a x^\alpha_a & \conj y^\alpha_a x^\alpha_a
\\
\conj x^\alpha_a y^\alpha_a & \conj y^\alpha_a y^\alpha_a
\end{pmatrix}
\propto \1_2,
\end{equation}
which implies that
\begin{equation}
\vec{V}:=\sum_{a=1}^{n}\vec{V}^{(a)}=0;
\end{equation}
as such, we can interpret the  \( n \) vectors as being the normal vectors to the faces of a polyhedron by means of the Minkowski theorem\footcite[The proof can be found in][chap.~7]{minkowski}
\begin{theorem}[Minkowski theorem]
Let \( \vec{v}_1,\dotsc\vec{v}_n \in \R^3 \) be vectors spanning \( \R^3 \) satisfying
\[
\vec{v}_1+ \vec{v}_2+\dotsb+\vec{v}_n=0. 
\]
Then there exist a unique (up to translation) convex polyhedron with \( n \) faces \( f_1,\dotsc,f_n \) such that \( \vec{v}_a \) is the normal vector to \( f_a \).
\end{theorem}

This construction is similar of the usual one in terms of spinors, which can be found for example in \cite{return_spinor}, and in fact coincides with it if \( \rank(\zeta)=2 \). Note that in the rank \( 2 \) case, as one can easily show,
\begin{equation}
\vec{V}^{(a)}\cdot \vec{V}^{(a)} = \tfrac{1}{4}\braket{z_a|z_a}^2 = \tfrac{1}{4}e_{aa}^2 \equiv \braket{\mathcal{A}_a}^2,
\end{equation}
that is the areas associated to the faces of the polyhedron, which are given by the length of their normals, are exactly the expectation values of the area operators. However, when \( \rank(\zeta)>2 \), due to the summation over \( \alpha \), such a relation between the norm of \( \vec{V}^{(a)} \) and \( e_{aa} \) is not available, namely
\begin{equation}
\vec{V}^{(a)}\cdot \vec{V}^{(a)} = \tfrac14 e_{aa}^2 + \sum_{\alpha,\beta=1}^{k} \paren[\big]{ \conj x^\alpha_a y^\alpha_a \conj y^\beta_a x^\beta_b - \conj x^\alpha_a x^\alpha_a \conj y^\beta_a y^\beta_a},
\end{equation}
so that at this stage the full relationship between the polyhedron we constructed and the quantum theory we started from is non fully understood. One should note that

\begin{equation}
\braces[\big]{V^{(a)}_i,V^{(a)}_j}=\tensor{\varepsilon}{_{ij}^k}V^{(a)}_k
\end{equation}
and
\begin{equation}
\braces{\vec{V},e_{ab}}=\braces{\vec{V},f_{ab}}=\braces {\vec{V}, \conj f_{\!ab}} =0,
\end{equation}
so that upon quantisation of the vectors we get
\begin{equation}
\vec{V}^{(a)} \rightarrow \vec{J}^{(a)},
\end{equation}
where
\begin{subequations}
\begin{align}
J^{(a)}_z &:=  \sum_{\alpha=1}^{k}\half\paren[\big]{{A^\alpha_a}\Dagger A^\alpha_a - {B^\alpha_a}\Dagger B^\alpha_a}
\\
J^{(a)}_+ &:= \sum_{\alpha=1}^{k}  {A^\alpha_a}\Dagger B^\alpha_a
\\
J^{(a)}_- &:= \sum_{\alpha=1}^{k}  {B^\alpha_a}\Dagger A^\alpha_a,
\end{align}
\end{subequations}
which we can regard as a generalisation of the \JS\ representation with \( 2k \) spinors instead of \( 2 \). An unexpected feature of this generalisation is that each \( \SU(2) \) representation \( F_j \) appears more than once in the Heisenberg group \( \mathrm{H}_{2k}(\R) \) representation generated by the harmonic oscillators: for example, when \( k=2 \) and \( j\in \half \N_0 \), both
\begin{equation}
\ket{(2j,0),(0,0)}_\text{HO} \quad \mbox{and} \quad \ket{(0,2j),(0,0)}_\text{HO}
\end{equation}
describe the highest weight vector \( \ket{j,j}\in F_j \),
where
\begin{equation}
\ket{(n_{A^1},n_{B^1}),(n_{A^2},n_{B^2}),\dotsc,(n_{A^k},n_{B^k})}_\text{HO} : = \bigotimes_{\alpha=1}^k \ket{n_{A^\alpha},n_{B^\alpha}}_\text{HO},
\end{equation}
while if \( j \in \N_0 \) the vector
\begin{equation}
\ket{(j,0),(j,0)}_\text{HO}
\end{equation}
works as well. It is likely that this property will play a key role in the full understanding of the semi-classical limit.

\subsection[Relationship with \texorpdfstring{\( \Sp(4n,\R) \)}{Sp(4n,R)} and Bogoliubov transformations]{Relationship with \texorpdfstring{\( \Sp(4n,\R) \)}{Sp(4n,R)} coherent states and Bogoliubov transformations}

The coherent states we have defined can be introduced can be reinterpreted in terms of \emph{Bogoliubov transformations} by making use of the connection between \( \SO\Star(2n) \) and the \emph{symplectic group} \( \Sp(4n,\R) \). Recall that, if we have a set of \( N \) decoupled harmonic oscillators
\begin{equation}
[C_a,C\Dagger*_b]=\delta_{ab},\quad [C_a,C_b]=[C\Dagger*_a,C\Dagger*_b]=0,
\end{equation}
a Bogoliubov transformation is a a canonical transformation which maps them to a new set of harmonic oscillators,
\begin{align}
\widetilde C_{a} &= U^{ab}C_b + V^{ab}C\Dagger*_b\\
\widetilde C\Dagger*_{a} &= \conj U^{ab}C\Dagger*_b + \conj V^{ab}C_b,
\end{align}
satisfying the usual commutation relations; we can write in a compact form
\begin{equation}
\begin{pmatrix*}[l]
\widetilde C \\ \widetilde C\Dagger
\end{pmatrix*}
=
\begin{pmatrix}
U & V \\ \conj V & \conj U
\end{pmatrix}
\begin{pmatrix*}[l]
C \\ C\Dagger
\end{pmatrix*}.
\end{equation}
The conditions on \( U \) and \( V \) such that
\begin{equation}
[\widetilde C_a,\widetilde C\Dagger*_b]=\delta_{ab},\quad [\widetilde C_a,\widetilde C_b]=[\widetilde C\Dagger*_a,\widetilde C\Dagger*_b]=0
\end{equation}
are
\begin{equation}\label{eq:sp_conditions}
UU\Dagger - V V\Dagger = \1,\quad U V\Transpose = V U\Transpose,
\end{equation}
which automatically ensure that \( U \) is invertible and that\footnote{See \cref{app:bounded_symmetric_domains}.}
\begin{equation}
\begin{pmatrix}
U & V \\ \conj V & \conj U
\end{pmatrix} \in \Sp(2N,\R);
\end{equation}
as such, we can interpret \( \Sp(2N,\R) \) as the group of Bogoliubov transformations of \( N \) harmonic oscillators. The vacuum for the set of new harmonic oscillators is given by
\begin{equation}
\ket{\widetilde 0 } := \mathcal{N} \exp\paren*{\half S^{ab} C\Dagger*_a C\Dagger*_b}\ket{0},
\end{equation}
also known as the \emph{squeezed vacuum}, where \( S \) is the symmetric matrix\footnote{It follows from \eqref{eq:sp_conditions}.}
\begin{equation}
S= - U^{-1}V;
\end{equation}
in fact, it is easy to see that
\begin{equation}
\begin{split}
C_d \ket{\widetilde 0} &= \mathcal{N} \sum_{k=0}^\infty \frac{1}{k!}\bracks*{C_d,\paren*{\half S^{ab} C\Dagger*_a C\Dagger*_b}^k}\ket{ 0}
\\
&= \mathcal{N} \sum_{k=0}^\infty \frac{1}{k!} k \paren*{\half S^{ab} C\Dagger*_a C\Dagger*_b}^{k-1}\paren*{\half S^{cd} C\Dagger_c + \half S^{dc}C\Dagger_c} \ket{ 0}
\\
&= \mathcal{N} \sum_{k=0}^\infty \frac{1}{k!} \paren*{\half S^{ab} C\Dagger*_a C\Dagger*_b}^{k} S^{dc} C\Dagger_c \ket{ 0}
\\
&= S^{dc} C\Dagger_c \ket{\widetilde 0},
\end{split}
\end{equation}
from which it follows that
\begin{equation}
\widetilde C_a \ket{\widetilde 0 } = 0.
\end{equation}
The fact that \( \ket{\widetilde 0} \) has finite norm can be proven by evaluating \( \braket{\widetilde 0|\widetilde 0} \) as a Gaussian integral, making use of the resolution of the identity in terms of the coherent states for the harmonic oscillators \( C_a \).

To connect \( \SO\Star(2n) \) to Bogoliubov transformations, note that \( \SO\Star(2n) \) can be embedded into \( \Sp(4n,\R) \) as\footnote{It is a simple exercise to show that the conditions \eqref{eq:so*_conditions} ensure that \eqref{eq:sp_conditions} hold.}
\begin{equation}
\varphi:
\begin{pmatrix}
X & Y \\ -\conj Y & \conj X
\end{pmatrix}
\in \SO\Star(2n)
\mapsto
\begin{pmatrix}
X & 0 & 0 & -Y \\
0 & X & Y & 0 \\
0 & -\conj Y & \conj X & 0\\
\conj Y & 0 & 0 &\conj X
\end{pmatrix}
\in \Sp(4n,\R),
\end{equation}
so that we can interpret \( \SO\Star(2n) \) as a subgroup of Bogoliubov transformations of the \( 2n \) harmonic oscillators \( A_a \), \( B_b \) that we use to construct the \JS\ representation. In particular, for the Bogoliubov transformation \( \varphi(g_\zeta^{-1}) \), with \( \zeta \in \Omega_n \) we get
\begin{equation}
S=\begin{pmatrix}
0 & -\zeta \\ \zeta & 0
\end{pmatrix},
\end{equation}
so that the associated squeezed vacuum is
\begin{equation}
\mathcal{N} \exp\paren*{z^{ab} B\Dagger*_a A\Dagger*_b}\ket{0} \equiv \mathcal{N} \exp\paren*{\half z^{ab} \widetilde F_{ab}}\ket{0},
\end{equation}
which is exactly the coherent state \( \ket{\zeta} \).
To summarise, we can regard the coherent intertwiners we defined in this chapter as the squeezed vacua associated to a subgroup of Bogoliubov transformations, isomorphic to \( \SO\Star(2n) \). The particular Bogoliubov transformations are exactly those for which the squeezed vacuum is still \( \SU(2) \) invariant (i.e, an intertwiner), so that we can essentially regard \( \SO\Star(2n) \) as the group of canonical transformations of \( 2n \) harmonic oscillators preserving \( \SU(2) \) invariance, where the \( \SU(2) \) action is implemented through the \JS\ representation.

\section{Concluding remarks}

We have seen in this chapter how, even when working in the Euclidean regime, i.e., with a compact gauge group, the spinorial framework induces an action of the non-compact group \( \SO\Star(2n) \) on the space of all \( n \)-valent intertwiners. The reason why this additional structure was overlooked until now, despite the fact that a similar result was known for the maximal compact subgroup \( \mathrm{U}(n) \subset \SO\Star(2n)\), essentially lies in the way the operator \( E_{ab} \) is defined: our definition differs from the usual one found in the literature\footnote{See for example \cite{FreidelLivine2011}.} in that it includes a \( \delta_{ab}\1 \) term, which ensures that the commutation relations for the \( E \), \( F \) and \( \widetilde F \) operators form a closed algebra, namely \( \so\Star(2n)_\C \); without it, the commutator \( [F_{ab},\widetilde F_{cd}] \) has some terms proportional to \( \1 \) appearing in it, which prevent the interpretation of the intertwiner space as a representation of \( \SO\Star(2n) \).

We have seen how this new \( \SO\Star(2n) \) action can be used to construct a set of coherent intertwiners, using the Gilmore--Perelomov generalised coherent states; as we noted, although some of these were already known and used, the vast majority of them are new, namely all those labelled by a matrix \( \zeta \) of rank greater than \( 2 \). As part of the analysis of the properties of these coherent states, we have shown that, in the semi-classical limit of large areas, each coherent state is peaked around a classical phase space which we can interpret as the classical geometry given by a convex polyhedron with \( n \) faces. Some work is still required to achieve the full understanding of the semi-classical limit, as there are some issues in the connection between the expectation values of the area operators and the areas of the faces of the polyhedron when \( \rank(\zeta)>2 \), i.e., for the previously unconsidered coherent states.

\chapter{Conclusions and outlook}
In the past three chapters we investigated a number of results, all related to each other, with non-compact groups as their common thread.
The first few of them have been mathematical in nature. First we saw how the \WE\ theorem can be generalised to arbitrary Lie groups, with the introduction of the new concept of weak tensor operators to make the treatment in the case of non-compact groups rigorous, then proceeded to construct the main ingredient of the theorem, the \CG\ decomposition of the product of finite and infinite-dimensional representations, for the specific cases of the Lorentz groups; finally, we were able to use these results to construct an analogue of the \JS\ representation for \emph{all} representation classes of \( \Spin(2,1) \) and \( \Spin(3,1) \). Although the results of representation theory were far more difficult to prove and have a much broader scope, the \JS\ representations are arguably the most important results of \cref{chap:wigner-eckart}: even though they are just a simple application of the \WE\ theorem, the importance of their applications to physics, and the fact that they were completely unknown for continuous series representations makes them of considerable value, and in fact it is only thanks to these results that it was possible to write \cref{sec:lorentzian_LQG} at all.

The rest of the thesis focused on applications to physics, in particular to quantum gravity. In \Cref{sec:lorentzian_LQG} we saw how the mathematical results discovered in the previous chapter can be used to implement an equivalent of the spinorial approach to loop quantum gravity in the \( (2+1) \) Lorentzian case. Although the results we found are very similar to the Euclidean ones, the Lorentzian case has several key differences, caused by the higher complexity of \( \Spin(2,1) \): for example, the \( E \), \( F \) and \( \widetilde F \) operators we defined in some cases take intertwiners between unitary representations to intertwiners between non-unitary representations, and as such cannot be  considered proper observables. Nevertheless, they can still be used to generate all of the geometric observables, and we were able to use them to construct a solvable Hamiltonian constraint, with the Lorentzian Ponzano--Regge amplitude in its kernel.
It is important to note that, although the \JS\ representation was already known in the case of discrete series representation, the action of the \( E \), \( F \) and \( \widetilde F \) operators involves Racah coefficients where one of the representations is \( F_{\shalf} \), which still require the knowledge of the recoupling theory results of \cref{chap:wigner-eckart}: as a consequence, the entirety of \cref{sec:lorentzian_LQG} is new, not only the results involving continuous representations.

Finally, in \cref{chap:so*} we switched gears and analysed some new properties of \emph{Euclidean} LQG. We saw that, although \( \SU(2) \) is compact, when working with the spinorial formalism the non-compact group \( \SO\Star(2n) \) appears naturally in the theory; specifically, the Hilbert space of all \( n \)-valent intertwiners provides an irreducible representation of \( \SO\Star(2n) \), which can be interpreted physically as the subgroup of Bogoliubov transformations  of \( 2n \) harmonic oscillators---the ones appearing in the \JS\ representation---that preserves the \(\SU(2) \) invariance of intertwiners. This new structure complements the known fact that the space of \( n \)-valent intertwiners with fixed area is a representation of \( \mathrm{U}(n) \), which incidentally is the maximal compact subgroup of \( \SO\Star(2n) \). Analogously to what was done for fixed-area intertwiner space in \cite{FreidelLivine2011}, we used the \( \SO\Star(2n) \) structure to construct a new kind of coherent intertwiners, making use of Perelomov's construction of coherent states for arbitrary Lie groups. Although some of these states were already considered in the literature, the ones we constructed are more general, as there is no requirement on the matrices labelling them to be of rank \( 2 \). In the end we have shown that, in the semi-classical limit, each of these coherent states is peaked around what can be interpreted as the classical geometry described by a convex polyhedron in \( \R^3 \), although some work remains to be done to fully understand this link.

\subsection*{Future work}

As we have seen in \cref{chap:wigner-eckart}, the techniques used to investigate the \CG\ decomposition of the product of a finite and an infinite-dimensional representation are similar for both the \( 3 \)D and \( 4 \)D Lorentz group, although the treatment is more convoluted in the higher dimensional case; it is therefore likely that these techniques can be used as a guideline for the study of more generic non-compact groups. In particular, it would be interesting to consider the quantum groups associated to \( \Spin(2,1) \), i.e., the \( q \)-deformed enveloping algebra \( \mathcal{U}_q(\spin(2,1)) \). The reason why this particular example would be worth studying is that it may be used to introduce a cosmological constant \( \Lambda\neq 0 \) in the Lorentzian LQG kinematical Hilbert space, similarly to \( \mathcal{U}_q(\su(2)) \) in the \( 3 \)D Euclidean case\footcite{Dupuis:2013lka,Dupuis:2014fya}. The knowledge of the \WE\ theorem in these cases would likely lead to a deformed \JS\ representation, which could be used to generalise the formulation of the Lorentzian spinorial framework of \cref{sec:lorentzian_LQG} in the presence of a non-zero cosmological constant. It is worth noting that, when \( \Lambda \neq 0 \), new interesting features appear, such as the Bañados\Endash{}Teitelboim\Endash{}Zanelli (BTZ) black hole\footcite{Banados:1992wn}, so that a generalisation to the quantum group could shed new light on the physics happening when \( \Lambda \) is non-zero.

In addition to the use of \( \mathcal{U}_q(\spin(2,1)) \), another line of future research spawning from the results of \cref{sec:lorentzian_LQG} could be the extension to the \( 4 \)D case. Using the results of \cref{sec:4D_JS} (\JS\ representation) for \( \Spin(3,1) \), the introduction of a spinorial formalism in theories using the \( 4 \)D Lorentz group as a gauge group should not be difficult. The quantum theory in this case is an interesting open problem: although a model exists in the \( \Spin(4) \) gauge group\footcite{Dupuis:2011fz}, the generalisation to the Lorentzian case was only developed at the classical level\footcite{Dupuis:2011wy}. It is not surprising that the spinorial framework is missing here, as, unlike the \( 3D \) Lorentz group, in \( 4 \)D the \JS\ representation was completely unknown for \emph{all} unitary representations. Hopefully, the results presented in this thesis will help bridge this gap.

The results of \cref{chap:so*} are the ones that better lend themselves to future research, since some of them are preliminary and require additional work. The next step will definitely be the understanding of the nature of the semi-classical limit of the coherent states and its connection to a classical geometry. The pursuit of this topic has great importance, as the states labelled by a matrix with rank greater than 2---those for which the understanding of the semi-classical limit is incomplete---are exactly those that have never been considered before; finding out their connection to a classical geometry is thus necessary to figure out the role that they will play in loop quantum gravity. 

Another interesting research topic related to the \( \SO\Star(2n) \) formulation of intertwiner space is its generalisation to \( \Spin(2,1) \) intertwiner, and is closely connected with the spinorial approach to \( 3 \)D Lorentzian LQG; in fact, the spinorial observables \( E \), \( F \) and \( \widetilde F \) introduced in \cref{sec:lorentzian_LQG} form an \( \so(2n,\C) \) algebra just like in the Euclidean case, i.e., intertwiner space can be seen as a representation of \( \SO(2n,\C) \). 
Although this preliminary result is easy to obtain, as we have seen multiple times in this thesis dealing with \( \Spin(2,1) \) makes things considerably more difficult than the \( \SU(2) \) case.
First of all, there is no single intertwiner space, as the spinorial observables do not change the \emph{class} of a representation, so that for example an intertwiner of the kind \( D^+\otimes D^+ \rightarrow D^+\) can never become one of the kind \( D^-\otimes D^- \rightarrow D^-\) under the action of the \( \SO(2n,\C) \) generators.
Moreover, it is unclear which \emph{real form} of \( \SO(2n,\C) \) should be used to make each kind of intertwiner space a unitary representation: it is entirely possible that different real forms may be needed, depending on the classes of the representations appearing in the intertwiners.

\appendixpage
\appendix
\chapter{Some facts about matrices}\label{app:matrices}
\section{Tridiagonal matrices}\label{app:tridiagonal}
\emph{Tridiagonal} matrices are square matrices whose only non-zero entries are on the main diagonal, the diagonal below it (\emph{subdiagonal}) and the diagonal above it (\emph{superdiagonal}). They can be visualised as 
\begin{equation}
A=
\begin{pmatrix}
b_1 & c_1\\
a_2 & b_2 & c_2\\
& \ddots & \ddots & \ddots\\
&& a_{n-1} &b_{n-1} & c_{n-1}\\
& & & a_n & b_n
\end{pmatrix},
\end{equation}
with the generic entry given by
\begin{equation}
\label{eq:tridiagonal}
A_{ij}=\delta_{i-1,j}\, a_i + \delta_{i,j}\,b_i + \delta_{i+1,j}\, c_i,
\end{equation}
where
\begin{equation}
a_1:=0 \quad \textnormal{and}\quad c_n:=0.
\end{equation}
A result holding for a certain class of tridiagonal matrices\footcite[Originally presented in][]{wigner_eckart}, will be proved here.
\begin{proposition}\label{prop:tridiagonal}
Let $A$ be a $n\times n$ tridiagonal matrix over a field $\K$.
If the superdiagonal (subdiagonal) entries of $A$ are all non-vanishing, its eigenspaces are all $1$-dimensional.
\end{proposition}
\begin{proof}
Consider the case of non-zero superdiagonal entries.
Recall that, if $\lambda\in\K$ is an eigenvalue of $A$, the associated eigenspace is $\ker(A-\lambda\1)$, the vector space of solutions to the equation
\begin{equation}
Ax=\lambda x,\quad x\in \K^n;
\end{equation}
with the notation introduced in~\eqref{eq:tridiagonal}, this is equivalent to the system of $n$ equations
\begin{equation}
\begin{cases}
\left(b_1-\lambda\right)x_1 + c_1\,x_2 = 0\\
a_i\, x_{i-1} + \left(b_i-\lambda\right)x_i + c_i\,x_{i+1}=0,\casestext{} i=2,\dotsc,n-1\\
a_n\,x_{n-1} + \left(b_n-\lambda\right)x_n=0.
\end{cases}
\end{equation}
If $x_1=0$ the first equation reduces to
\begin{equation}
c_1\,x_2=0,
\end{equation}
which implies $x_2$ is zero as well, since all the $c$'s are non-vanishing. In general, the $k$th equation will be
\begin{equation}
c_k\,x_{k+1}=0,
\end{equation}
i.e., the only solution with $x_1=0$ is the null vector.

Let then $x_1$ be an arbitrary non-zero value. Substituting each equation in the next one, the first $n-1$ equations reduce to a system of equations of the form
\begin{equation}
\label{eq:tridiagonal_solution}
c_i\,x_{i+1}=\alpha_{i+1}\,x_1,\qquad i=1,\dotsc,n-1,
\end{equation}
with each $\alpha$ depending solely on $\lambda$ and on the matrix entries.
These always have solution, since one can safely divide by the $c$'s;
as a consequence, the solution is \emph{completely} specified by the value of $x_1$, which can be factored out as a scalar coefficient.
The $n$th equation is automatically satisfied, as it was assumed that $\lambda$ is an eigenvalue.
By virtue of eqs. (\ref{eq:tridiagonal_solution}), all the non-zero solutions of the eigenvalue equation are proportional to each other, so that
\begin{equation}
\dim \ker(A-\lambda\1)=1.
\end{equation}
The proof for the case of non-zero subdiagonal entries proceeds analogously.
\end{proof}

\section{Anti-symmetric matrices}\label{app:antisymmetric}

An anti-symmetric matrix $X$ is one which satisfies $X\Transpose = -X$. In the specific case of complex matrices, one can prove the following result\footcite[corollary~2]{Youla_1961}.

\begin{lemma}[Decomposition of anti-symmetric matrices]\label{lem:antisymmetric_decomposition}
Any anti-symmetric matrix $X\in M_n(\C)$ can be decomposed as $U M U\Transpose$, where $U$ is a unitary matrix and\footnote{Here, for two square matrices (possibly of different dimensions) $A$ and $B$, $A \oplus B$ denotes the block matrix $\begin{smallpmatrix}
A & 0\\ 0 & B
\end{smallpmatrix}$.}
\begin{equation}
M=
\begin{cases}
\bigoplus_{a=1}^{\frac{n}{2}}
\begin{smallpmatrix}
0 & -\lambda_a\\ \lambda_a &0
\end{smallpmatrix}
&\casesif \mbox{$n$ is even}
\\\addlinespace
\bigoplus_{a=1}^{\frac{n-1}{2}}
\begin{smallpmatrix}
0 & -\lambda_a\\ \lambda_a &0
\end{smallpmatrix}
\oplus (0)
&\casesif \mbox{$n$ is odd},
\end{cases}
\end{equation}
with
\begin{equation}
\lambda_1\geq \lambda_2 \geq\dotsm \geq\lambda_\nu\geq 0,\quad\nu=\floor*{\half[n]}.
\end{equation}
It follows that $\rank(X)=\rank(M)$ is necessarily even.
\end{lemma}

We can use this \namecref{lem:antisymmetric_decomposition} to prove a useful result in the case of anti-symmetric matrices of rank \(2\).

\begin{corollary}
Let \(X\in M_n(\C)\) be an anti-symmetric matrix of rank \(2\). Then
\[
XX\Star X = \half \tr(X\Star X) X
\quad\mbox{and}\quad\det(\1-X\Star X)=\paren*{1-\half \tr(X\Star X)}^2.
\]
\end{corollary}
\begin{proof}
Let $X=U M U\Transpose$ be the decomposition of $X$ given by \cref{lem:antisymmetric_decomposition}. Since the rank is $2$, it will be
\begin{equation}
M=\begin{pmatrix}
0 &-\lambda \\ \lambda & 0
\end{pmatrix}
\oplus \0_{n-2}
\end{equation}
so that
\begin{equation}
M\Star M=\begin{pmatrix}
\lambda^2 & 0 \\ 0 & \lambda^2
\end{pmatrix}
\oplus \0_{n-2};
\end{equation}
it follows that
\begin{equation}
\tr(X\Star X)=\tr(\conj{U}M\Star M U\Transpose)=\tr(M\Star M)=2\lambda^2
\end{equation}
and
\begin{equation}
X X\Star X= U M M\Star M U\Transpose=\lambda^2 U M U\Transpose = \half \tr(X\Star X) X.
\end{equation}
Moreover,
\begin{equation}
\begin{split}
\det(\1-X\Star X)&=\det(\conj{U}U\Transpose-\conj{U}M\Star M U\Transpose)
\\
&=\det(\1-M\Star M)
\\
&=\paren[\big]{1-\lambda^2}^2
\\
&=\paren*{1-\half \tr(X\Star X)}^2.
\end{split}
\end{equation}
\end{proof}
\chapter{\CG\ coefficients}\label{app:CG}

\section{\texorpdfstring{\(\Spin(2,1)\)}{Spin(2,1)} \CG\ coefficients}\label{app:3D_CG}
\begin{table}[h!]
\begin{tabularx}{\textwidth}{p{0.75in}*{2}{>{\centering\arraybackslash}X}} \toprule
& \(J=j-\frac{1}{2}\) &  \(J=j+\frac{1}{2}\)  \\
\midrule
\(\mu=-\frac{1}{2}\) &\(-\frac{\sqrt{j+M+\frac{1}{2}}}{\sqrt{2j+1}}\) & \(\frac{\sqrt{j-M+\frac{1}{2}}}{\sqrt{2j+1}}\) \\
\addlinespace
\(\mu=+\frac{1}{2}\) & \(\frac{\sqrt{j-M+\frac{1}{2}}}{\sqrt{2j+1}}\) & \(\frac{\sqrt{j+M+\frac{1}{2}}}{\sqrt{2j+1}}\) \\
\bottomrule
\end{tabularx}
\caption{Clebsch--Gordan coefficients \(B(J,M|\gamma,\mu;j,M-\mu)\) for \(\gamma=\frac{1}{2}\).}
\label{tab:3d-1/2}
\end{table}
\begin{table}[h!]
\begin{tabularx}{\textwidth}{p{0.75in}*{3}{>{\centering\arraybackslash}X}} \toprule
& \(J=j-1\) &  J=j & \( J=j+1 \)\\
\midrule
\(\mu=-1\) &\(\frac{\sqrt{j+M}\sqrt{j+M+1}}{\sqrt{2j}\sqrt{2j+1}} \) &
\(-{\scriptstyle\sqrt{2}}\frac{\sqrt{j-M}\sqrt{j+M+1}}{\sqrt{2j}\sqrt{2j+2}}\) & 
\(\frac{\sqrt{j-M}\sqrt{j-M+1}}{\sqrt{2j+1}\sqrt{2j+2}}\) \\
\addlinespace
\(\mu=0\) & \(-{\scriptstyle\sqrt{2}}\frac{\sqrt{j-M}\sqrt{j+M}}{\sqrt{2j}\sqrt{2j+1}}\) & \(-\frac{2M}{\sqrt{2j}\sqrt{2j+2}}\) &
\({\scriptstyle\sqrt{2}}\frac{\sqrt{j-M+1}\sqrt{j+M+1}}{\sqrt{2j+1}\sqrt{2j+2}}\)\\
\addlinespace
\(\mu=+1\) & \(\frac{\sqrt{j-M}\sqrt{j-M+1}}{\sqrt{2j}\sqrt{2j+1}}\) & 
\({\scriptstyle\sqrt{2}} \frac{\sqrt{j+M}\sqrt{j-M+1}}{\sqrt{2j}\sqrt{2j+2}}\) &
\(\frac{\sqrt{j+M}\sqrt{j+M+1}}{\sqrt{2j+1}\sqrt{2j+2}}\) \\ \bottomrule
\end{tabularx}
\caption{Clebsch--Gordan coefficients \(B(J,M|\gamma,\mu;j,M-\mu)\) for \(\gamma=1\).}
\label{tab:3d-1}
\end{table}
\noindent Explicit values for the Clebsch--Gordan coefficients are presented here, for the small values $\gamma=\frac{1}{2}$ (table \ref{tab:3d-1/2}) and $\gamma=1$ (table \ref{tab:3d-1}), with arbitrary $j$. The tables are valid for $D^\pm_j$, $C^\varepsilon_j$ and $F_j$, provided only the allowed values of $j$, $J$ and $M$ are considered. The coefficients are normalized in such a way that
\begin{equation}
A(\gamma,\mu;j,m|J,M)=B(J,M|\gamma,\mu;j,m)
\end{equation}
and that, for the finite-dimensional series (with $j\geq \gamma$), they coincide with the $\mathfrak{su}(2)$ ones. Moreover, in analogy with the $\mathfrak{su}(2)$ case, the Clebsch--Gordan coefficients for the coupling $V_j\otimes F_\gamma\cong F_\gamma\otimes V_j$ are chosen to be
\begin{equation}
B(J,M|j,m;\gamma,\mu):=(-1)^{J-j-\gamma}B(J,M|\gamma,\mu;j,m).
\end{equation}

Some properties of the Clebsch--Gordan coefficients ar also listed in this section. Assume that the coupling $F_\gamma\otimes V_j$, with $V_j$ an arbitrary irreducible \((\mathfrak{g},K)\)-module, is decomposable, and consider the Clebsch--Gordan coefficients in the form presented in \cref{sec:3d-lorentz-group}, that is such that the diagonalized basis vectors are
\begin{equation}
\label{eq:CG_recur0}
\ket{J,M}=\sum_{\mu}\sum_{m}A(\gamma,\mu;j,m|J,M)\ket{\gamma,\mu;j,m}.
\end{equation}
One can always rescale these vectors so that
\begin{equation}
\mathcal{J_\pm}\ket{J,M}=C_\pm(J,M)\ket{J,M\pm 1}.
\end{equation}
By acting with $\mathcal{J_\pm}$ on both sides of (\ref{eq:CG_recur0}) and equating the coefficients of each basis vector we find that the Clebsch--Gordan coefficients must obey the recursion relation
\begin{multline}\label{eq:CG_recur1}
C_\pm(J,M)A(\gamma,\mu;j,m|J,M\pm 1)=
C_\pm(\gamma,\mu \mp 1)A(\gamma,\mu \mp 1;j,m|J,M) \\
+C_\pm(j,m\mp 1)A(\gamma,\mu;j,m\mp 1|J,M);
\end{multline}
analogously, we fins for the inverse coefficients
\begin{multline}\label{eq:CG_recur2}
C_\pm(J,M)B(J,M\pm 1|\gamma,\mu;j,m)=
C_\pm(\gamma,\mu \mp 1)B(J,M|\gamma,\mu \mp 1;j,m) \\
+ C_\pm(j,m\mp 1)B(J,M|\gamma,\mu;j,m\mp 1).
\end{multline}
Since both the coefficients and their inverse, for each fixed $J$, are solutions the same homogeneous linear system, they must be proportional to each other: we will choose their normalization so that
\begin{equation}
A(\gamma,\mu;j,m|J,M)=B(J,M|\gamma,\mu;j,m).
\end{equation}
Since the recursion relations only relate coefficients with the same $J$, one could think \emph{a priori} that coefficients with different $J$ are independent. It will be shown in the following that this is not true.

\begin{proposition}
Consider the coupling \( F_\gamma \otimes V_j \) of a finite-dimensional module and an irreducible one, of any class. Whenever the denominator is defined, the \CG\ coefficients satisfy
\begin{equation*}
\frac{B(J+1,M|\gamma,-\gamma;j,m)}{B(J,M|\gamma,-\gamma;j,m)}\propto\frac{\sqrt{J-M+1}}{\sqrt{J+M+1}},
\end{equation*}
where the proportionality factor is fixed by the normalisation of the \CG\ coefficients and does not depend on \( M \) or \( m \).
\end{proposition}
\begin{proof}
Consider the particular case of (\ref{eq:CG_recur1})
\begin{equation}
C_+(J,M)B(J,M+1|\gamma,-\gamma;j,m+1)=C_+(j,m)B(J,M|\gamma,-\gamma;j,m),
\end{equation}
where we used the fact that
\begin{equation}
C_+(\gamma,-\gamma-1)=0.
\end{equation}
By considering the same equation for $J+1$ and dividing by the first one, we obtain
\begin{equation}
D_J(M+1):=\frac{B(J+1,M+1|\gamma,-\gamma;j,m+1)}{B(J,M+1|\gamma,-\gamma;j,m+1)}=\frac{C_+(J,M)}{C_+(J+1,M)}\frac{B(J+1,M|\gamma,-\gamma;j,m)}{B(J,M|\gamma,-\gamma;j,m)},
\end{equation}
i.e.,
\begin{equation}
D_J(M+1)=\frac{\sqrt{J-M}\sqrt{J+M+1}}{\sqrt{J-M+1}\sqrt{J+M+2}}D_J(M).
\end{equation}
It is easy to see by recursion that
\begin{equation}
D_J(M+n)=\frac{\sqrt{J-M-n+1}\sqrt{J+M+1}}{\sqrt{J+M+n+1}\sqrt{J-M+1}}D_J(M),\quad n\in\mathbb{N},
\end{equation}
from which it follows that
\begin{equation}
D_J(M)\equiv \frac{B(J+1,M|\gamma,-\gamma;j,m)}{B(J,M|\gamma,-\gamma;j,m)}=\alpha(J)\frac{\sqrt{J-M+1}}{\sqrt{J+M+1}},
\end{equation}
where $\alpha$ is arbitrary and depends on the normalization.

\end{proof}

\section{\texorpdfstring{\(\Spin(3,1)\)}{Spin(3,1)} \CG\ coefficients}\label{app:4D_CG}

\begin{table}[h!]
\begin{tabularx}{\textwidth}{p{1in}*{2}{>{\centering\arraybackslash}X}} \toprule
& \((\Lambda,\Rho)=(\lambda-\shalf,\rho-\tfrac{A}{2})\) & \((\Lambda,\Rho)=(\lambda+\shalf,\rho+\tfrac{A}{2})\)  \\
\midrule
\(j=J-\half\qquad\) & \(\ii A\frac{\sqrt{J-\lambda+\shalf}\sqrt{J-A\rho+\shalf}}{\sqrt{2J+1}\sqrt{\lambda+A\rho}}\) & \(\frac{\sqrt{J+\lambda+\shalf}\sqrt{J+A\rho+\shalf}}{\sqrt{2J+1}\sqrt{\lambda+A\rho}}\) \\
\addlinespace
\(j=J+\half\) & 
\(\frac{\sqrt{J+\lambda+\shalf}\sqrt{J+A\rho+\shalf}}{\sqrt{2J+1}\sqrt{\lambda+A\rho}}\) & \(-\ii A\frac{\sqrt{J-\lambda+\shalf}\sqrt{J-A\rho+\shalf}}{\sqrt{2J+1}\sqrt{\lambda+A\rho}}\) \\
\bottomrule
\end{tabularx}
\caption{\CG\ coefficients \(\mathrm{B}\lbrace(\Lambda,\Rho)J|\gamma_A;(\lambda,\rho)j\rbrace\) for \(\gamma=\half\).}
\label{tab:4d-1/2}
\end{table}

\noindent Some notions about the \CG\ coefficients of the coupling \(F^A_\gamma\otimes V_{\lambda,\rho}\) are presented here. In particular, explicit values for some \CG\ coefficients, namely those with \(\gamma=\half\) are listed in \cref{tab:3d-1/2}; the normalisation is chosen so that
\begin{equation}
\mathrm{B}\big\lbrace(\Lambda,\Rho)J\big|\half_A;(\lambda,\rho)j\big\rbrace \equiv \mathrm{A}\big\lbrace\half_A;(\lambda,\rho)j\big|(\Lambda,\Rho)J\big\rbrace.
\end{equation}
Moreover, we will prove the following useful property:
\begin{proposition}\label{prop:4d_CG_property}
Consider the product \(F^A_\gamma\otimes V_{\lambda,\rho}\), with \(\gamma\geq\shalf\) and \(V_{\lambda,\rho}\) infinite\Hyphdash{}dimensional. If a \CG\ decomposition exists, when \(J\geq \abs{\lambda}+\gamma\) the \CG\ coefficients satisfy, for all \((\Lambda,\Rho)\in \cC^A_J(\lambda,\rho,\gamma)\),
\begin{equation*}
\frac{\mathrm{B}\lbrace(\Lambda+1,\Rho+A)J|\gamma_A;(\lambda,\rho)J-\gamma\rbrace}{\mathrm{B}\lbrace(\Lambda,\Rho)J|\gamma_A;(\lambda,\rho)J-\gamma\rbrace}\propto\frac{\sqrt{J+\Lambda+1}\sqrt{J+A\Rho+1}}{\sqrt{J-\Lambda}\sqrt{J-A\Rho}},
\end{equation*}
where the proportionality factor is fixed by the normalisation of the \CG\ coefficients and does not depend on \(J\).
\end{proposition}
\begin{proof}
When \(J\geq \abs{\lambda}+\gamma\) the \((J_0,J^2)\)-eigenspace \(V^J=V^J_J\) is spanned by the \(2\gamma+1\) vectors\footnote{Requiring that \(J\geq\abs{\lambda}+\gamma\) is needed to ensure \(j=J-\gamma\) is allowed, i.e., \(J-\gamma\in \abs{\lambda}+\N_0\).}
\begin{equation}
\label{eq:CGapp_basis1}
\ket{(j)J}=\sum_{\mu\in\cM_\gamma}\braket{\gamma,\mu;j,J-\mu | J,J}\ket{\gamma_A,\mu}\otimes\ket{(\lambda,\rho)j,J-\mu},
\end{equation}
where \(j\in \lbrace J-\gamma,\dotsc,J+\gamma\rbrace\).
Since \(F^A_\gamma\otimes V_{\lambda,\rho}\) is decomposable, \(\cC^A_J(\lambda,\rho,\gamma)=\cC^A(\lambda,\rho,\gamma)\) does not depend on \(J\) when \(J\geq \abs{\lambda}\) and
\begin{equation}
\label{eq:CGapp_basis2}
\ket{(j)J}
=\sum_{(\Lambda,\Rho)\in\cC^A}\mathrm{B}\lbrace(\Lambda,\Rho)J|\gamma_A;(\lambda,\rho)J \rbrace\ket{(\Lambda,\Rho)J}.
\end{equation}
Equating \eqref{eq:CGapp_basis1} and \eqref{eq:CGapp_basis2} in the particular case \(j=J-\gamma\) gives\footnote{Recall that in this case the only non-zero coefficient in \eqref{eq:CGapp_basis1} is  \(\braket{\gamma,\gamma;J-\gamma,J-\gamma|J,J}=1\).} 
\begin{equation}
\label{eq:CGapp_vector}
\ket{\gamma_A,\gamma}\otimes\ket{(\lambda,\rho)J-\gamma,J-\gamma}
=
\sum_{(\Lambda,\Rho)\in\cC^A}\mathrm{B}\lbrace(\Lambda,\Rho)J|\gamma_A;(\lambda,\rho)J-\gamma\rbrace\ket{(\Lambda,\Rho)J}.
\end{equation}
Acting with \(K_+\) on both sides of~\eqref{eq:CGapp_vector} we get respectively
\begin{equation}
\begin{split}
&-P^+_{\lambda,\rho}(J-\gamma)\sqrt{2J-2\gamma+1}\sqrt{2J-2\gamma+2}\ket{\gamma_A,\gamma}\otimes\ket{(\lambda,\rho)J+1-\gamma,J+1-\gamma}
\\
=&-P^+_{\lambda,\rho}(J-\gamma)\sqrt{2J-2\gamma+1}\sqrt{2J-2\gamma+2}\ket{(J+1-\gamma)J+1}
\\
=&\!\begin{multlined}[t][0.86\linewidth]
-P^+_{\lambda,\rho}(J-\gamma)\sqrt{2J-2\gamma+1}\sqrt{2J-2\gamma+2}
\\[0.5\defaultaddspace]
\sum_{(\Lambda,\Rho)\in\cC^A}\mathrm{B}\lbrace(\Lambda,\Rho)J+1|\gamma_A;(\lambda,\rho)J+1-\gamma\rbrace\ket{(\Lambda,\Rho)J+1}
\end{multlined}
\end{split}
\end{equation}
for the LHS and
\begin{equation}
-\sum_{(\Lambda,\Rho)\in\cC^A}\mathrm{B}\lbrace(\Lambda,\Rho)J|\gamma_A;(\lambda,\rho)J-\gamma\rbrace P^+_{\Lambda,\Rho}(J)\sqrt{2J+1}\sqrt{2J+2}\ket{(\Lambda,\Rho)J+1}
\end{equation}
for the RHS; it follows that, for each \((\Lambda,\Rho)\in\cC^A(\lambda,\rho,\gamma)\),
\begin{multline}
\label{eq:CGapp_equality}
\mathrm{B}\lbrace(\Lambda,\Rho)J|\gamma_A;(\lambda,\rho)J-\gamma\rbrace P^+_{\Lambda,\Rho}(J)\sqrt{2J+1}\sqrt{2J+2}=\\
\mathrm{B}\lbrace(\Lambda,\Rho)J+1|\gamma_A;(\lambda,\rho)J+1-\gamma\rbrace P^+_{\lambda,\rho}(J-\gamma)\sqrt{2J-2\gamma+1}\sqrt{2J-2\gamma+2}.
\end{multline}
Now let
\begin{equation}
f_J(\Lambda,\Rho):=\frac{\mathrm{B}\lbrace(\Lambda+1,\Rho+A)J|\gamma_A;(\lambda,\rho)J-\gamma\rbrace}{\mathrm{B}\lbrace(\Lambda,\Rho)J|\gamma_A;(\lambda,\rho)J-\gamma\rbrace},\quad (\Lambda,\Rho)\in\cC^A(\lambda,\rho,\gamma),
\end{equation}
where the numerator may vanish if \((\Lambda+1,\Rho+A)\not\in\cC_{J}(\lambda,\rho,\gamma)\); it follows from~\eqref{eq:CGapp_equality} that
\begin{equation}
\begin{split}
f_{J+1}(\Lambda,\Rho)&=
\frac{P^+_{\Lambda+1,\Rho+A}(J)}{P^+_{\Lambda,\Rho}(J)}f_J(\Lambda,\Rho)\\
&=
\frac{\sqrt{J+\Lambda+2}\sqrt{J+A\Rho+2}}{\sqrt{J-\Lambda+1}\sqrt{J-A\Rho+1}}\frac{\sqrt{J-\Lambda}\sqrt{J-A\Rho}}{\sqrt{J+\Lambda+1}\sqrt{J+A\Rho+1}}f_J(\Lambda,\Rho).
\end{split}
\end{equation}
One can check recursively that it must be, for each \(n\in\N\),
\begin{equation}
f_{J+n}(\Lambda,\Rho)=\frac{\sqrt{J+n+\Lambda+1}\sqrt{J+n+A\Rho+1}}{\sqrt{J+n-\Lambda}\sqrt{J+n-A\Rho}}\frac{\sqrt{J-\Lambda}\sqrt{J-A\Rho}}{\sqrt{J+\Lambda+1}\sqrt{J+A\Rho+1}}f_J(\Lambda,\Rho);
\end{equation}
the solution of this \emph{recurrence relation} in \(J\) is 
\begin{equation}
f_J(\Lambda,\Rho)\propto\frac{\sqrt{J+\Lambda+1}\sqrt{J+A\Rho+1}}{\sqrt{J-\Lambda}\sqrt{J-A\Rho}},
\end{equation}
where the proportionality constant does not depend on \(J\).
\end{proof}

\chapter{Bounded symmetric domains}
\label{app:bounded_symmetric_domains}

This appendix contains definitions and results related to the groups \( \SO\Star(2n) \) and \( \Sp(2n,\R) \) and their action on their respective bounded symmetric domains. The exposetion closely follows \cite{boothby_symmetric_1972}.

In order to treat both groups at the same time, we will define
\begin{equation}
G^\varepsilon(2n) := \set*{g \in \SU(n,n) \setstx g\Transpose
\begin{pmatrix}
0  & \1_n \\ -\varepsilon\1_n & 0
\end{pmatrix}
g = 
\begin{pmatrix}
0  & \1_n \\ -\varepsilon\1_n & 0
\end{pmatrix}
}, \quad \varepsilon=\pm1,
\end{equation}
where
\begin{equation}
\SU(n,n)=\set*{g\in \SL(2n,\C)  \setstx g\Star
\begin{pmatrix}
\1_n & 0\\ 0 & -\1_n
\end{pmatrix}
g=
\begin{pmatrix}
\1_n & 0\\ 0 & -\1_n
\end{pmatrix}
}
\end{equation}
and \( SL(2n,\C) \) is the group of \( 2x\times 2n \) complex matrices with determinant \( 1 \). Then we have
\begin{equation}
\SO\Star(2n):=G^-(2n),\qquad \Sp(2n,\R):=G^+(2n).
\end{equation}
In both cases the maximal compact subgroup is
\begin{equation}
K:=\set*{
\begin{pmatrix}
U & 0 \\ 0 & \conj U
\end{pmatrix}
\setstx U\in \mathrm{U}(n)
} \cong \mathrm{U}(n).
\end{equation}

\section{Parametrisation of the group \texorpdfstring{\(\displaystyle G^\varepsilon(2n)\)}{Ge(2n)}}

Elements of \(G^\varepsilon(2n)\) can be parametrised as \(2\times 2\) block matrices
\begin{equation}
\begin{pmatrix}
A&B\\C&D
\end{pmatrix}
,\quad A,B,C,D\in M_n(\C),
\end{equation}
satisfying the conditions
\begin{subequations}
\label{eq:G_conditions}
\begin{align}
A\Star A - C\Star C &=\1 \\
D\Star D - B\Star B &=\1 \\
A\Transpose D -\varepsilon C\Transpose B &=\1
\end{align}
\begin{align}
A\Star B &= C\Star D \\
A\Transpose C &= \varepsilon C\Transpose A \\
B\Transpose D &= \varepsilon D\Transpose B
\end{align}
\begin{equation}
\det
\begin{pmatrix}
A&B\\C&D
\end{pmatrix}
=1.
\end{equation}
\end{subequations}
This parametrisation can be greatly simplified, owing to the following propositions.
\begin{proposition}
\label{prop:AD_invertible}
Let \(g=\begin{pmatrix}
A & B \\ C & D
\end{pmatrix}\in G^\varepsilon(2n)\). Then \(A\) and \(D\) are necessarily invertible and
\begin{equation*}
\det (g) =\frac{\det (D)}{\det (A\Star)}.
\end{equation*}
\end{proposition}
\begin{proof}
\begin{proofenumerate}
\item 
Suppose that \(A\) is not invertible; then there must be a non-zero \(v\in\C^n\) such that~\(Av=0\). It follows from \eqref{eq:G_conditions} that
\begin{equation}
v=(A\Star A-C\Star C)v=-C\Star Cv;
\end{equation}
however, \(C\Star C\) is a positive semi-definite matrix, i.e., all of its eigenvalues are non-negative, hence a contradiction. An analogous argument shows that \(D\) is invertible as well.

\item Recall that, if \(D\) is invertible,
\begin{equation}
\det
\begin{pmatrix}
A & B \\ C & D
\end{pmatrix}
=\det(D)\det(A-BD^{-1}C).
\end{equation}
As \(A\Star\) is invertible, we have\footnote{Recall that \(A\Star B=C\Star D\) and \(A\Star A-C\Star C=\1\).}
\begin{equation}
\begin{split}
\det (A-B D^{-1}C) &=\frac{\det (A\Star A-A\Star B D^{-1}C)}{\det(A\Star)}\\
&= \frac{\det (A\Star A-C\Star D D^{-1}C)}{\det(A\Star)}\\
&= \frac{\det (A\Star A-C\Star C)}{\det(A\Star)}\\
&= \frac{1}{\det(A\Star)},
\end{split}
\end{equation}
which concludes the proof.
\end{proofenumerate}
\end{proof}
\begin{lemma}\label{lem:inverse}
Let \(g=\begin{pmatrix}
A & B \\ C & D
\end{pmatrix}\in G^\varepsilon(2n)\). The inverse of \(g\) has the form
\begin{equation*}
g^{-1}=\begin{pmatrix}
D\Transpose & -\varepsilon B\Transpose \\ -\varepsilon C\Transpose & A\Transpose
\end{pmatrix};
\end{equation*}
moreover, in addition to the constraints \eqref{eq:G_conditions}, it must be
\begin{equation*}
AD\Transpose-\varepsilon B C\Transpose=\1,\quad B A\Transpose=\varepsilon AB\Transpose  \quad\mbox{and}\quad C D\Transpose =\varepsilon D C\Transpose.
\end{equation*}
\end{lemma}
\begin{proof}
\begin{proofenumerate}
\item
One can check explicitly that
\begin{equation}
\begin{pmatrix}
D\Transpose & -\varepsilon B\Transpose \\ -\varepsilon C\Transpose & A\Transpose
\end{pmatrix}
\begin{pmatrix}
A & B \\ C & D
\end{pmatrix}=
\begin{pmatrix}
D\Transpose A - \varepsilon B\Transpose C & D\Transpose B -\varepsilon B\Transpose D\\
A\Transpose C -\varepsilon C\Transpose A & A\Transpose D - \varepsilon C\Transpose B
\end{pmatrix}
=\begin{pmatrix}
\1 & 0 \\ 0 &\1
\end{pmatrix},
\end{equation}
as it follows directly from \eqref{eq:G_conditions}, hence
\begin{equation}
\begin{pmatrix}
D\Transpose & -\varepsilon B\Transpose \\ -\varepsilon C\Transpose & A\Transpose
\end{pmatrix}=g^{-1}.
\end{equation}
\item
Since, as \(g\) is a square matrix, it must be \(g g^{-1}=\1\) as well, one has
\begin{equation}
\begin{pmatrix}
\1 & 0 \\ 0 &\1
\end{pmatrix}
=
\begin{pmatrix}
A & B \\ C & D
\end{pmatrix}
\begin{pmatrix}
D\Transpose & -\varepsilon B\Transpose \\ -\varepsilon C\Transpose & A\Transpose
\end{pmatrix}
=\begin{pmatrix}
AD\Transpose - \varepsilon B C\Transpose & BA\Transpose - \varepsilon A B\Transpose\\
CD\Transpose - \varepsilon D C\Transpose & DA\Transpose - \varepsilon C B\Transpose
\end{pmatrix}
\end{equation}
from which the additional constraints follow.
\end{proofenumerate}
\end{proof}
\begin{proposition}
Let \(
\begin{pmatrix}
A & B \\ C & D
\end{pmatrix}
\in G^\varepsilon(2n).
\)
Then it must necessarily be
\begin{equation}
C=\varepsilon \conj B,\quad D=\conj A,
\end{equation}
which automatically ensures the \(\det(g)=1\).
\end{proposition}
\begin{proof}
From \eqref{eq:G_conditions} and the results of \cref{lem:inverse} follows that
\begin{equation}
\begin{split}
A\conj D^{-1}&=A(D\Transpose-B\Transpose\conj B \conj{D}^{-1})\\
&=\1+\varepsilon B C\Transpose - AB\Transpose \conj B  \conj{D}^{-1}\\
&=\1+\varepsilon B C\Transpose - \varepsilon B A\Transpose  \conj B  \conj{D}^{-1}\\
&=\1+\varepsilon B C\Transpose-\varepsilon B C\Transpose,
\end{split}
\end{equation}
thus \(D=\conj A\). Then \(A\Star B=C\Star \conj A\) and \( A\Transpose  C=\varepsilon C\Transpose A\) so that
\begin{equation}
C=\varepsilon (A^{\transpose})^{-1} C\Transpose A = \varepsilon \conj B.
\end{equation}
The fact that \(\det(g)=1\) follows directly from \cref{prop:AD_invertible}.
\end{proof}
Putting everything together: the elements of \(G^\varepsilon(2n)\) are parametrised by block matrices of the form
\begin{equation}
g=\begin{pmatrix}
A & B\\ \varepsilon \conj B & \conj A
\end{pmatrix}
\end{equation}
with \(\det(A)\neq0\) and \(A\), \(B\) satisfying
\begin{subequations}
\label{eq:G_conditions2}
\begin{gather}
AA\Star-BB\Star=\1\\
A\Star A-B\Transpose\conj B =\1
\\
A\Star B=\varepsilon B\Transpose \conj A\\
BA\Transpose = \varepsilon A B\Transpose,
\end{gather}
\end{subequations}
with inverse
\begin{equation}
g^{-1}=
\begin{pmatrix}
A\Star & -\varepsilon B\Transpose\\
-B\Star & A\Transpose
\end{pmatrix}.
\end{equation}

\section{The Lie algebra \texorpdfstring{\(\displaystyle \mathfrak{g}^\varepsilon(2n)\)}{ge(2n)}}
The Lie algebra of \( G^\varepsilon(2n) \) is given by
\begin{equation}
\mathfrak{g}^\varepsilon(2n) = \set*{V \in \su(n,n) \setstx V\Transpose
\begin{pmatrix}
0  & \1_n \\ -\varepsilon\1_n & 0
\end{pmatrix}
= -
\begin{pmatrix}
0  & \1_n \\ -\varepsilon\1_n & 0
\end{pmatrix}
V
},
\end{equation}
where
\begin{equation}
\su(n,n)=\set*{V\in \mathfrak{sl}(2n,\C)  \setstx V\Star
\begin{pmatrix}
\1_n & 0\\ 0 & -\1_n
\end{pmatrix}
= -
\begin{pmatrix}
\1_n & 0\\ 0 & -\1_n
\end{pmatrix}
V
}
\end{equation}
and
\begin{equation}
\mathfrak{sl}(2n,\C) =\set*{V\in M_n(\C)  \setstx \tr V=0 }.
\end{equation}
It is easy to see that the elements of \( \mathfrak{g}^\varepsilon(2n) \) can be parametrised by \( 2n\times 2n \) matrices
\begin{equation}
V=
\begin{pmatrix}
X & Y \\ \varepsilon \conj Y & \conj X,
\end{pmatrix}
\end{equation}
with
\begin{equation}
X\Star=-X,\quad Y\Transpose =\varepsilon Y.
\end{equation}
As \( X \in M_n(\C)\) is anti-hermitian, it has \( n^2 \) real degrees of freedom, while \( Y \) is symmetric if \( \varepsilon=1 \) and anti-symmetric if \( \varepsilon=-1 \), so that it has \( n^2 + \varepsilon n \) real degrees of freedom; it follows that \( \dim \mathfrak{g}^\varepsilon(2n)=n(2n+\varepsilon) \) as a real Lie algebra.

A basis for \(\mathfrak{g}^\varepsilon(2n)_\C\) is given by the matrices
\begin{equation}
E_{ab}=
\begin{pmatrix}
\Delta_{ab} & 0 \\ 0 & -\Delta_{ba}
\end{pmatrix},\quad
F_{ab}=
\begin{pmatrix}
0 & 0 \\ \Delta_{ab} +\varepsilon \Delta_{ba} & 0
\end{pmatrix},\quad
\widetilde F_{ab}=
\begin{pmatrix}
0 & \Delta_{ab} +\varepsilon \Delta_{ba} \\ 0 & 0
\end{pmatrix},\quad
\end{equation}
where \( a,b=1,\dotsc,n \) and \( \Delta_{ab}\in M_n(\C) \) is the matrix with entries
\begin{equation}
(\Delta_{ab})_{cd}=\delta_{ac}\delta_{bd};
\end{equation}
the \( E_{ab} \) matrices span the complexification of the subalgebra \( \mathfrak{u}(n) \).
Using the fact that
\begin{equation}
\Delta_{ab}\Delta_{cd}=\delta_{bc}\Delta_{ad},
\end{equation}
we can easily compute the commutation relations of the complexified generators, which are
\begin{subequations}
\begin{align}
[E_{ab},E_{cd}]		&= \delta_{cb}E_{ad} - \delta_{ad}E_{cb} \\
[E_{ab},\widetilde F_{cd}]	&= \delta_{bc}\widetilde F_{ad} + \varepsilon \delta_{bd}\widetilde F_{ac} \\
[E_{ab},F_{cd}]		&=  - \delta_{ac}F_{bd} -\varepsilon \delta_{ad}F_{bc} \\
[F_{ab},\widetilde F_{cd}]	&=  - \delta_{cb}E_{da} -\delta_{da}E_{cb} -\varepsilon \delta_{db}E_{ca} - \varepsilon \delta_{ca}E_{db}  \\
[F_{ab},F_{cd}]		&= [\widetilde F_{ab},\widetilde F_{cd}] = 0.
\end{align}
\end{subequations}

\section{Bounded symmetric domains}

Bounded symmetric domains are a class of domains in \( \C^N \) of the form \( G/K \), where \( G \) is a non-compact semi-simple and \( K \) is its maximal compact subgroup; here \( G/K \) denotes the \emph{left coset space}
\begin{equation}
G/K:=\set{gK \setst g\in G},\quad gK:=\set{gk\setst k\in K}.
\end{equation}
For the particular case of \( G=G^\varepsilon(2n) \), \( K=U(n) \), the bounded symmetric domain is isomorphic to the domain
\begin{equation}
\Omega_n^\varepsilon:=
\set{
\zeta\in M_n(\C)\setst\1-\zeta\Star\zeta>0,\zeta\Transpose=\varepsilon\zeta
},
\end{equation}
on which \(  G^\varepsilon(2n)\) operates holomorphically  as
\begin{equation}
g(\zeta):=(A\zeta+B)(C\zeta+D)^{-1},\quad
g=
\begin{pmatrix}
A&B\\C&D
\end{pmatrix}.
\end{equation}
This action is well defined: in fact we have\footcite{boothby_symmetric_1972}
\begin{equation}\label{eq:bounded_action_proof}
\begin{split}
(C\zeta+D)\Star(C\zeta+D) - (A\zeta+B)\Star(A\zeta+B) &=
\begin{pmatrix}
\zeta\Star & \1 
\end{pmatrix}
g\Star
\begin{pmatrix}
\1 & 0 \\ 0 & -\1
\end{pmatrix}
g
\begin{pmatrix}
\zeta \\ \1
\end{pmatrix}
\\
&=
\begin{pmatrix}
\zeta\Star & \1 
\end{pmatrix}
\begin{pmatrix}
\1 & 0 \\ 0 & -\1
\end{pmatrix}
\begin{pmatrix}
\zeta \\ \1
\end{pmatrix}
\\
&=
\zeta\Star\zeta -\1 < 0;
\end{split}
\end{equation}
if \( (C\zeta+D) \) were not invertible, there would be a non-zero vector \( v \) in its kernel, so that
\begin{equation}
\begin{split}
0 &>  v\Star(C\zeta+D)\Star(C\zeta+D)v - v\Star(A\zeta+B)\Star(A\zeta+B)v
\\ & = - v\Star(A\zeta+B)\Star(A\zeta+B)v \leq 0,
\end{split}
\end{equation}
which leads to a contradiction. Moreover, it follows from \eqref{eq:bounded_action_proof} that
\begin{equation}
\1 - g(\zeta)\Star g(\zeta) = \bracks[\big]{(C\zeta+D)^{-1}}\Star (\1 - \zeta\Star \zeta)(C\zeta+D)\geq 0,
\end{equation}
so that indeed \( g(\Omega\varepsilon_n)\subseteq \Omega\varepsilon_n\). The fact that \( \Omega\varepsilon_n \cong G^\varepsilon(2n)/\mathrm{U}(n) \) is a consequence of the following propositions.
\begin{proposition}
\label{prop:transitive_action}
The action of \(G^\varepsilon(2n)\) on \(\Omega^\varepsilon_n\) is \emph{transitive}, i.e., for all \(\zeta,\omega\in \Omega^\varepsilon_n\) there is \(g\in G^\varepsilon(2n)\) such that \(\omega=g(\zeta)\).
\end{proposition}
\begin{proof}
First notice that for each \(\zeta\in\Omega^\varepsilon_n\) there is a group element that sends \(\zeta\) to \(0\). In fact,
\begin{equation}
g(0)= B \conj A^{-1}=\zeta \quad\Leftrightarrow\quad
B=\zeta \conj A,
\end{equation}
where, owing to \eqref{eq:G_conditions2}, \(A\) has to satisfy
\begin{equation}
A A\Star=(\1-\zeta\zeta\Star)^{-1},
\end{equation}
i.e., \(A\) is a \emph{square root} of the positive-definite matrix \((\1-\zeta\zeta\Star)^{-1}\); in particular one can choose the \emph{unique} positive-definite square root, denoted by \(\sqrt{(\1-\zeta\zeta\Star)^{-1}}\). One can check explicitly that\footnote{Recall that since the positive-definite square root is unique, one has \((\sqrt{A})\Transpose\equiv \sqrt{A\Transpose}\) and analogous expressions for \(\conj A\), and~\(A\Star\).}
\begin{equation}
g_{\zeta}:=
\begin{pmatrix}
\sqrt{(\1-\zeta\zeta\Star)^{-1}} & \zeta\sqrt{(\1-\zeta\Star\zeta)^{-1}}\\
\zeta\Star\sqrt{(\1-\zeta\zeta\Star)^{-1}} & \sqrt{(\1-\zeta\Star\zeta)^{-1}}
\end{pmatrix}
\end{equation}
satisfies all the constraints \eqref{eq:G_conditions2}, so it belongs to \(G^\varepsilon(2n)\). Then for any \(\zeta,\omega\in \Omega^\varepsilon_n\) one has
\begin{equation}
(g_\omega g^{-1}_\zeta)(\zeta)=g_\omega(0)=\omega,
\end{equation}
so the action is transitive.
\end{proof}

\begin{proposition}
The \emph{isotropy subgroup} of \(0\in\Omega^\varepsilon_n\) is \(K_n\cong U(n)\), the maximal compact subgroup of \(G^\varepsilon(2n)\).
\end{proposition}
\begin{proof}
Let
\begin{equation}
g=\begin{pmatrix}
A & B \\ \varepsilon \conj B & \conj A
\end{pmatrix}
\end{equation}
a generic \( G^\varepsilon(2n) \) element. We have
\begin{equation}
g(0)= B D^{-1},
\end{equation}
which vanishes if and only if \( B \) is the zero matrix. It follows that
\begin{equation}
\1 = A A\Star - B B\Star \equiv A A\Star,
\end{equation}
so that the subgroup that leaves \( 0\in \Omega^\varepsilon_n \) invariant is \( K \).
\end{proof}

\backmatter

\printbibheading
\printbibliography[heading=none]

\end{document}